\documentclass{ws-ijmpa}
\pdfoutput=1
\usepackage{graphicx}
\usepackage{bm}
\usepackage{mathrsfs}
\usepackage{amsmath}
\usepackage{amssymb}
\usepackage{amstext}
\usepackage{epsfig,latexsym}
\newcommand{\DFIGURE}[1]{\begin{figure}[htbp]#1\end{figure}}
\newcommand{\HFIGURE}[3]{\begin{figure}\center{\includegraphics[width=0.49\textwidth]{#1}}\caption{#2}\label{#3}\end{figure}}

\newcommand{\cQ}{q_\mt{7}} 

\newcommand{\qe}{\mathfrak{e}}
\newcommand{\qnn}{\mathfrak{n}}

\newcommand{\be}{\begin{equation}}
\newcommand{\ee}{\end{equation}}
\newcommand{\bea}{\begin{eqnarray}}
\newcommand{\eea}{\end{eqnarray}}
\newcommand{\ba}{\begin{eqnarray}}
\newcommand{\ea}{\end{eqnarray}}

\newcommand{\beq}{\begin{equation}}
\newcommand{\eeq}{\end{equation}}
\newcommand{\beqa}{\begin{eqnarray}}
\newcommand{\eeqa}{\end{eqnarray}}
\newcommand{\beqar}{\begin{eqnarray*}}
\newcommand{\eeqar}{\end{eqnarray*}}

\newcommand{\eg}{{\it e.g.,}\ }
\newcommand{\ie}{{\it i.e.,}\ }

\newcommand{\tlt}{{\tilde{t}}} 
\newcommand{\tx}{{\tilde{x}}}

\newcommand{\ts}{{\tilde{\sigma}}}
\newcommand{\tlnu}{{\tilde{\nu}}}
\newcommand{\tlgam}{{\tilde{\gamma}}}

\newcommand{\tlc}{{\tilde{c}}}
\newcommand{\tlm}{{\tilde{m}}}
\newcommand{\tlb}{{\tilde{B}}}
\newcommand{\tlrho}{{\tilde{\rho}}}

\newcommand{\tlq}{\tilde{q}}
\newcommand{\tom}{{\tilde{\omega}}}

\def\nc {N_\mt{c}}
\def\nf {N_\mt{f}}

\def\t6 {T_\mt{D6}}

\newcommand{\ls}{\ell_\mt{s}}

\newcommand{\mt}[1]{\textrm{\tiny #1}}

\newcommand{\trho}{{\tilde{\rho}}}  

\renewcommand{\Re}{\mathrm{Re}\,}
\renewcommand{\Im}{\mathrm{Im}\,}

\newcommand{\ei}{\end{itemize}}
\newcommand{\ben}{\begin{enumerate}}
\newcommand{\een}{\end{enumerate}}

\newcommand{\bmtx}{\left[ \begin{array}{cc}}
\newcommand{\emtx}{\end{array} \right]}
\newcommand{\bvec}{\left[ \begin{array}{c}}
\newcommand{\evec}{\end{array} \right]}

\renewcommand{\theequation}{\thesection.\arabic{equation}}
\renewcommand{\Re}{{\mbox{Re }}}
\renewcommand{\Im}{{\mbox{Im }}}

\newcommand{\bfig}{\begin{figure}}
\newcommand{\efig}{\end{figure}}

\newcommand{\order}{{\mathcal{O}}}
\newcommand{\myg}{{\mathcal{G}}}
\newcommand{\gperp}{{\mathcal{G}^{\bot}}}

\newcommand{\phipsi}{{\phi_{\!\Psi}\!}}
\newcommand{\phiz}{{\phi_{\!z}\!}}
\newcommand{\Gs}{{G_{\!S}\!}}
\newcommand{\Ga}{{G_{\!A}\!}}
\newcommand{\Fnaught}{{F^{\! (0)}\!\!}}
\newcommand{\gone}{{g^{\! (1)}\!\!\!\!\!}}
\newcommand{\gtwo}{{g^{\! (2)}\!\!\!\!\!}}
\newcommand{\phione}{\phi}

\newcommand{\labell}[1]{\label{#1}}
\newcommand{\reef}[1]{(\ref{#1})}
\newcommand{\mref}[1]{\ref{#1}}
\newcommand{\mlabel}[1]{\label{#1}}
\begin{document}%
\title{HOLOGRAPHIC EXPERIMENTS ON DEFECTS}%
\author{MATTHIAS C WAPLER}%
\address{Perimeter Institute for Theoretical Physics,
Waterloo, Ontario N2L 2Y5, Canada \\and\\
Department of Physics \& Astronomy and Guelph-Waterloo Physics
Institute,\\
\ \ \ University of Waterloo,
Waterloo, Ontario N2L 3G1, Canada \\and\\
Center for Quantum Spacetime, Sogang University, Seoul, Korea \\
 E-mail: 
 wapler@sogang.ac.kr}%
%
%
%
%
%
%
\maketitle%
\begin{abstract}%
Using the AdS/CFT correspondence, we study the anisotropic charge transport properties of both supersymmetric and non-supersymmetric matter fields on (2+1)-dimensional defects coupled to a (3+1)-dimensional ${\cal N}=4$ SYM ``heat bath''.

We focus on the cases of a finite external background magnetic field, finite net charge density and finite mass and their combinations. In this context, we also discuss the limitations due to operator mixing that appears in a few situations and that we ignore in our analysis.

At high frequencies, we discover a spectrum of quasiparticle resonances due to the magnetic field and finite density and at small frequencies, we perform a Drude-like expansion around the DC limit. Both of these regimes display many generic features and some features that we attribute to strong coupling, such as a minimum DC conductivity and an unusual behavior of the ``cyclotron'' and plasmon frequencies, which become related to the resonances found in the conformal case in an earlier paper.
We further study the hydrodynamic regime and the relaxation properties, from which the system displays a set of different possible transitions to the collisionless regime. The mass dependence can be cast in two regimes: a generic relativistic behavior dominated by the UV and a non-linear hydrodynamic behavior dominated by the IR. In the massless case, we furthermore extend earlier results from the literature to find an interesting selfduality under a transformation of the conductivity and the exchange of density and magnetic field.
\end{abstract}

\section{Introduction}
The AdS/CFT correspondence \cite{juan}\cdash\cite{bigRev} is a very important tool to study strongly coupled field theories using both string theory based and other gravitational setups. The first, and most common, example is the duality between a (black) stack of $N_c$ D3 branes generating an $AdS_5 \times S^5$ geometry in the decoupling limit and a thermal $\mathcal{N} = 4$ $SU(N_c)$ super-Yang-Mills theory on its boundary.
Of particular importance to experiment has been the introduction of fields transforming in the fundamental representation of the $SU(N_c)$ by introducing a small number of $N_f$ ``probe'' D7-branes into the background, covering all of the AdS directions \cite{lisa1}\cdash\cite{karchkatzmaldacena}. The $U(N_f)$ symmetry of the stack of probe branes then corresponds to the flavor symmetry.
Obviously this setup is still significantly different from the standard model QCD, but studying the physical properties of both this model and the T-dual D4-D6 setup have received great interest \cite{recent1}\cdash\cite{long2}. 
Particular activity has been related to the thermodynamics and the phase structure \cite{long1,long2} and to the ``meson spectrum'' \cite{meson,fast}.
The hope for experimental matchings is that some results obtained from AdS/CFT may be sufficiently generic, such that they also apply to QCD.

Recently there has also been great effort on applying the AdS/CFT correspondence to condensed matter physics. One particular motivation for why this may be successful is the fact that while there is only one QCD, there are on the one hand many different strongly coupled effective field theories in condensed matter physics and on the other hand there exist a large number AdS string vacua. Hence there is a significant potential for close matchings of gravitational duals to experimentally relevant CFTs.
Starting with the study of transport properties of 2+1 dimensional field theories dual to an $AdS_4 \times S^6$ geometry obtained from an M2-brane setup in ref. \refcite{pavel}, there has been a huge amount of activity studying 2+1 dimensional systems. Those systems display a variety of interesting properties, such as Hall conductivity \cite{hallo}\cdash\cite{mmore5}, superconductivity and superfluidity \cite{more1}\cdash\cite{more12}. While most constructions are not based on string theory, there are some string theory based constructions that ensure that the systems are pathology-free. A good example for how realistic those constructions are is e.g. the reproduction of the quantization of the magnetic flux in superconductors \cite{magsup1}\cdash\cite{magsup3} or the Nernst effect \cite{nernst}. Another interesting example of recent progress is the construction of duals of non-relativistic CFTs \cite{nonrel2}\cdash\cite{nonrel16}. A yet not satisfactorily addressed question is what the implications of the Fermi-Dirac distribution is in purely fermionic systems \cite{fermi}.

The most common example of how systems with conformal symmetry arise in condensed matter physics is the quantum critical phase. This phase arises in the context of a phase transition at zero temperature, a so-called ``quantum critical phase transition'' at a ``quantum critical point''. At this point, the system displays scale-invariant behavior as also in other phase transitions, but in contrast to phase transitions at finite temperature, it extends into a whole region in the phase diagram that may be described by a conformal field theory, the so-called quantum critical phase; shown in fig. \ref{surfdi}. As an example, the conductivity in the quantum critical phase is thought to be controlled by a universal function $\Sigma$ that depends on the ratio $\omega/T$ and some dimensional temperature scaling, $\sigma(\omega)  = Q^2(T/c)^{d-2}\Sigma(\omega/T)$ for some microscopically determined velocity $c$. Obtaining $\Sigma$ however is a difficult task and can only be done in certain limits. For instance for large frequencies, one expects $\sigma \sim \omega^{d-2}$ \cite{subir,sachdev}.

All of the above-mentioned systems have in common that they are described by 2+1 dimensional field theories. In our 3+1 dimensional world however, all 2+1 dimensional systems are strictly speaking defects. In some cases this fact may be less relevant and in other cases more relevant. Hence, it is interesting to study the physics of a 2+1 dimensional defect in order to explore what difference there is to purely 2+1 dimensional systems. Defect field theories are basically field theories in which matter that is confined to some hypersurface interacts via a field theory in the bulk. 
While there is some review literature in a soft condensed matter context, related to aspects like the statistical mechanics of crystal defects in the context of melting behaviors, defects in polymers or flux tubes in superconductors (for a review see ref. \refcite{soft}) there seems to be not much review literature related to the defects and their aspects that we are interested in.
Hence, a motivated guess may be that they have many properties in common with surfaces, which have been studied extensively. 
To illustrate their properties, we can look in fig. \mref{surfdi} at a generic surface phase diagram of some system described by a bulk coupling $J_b$ and a  surface coupling $J_s$ -- for a review on the subject see refs. \refcite{surftrans1,surftrans2}. There we see that over most of the parameter range the surface and the bulk are in the same phase and display a simultaneous ``ordinary'' phase transition. As we tune the surface coupling beyond a ``special point'', which is some critical multiple of the bulk coupling, the phase transitions on the surface and in the bulk separate into a surface phase transition and an ``extraordinary'' phase transition in which there is a phase transition only in the bulk. It is obvious from the ratio $J_s/J_b$ in this regime that the ordered phase on the surface extends to higher temperatures than the ordered phase in the bulk. However it is quite interesting that this splitting of phase transitions typically occurs as $J_s$ becomes greater than $J_b$ and hence there is no ``mirror symmetric'' version of this plot.

\DFIGURE{
\includegraphics[width=0.49\textwidth]{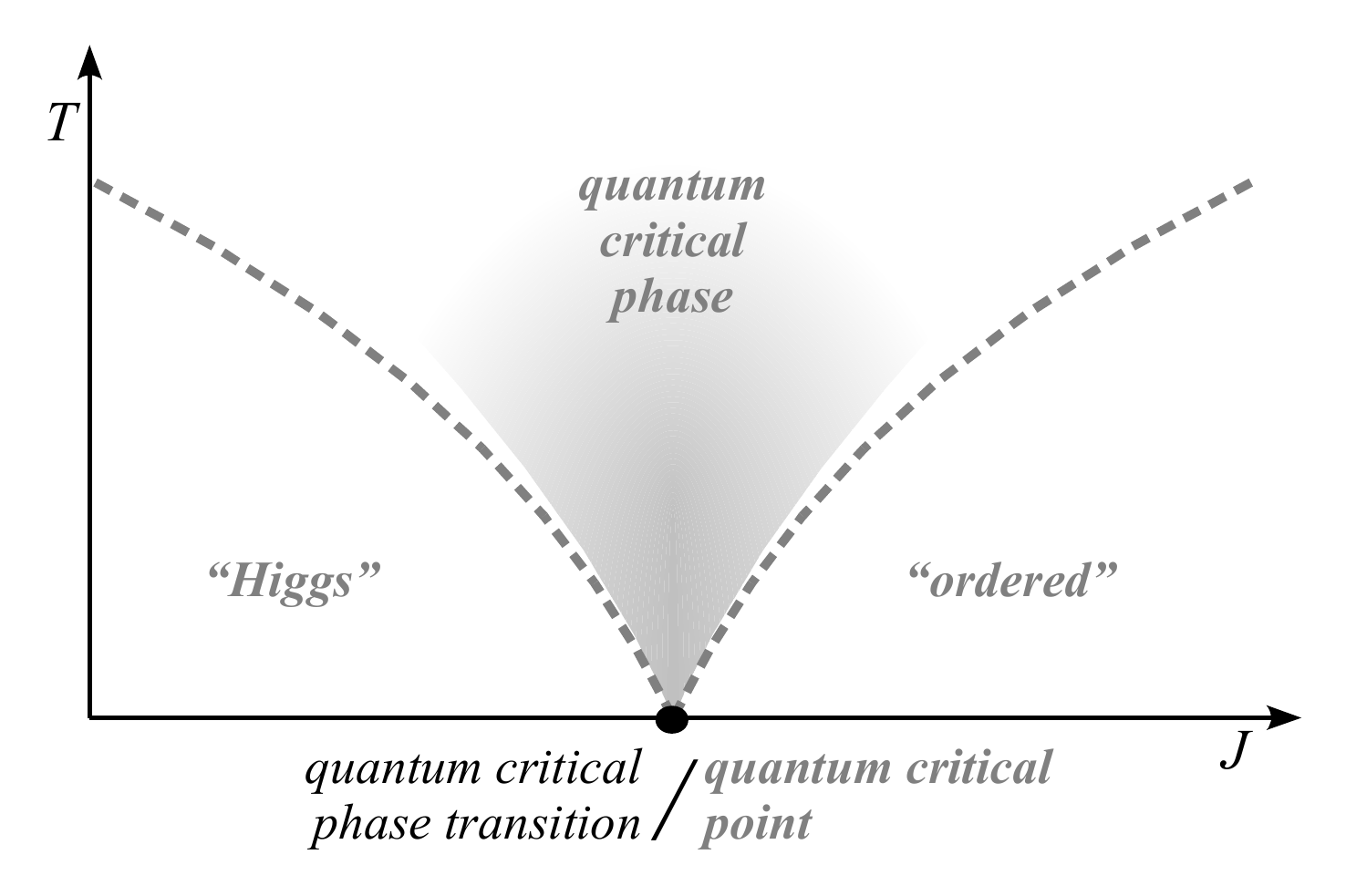}
\includegraphics[width=0.49\textwidth]{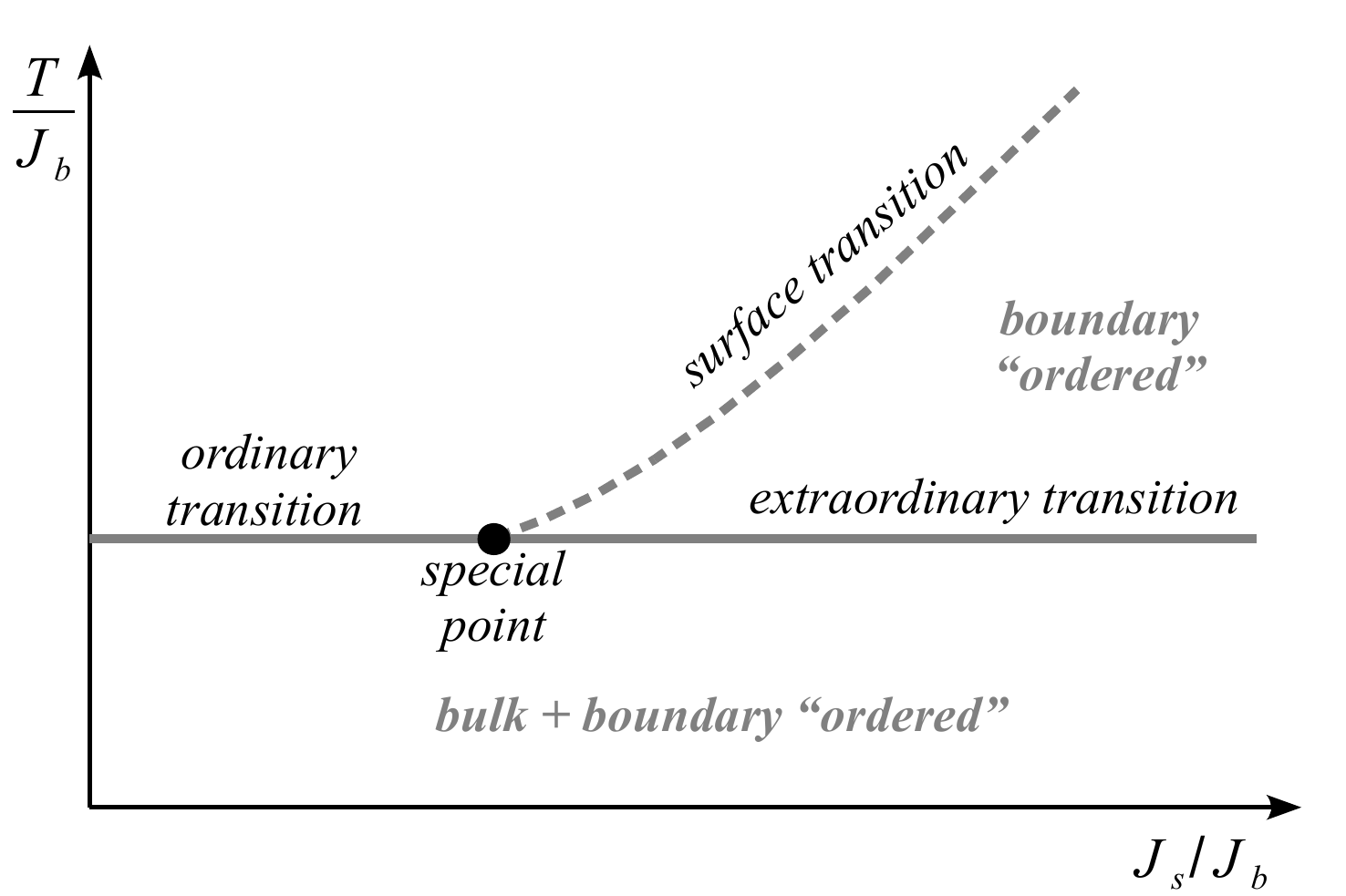}
\caption{Left: A generic quantum critical phase transition and quantum critical phase. Right: A generic surface phase diagram.}\mlabel{surfdi}}

In this paper, we will study the transport properties of the matter along the defect in order to infer on its physical properties. In that sense we will be put sometimes in the perspective of an experimentalist, trying to interpret our results. 
In order to identify the characteristics that are due to having a defect rather than just a plain 2+1 dimensional field theory, a major theme in this research is the inclusion of a parameter that is related to a difference in the level of the gauge group $N_c \rightarrow \delta N_c$ between both sides of the defect. The particular focus is then to extend earlier results obtained in the purely conformal case to the case of finite background parameters.

Also, we will try to link the properties obtained in different regimes rather than studying only one particular limit or one particular effect. In this spirit, we will study all frequency and wavenumber scales, from the DC limit and the hydrodynamic regime (the ``collision-dominated'' regime at small frequencies and small wavenumbers) up to the quasiparticle limit (the ``collisionless'' regime at large frequencies) -- and we will demonstrate in several cases how a length scale obtained in one regime, direction or context will govern some properties also in another regime. 
Furthermore, we will be interested how effects that we are familiar with from ordinary weakly coupled free electron gas type physics will manifest themselves in this strongly coupled system, as we turn on various ``condensed matter'' parameters, i.e. the net baryon number density, background magnetic field and ``quark'' mass. Certainly, since we are working at finite temperatures, there will aways be a finite total ``quark'' density, and the net baryon number density in some sense corresponds to the difference of the number of ``quarks'' and ``anti-quarks''. While those quantities may move us away from the quantum critical point in the defect field theory, the bulk theory will still remain $\mathcal{N}=4$ SYM. In terms of the phase diagram of surface phase transitions, this would move us along the direction of surface coupling towards the extraordinary phase transition. 

One interest is furthermore to explore what happens in this defect setup to the result of the constant conductivity due to the electromagnetic duality that was studied in ref. \refcite{pavel}, in particular at finite background quantities.

Our defect CFT is realized by inserting $N_f$ probe D5- or D7-branes into the background of a black D3-brane. In either construction, the difference $\delta N_c$ in the level of the gauge group of the 3+1 SYM will be introduced by an additional flux on the probe brane in the compact sphere.
The defect CFT
constructed with the D5-branes is certainly well known
\cite{lisa1,lisa2,hirosi}\cdash\cite{jaume2}. Certain aspects of the D7-brane
construction have also been studied previously
\cite{rey1,rey2,quantumhall} but we should note that the internal flux
introduced here is essential to remove an instability that would
otherwise appear in this construction. 
The finite magnetic field and net density are introduced using the well-known duals of a magnetic field and an electric field, respectively, in the world volume of the probe brane.
The finite quark mass will be obtained by a deformation of the embedding in the compact sphere in the same fashion in which it was done in the duals for 3+1 dimensional QCD-like systems \cite{johanna1}\cdash\cite{long2,findens1}\cdash\cite{us}.
To obtain the transport properties, we will use linear response
 theory to study the conductivity on the defect at finite frequency
 and temperature and at finite wave-number, \ie the conductivity of
 an anisotropic current. In the gravity side, this corresponds to studying the gauge field on the probe brane world-volume. For anisotropic perturbations in the presence of a combination of both scalar (mass, $\delta N_c$) and vector (density, magnetic field) backgrounds, some of the modes of the gauge field couple to the perturbations of the scalar sector -- corresponding to operator mixing. We will, however ignore this mixing at the expense that in some cases our results may not be accurate.
 
 The outline of this paper is as follows:
In section \ref{fancybgg}, we describe the construction of the background, where we first review in\ref{fancysetup} the (gravity dual of) the $\mathcal{N}=4$ SYM, then
describe how to introduce the background quantities in the defect system in the supersymmetric (\ref{adscmt_susy}) and non-supersymmetric (\ref{adscmt_nonsusy}) cases and also describe several problems that arise in the non-supersymmetric D3-D7 setup in section \ref{adscmt_nonsusy} - which motivate us not to pursue the massive D7 case. We then show in section \mref{fancygetcon}, how to obtain the conductivity from linear response theory in our case, and also demonstrate the effects of electromagnetic duality at finite density and finite magnetic fields in sec. \ref{femdual}. The necessary steps of explicit computations are mentioned in \ref{explicit} and in section \ref{checkapp} we also discuss the limitations that arise from ignoring the mixing of the gauge field to the scalars. 
In section \mref{analcon}, we then derive analytic results in various limits, first in the isotropic DC limit $\omega \rightarrow 0$ (sec. \ref{DClimit}) and in the small frequency expansion beyond the DC limit (sec. \ref{smallfreq}). Then, we obtain the diffusion constant and ``electric'' permittivity and consider the hydrodynamic regime ($\omega ,\,  q \ll 1$) in section \ref{diffanal} and finally, in section \ref{tzero}, we consider the small temperature limit $q\gg T$ at small frequencies $\omega \ll q$ in various regimes of the density and magnetic field. 
The numerically-obtained results are shown in section \ref{numcon}. First, the full spectral curves are presented in section \mref{fancyspectral}. Then we present and discuss the purely dissipative poles that we obtain numerically in section \ref{diffnrelax} and finally in section \ref{llplasmon} we study the quasiparticle poles in the correlator that we obtain numerically both directly and from the spectral curves.
We present the explicit form of the induced metric on the brane in appendix \mref{gform}, and in appendix \mref{solidrev}, we review some basic properties of weakly coupled systems in order to introduce the terminology and remind the reader of the reader of some intuition and generic expectations.
Our results are summarized and discussed in section \ref{discuss}
%
%
%
\section{Turning on the Background Parameters}\mlabel{fancybgg}
\subsection{$D3$ = ({\cal{N}}=4 SYM) background}\mlabel{fancysetup}
Let us remind ourselves of the super Yang-Mills background.
We start off with the well-known $AdS_5 \times S^5$ background of $N_c$ D3 branes in the decoupling limit corresponding to an ${\mathcal{N}} = 4$ SYM theory on the boundary \cite{juan}\cdash\cite{bigRev} with $U(N_c)$ gauge group. We work in the limit of $N_c \rightarrow \infty$, at Yang-Mills coupling $g_{YM}^2 = 2\pi g_s \rightarrow 0$ in the field theory, such that we consider the large t'Hoft coupling limit $\lambda = g^2_{YM} N_c \to \infty$ and we can use the supergravity limit as $L^4 = 4\pi g_s N_c l_s^4 \rightarrow \infty$. At finite temperature $T = \frac{r_0}{\pi L^2}$, the background metric is written as 
\begin{equation}
ds^2 \, = \, \frac{r^2}{L^2}\left(- (1-r_0^4/r^4) dt^2 + d\vec{x}_3^2  \right) + \frac{L^2}{r^2}\left(\frac{dr^2}{(1-r_0^4/r^4)} + r^2 d\Omega_5^2 \right)\ , ~~~~~ C^{(4)}_{\!\! 0123}= -\frac{r^4}{L^4} \, .
\end{equation}
Considering only $T > 0$ allows us to go to dimensionless coordinates $u = \frac{r_0}{r}, \ \tlt = \frac{ r_0 t }{L^2}, \ \vec{\tx} = \vec{x}\frac{ r_0}{L^2}$:
\begin{equation}
ds^2 \, = \, \frac{L^2}{u^2} \left(- (1-u^4)d\tlt^2 + d\vec{\tx}_3^2 + \frac{du^2}{1 - u^4} + u^2 d\Omega_5^2 \right) .
\end{equation}

In
the field theory side, all fields in the SYM theory transform in the adjoint
representation of the gauge group. 
The most well-known approach to introducing
matter fields transforming in the fundamental representation is to
insert probe D7-branes in the radial direction into the supergravity background
\cite{karchkatzkarch,karchkatzmaldacena}, intersecting in the ``flat'' (SYM) directions with the D3-branes. 
If the intersection overs only part of the flat directions, 
this will create a 
defect field theory, where the fundamental fields are
only supported on a subspace within the four-dimensional spacetime
of the gauge theory. As in ref. \refcite{baredef}, we will consider a
$(2+1)$-dimensional defect by inserting $\nf$ D$p$-branes, with
three dimensions parallel to the SYM directions and $p-3$ directions
wrapped on the $S^5$, considering both probe D5-
and D7-branes. If we consider the supergravity background as the
throat geometry of $\nc$ D3-branes, our defect constructions are
described by the following array:
\begin{equation}
\begin{array}{rcccccccccccl}
  & & 0 & 1 & 2 & 3 & 4& 5 & 6 & 7 & 8 & 9 &\\
\mathrm{background\,:}& D3 & \times & \times & \times & \times & & &  & & & & \\
\mathrm{probe\,:}& D5 & \times & \times & \times &  & \times  & \times & \times & &  & &  \\
& D7 & \times & \times & \times &  & \times  & \times & \times & \times & \times & &   \\
\end{array}
\labell{array}
\end{equation}

The D5-brane construction is supersymmetric and the dual field
theory is now the SYM gauge theory coupled to $\nf$ fundamental
hypermultiplets, which are confined to a (2+1)-dimensional defect.
Note that the supersymmetry has been reduced from ${\cal N}=4$ to
${\cal N}=2$ by the introduction of the defect. In the D7-brane
case, we have lost supersymmetry altogether and the defect supports
$\nf$ flavors of fermions, again in the fundamental representation
\cite{rey1,rey2}. One should worry that the lack of supersymmetry in the
latter case will manifest itself with the appearance of
instabilities. In ref. \refcite{baredef}, we showed how this can be avoided, and we discuss in section \mref{adscmt_nonsusy} how this instability becomes apparent in the scaling dimension of the scalar field that corresponds to the deformation of the $S^4$ of the D7-worldvolume inside the $S^5$ background. There, we also discuss some problems related to the reliability of the quenched approximation that we consider in this paper.
In this limit, $\nf
\ll \nc$, the D5-branes may be treated as probes in the
supergravity background, \ie we may ignore the gravitational
back-reaction of the branes. For the D7-branes, however, this is only true locally and not in the asymptotic regime.

As we commented above, a similar holographic framework has been used
extensively to study the properties of the ${\cal N}=2$ gauge theory
constructed with parallel D7- and D3-branes, \ie the fundamental
fields propagate in the full four-dimensional spacetime -- \eg see
refs. \refcite{recent1}--\refcite{long2}. There, it was found that if a ``quark'' mass $M_q$ is introduced for the
hypermultiplets, it was found that the scale
$M_\mt{fun}\sim M_q/\sqrt{\lambda}$ plays a special role in this
theory. First, the ``mesons", bound states of a fundamental and an
anti-fundamental field, are deeply bound with their spectrum of
masses characterized by $M_\mt{fun}$ \cite{meson}. Next at a
temperature $T\sim M_\mt{fun}$, the system undergoes a phase
transition characterized by the dissociation of the mesonic bound
states \cite{long1,long2}. The meson spectrum is
characterized by the same mass scale $M_\mt{fun}$
\cite{ramallo1,ramallo2,holomeson} and these states are completely dissociated
in a phase transition at $T\sim M_\mt{fun}$.
A similar behavior can be observed for defects in the D5 case that are T-dual to the mentioned D3-D7 configuration. 
We will discuss this briefly in section \mref{adscmt_susy} and more in detail in ref. \refcite{thermpaper}, where we also find some properties that are new in the case of a defect. 

\subsection{Introducing the defect}\mlabel{fancyintro}
In the supergravity limit, the D5 brane action of the $U(1)$ subgroup of the $U(N_f)$ is just the DBI action plus a Chern-Simons term 
\begin{equation}
S \, = \, - T_5 N_f \int_{D5} \sqrt{- det(P[G] + 2\pi l_s^2 F )} \, +\, T_5 N_f \int_{D5} C^{(4)} \wedge 2\pi l_s^2 F  \ ,
\end{equation}
where the factors of $N_f$ arise from taking the trace over the flavor degrees of freedom, arising from the stack of $N_f$ coincident branes.
To simplify things further, we work in the quenched approximation $N_f \ll N_c$, such that we can ignore the backreaction of the probe branes.

Preserving translational invariance in the flat directions and rotational invariance on the sphere, together with the choice of the embedding \reef{array} dictates the induced metric on the D5 brane to be of the form
\begin{eqnarray}\labell{branemet}
ds^2 &=& \frac{L^2}{u^2} \left( -(1- u^4)d\tlt^2  + d\vec{\tx}_2^2  + \left(1+(1-u^4)\left(z'(u)^2   + u^2\frac{\Psi'(u)^2}{1-\Psi(u)^2} \right) \right) \frac{du^2}{1-u^4}  \nonumber \right. \\ & & \left. \ \ \ \ \ \ \ \ \ \ \, + \, u^2 (1-\Psi(u)^2) d\Omega_2^2\right) \ ,
\end{eqnarray} 
where $z$ is the position of the brane in normal flat direction and $\Psi$ describes the size of the $S^2$ through the embedding
\begin{equation}
d\Omega_5 ^2\  =\ d\psi ^2\, +\, \cos^2 \psi
\, d\Omega_2^2 \,+\,\sin^2 \psi \, d{\Omega}_2^2 \ , \ \ \ \sin \psi  =:  \Psi \ .
\end{equation}

In a previous paper, ref. \refcite{baredef}, we discussed the case of the trivial solution $\Fnaught=0$ for the $U(1)$ background in the flat directions and for the $S^2$ radius $\Psi = 0$.
In other words, the discussion of the setup was limited to the case of vanishing ``quark'' mass for the matter on the defect, vanishing net density of matter and antimatter on the defect (net baryon number density) and vanishing external magnetic field applied to the defect. However, a flux $\Fnaught=\frac{q}{N_f} d\Omega_2$
on the compact sphere was turned on. This corresponds on the gravity side to having an extra set of $q$ D3 branes pulling on the D5 from one side of the defect and on the field theory side to having an extra number of colors, $\delta N_c = q=: f\frac{L^2 N_f}{\pi l_s^2} = f\frac{\sqrt{\lambda} N_f}{\pi} \in \mathbb{Z}$ on that side of the defect. Both from the embedding geometry $z(u)$, and from the resulting quasiparticle spectrum in the field theory, it was argued that this flux also introduced a finite width, $\Delta z$, of the defect. 
The embedding was found to be
\begin{equation}\label{ozembed}
z(u)' = \frac{- f}{\sqrt{1+f^2 u^4}} \ ,
\end{equation}
which has in principle some analytical solution.
Here, as everywhere in this paper, we use the notation $(\cdot)' := \partial_u (\cdot)$.
Even though $\Psi$ has a tachyonic mode, corresponding to shrinking to zero size, it was shown that its mass lies above the Breitenlohner-Friedmann bound $m_\Psi =  \frac{-2}{L_{AdS_4}^2} > \frac{-9}{L_{4 AdS_4}^2} $ \cite{BF1,BF2}, such that it does not cause an instability.

\subsection{AdS/CMT Dictionary (Supersymmetric Case)}\mlabel{adscmt_susy}
Now the situation is slightly more non-trivial, as we wish to introduce finite values for the mass, baryon density and magnetic field. 
Using the $AdS/CFT$ dictionary in ref. \refcite{bigRev} in analogy with the $3+1$ dimensional system, e.g. refs. \refcite{hallo,findens1,findens2,magnetic1,magnetic2}, we find 
the gravity dual of the baryon density
\begin{equation}
\rho_0  \ = \ - 2\pi^2\frac{\delta S}{\delta \Fnaught_{tr}}\ =\ 2\pi^2 (\pi T)^2 \varepsilon_0 \lim_{u\rightarrow 0} A'(u)\ =:\ - 2\pi^2 (\pi T)^2 \lim_{u\rightarrow 0} E(u)
\end{equation}
and magnetic field $B$ to be related to a non-trivial $U(1)$ background on the brane:
\begin{equation}\labell{eback}
\Fnaught|_{u \rightarrow 0} \, = \, - E(u) dt\wedge dr  + B dx\wedge dy \,  =: \, F_E + F_B \ .
\end{equation}
The factor $\varepsilon_0 := \frac{r_0}{g_4^2 L^2} = \frac{\pi\,T}{
g_4^2}$ arises from the spherical factor and the overall factors in the DBI action.
We can also define the (asymmetric) background metric
\beq \labell{gdef}
G = g+ \Fnaught \ .
\eeq

By analogy with the $3+1$ dimensional $D3-D7$ system, we can repeat the arguments in refs. \refcite{johanna1}--\refcite{long2,findens1,findens2,us}, and associate a non-trivial embedding $\Psi(u)$ with a finite quark mass $M_q$ and dual condensate $C$. This condensate has on the one hand an interpretation as a chemical potential for $M_q$ and on the other hand is considered in QCD contexts considered as the order parameter of chiral symmetry breaking.

In the parametrization \reef{eback} the DBI-CS action becomes then
\begin{eqnarray}\labell{backact}
S &=& 4\pi L^2  T_5  \int\! d\mathfrak{x}^4 \left(\sqrt{-\det G}\sqrt{(1-\Psi^2)^2 + f^2} + f u^4  z' \right) \\ \nonumber &=& 4 \pi L^2 T_5 \int d\sigma^4\left( \sqrt{-\det G}\sqrt{1+F_{E}^2}\sqrt{1+F_{\tlb}^2}\sqrt{(1-\Psi^2)^2 + f^2}  + f u^4 z'\right),
\end{eqnarray}
and one trivially finds the background solution
\begin{eqnarray}\labell{backgd}
\!\!\!\!\!\!\!\!\!\!\!\!\!\! B &=& const. \\ 
\!\!\!\!\!\!\!\!\!\!\!\!\!\! E(u) &=&  \frac{\tlrho \sqrt{1+f^2}\sqrt{1- \Psi^2 + u^2 (1-u^4) \Psi'^2}}{\sqrt{1-\Psi^2}\sqrt{1+\big(  f^2 +(\tlrho^2+ \tlb^2)(1+f^2)\big)u^4 + (1+\tlb^2 u^4)\Psi^2(\Psi^2-2)}}  \\
\!\!\!\!\!\!\!\!\!\!\!\!\!\! \partial_u z & = &  \frac{-f \sqrt{1- \Psi^2 + u^2 (1-u^4) \Psi'^2}}{\sqrt{1-\Psi^2}\sqrt{1+\big( f^2+(\tlrho^2 + \tlb^2)(1+f^2)\big)u^4 + (1+\tlb^2 u^4)\Psi^2(\Psi^2-2)}} 
\end{eqnarray} 
for all the physically relevant fields, except for $\Psi(u)$, because that one enters the action both directly and with one derivative. 
For convenience, we defined the dimensionless parameters $\tlrho := \frac{\rho_0}{2\pi^2 (\pi T)^2 \varepsilon_0}$ and $\tlb = \frac{B}{(\pi T)^2}$.
We see that the width of the defect from the brane picture, $z_{max} := \lim_{u->1} z(u)$, decreases as we increase $B$ and $\rho_0$ as they appear only in the denominator. This may appear somewhat counter-intuitive from a weakly coupled point of view, but it is what we should expect, as the system is strongly coupled, or the correlation length diverges, and hence the ``contractive force'' scales with the total number of particles.

The equation of motion for $\Psi(u)$ becomes
\begin{eqnarray}\labell{psieom}
\textstyle{\frac{2(1+\tlb^2u^4)(1-\Psi^2)^3 +u^2 (1-u^4)\big(1-(f^2 + (\tlrho^2+\tlb^2)(1+f^2))u^4 + (1+\tlb^2 u^4)\Psi^2(\Psi^2-2) \big)\Psi'^2}{u^4 (1\!-\!\Psi^2)\sqrt{(1\!-\!\Psi^2)\big(1\!-\!\Psi^2+(u^2\!-u^6)\Psi'^2 \big)\big(1+( f^2\!+(\tlrho^2\! + \tlb^2)(1+f^2))u^4 + (1+\tlb^2 u^4)\Psi^2(\Psi^2\!-2)\big)}}} \nonumber \\
\textstyle{\!\!\!\!\!\!\!\!\! = \
\partial_u  \left(\Psi' \frac{1-u^4}{u^2}\sqrt{\frac{1+( f^2+(\tlrho^2 + \tlb^2)(1+f^2))u^4 + (1+\tlb^2 u^4)\Psi^2(\Psi^2-2)}{(1-\Psi^2)(1-\Psi^2+(u^2-u^6)\Psi'^2)}} \right)}
 \ , ~~~~~~~~
\end{eqnarray}
which has no analytical solution, except for some limiting cases. For $u \rightarrow 0$, it is easy to see that the solution becomes
\begin{equation}\labell{psiasym}
\Psi \, \sim \, \tlm \, u \, + \, \tlc \, u^2 \ ,
\end{equation}
where $\tlm$ and $\tlc$ are dimensionless free parameters that are determined by the boundary conditions. Now, we see that the argument of the T-dual case of the D3-D7 system with $3+1$ intersecting directions \cite{johanna1}\cdash\cite{long2,findens1}\cdash\cite{us} also applies to our case, and the quark mass $M_q$ and condensate $C$ are given by 
\begin{equation}
M_q \, = \, \frac{r_0 \,\tlm}{2^{3/2}\pi l_s^2} \, = \, \sqrt{\lambda} \frac{T}{2^{3/2}}  \tlm\ \ \  \mathrm{and}  \ \ \  C\, = \, \sqrt{2}4\pi^2 \tlc \, r_0^2 N_f l_s^2 T_5 \, = \,  \frac{1}{4\pi} \tlc T^2 N_f N_c\ .
\end{equation}
This can be straightforwardly obtained from the results in ref. \refcite{long1,long2}, but also in our case we see that this relates to the length of a string spanning on the sphere from the D3 branes to the D5 branes. In ref. \refcite{thermpaper}, we discuss this more in detail and verify that $C$ is indeed the dual chemical potential to the mass.

In order to find the solution for the full geometry for a given mass however, we need consider the equation near the horizon, where \reef{psieom} reduces to first order,
\begin{equation}\label{bcdic}
\Psi'|_{u\rightarrow 1} \, =\, \frac{1}{2}\frac{(1+\tlb^2)\Psi_0 (1-\Psi_0^2)^2}{(1-\Psi_0^2)^2+f^2+\tlrho^2+\tlb^2\big(1 + (1-\Psi_0^2)^2\big)} \ ,
\end{equation}
effectively relating $\tlm$ and $\tlc$. The only remaining boundary condition at the horizon is then $\Psi_{u\rightarrow 1} = \Psi_0$. 
Because the boundary condition \reef{bcdic} is a consequence equation of motion, we cannot use $\tlm$ instead as a boundary condition, but we have to find recursively $\Psi_0$ for a given value of $\tlm$. This is because implicitly, on-shell, $\tlc$ is a function of $\tlm$ and starting to integrate at $u=0$ with some random combination of $\tlc$ and $\tlm$ means the equations of motion cannot be on-shell as we approach the horizon.

At vanishing density $\rho_0=0$ and vanishing compact flux $f$, we find again that the black hole embedding which gives us free quarks, limits $M_q < M_{crit}$. At $M_{crit}$, we have the $2+1$ analogue of the phase transition that was found for the $3+1$ system in refs. \refcite{johanna1}--\refcite{long2}. It turns out that the critical mass decreases as we turn on the magnetic field in the $2+1$ field theory. This is discussed in detail in ref. \refcite{thermpaper}.
%
%
%
This phase transition disappears (at least in our case where we consider only the $U(1)$ background) as we turn on either a finite baryon density $\rho_0$, or as we choose the compact magnetic flux $f$ to be non-zero. Essentially, this happens because the charge can only be supported by the blackhole embedding and the action becomes singular when $\Psi(u) =0$ at finite flux just as in the $3+1$ dimensional case in ref. \refcite{long1,long2}. 

In some of our studies of the effects of finite masses, we will have to consider finite $\rho_0$ or $f$, to allow for sufficiently large masses. In the limit of very large masses (i.e. $\Psi \rightarrow 1_-$ near the horizon), one can see that over $u\in ]0,1]$, the equation of motion for $\Psi$ is also solved by $\Psi \sim 1$ -- to see this in eq. \reef{psieom}, one needs to take $\Psi' \rightarrow 0$ first. This demonstrates how a new length scale arises for large masses, $\tlm\gg 1$ as the profile splits approximately into two parts -- one with $u > u_m\sim\frac{1}{\tlm}$ and $\Psi \sim 1$ for some value $u_m$ and one given approximately by the asymptotic solution \reef{psiasym}. Around $u_m$, $\Psi''$ diverges. It would be interesting to see whether this limit has any relation to the recent discussion of non-relativistic AdS/CFT \cite{nonrel2}\cdash\cite{nonrel16}.

A more thorough discussion of the thermodynamics and the phase structure can be found in ref. \refcite{thermpaper}.
%
\subsection{Non-supersymmetric $D3-D7$ intersection}\mlabel{adscmt_nonsusy}
The non-supersymmetric case is very similar to the D5 case as it differs in the massless case only by the geometry and field configuration in the $S^5$ factor. Now we parametrize the $S^5$ in the bulk space as $d\Omega_5 ^2 = d\psi ^2 + \cos^2 \psi
\, d\Omega_4^2$, such that we have the induced metric
\begin{eqnarray}\labell{branemettt}
ds^2 &=& \frac{L^2}{u^2} \left( -(1- u^4)d\tlt^2  + d\vec{\tx}_2^2  + \left(1+(1-u^4)\left(z'(u)^2   + u^2\frac{\Psi'(u)^2}{1-\Psi(u)^2} \right) \right) \frac{du^2}{1-u^4}  \nonumber \right. \\ & & \left. \ \ \ \ \ \ \ \ \ \ \, + \, u^2 (1-\Psi(u)^2) d\Omega_4^2\right) \ .
\end{eqnarray}
and we set up an instanton on the $S^4$ instead of the magnetic charge on the $S^2$.
The coupling to the five-form flux comes now via the Chern-Simons term
\begin{equation}
\frac{(2\pi l_s^2)^2}{2}T_7 N_f \int_{D7} C^{(4)} \wedge  F \wedge  F \ .
\end{equation}
 This CS term however also causes the $D7$ setup to differ from the $D5$ setup in the massive case, as can be seen most easily by integrating this term by parts to give us (modulo a total derivative) 
 \begin{equation}
\frac{(2\pi l_s^2)^2}{2}T_7 N_f \int_{D7} F^{(5)} \wedge  A \wedge  F  \ \rightarrow \ 
8 T_7 N_f \frac{\pi^5 l_s^4}{L^4} \int_0^1 du \frac{\Psi'}{\sqrt{1-\Psi^2}}  \int_{\mathbb{R}^{(2,1)}} A \wedge  F  \ .
 \end{equation}
 The term on the right hand side with the factor $\Psi'$ arises from the fact that the deformed embedding of the $S^4$ inside the $S^5$ causes the dual $F^5$ on the sphere to pull back to the brane worldvolume. Integrating out the $S^4$ of the worldvolume gives us then the right hand side,
which is just a Chern-Simons term with radius-dependent coupling. This was used in ref. \refcite{hall1} to obtain a Hall effect in a setup that would be considered from the perspective of this paper to be unstable.
This term will obviously modify the two-point functions, and will be interesting to consider in further work, but there are some problems with the massive D7 case that we will outline below, so we will here only consider the massless $D7$ defect.

The instanton solution was found in refs. \refcite{neil2,ConstableRobEtc}  and is outlined  also in ref. \refcite{baredef} and yields  
\beq
\frac{1}{8\pi^2}\oint_{S^4} Tr \Fnaught\wedge \Fnaught\ =:\ \cQ\ \in\ \mathbb{Z}
\eeq
and the $S^4$ factor 
\beq
\oint_{S^4} d^4\Omega\  =\ \frac{8 \pi^2}{3}\left(\nf\, L^4 (1 - \Psi^2) + 6 \pi^2 \ls^4 |\cQ| \right)  \ ,
\eeq
such that the Ansatz \reef{eback} puts the action into the form
\begin{equation}
S \ = \ \frac{8 \pi^2}{3} N_f\, L^4  T_7  \int d\sigma^4 \left(\sqrt{\det G}((1 - \Psi^2) + |Q|) + Q u^4 \partial_u z \right) \ 
\end{equation}
where we defined $Q = 6 \pi^2\frac{\ls^4}{L^4}
\frac{\cQ}{\nf}=\frac{6\pi^2}{\lambda}\frac{\cQ}{\nf}$. Having an isotropic solution on the $S^{4}$, i.e. having the corresponding symmetries unbroken requires $\frac{N_f(N_f^2-1)}{6} \ge \cQ$.
In the general case ($\Psi \neq 0$), this action gives different solutions than \reef{backgd}, however in the case $\Psi=0$ the solutions are given precisely by \reef{backgd}, provided we replace the flux parameter with $f_7 \equiv \frac{Q}{\sqrt{1+ 2|Q|}}$. 

%
In ref. \refcite{baredef} it was found that the mass of the tachyonic mode of the $S^4$ radius of the D7 probe brane satisfies the BF bound only for $f^2 > 49/32$, and a quick calculation shows that this also happens to apply in this background -- independent of  $\tlrho$ and $\tlb$.
This, we will see, is reflected in the asymptotic behavior of $ \Psi(u)$. 
We will not bother the reader with the lengthy form of the equations of motion for $\Psi$, however, we note that the asymptotic solution takes the form
\begin{equation}
\Psi(u)\ \sim \  \tlm_7 u^{\alpha_{-}} \, + \, \tlc_7 u^{\alpha_{+}} ~~ , ~~~~\alpha_\pm = \frac{3}{2}\pm \sqrt{\frac{4Q^2 -7 -12Q}{2+4Q}}
\end{equation} 
which implies that above the BF bound, the solution will be a power law, and below the BF bound it will be oscillatory -- indicating the instability. Above the BF bound, we could, in principle, identify those two modes with a ``mass-like''  operator (and a ``condensate'' operator) of non-integer conformal dimension, motivated by the fact that this is related to the separation of the D3 and D7 branes in the sphere. Possibly one could interpret this behavior with a ``running'' mass. However, there is a very significant problem that may be more worrying than the instability at vanishing mass: A solution of $N_7$ D7 branes causes an asymptotic deficit angle of $\frac{N_f}{12}$ because of backreaction (see e.g. ref. \refcite{Ortin1}). At the necessary finite $f_7$, and hence $\cQ$ of order $\lambda$, we need at stack of D7 branes as determined by the limit $\frac{N_f(N_f^2-1)}{6} \ge \cQ$ for spherically symmetric solution from ref. \refcite{ConstableRobEtc}. In order to still have the same kind of field theory with the same symmetries, we require $N_f \gg 12$, which implies that the solution cannot connect to an asymptotic space-time. Note that this was exactly the $S^4$ factor which broke supersymmetry, and this factor would be highly modified for solutions with $\cQ$ below this bound. Below the BF bound the oscillatory asymptotic solution is non-physical in our setup and furthermore the corresponding operator would have complex conformal dimension. It seems that there are some non-standard ways to interpret the $f_7=0$ case in the context of the quantum Hall effect \cite{quantumhall}, but we will not pursue the $\Psi \neq 0$ case in the rest of this paper. It is a noteworthy curiosity, that in the absence of the pullback of the CS term to the flat directions, the resulting spectral functions are identical to the D5 case under an appropriate identification of the mass-like operators.
\section{Computing the Conductivity}\mlabel{fancygetcon}
One common approach to compute the conductivity that we will pursue here is
linear response theory, i.e. applying the Kubo formula
\begin{equation}
\pi^{-1} T \tilde{\sigma}_{ij} = \sigma_{ij} \, = \, \frac{i}{\omega} C_{ij} ~~~ \ \, \mathrm{or\ for\
convenience} \ \ \tilde{\sigma}_{ij} \, :=  \, \pi T \sigma_{ij} \,
= \, \frac{i}{\tom} C_{ij}\  \ ,
\end{equation}
that gives the conductivity for currents resulting from small perturbations in terms of the retarded Green's function, which is given in terms of the correlator as
\begin{equation}
 C_{ij}(x-y)=-i\,\theta(x^0-y^0)\,\langle\, [J_i(x),J_j(y)]\, \rangle \ .
\end{equation}
We define for later convenience $\tilde{\sigma} = \frac{\sigma}{\pi T}$.
Since the baryon number current $J_i$ is dual to the gauge field of the $U(1)$ subgroup of the $U(N_f)$, $A_i$, the correlator is given by the variation of the on-shell action
\begin{equation}\labell{correlact}
C_{i j} = \frac{\delta^2 S}{\delta A^\star_{i, 0}\delta A_{j, 0}} ,
\end{equation}
where $A_0$ is the boundary value of the gauge field at the asymptotic boundary $u=0$.

In order to obtain this correlator, we have to consider second order perturbations of the action with respect to the gauge field and obtain the equations of motion for this pertubation, denominated now with $F$. Strictly speaking, we also have to take into account the pertubation of the scalars $\phione_i$ up to the same order. They appear in the induced metric as 
$\delta g_{\mu\nu} = \partial_\mu \phione_i \, \partial_\nu \phione_j \, g^{ij}\partial_\mu \phione_i \, \partial_\nu \phi_j \, g^{ij} \, + \,\partial_\mu \phi_i \, \partial_\nu \phione_j \, g^{ij}\,+\,\partial_\mu \phione_i \, \partial_\nu \phione_j \, g^{ij}\, =: \,\gone_{\mu\nu}\,+\,\gtwo_{\mu\nu}$, where the linear term appears only at finite mass or finite $f$ and may induce mixing with the vector perturbations.
%
%
%

Considering only the scalars that mix with the vectors, the action at second order can be written most conveniently as
\begin{eqnarray}\labell{robbackactF}
S& =& - \frac{1}{4 g_4^2} \int d\mathfrak{x}^4  \,\sqrt{- \det \, G}\,\left( \frac{\sqrt{(1-\Psi^2)^2 + f^2}}{\sqrt{1+f^2}} \left(F_{\alpha \beta}G^{\alpha \gamma}F_{\gamma \delta} G^{\delta \alpha} - \frac{1}{2}(F_{\alpha \beta}\Ga^{\alpha \beta})^2 \right.\right. \nonumber \\
 & & ~~~~~~~~~~~~~~~~~~~~~~~~~  +4 \gone_{\alpha \beta}\Gs^{\alpha \gamma}F_{\gamma \delta} \Ga^{\delta \alpha}
 - \gone_{\alpha \beta}\Gs^{\alpha \beta}F_{\gamma\delta}\Ga^{\gamma\delta} \nonumber \\
& & ~~~~~~~~~~~~~~~~~~~~~~~ \left.
 -\,  \frac{1}{2}(\gone_{\alpha \beta}\Gs^{\alpha \beta})^2 \, + \, \gone_{\alpha \beta}G^{\alpha \gamma}\gone_{\gamma \delta} G^{\delta \alpha} \, - \, 2 \gtwo_{\alpha \beta}\Gs^{\alpha \beta}
\right) \nonumber \\
& & ~~~~~~~~~~~~~~~~
- 2 \frac{(1-\Psi^2)\Psi}{\sqrt{1+f^2}\sqrt{(1-\Psi^2)^2 + f^2}} \phipsi \left(F_{\alpha \beta}\Ga  ^{\alpha \beta} +\gone_{\alpha \beta}\Gs^{\alpha \beta}\right) \nonumber \\
& & ~~~~~~~~~~~~~~~~~~~\left.\left.
 - 2 \frac{(1-\Psi^2)^3 + (1-3 \Psi^2)f^2 }{\sqrt{1+f^2}((1-\Psi^2)^2 + f^2)^{3/2}} \phipsi^2
 \right.\right) \ ,
\end{eqnarray}
where $\Gs$ and $\Ga$ are the symmetric and antisymmetric parts of the inverse asymmetric metric.

For simplicity of the computations however, we will ignore the interaction terms between the scalar and vector sector. We will discuss the limitations on the reliability of our results later, but first let us look at the simplified action:
\begin{equation}\labell{robbackact}
S = - \frac{1}{4 g_4^2} \int d\mathfrak{x}^4 \frac{\sqrt{(1\!-\!\Psi^2)^2 + f^2}}{\sqrt{1\!+\!f^2}} \sqrt{- \det \, G} \left(F_{\alpha \beta}G^{\alpha \gamma}F_{\gamma \delta} G^{\delta \alpha} - \frac{1}{2}F_{\alpha \beta}G^{\alpha \beta}F_{\gamma \delta} G^{\gamma \delta } \right)  ,
\end{equation}
where  $G$ is the asymmetric combined metric \reef{gdef}, and $g_4^2$ is defined as 
\beq
\frac{1}{g_4^2}\  =\   4\pi (2\pi l_s^2)^2 L^2 \sqrt{1+f^2}T_5\ =\ \sqrt{1+f^2}\frac{2}{\pi}\frac{N_c}{\sqrt{2\lambda}} \ .
\eeq
In some sense, there is now a radius dependent coupling $\frac{\sqrt{(1-\Psi^2)^2 + f^2}}{\sqrt{1+f^2}}$, that always goes to unity asymptotically or obviously everywhere in the massless case.
Surprisingly, the gauge field background dies off sufficiently fast asymptotically, such that as $u\rightarrow 0$, the action just becomes the Maxwell action with coupling $g_4$ and in a suitable gauge $A_u=0$ the correlator is still given by the asymptotic mode function
\begin{equation}\labell{condmode}
C^{i j} =  \frac{\varepsilon_0}{\sqrt{1+f^2}}\left. \frac{\delta \left(\partial_u A_j\right)}{\delta A_{i }}\right|_{u\rightarrow0} \ ,
\end{equation}
where $\varepsilon_0$ is defined as $\varepsilon_0 = \frac{\pi T}{g_4^2}$.

In the rest of our analysis, we will consider the Fourier-transformed fields in the flat directions, \eg 
\begin{equation}
A_\mu(\mathfrak{x}) \ = \ \int \frac{d^3 x}{(2\pi)^3} e^{i k_a x^a}A_\mu(k,u)  \ , \ \ \ k = (\omega,q,0)\ ,
\end{equation}
where we considered without loss of generality momentum to be carried in the $x$ direction such that the $y$ direction is ``transverse''.

\subsection{Electromagnetic duality}\mlabel{femdual}
In this background, we see that the effective action for the gauge field is not invariant under electromagnetic duality $F\rightarrow \star F$. Hence, the relation $C_{xx} \, = \, - \frac{\varepsilon_0^2 \omega^2}{C_{yy}}$ that was found in ref. \refcite{pavel} does not apply in this case. Since the DBI action at constant coupling, i.e. in the massless case, however still obeys this duality, one would expect that it survives in some form under the exchange of the magnetic and electric charges on the probe brane, i.e. under the exchange of the density and magnetic field in the field theory side.
To quantify this further, let us look at the transformations under $F\rightarrow \star F$ of the Fourier-transformed gauge field in the gauge $A_u = 0$ that led to \reef{condmode}, obviously at $M_q=0$. The relevant components are at asymptotic infinity:
\begin{equation}
(\star F)_{tx}|_{u=0} \, =\, -\sqrt{1+f^2}F_{uy}|_{u=0} ~~\mathrm{and} ~~~
(\star F)_{ty}|_{u=0} \, =\, \sqrt{1+f^2} F_{ux}|_{u=0} \ ,
\end{equation}
such that the variation w.r.t. the gauge field becomes in terms of the transformed gauge field, denoted in abusive notation as $\star A$:
\begin{eqnarray}
\!\!\!\!\!\!\! \left.\frac{\partial}{\partial A_x}\right|_{\!u=0} \!\!\!\!\!\!\!\! & =\!\! & - \frac{\sqrt{1+f^2}}{i \tom}\left(\frac{\partial A_y'}{\partial A_x}\frac{\partial}{\partial (\star A)_x} - \frac{\partial A_x'}{\partial A_x}\frac{\partial}{\partial (\star A)_y} \right)_{\!\!u=0} \!\!\!\!\!\! =  \frac{\partial}{\varepsilon_0}\left(\tilde{\sigma}_{xy}\frac{\partial}{\partial (\star A)_x} - \tilde{\sigma}_{xx}\frac{1}{\partial (\star A)_y} \right)_{\!\!u=0}~~~~~ { } \\
\!\!\!\!\!\!\! \left.\frac{\partial}{\partial A_y}\right|_{\!u=0} \!\!\!\!\!\!\!\! & = \!\! & - \frac{\sqrt{1+f^2}}{i \tom}\left(\frac{\partial A_y'}{\partial A_y}\frac{\partial}{\partial (\star A)_x} - \frac{\partial A_x'}{\partial A_y}\frac{\partial}{\partial (\star A)_y} \right)_{\!\!u=0} \!\!\!\!\!\! =  \frac{\partial}{\varepsilon_0}\left(\tilde{\sigma}_{yy}\frac{1}{\partial (\star A)_x} + \tilde{\sigma}_{xy}\frac{\partial}{\partial (\star A)_y} \right)_{\!\!u=0} \!\!\!\!\!\!\! . 
\end{eqnarray}
Rewriting the conductivity obtained from \reef{correlact} then in terms of the transformed fields gives us 
\begin{eqnarray}
\tilde{\sigma}_{xx} & = & \frac{1}{\varepsilon_0^2}\left(\tilde{\sigma}_{xy}\tilde{\sigma}_{xy} [\star \tilde{\sigma}]_{xx}  + \tilde{\sigma}_{xx}\tilde{\sigma}_{xx} [\star \tilde{\sigma}]_{yy}\right)  \\
\tilde{\sigma}_{yy} & = & \frac{1}{\varepsilon_0^2}\left(\tilde{\sigma}_{yy}\tilde{\sigma}_{yy} [\star \tilde{\sigma}]_{xx}  + \tilde{\sigma}_{xy}\tilde{\sigma}_{xy} [\star \tilde{\sigma}]_{yy}\right)  \\
\tilde{\sigma}_{xy} & = & \frac{1}{\varepsilon_0^2}\left(\tilde{\sigma}_{xy}\tilde{\sigma}_{yy} [\star \tilde{\sigma}]_{xx}  + \tilde{\sigma}_{xx}\tilde{\sigma}_{xy} [\star \tilde{\sigma}]_{yy} + (\tilde{\sigma}_{xy}\tilde{\sigma}_{xy}-\tilde{\sigma}_{xx}\tilde{\sigma}_{yy}) [\star \tilde{\sigma}]_{xy}\right)  
\end{eqnarray}
where we used \reef{condmode} and $S[  F] = S[\star F]$, and defined $[\star \sigma]_{\mu \nu} = \frac{\delta^2 S[\star F]}{\delta [\star A]_{i, 0}\delta [\star A]_{j, 0}}$. Since $F \leftrightarrow \star F$ exchanges the electric and magnetic charges on the probe brane defined at infinity -- i.e. exchanges density and magnetic field in the field theory side - $\star \sigma$ is just the conductivity and the exchange of $\tlrho$ and $\tlb$. 
Finally, we can solve for $\star \sigma$ and obtain
\begin{equation}\labell{emdualcon}
\sigma[\tlb,\tlrho]_{a b} \ = \ \left(\star \sigma[\tlrho,\tlb] \right)_{ab} \ = \ \frac{1}{\varepsilon_0^2}\left(\left(\tilde{\sigma}[\tlrho,\tlb]\right)^{-1} \right)_{cd}\varepsilon^c_a \varepsilon^d_b \ 
\end{equation}
where $a,b \in \{ x, y\}$.

This result is remarkable, since it relates the transport properties under the exchange of two quantities that are completely distinct in nature from the condensed matter point of view. Furthermore, it applies to a whole class of strongly coupled $2+1$ dimensional systems, whose gravity dual obeys the electromagnetic duality. Hence, such a relation is a generic prediction from AdS/CFT for a quantum critical 2-dimensional system. For theories not obeying \reef{condmode}, there may potentially be additional terms in \reef{emdualcon}. It seems that this is an implication of the ``particle-vortex duality'' found in refs. \refcite{dolan,sl2z}, extended to finite frequencies and accordingly to a complex conductivity tensor. Certainly, this duality does not generate the full $SL(2,\mathbb{Z})$. A candidate for the second generator is simply a shift in the theta angle and the corresponding Hall conductivity from the Wess-Zumino term in the action that was outlined in ref. \refcite{baredef}. For another discussion of $SL(2,\mathbb{Z})$ symmetry in the context of AdS/CFT in a somewhat different limit that appeared very recently on the arXiv see ref. \refcite{myfriendsincalifornia}.

This duality holds always in the massless case to numeric accuracy. Hence it is not possible to visually ``compare'' the result in a plot.
\subsection{Explicit Computations}\label{explicit}
To proceed further let us start by writing out the equations of motion explicitly:
\begin{eqnarray}\labell{eoms}
\labell{fyeq}A_y\!\!:\!\! & & 0\! =  ({\scriptstyle{\sqrt{-G}}}G^{yy} G^{ u u} A_y')' \, +\, {\scriptstyle{\sqrt{-G}}}G^{yy}( G^{ xx} \partial_x^2 +  G^{tt}\partial_t^2)A_y \,\nonumber \\
\!\!\!\!\!\!\! & & ~~~~~ ~~~~~ + \, \left({\scriptstyle{\sqrt{-G}}}G^{tu} G^{xy} \right)' \left(\partial_x A_t - \partial_t A_x \right)   \\
\labell{fueq}A_u\!\!:\!\! & & 0\! = G^{tt} \partial_t A_t' +  G^{xx} \partial_x A_x' \\
\labell{fteq}A_t\!\!:\!\! & & 0\! = ({\scriptstyle{\sqrt{-G}}}G^{tt} G^{ u u} A_t')'  +  {\scriptstyle{\sqrt{-G}}}G^{tt} G^{xx} \left(\partial_x^2 A_t -  \partial_t \partial_x A_x \right) - \left({\scriptstyle{\sqrt{-G}}}G^{tu} G^{xy}\right)' \partial_x A_y ~~~~~~~~ { } \\
\labell{fxeq} A_x\!\!:\!\! & & 0\! = ({\scriptstyle{\sqrt{-G}}}G^{xx} G^{ u u} A_x')' +  {\scriptstyle{\sqrt{-G}}}G^{tt} G^{xx} \left(\partial_t^2 A_x - \partial_t\partial_x A_t \right)  +  \left({\scriptstyle{\sqrt{-G}}}G^{tu} G^{xy}\right)' \partial_t A_y  . ~~~~~~~~ { }
\end{eqnarray}
For convenience of the reader, we stick here to the concise notation in terms of $G$ and summarize the exact form of the components $G$ in appendix \mref{gform}. Also, in this expression, and for the rest of this paper, we absorbed the radius-dependent coupling into the determinant of the metric, in somewhat abusive notation: 
\begin{equation}
\frac{\sqrt{(1-\Psi^2)^2 + f^2}}{\sqrt{1+f^2}} \sqrt{-\det\,G} \, \rightarrow \, \sqrt{- G} \ .
\end{equation}
Finally, we also remind ourselves that $G^{yy} = G^{xx}$, so while keeping them distinctively for didactic reasons in most places, in some places they will be interchanged to simplify expressions.

It can be easily verified by using the equation for $A_u$, that the equations for $A_t$ and $A_x$ are degenerate. Hence, our strategy will be to eliminate $A_t$ from the equation for $A_x$ and produce an equation for $A_x'$, by multiplying \reef{fteq} with $\sqrt{-G}G^{yy} G^{ u u}$ and differentiating with respect to $u$.
This gives us:
\begin{eqnarray}\labell{fyyeq}
0 & = & \left(\sqrt{-G}G^{yy} G^{ u u} A_y'\right)' \ +\ \left(\sqrt{-G}G^{yy} G^{ xx} \partial_x^2 + \sqrt{-G}G^{yy} G^{tt}\partial_t^2\right)A_y \  \\ \nonumber 
& & ~~~~~~~~~~~ + \ \frac{\left(\left(\sqrt{-G}G^{tu} G^{xy}\right)' \right)^2}{\sqrt{-G}G^{tt} G^{xx}} A_y \ + \ \frac{\left(\sqrt{-G}G^{tu} G^{xy}\right)' }{\sqrt{-G}G^{tt} G^{xx}} \mathcal{A}_x' \\
0 & = & \left(\sqrt{-G}G^{yy} G^{ u u} \mathcal{A}_x'\right)' \ +\ \left(\sqrt{-G}G^{yy} G^{ xx} \partial_x^2 + \sqrt{-G}G^{yy} G^{tt}\partial_t^2\right)\mathcal{A}_x \labell{fxxeq} \\ \nonumber
& & ~~ -  \frac{\left(-G\, G^{tt}G^{uu} G^{xx} G^{yy}\right)'}{\sqrt{-G} G^{tt} G^{xx}} \mathcal{A}_x'\ +  (-G)\, G^{tt}G^{uu} G^{xx} G^{yy} \left(\frac{\left(\sqrt{-G}G^{tu} G^{xy}\right)' }{\sqrt{-G}G^{tt} G^{xx}}A_y\right)'  ,
\end{eqnarray}
where $\mathcal{A}_x = \frac{\sqrt{-G}G^{uu} G^{xx} A_x'}{i \tom}$. These equations separate at vanishing density or vanishing magnetic field -- as they should, because we do not expect a Hall effect in this case.

Using \reef{fxeq}, we can recover $A_x|_{u=0} =- i \frac{\mathcal{A'}_x}{\omega (1+f^2)}$, which will allow us to compute the conductivity tensor. Near the horizon, the solutions become approximately
\begin{equation}\labell{fbdysol}
A_y \, = \, (1-u)^{i \omega/4} \left(A_{y}^{(0)}+ A_{y}^{(1)} (1-u) \right)  , ~~ \mathcal{A}_x \, = \, (1-u)^{i \omega/4} \left(\mathcal{A}_{x}^{(0)}+ \mathcal{A}_{x}^{(1)} (1-u) \right) ,
\end{equation}
where $A_{y}^{(0)}$ and $\mathrm{A}_{x}^{(0)}$ are arbitrary constants and $A_{y}^{(1)}$ and $\mathrm{A}_{x}^{(1)}$ are determined straightforwardly in terms of $\rho_0$, $B$, $\Psi_0$, $A_{y}^{(0)}$ and $\mathcal{A}_{x}^{(0)}$, but somewhat lengthy and without physical insight and hence omitted here. To compute the conductivity, we could then fix the boundary conditions for $A_y \in \{1,0\}$ and $\mathcal{A}_x\in \{1,0\}$ at $u \rightarrow 0$ and enforce the leading behavior of \reef{fbdysol} as a boundary condition at $u\rightarrow 1$ as done in ref. \refcite{baredef}. This is however a numerically non-trivial boundary value problem. Hence, it is more reliable and less time-intensive to simply enforce \reef{fbdysol} for two independent choices of $\{A_{y}^{(0)},\, \mathrm{A}_{x}^{(0)}\}$, labeled $\mathfrak{a}$ and $\mathfrak{b}$, to obtain $\{A_{y}(\mathfrak{a}),A_{y}'(\mathfrak{a}),\mathrm{A}_{x}(\mathfrak{a}),\mathrm{A}_{x}'(\mathfrak{a})\}$ and $\{A_{y}(\mathfrak{b}),A_{y}'(\mathfrak{b}),\mathrm{A}_{x}(\mathfrak{b}),\mathrm{A}_{x}'(\mathfrak{b})\}$ at $u\rightarrow 1$ and then use the linearity of the problem to compute the variation in \reef{condmode} exactly. Furthermore, this strategy is very suitable from a computational point of view, as it allows us to naturally the parallelize solving the equations of motion, i.e. the most time consuming step, on a dual-core processor.

Putting everything together, we finally obtain:
\begin{displaymath}
\tilde{\sigma} = \varepsilon_0 \left[\begin{array}{cc}- i \sqrt{1+f^2}\tom\frac{\mathcal{A}_x\!(\mathfrak{a}) A_y\!(\mathfrak{b}) - \mathcal{A}_x\!(\mathfrak{b}) A_y\!(\mathfrak{a})}{\mathcal{A}_x\!'(\mathfrak{a}) A_y\!(\mathfrak{b}) - \mathcal{A}_x\!'(\mathfrak{b}) A_y\!(\mathfrak{a})} & 
-\frac{1}{\sqrt{1+f^2}}\frac{\mathcal{A}_x\!(\mathfrak{b}) \mathcal{A}_x\!'(\mathfrak{a}) - \mathcal{A}_x\!(\mathfrak{a}) \mathcal{A}_x\!'(\mathfrak{b})}{\mathcal{A}_x\!'(\mathfrak{a}) A_y\!(\mathfrak{b}) - \mathcal{A}_x\!'(\mathfrak{b}) A_y\!(\mathfrak{a})} \\
- \sqrt{1+f^2}\frac{A_y\!(\mathfrak{b}) A_y\!'(\mathfrak{a}) - A_y\!(\mathfrak{a}) A_y\!'(\mathfrak{b})}{\mathcal{A}_x\!'(\mathfrak{a}) A_y\!(\mathfrak{b}) - \mathcal{A}_x\!'(\mathfrak{b}) A_y\!(\mathfrak{a})} &
\frac{i}{\sqrt{1+f^2}\tom}\frac{\mathcal{A}_x\!'(\mathfrak{a}) A_y\!'(\mathfrak{b}) - \mathcal{A}_x\!'(\mathfrak{b}) A_y\!'(\mathfrak{a})}{\mathcal{A}_x\!'(\mathfrak{a}) A_y\!(\mathfrak{b}) - \mathcal{A}_x\!'(\mathfrak{b}) A_y\!(\mathfrak{a})} 
\end{array}\right]_{u\rightarrow 0} \ .
\end{displaymath}
Formally, this is asymmetric, such that the (anti)symmetry of the numerical result is a check for the consistency and the accuracy of the numerical solutions for $\mathcal{A}_x$ and $A_y$. We also note that in the limit of $\{\rho_0,B,\Psi\} = 0$, we just recover the equations that were found in the conformal case in ref. \refcite{baredef}. In principle, the duality from \mref{femdual} suggests that there may exist a field redefinition for $A_y$ and $\mathcal{A}_x$, such that the asymptotic solutions for the resulting fields are exchanged under $\tlrho\leftrightarrow \tlb$.
However, there is no guarantee that this redefinition can be written analytically.
\subsection{Validity of the diagonal approximation}\label{checkapp}
Now, let us look at the limitations of the reliability of our results that arise when we ignore the interaction terms.
The relevant contracted terms involving fields at first order are 
\begin{eqnarray}
\!\!\! \gone_{\alpha \beta}\Gs^{\alpha \beta} & = & \textstyle{2\frac{u^2(1-u^4)(1-\Psi^2)}{1+\tlb^2 u^4}\frac{1+(f^2 +\tlrho^2 + \tlb^2)u^4-\left(1+\frac{\tlb^2 u^2}{1+f^2}\right)(1-(1-\Psi^2)^2)}{(f^2+(1-\Psi^2)^2)((1-\Psi^2)+u^2(1-u^4)\Psi')}\left(u^2 z' \partial_u\phiz + \frac{\Psi'}{1-\Psi^2}\partial_u\phipsi\right)}~~~~~~~~{}\\
\!\!\! F_{\alpha \beta}\Ga F ^{\alpha \beta} & = & \textstyle{ 2\frac{-\tlb u^4}{1+\tlb^2 u^4}F_{xy} + \tlrho u^4\frac{\sqrt{1-\Psi^2}\sqrt{1+(f^2 +\tlrho^2 + \tlb^2)u^4-\left(1+\frac{\tlb^2 u^2}{1+f^2}\right)(1-(1-\Psi^2)^2)}}{(1+\tlb^2 u^4)(f^2+(1-\Psi^2)^2)\sqrt{(1-\Psi^2)+u^2(1-u^4)\Psi'}}F_{tr}} ~~~~~{}
\end{eqnarray}
and the relevant mixing term that contracts directly the different sectors:
\begin{eqnarray}\nonumber
\!\!\!\!& \!\! &\gone_{\alpha \beta}\Gs^{\alpha \gamma}F_{\gamma \delta} \Ga^{\delta \alpha} = \textstyle{
\frac{u^6 \sqrt{1+(f^2 +\tlrho^2 + \tlb^2)u^4-\left(1+\frac{\tlb^2 u^2}{1+f^2}\right)(1-(1-\Psi^2)^2)}}{(1+\tlb^2 u^4)^2\sqrt{f^2+(1-\Psi^2)^2} \sqrt{(1-\Psi^2)+u^2(1-u^4)\Psi'} }}\times \\ 
\!\!\!\!& \!\! &  ~~~ \Bigg(\textstyle{
\frac{\tlb (1-u^4)\sqrt{1-\Psi^2}\sqrt{1+(f^2 +\tlrho^2 + \tlb^2)u^4-\left(1+\frac{\tlb^2 u^2}{1+f^2}\right)(1-(1-\Psi^2)^2)}}{\sqrt{(1-\Psi^2)+u^2(1-u^4)\Psi'}} \,} \times\nonumber\\ 
\!\!\!\!& \!\! & ~~~~~~~~~~~~~~~~~~~\textstyle{\left(\left(u^2 z' \partial_x\phiz + \frac{\Psi'}{1-\Psi^2}\partial_x\phipsi\right)F_{ry}\left(u^2 z' \partial_y\phiz + \frac{\Psi'}{1-\Psi^2}\partial_y\phipsi\right)F_{rx}\right)  } \nonumber\\
\!\!\!\!& \!\! &  \textstyle{-2 \tlrho\! \left(u^2 z' \partial_u\phiz +\! \frac{\Psi'}{1-\!\Psi^2}\partial_u\phipsi\right)\!\!F_{tr}-\frac{\tlb (1-u^4)(1-\Psi^2)\left(1+(f^2 +\tlrho^2 + \tlb^2)u^4-\left(1+\frac{\tlb^2 u^2}{1+f^2}\right)(1-(1-\Psi^2)^2)\right)}{\left(f^2+(1-\Psi^2)^2\right)\left((1-\Psi^2)+u^2(1-u^4)\Psi'\right)}} \times \nonumber \\
\!\!\!\!& \!\! & ~~~~~~~~~~~~~~~~~~~~~~~\textstyle{\left(\left(u^2 z' \partial_x\phiz + \frac{\Psi'}{1-\Psi^2}\partial_x\phipsi\right)F_{tx}+\left(u^2 z' \partial_y\phiz + \frac{\Psi'}{1-\Psi^2}\partial_y\phipsi\right)F_{ty}\right)}
\Bigg)
\end{eqnarray}

Inspecting these cross terms and also the other terms in the action \reef{robbackactF} and in the equations of motion \reef{eoms}, we find in general that generically the coupling of the vectors to the scalars is proportional to the density or magnetic field and the corresponding background scalar. While $\phiz$ is massless and couples only via its derivatives, proportional to $z'$, $\phipsi$ is massive and couples in addition via its magnitude proportional to $\Psi$. Also, generically the interaction terms are suppressed compared to the other terms at asymptotic infinity, and the coupling to $\phiz$ is also suppressed at small $u$, and near the horizon. 

For the individual equations of motion, the form of the interaction terms implies that the transverse component $A_y$ couples proportional to $k \tlb$, so the coupling vanishes both in the case of vanishing magnetic field and in the isotropic case. For the longitudinal field $A_x$, and the equation for the related $\mathcal{A}_x$, the coupling is proportional to $\tlq \tlrho$, so it vanishes in the isotropic case and at vanishing density. This is actually an interesting example of the outcome of the electromagnetic duality, since it relates the properties of the conductivity under the exchange of $x\leftrightarrow y$ and $\tlrho \leftrightarrow \tlb$.

In the rest of this paper, we will continue to ignore the coupling of the vector field to the scalars, so we have to keep in mind that the details of the results for $\sigma_{yy}$ may not be accurate in the rigorous in the case of finite mass or $f$ in the presence of both a magnetic field and finite wavenumber,  $\sigma_{xx}$ in the case of finite $\tlm$ or $f$ at finite density and finite wavenumber and obviously $\sigma_{xy}$ always at finite $\tlm$, finite $f$ and non-vanishing wavenumber $\tlq$. Even in those cases, one can expect however that many of the features that arise from the UV behavior, i.e. from the asymptotic region, ``survive'' and also at least in the case of finite $f$ also qualitative features that arise from the IR in the near-horizon region. Hence we will show all the results uniformly, also in regions of the parameter space that are not fully accurate in the rigorous top-down point of view. In some situations, we also turn on a finite $f$ in order to lower the ``effective temperature'' that we discuss in section \ref{tzero}, and make some qualitative results more apparent.
\section{Analytic results}\mlabel{analcon}
\subsection{Isotropic perturbations with small frequencies}\mlabel{DCsmall}
\subsubsection{DC Limit}\mlabel{DClimit}
One limit of obvious interest is the isotropic DC limit (i.e. $\tom,\tlq \ll\{ 1,f^{-1},b^{-1},\tlrho^{-1}\}$). To analyze this case, we define a new radial coordinate $s$, $\frac{\partial s}{\partial u}  = \left(\sqrt{-G}G^{yy} G^{ u u} \right)^{-1}$ and study the original equations of motion. Now, they just read
\begin{eqnarray}
A_y\!\!: & 0 = & \partial_s^2 A_y \ +\ \sqrt{-G}G^{yy} G^{uu}(\sqrt{-G}G^{yy} G^{ xx} \partial_x^2 + \sqrt{-G}G^{yy} G^{tt}\partial_t^2)A_y \ \nonumber \\
 & & ~~~~~~ + \ \sqrt{-G}G^{yy} G^{uu}\left(\sqrt{-G}G^{tu} G^{xy} \right)' \left(\partial_x A_t - \partial_t A_x \right) \labell{DC_Ay}\\
A_u\!\!: & 0 = & G^{tt} \partial_t \partial_s A_t \ + \ G^{x} \partial_x \partial_s A_x \labell{DC_Au}\\
A_t\!\!: & 0 = & \frac{G^{tt}}{G^{yy}}\partial_s^2 A_t \ + \sqrt{-G}G^{yy} G^{uu}\left(\frac{G^{tt}}{G^{yy}}\right)' \partial_s A_t \ +\\ \nonumber 
 & &\!\!\!\!\!\!\!\!  (-G)G^{yy} G^{uu}G^{tt} G^{xx} \left(\partial_x^2 A_t - \partial_t \partial_x A_x \right) - \sqrt{-G}G^{yy} G^{uu}\left(\sqrt{-G}G^{tu} G^{xy}\right)' \partial_x A_y \labell{DC_At}\\
A_x\!\!: & 0 = & \partial_s^2 A_x\, + \, (-G)G^{yy} G^{uu} G^{tt} G^{xx} \left(\partial_t^2 A_x - \partial_t\partial_x A_t \right) \, \nonumber \\
& & ~~~~~~~~~ + \, \sqrt{-G}G^{yy} G^{uu}\left(\sqrt{-G}G^{tu} G^{xy}\right)' \partial_t A_y \ . \labell{DC_Ax}
\end{eqnarray}
First, we consider the equations in the limit $u\rightarrow 1$. In this limit, we have \\$\frac{\partial s}{\partial u} = \frac{1+\tlb^2+F^2}{\sqrt{(1+f^2) \tlb^2 \left(f^2 + (1-\Psi_0^2)^2\right) +(1+f^2)^2  \left(f^2+\tlrho^2 + (1-\Psi_0^2)^2\right) }}\frac{1}{(1-u^4)} + \order (1) =: \frac{4 c_s}{(1-u^4)}+ \order (1)$, such that $s\sim - c_s \ln(1-u^4)$ or $(1-u^4) \sim e^{-\frac{s}{c_s}}$. Note that in this coordinate, the horizon is located at $s\rightarrow \infty$ and asymptotic infinity is at $s=0$. Now, the equations of motions reduce to 
\begin{eqnarray}
0 &=& \partial_s^2 A_{y,x}\! + (1\!+\!f^2)\frac{\tlb^2\left(f^2+(1\!-\!\Psi_0^2)^2\right) +(1\!+\!f^2)(\tlrho^2+ f^2 +(1\!-\!\Psi_0^2)^2)}{(1 + \tlb^2+f^2)^2}\partial_t^2   A_{y,x} ~~~~~~~~ { }  \\
0 & =& \partial_s A_t ~~, ~~~~ 0\, = \, \partial_s^2 A_t \ ,
\end{eqnarray}
up to order $e^{-\frac{s}{c_s}}$. This has the solution 
\begin{eqnarray}
A_t & = & A_t^0 ~~~, ~~~~ A_{x,y} = A_{x,y}^0 e^{i\nu s} ~ ~; \\ \nonumber 
\nu & = & \tom 
\sqrt{1+f^2}\frac{\sqrt{\tlb^2\left(f^2+(1-\Psi_0^2)^2\right) +(1+f^2)(\tlrho^2+ f^2 +(1-\Psi_0^2)^2)}}{1+\tlb^2+f^2} \ .
\end{eqnarray}
Then, we consider the region of $s \in [0,\order(1)\times c_s]$. To obtain the diagonal and Hall conductivities, we set e.g. $A_x=0$ and $A_t=0$ at $s=0$, and study the linear dependence of $\partial_s A_y$ and $\partial_s A_x$ on $A_y$. Combining the equations \reef{DC_Ay} to \reef{DC_Ax}, we find that $\partial_s^2 A_y \lesssim \order \left((\tom^2+\tlq^2) (1 + f^2+\tlb^2+\tlrho^2)\right) A_y$, such that the change in $\partial_s A_y$ over this region is $\delta(\partial_s A_y) \lesssim \order\left((\tom^2+\tlq^2)c_s (1 + f^2+\tlb^2+\tlrho^2)\right) A_y$. Hence, we have to leading order 
\begin{equation}
\partial_s A_y \  = \ i  \nu A_y \ ,
\end{equation}
which gives us the isotropic DC diagonal conductivity and also allows us to obtain the Hall conductivity.

Looking at $A_x$, we find that to leading order in $\tom, \tlq$, we have 
\begin{eqnarray}
0&=&\partial_u \partial_s A_x + i \tom A_y \partial_u \left(\sqrt{-G}G^{tu} G^{xy}\right) ~~~ \mathrm{or} \\ \nonumber
0&=&\partial_s^2 A_x + i \tom A_y \sqrt{-G}G^{uu} G^{yy} \partial_u \left(\sqrt{-G}G^{tu} G^{xy}\right) \ ,
\end{eqnarray}
and hence $A_x^{0}$ is a consistent solution near the horizon at large $s$.
In the asymptotic region at small $s$ and in the limit of $\tom \ll 1$ the first integral of $u$ can be done analytically, such that 
we obtain 
\begin{equation}
\partial_s A_x \, = \, - i \tom A_y^0 \left[\sqrt{-G}G^{tu} G^{xy} \right]_{u=1}^{u=u} 
\end{equation}
since we should have by consistency $\partial_s A_x \rightarrow 0$ as $s \rightarrow 0$. Hence $A_x|_{u=0} \sim A_y \times \order(\tom) \ll A_y|_{u=0}$
%

Finally, we note that $\frac{\partial s}{\partial u} = 1$ as $u \to 0$, such that we can write down the DC conductivity tensor
\begin{displaymath}\labell{DCtensor}
\tilde{\sigma}^{DC}\!\! = \varepsilon_0 \! \left[\begin{array}{cc}
\frac{\sqrt{\tlb^2\left(f^2+(1-\Psi_0^2)^2\right) +(1+f^2)(\tlrho^2+ f^2 +(1-\Psi_0^2)^2)}}{1+\tlb^2+f^2} &
\frac{\tlb \tlrho }{1+\tlb^2+f^2}\\
- \frac{\tlb \tlrho }{1+\tlb^2+f^2}&\!\!\!\!\!\!\!\!\!\!\!
\frac{\sqrt{\tlb^2\left(f^2+(1-\Psi_0^2)^2\right) +(1+f^2)(\tlrho^2+ f^2 +(1-\Psi_0^2)^2)}}{1+\tlb^2+f^2}
\end{array}\right] \ .
\end{displaymath}
It is straightforward to verify that at $\Psi_0 = 0$, i.e. $M_q=0$, this expression obeys the duality under exchange of $\tilde{\sigma} \leftrightarrow \frac{1}{\varepsilon_0}\tilde{\sigma}^{-1}$, $x\leftrightarrow y$, $\tlrho\leftrightarrow \tlb$.
Obviously, to obtain the full dependence at finite mass, we have to invert $M_q(f,\tlrho,\tlb,\Psi_0)$, in order to obtain $\Psi_0$ as a function of $\{M_q,f,\tlrho,\tlb\}$ -- but this is not possible in closed-form. At small masses $\tlm \ll 1$ and small $f$, $\tlrho$ and $\tlb$, however, one can use $\tlm \sim \Psi_0$, and at large quark mass $\tlm \gg 1$ at finite $\tlrho$ and $f$, we have $\Psi_0 \sim 1$ and the result becomes independent of the quark mass.

If we compare this result with the Drude conductivity \reef{drudecon}, we find that this is qualitatively what one would expect. We can identify $\frac{\tlb}{\sqrt{1+f^2}} = \omega_c \tau$, $\mu \, \Delta \qnn = \frac{\varepsilon_0}{\pi T}  \tlrho $ and $\mu \qnn = \frac{\varepsilon_0 }{\pi T}\frac{\sqrt{\tlb^2\left(f^2+(1-\Psi_0^2)^2\right) +(1+f^2)(\tlrho^2+ f^2 +(1-\Psi_0^2)^2)}}{1+f^2}$. The fact that the expression for $\mu \qnn$ is somewhat complicated is not surprising, since it results from the density of quark-antiquark pairs in thermal equilibrium. What is somewhat surprising is the fact that at finite $f$, there is only very limited dependence on the quark mass $M_q$ -- because one might have thought that (at vanishing $\rho_0$) $\qnn$ is strongly suppressed at large $M_q$ -- but one should not interpret too much into this result. What comes as expected though is the fact that $\mu \qnn \propto \tlrho$ at large $\tlrho$. 

\subsubsection{Small frequencies}\mlabel{smallfreq}
Next, let us try to extract the subleading terms in the conductivity at small frequencies. To do so, we perturb the equations of motion for $A_y$ and $A_x$ \reef{DC_Ay},\reef{DC_Ax} around the DC solution by taking $A_{x,y} \rightarrow  A_{x,y}^{0}e^{i\nu s}+  A_{x,y}^{(1)}$.
The equations of motion for $A^{(1)}_{x,y}$ becomes then at $\tlq  = 0$
\begin{eqnarray}\labell{smfreqeq}
 \partial_s^2 A^{(1)}_{x} &=& \delta\left(\myg^{tu}\myg^{uy}\right)\, \tom^2 A^{(0)}_{x} e^{i\nu s} \ + \ i \tom \myg^{uy} \left(\gperp \right)' A_{y} e^{i\nu s} \ ,\\
 \partial_s^2 A^{(1)}_{y} &=& \delta\left(\myg^{tu}\myg^{uy}\right)\, \tom^2 A^{(0)}_{y} e^{i\nu s} \ - \ i \tom \myg^{uy} \left(\gperp \right)' A_{x} e^{i\nu s} \ ,\\ \nonumber
 \myg^{uy} & := & \sqrt{-G}G^{yy} G^{uu} \, , \, \ \ \myg^{ty}  :=  \sqrt{-G}G^{yy} G^{tt} 
\, , \, \ \ \gperp  :=  \sqrt{-G}G^{xy} G^{tu} \, , \, \ \ \\ \nonumber
\delta(\cdot)  & := & (\cdot) - (\cdot)_{u\rightarrow 1} \ .
\end{eqnarray}
We also write out the symbols $\myg$ in appendix \mref{gform}.
For simplicity, we choose as above $s|_{u=0} = 0$, and we use $u$ as a variable to work with. Before proceeding, we look at the correction to the conductivity:
\begin{equation}
\tilde{\sigma}_{yy} \ = \ - \varepsilon_0 \frac{i}{\omega}\left. \frac{\delta A'_y}{\delta A_y}\right|_{u \rightarrow 0} \nonumber \\
\ \sim \ \tilde{\sigma}_{yy}^{DC} \left(1 +\frac{1}{i \nu} \frac{\partial_s A^{(1)}_y}{A^{(0)}_y} - \frac{A^{(1)}_y}{A^{(0)}_y} \right)_{u\rightarrow 0, \ A^{(0)}_x = 0}  \ .
\end{equation}
Primarily, we are interested in the $\order(\omega^2)$ corrections to the real part of the conductivity, so we need to keep track of $A^{(1)}$ up to $\order (\omega^2)$ and $\partial_u A^{(1)}$ up to $\order(\omega^3)$, which coincides with the accuracy of the first perturbation, as the natural expansion parameter is $\omega^2$.
In the case of $\rho_0  B = 0$, only the diagonal term in the equations of motion contributes, so we find to the relevant order 
\begin{eqnarray}
\partial_s A^{(1)}_y & =& -\tom^2 A^{(0)}_y  \int_u^1 d\tilde{u} (1+ i\nu s(\tilde{u})) \frac{\delta\left(\myg^{tu}\myg^{uy}\right)}{\myg^{uy}} \labell{aone} \\ \labell{aoned}
 A^{(1)}_y & =& \tom^2 A^{(0)}_y  \int^1_u d\hat{u} \frac{1}{\myg^{uy}} \int_{\hat{u}}^1 d\tilde{u}  \frac{\delta\left(\myg^{tu}\myg^{uy}\right)}{\myg^{uy}} \ .
\end{eqnarray}
We remind ourselves that $s(u) = \int_0^u \frac{1}{\myg^{uy}}$, such that $s(u)$ and the first integral of $\frac{\delta\left(\myg^{tu}\myg^{uy}\right)}{\myg^{uy}}$ can be easily computed analytically at $M_q = 0$ and expressed in terms of hypergeometric functions. The second integrals however have to be computed numerically even in the massless case. To demonstrate convergence, we note that any combination of the form  $\sqrt{-G}G^{\mu \nu} G^{\alpha \beta}$ is finite at $u\rightarrow 0$ and at $u \rightarrow 1$ we have $s \propto - \ln (1-u^4)$, $\myg^{uy} \propto (1-u^4) \propto \delta\left(\myg^{tu}\myg^{uy}\right)$. The convergence of $\int \! du \ln (1-u^4)$ is also the reason why we could expand the exponential at sufficiently small $\omega$. 

Including the case of $B\times \rho_0 \neq 0$ is slightly more tedious. First, we compute $A_x$ up to $\order(\omega^2)$ under the condition that $A_x|_{u=0} = 0$. To do so, we first need to integrate $\partial_u (\partial_s A_x)$ in \reef{DC_Ax}. The condition $A_x|_{u=0} = 0$ implies then that $\partial_s A_x|_{u=1} \sim i\nu \int_{u=0}^{u=1} \partial_s A_x$, such that we can, in the limit of small $\tom$, use $\partial_s A_x|_{u=1}=0$. Hence, we get:
\begin{eqnarray}
A_x & = &  i \tom A_y^{(0)}\int_0^u  d\tilde{u} \frac{\delta\left(\gperp\right)}{\myg^{uy}} - \tom \nu A_y^{(0)}\int_0^u d\hat{u} \left(s(\hat{u}) \frac{\delta\left(\gperp\right)}{\myg^{uy}} - \frac{1}{\myg^{uy}} \int_0^{\hat{u}} d\tilde{u} \frac{\delta\left(\gperp\right)}{\myg^{uy}} \right) \nonumber \\
& & ~~~~~ - \nu \tom A_y^{(0)}\int_0^1  d u \frac{\delta\left(\gperp\right)}{\myg^{uy}}\ . \labell{axgot}
\end{eqnarray}
Here, as in the rest of this section, we performed the integration by parts in order to limit the number of consecutive integrals to two integrals.
Now, we can compute the additional contribution to $A^{(1)}_y$, which can still be written in terms of double integrals, with the first one computable analytically at $M_q=0$:
\begin{eqnarray}
\delta \partial_s A^{(1)}_y & = & \tom^2 A_y^{(0)}\left( \left[\gperp \int_0^{\hat{u}}  d\tilde{u} \frac{\delta\left(\gperp\right)}{\myg^{uy}}\right]_{\hat{u}=u}^{1}
- \int_u^{1}  d\tilde{u} \frac{\gperp \delta\left(\gperp\right)}{\myg^{uy}}\right)  \labell{delaoned} \\
& &\!\!\!\!\!\!\!\!\!\!\!\!\!\!\!\!\!\!\!\!\!\!\!\!\!\!\!- i \tom^2 \nu A_y^{(0)}\!\!\left(\left[\gperp
\int_0^{\bar{u}}\!\!\! d\hat{u} \left(s(\hat{u})\left( \frac{\delta\left(\gperp\right)}{\myg^{uy}}+ \int_0^1\!\!\!  d u \frac{\delta\left(\gperp\right)}{\myg^{uy}}\right)  - \frac{1}{\myg^{uy}} \int_0^{\hat{u}}\!\!\! d\tilde{u} \frac{\delta\left(\gperp\right)}{\myg^{uy}} \right)\right]_{\bar{u}=u}^{1} \right. \nonumber \\ \nonumber
& - & \left.
\int_u^{1}\!\!\! d\hat{u} \, \gperp \left(s(\hat{u})\left( \frac{\delta\left(\gperp\right)}{\myg^{uy}} + \int_0^1 \!\!\! d u \frac{\delta\left(\gperp\right)}{\myg^{uy}}\right) - \frac{1}{\myg^{uy}} \int_0^{\hat{u}}\!\!\! d\tilde{u} \, \frac{\delta\left(\gperp\right)}{\myg^{uy}} \right)\right)\\ \labell{delaone}
\delta  A^{(1)}_y & = &  \tom^2 A_y^{(0)}
\int_{u}^1 d\bar{u}\frac{1}{\myg^{uy}}\left( \left[\gperp \int_0^{\hat{u}}  d\tilde{u} \frac{\delta\left(\gperp\right)}{\myg^{uy}}\right]_{\hat{u}=\bar{u}}^{1}
- \int_{\bar{u}}^{1}  d\tilde{u} \frac{\gperp \delta\left(\gperp\right)}{\myg^{uy}}\right)
  .
\end{eqnarray}
The integral for $\delta  A^{(1)}_y$ might seem divergent to the reader, but by close inspection it is apparent that the integrand is finite as $u\rightarrow 1$.
Finally, we can write the correction to the diagonal conductivity which simplifies significantly after some simple algebra: After setting $u=0$, we can eliminate the first term in the first line and all of the second line in \reef{delaoned} and then it turns out that most of the terms in $A_y^{(1)}$ and $\partial_s A_y^{(1)}$ are pairwise equal, such that we obtain
\begin{eqnarray}
\tilde{\sigma}_{yy} & = & \tilde{\sigma}_{yy}^{DC} \left(1 -  \tom^2 \left(
2  \int_0^1 d\, u \,  s(u) \frac{\delta\left(\myg^{ty}\myg^{uy} \right)}{\myg^{uy}} 
\ - \ 
2 \int_0^{1} \! d u \, s(u) \frac{\delta(\gperp)^2}{\myg^{uy}} \right. \right. \nonumber \\
& & \left. \left. 
~~ -2 \int_0^{1} du \,  \frac{\delta(\gperp )}{\myg^{uy}} \int_0^{u} d\tilde{u} \, \frac{\delta\left(\gperp\right)}{\myg^{uy}}
\, - \int_u^{1} du \, \gperp \, s(u) \int_0^1  d u \frac{\delta\left(\gperp\right)}{\myg^{uy}}
 \right)
\right)  \nonumber \\
& + & 
i \varepsilon_0 \tom \left( \int_0^1 d\, u \frac{\delta\left(\myg^{ty}\myg^{uy} \right)}{\myg^{uy}} \ + \ 
\int_0^{1}  d\tilde{u}  \frac{ \delta(\gperp)^2}{\myg^{uy}}\right) \ .
\end{eqnarray}

For completeness, we can also compute the contribution to the Hall conductivity. To do so, we again consider a pertubation that keeps $A_x|_{u=0} = 0$.
The Hall conductivity will then be to order $\omega^2$
\begin{equation}
\tilde{\sigma}_{xy} \ = \ \tilde{\sigma}_{xy}^{DC}\left(1 - \left.\frac{A_y^{(1)}}{A_y^{(0)}}\right|_{u\rightarrow 0} + \left.\frac{\partial_s A_x^{(1)}}{\partial_s A_x^{(0)}}\right|_{u\rightarrow 0}\right) \ .
\end{equation}
We already know $\frac{A_y^{(1)}}{A_y^{(0)}}$, so we only need to compute $\partial_s A_x^{(0)}$.
There will be two contributions, from the diagonal and off-diagonal terms in the equation of motion for $A_x$. Using
as zeroth order the first term 
\begin{equation}
A_x^{(0)} \ = \  i \tom A_y^{(0)}\int_0^1  d\tilde{u} \frac{\delta\left(\gperp\right)}{\myg^{uy}}
\end{equation}
from \reef{axgot},
we find that the contributions from the diagonal term in \reef{smfreqeq} is 
\begin{equation}
\partial_s A_x^{(1)} \ = \- i \tom^3 A_y^{(0)}\int_u^1 d\bar{u}  \frac{\delta\left(\myg^{ty}\myg^{uy} \right)}{\myg^{uy}}
\int_0^{\bar{u}}  d\tilde{u}  \frac{\delta\left(\gperp\right)}{\myg^{uy}} 
%
\, -\, \tom \nu A_y^{(0)} \int_0^1  d\tilde{u} \frac{\delta\left(\gperp\right)}{\myg^{uy}}\labell{delax0} \ ,
\end{equation}
where the second term comes from the oscillatory behavior towards the horizon at large $s$.

In the off-diagonal term, let us first write the $\order (\omega^2)$ term
\begin{eqnarray}
\delta \partial_s A_x^{(1)} & = &  \tom \nu A_y^{(0)}\left(\gperp s(u) - \int_u^1\!\!d\tilde{u} \frac{\gperp}{\myg^{uy}} \right) \\
\delta A_x^{(1)} & = &  \tom \nu A_y^{(0)}\int_0^u  \frac{d\tilde{u}}{\myg^{uy}}\left(\gperp s(\tilde{u}) - \int_{d\tilde{u}}^1\!\!d\bar{u} \frac{\gperp}{\myg^{uy}} \right) \ ,
\end{eqnarray}
giving rise to an $\order (\omega^3)$ term
\begin{equation}\labell{delax1}
\delta_1\partial_s A_x^{(1)} \ = \ 2 i \tom \nu^2 A_y^{(0)}\int_0^1 \!\! d u\frac{\gperp s(\tilde{u})}{\myg^{uy}}  \ .
\end{equation}

Using the fact that $\gperp|_{u=0} =0$, the direct $\order (\omega^3)$ contributions from the cross-term are read off from \reef{aone} and \reef{delaone}:
\begin{eqnarray}
\delta_2 \partial_s A_x^{(1)} & = & i \tom \nu^2 A_y^{(0)} \gperp s(u)^2 + 2 \int_u^1\!\!\! d \tilde{u}\frac{\gperp}s(\tilde{u}){\myg^{uy}} \nonumber \\ \nonumber
& & ~~~ - i \tom^3A_y^{(0)}\left( \gperp \int_u^1\!\!\! d \tilde{u} \frac{1}{\myg^{uy}} \int_{\tilde{u}}^1\!\!\! d \bar{u} \frac{\delta(\myg^{tx}\myg^{uy})}{\myg^{uy}} 
+\int_u^1\!\!\! d \tilde{u} \frac{\gperp}{\myg^{uy}} \int_{\tilde{u}}^1\!\!\! d \bar{u} \frac{\delta(\myg^{tx}\myg^{uy})}{\myg^{uy}} \right. \\ \nonumber
& & ~~~ - \left.\gperp \int_u^1\!\! \frac{d\tilde{u}}{\myg^{uy}} \left(\delta(\gperp )\int^{\tilde{u}}_0\!\!\! d \bar{u} \frac{\delta(\gperp)}{\myg^{uy}} -\int^{\tilde{u}}_0\!\!\! d \bar{u} \frac{\left(\delta(\gperp)\right)^2}{\myg^{uy}} \right) \right. \nonumber \\
& & ~~~ + \, \left.
\int_u^1\!\! d\tilde{u}\frac{\gperp}{\myg^{uy}} \left(\delta(\gperp )\int^{\tilde{u}}_0\!\!\! d \bar{u} \frac{\delta(\gperp)}{\myg^{uy}} -\int^{\tilde{u}}_0\!\!\! d \bar{u} \frac{\left(\delta(\gperp)\right)^2}{\myg^{uy}} \right)
\right)
\end{eqnarray}
We can note that the first term in each line vanishes if we take $u\rightarrow 0$. In this case, also the contribution from \reef{delaone} and the last two ``sub-terms'' all combine into one term, the contributions from \reef{delax1} and the first line are equal, as are the second line and the contribution from \reef{delax0}. Hence, we the result can be written as :
%
\begin{eqnarray}
\tilde{\sigma}_{xy} & = &  \varepsilon_0\frac{\tlrho \tlb}{1+f^2+\tlb^2}
+ 2  i  \nu \varepsilon_0  \int_0^1  d\tilde{u} \frac{\delta\left(\gperp\right)}{\myg^{uy}} \ +
%
 \\ \nonumber 
& & \!\!\!\!\!\!\!\!\!\!\! \tom^2 \varepsilon_0\! \int_0^1 \! d u\frac{\delta (\myg^{tu} \myg^{xy})-\delta(\gperp)}{\myg^{uy}}
\int_0^{u} \! d\bar{u} \frac{\delta(\gperp)}{\myg^{uy}} 
  - 
 \nu^2 \varepsilon_0 \! \int_0^1\! d u\, s(u)\frac{\delta (\gperp)}{\myg^{uy}}
\int_0^{u}\!  d\bar{u} \frac{\delta(\gperp)}{\myg^{uy}}\ .
\end{eqnarray}
%
Let us now look at the fruits of this algebra. 
\DFIGURE{
\includegraphics[width=0.49\textwidth]{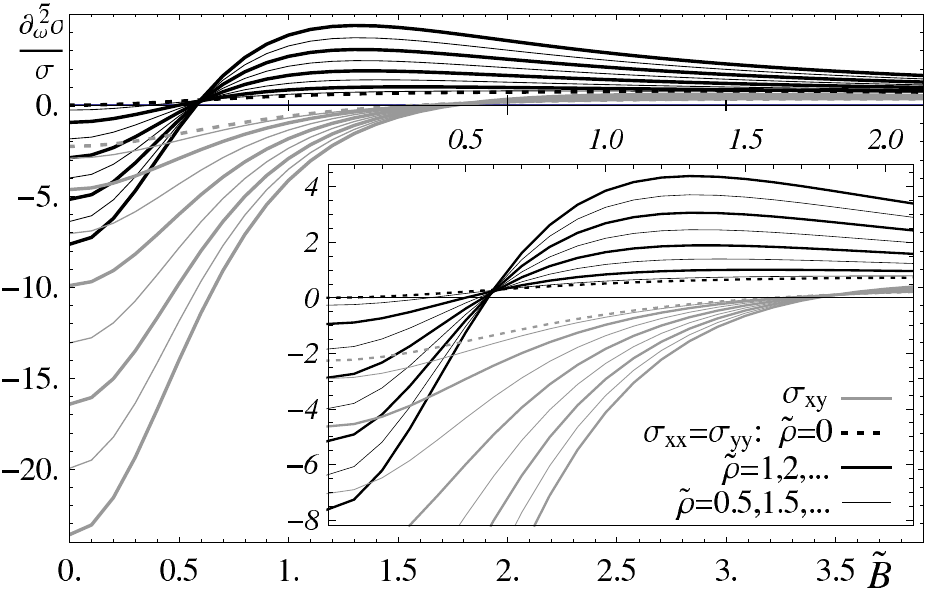}
\includegraphics[width=0.49\textwidth]{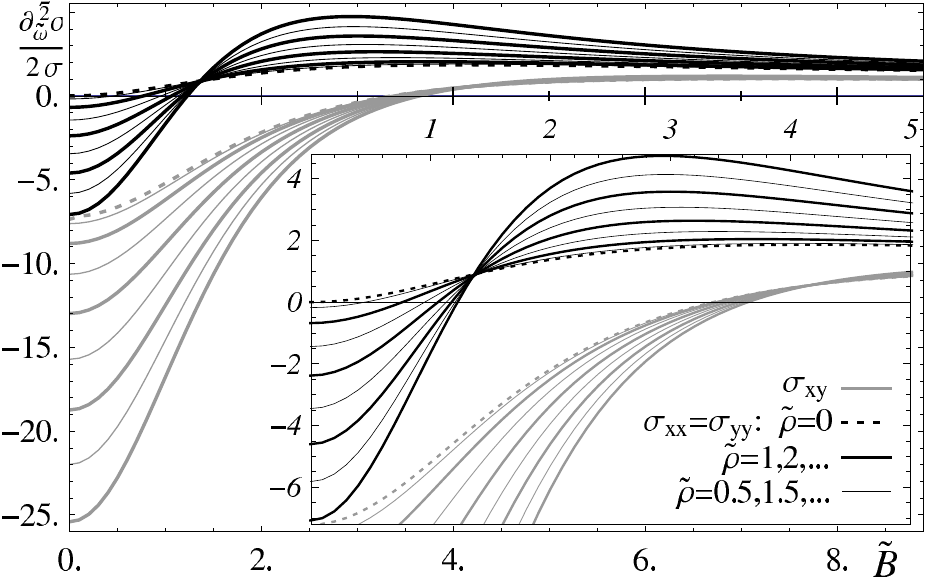}
\caption{The quadratic factor in the small-frequency expansion of the conductivity $\frac{\partial^2_\tom \sigma}{2 \sigma}$ as a function of the magnetic field for various values of the density. Left: $f=0$. Right: $f=2$.}
\mlabel{fruits_mag_f}}

In figure \mref{fruits_mag_f}, we show the behavior of quadratic term in relation to the magnetic field. We see that it behaves approximately as in the Simple Drude conductivity picture outlined in section \mref{metalcon}, with a few differences in the details. Essentially the second order terms for the diagonal and Hall conductivities start off at $B \rightarrow 0$ (In practice $\tlb = 10^{-5}$)  at some negative value that is approximately proportional to the density and represents the relaxation time $\tau^{-2}$ -- where we notice the diverging relaxation time at $\rho_0=0=B$ that gave rise to the constant DC conductivity $\ts(\tom)=i\frac{\Pi(\tom,0)}{\tom}=  \varepsilon_0 = \pi \,D\,T \,
\varepsilon$ or $\sigma(\omega) = D\,\varepsilon$
found in ref. \refcite{baredef} and for a similar system first in ref. \refcite{pavel}. At larger magnetic fields, it rises $\propto B^2$ and becomes positive and then tails off after some maximum. The coefficient for the diagonal conductivity approaches a constant at large magnetic fields and the one of the Hall conductivity tails off approximately $\propto B^{-1}$ whilst the expectation from the Drude picture at constant $\tau$ would have been $\propto B^{-2}$ -- indicating
at least a $B${}-dependence of relaxation time. The most striking feature is the ``node'' at which the correction term becomes independent of the density. In the Drude model, this would be the value of $\omega_c \tau$ at which the quadratic term vanishes.
\DFIGURE{
\includegraphics[width=0.49\textwidth]{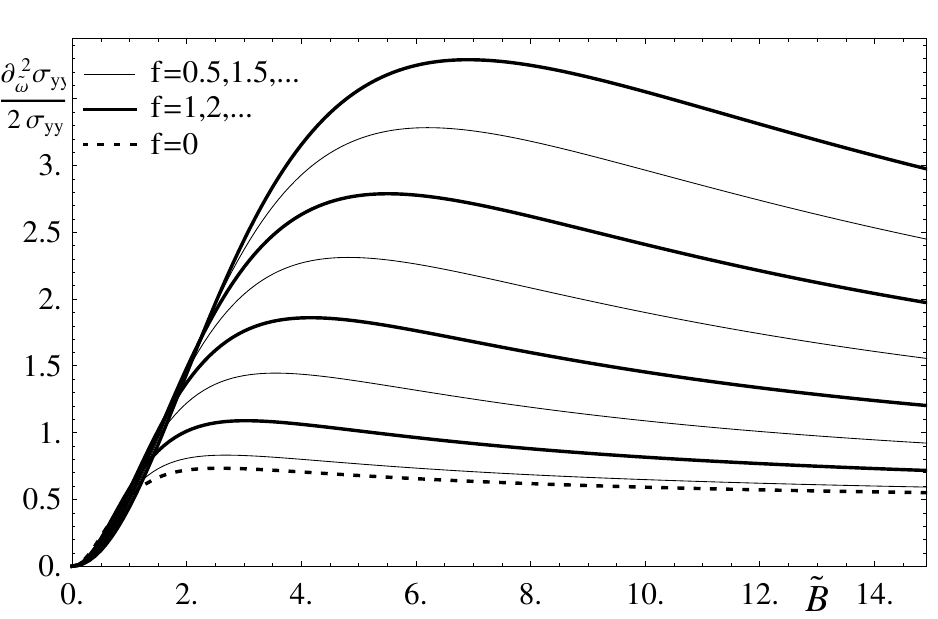}
\includegraphics[width=0.49\textwidth]{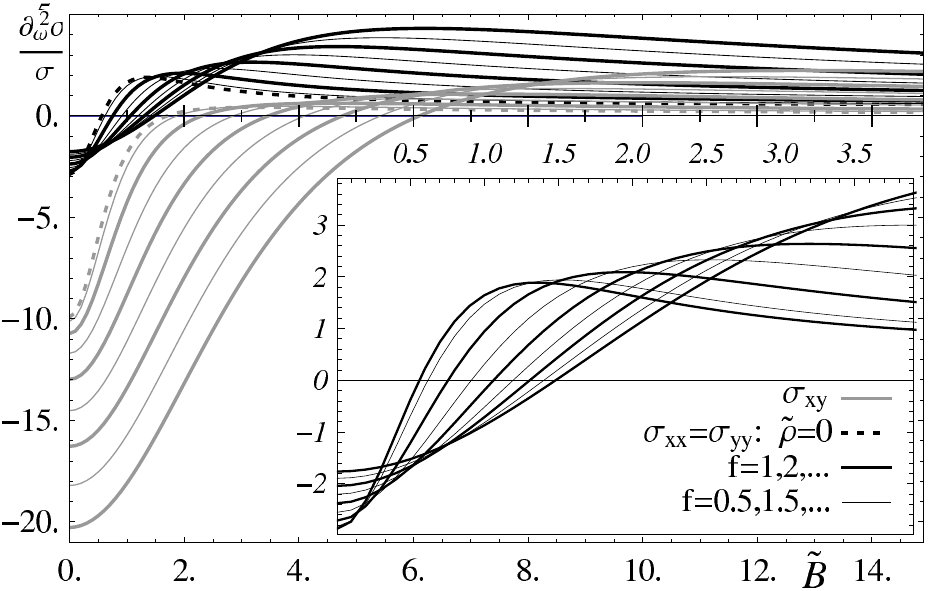}
\caption{The quadratic factor in the small-frequency expansion of the conductivity $\frac{\partial^2_\tom \sigma}{2 \sigma}$ as a function of the magnetic field for various values of $f$. Left: $\tlrho=0$. Right: $\tlrho=2$.}\mlabel{fruits_mag_e}}
Looking in fig. \mref{fruits_mag_e} at how $f$ shifts those curves, we find that at small $\tlb$, in the negative region in the case of finite $\tlrho$, that they are shifted towards $0$ for increasing $f$, implying that the relaxation time increases, whereas for large values of $\tlb$, they are shifted to larger values - which is simply an implication of the observation that $\omega_c \tau \sim \frac{\tlb}{\sqrt{1+f^2}}$.

Going a step further, we can check the generic predictions from section \mref{metalcon}. In fig. \mref{fruits_mag_stuff}, we see that the ratio $\frac{\partial^2_\omega \sigma^\parallel \ \sigma^\perp}{\sigma^\parallel \ \partial^2_\omega \sigma^\perp}$ at $\tlb \rightarrow 0$ approaches precisely the prediction value $\frac{1}{3}$ at large densities with a convergence rate that decreases with increasing $f$ -- even though we are in a completely different, i.e. strong coupling, regime. This also indicates that at large net densities and small $f$, we approach the classical Drude behavior, whereas for small densities or large $f$, we are in a completely different ``phase''. At large magnetic fields, however, this ratio does not become constant and depends significantly on $f$, but at least it seems that always $\frac{\partial^2_\omega \sigma^\parallel }{\sigma^\parallel} > \frac{\partial^2_\omega \sigma^\perp }{\sigma^\perp}$. Furthermore, we can look at the location of the node, $\tlb_{crit}$, which indicates the $\tlb$ value where $\frac{\partial^2_\omega \sigma}{2 \sigma} = 0$, i.e. where the peak turns into a minimum. For the diagonal conductivity, we find that $\frac{\tlb_{crit}^2}{1+f^2} \approx 0.342$ at $f=0$ which converges to $\frac{\tlb_{crit}^2}{1+f^2} \approx 0.397$ at large f. For the Hall conductivity, the value starts at $3.15$, has a maximum of $3.90$ around $f\sim 80$ and then converges to $3.88$. The variation in the ratio of those critical $\tlb^2$ values is even smaller - between $9.18$ and $9.84$. If we were to associate $\tau \omega_c = \frac{\tlb}{\sqrt{1+f^2}}$ as suggested from the DC conductivity in section \mref{DClimit}, this is reasonably close to the values from the Drude model of $\frac{1}{3}( \omega_c \tau)^2$ and $3 (\omega_c \tau)^2$.
%
\DFIGURE{
\includegraphics[width=0.49\textwidth]{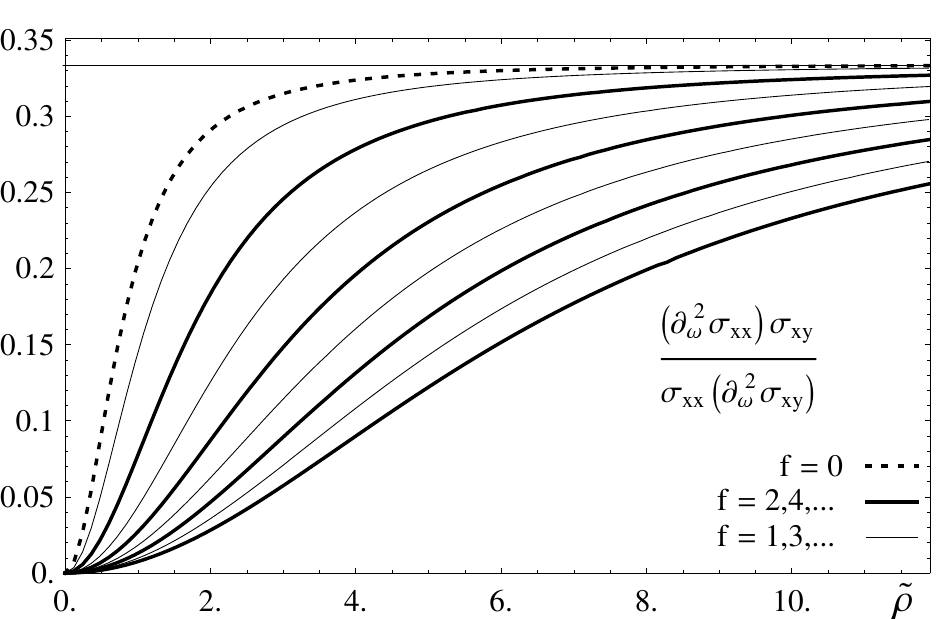}
\includegraphics[width=0.49\textwidth]{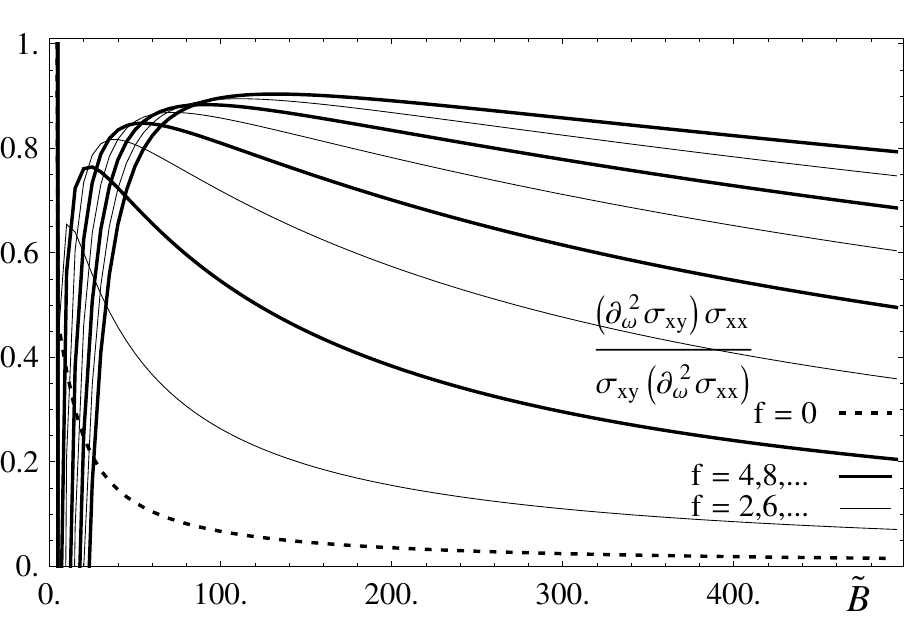}
\caption{The ratio of the quadratic factors  in the small-frequency expansion of the diagonal and Hall conductivities. Left: $\frac{\partial^2_\omega \sigma^\parallel \ \sigma^\perp}{\sigma^\parallel \ \partial^2_\omega \sigma^\perp}$ at $\tlb = 10^{-5}$ as a function of $\tlrho$ for various values of $f$. Right: $\frac{\partial^2_\omega \sigma^\perp \ \sigma^\parallel}{\sigma^\perp \ \partial^2_\omega \sigma^\parallel}$ at large magnetic fields as a function of $\tlb$ for $\tlrho=10$ and various values of $f$. The density $\tlrho$ changes only the behavior at small magnetic fields and leaves the large-$\tlb$ tail unchanged.}\mlabel{fruits_dens}}

Looking in figure \mref{fruits_dens} at the quadratic term of 
$\sigma_{xx}$ at $\tlb =0$, where 
$\frac{\partial^2_\tom \sigma}{2 \sigma}$ 
becomes 
$\frac{\partial^2_\tom \sigma}{2 \sigma} = -\tau^{-2}$,
we find in fig. \mref{fruits_dens}, that 
$\tau^{-2}$ 
is approximately proportional to the density, with a coefficient of
$\tau^{-2} \approx 2.6 \tlrho$. From another perspective, this means that the relaxation time is approximately proportional to the mean distance between ``quarks'', $\tau \propto \rho_0^{-1/2}$, but not the naive geometric mean free path in a system of weakly coupled particles. 
\HFIGURE{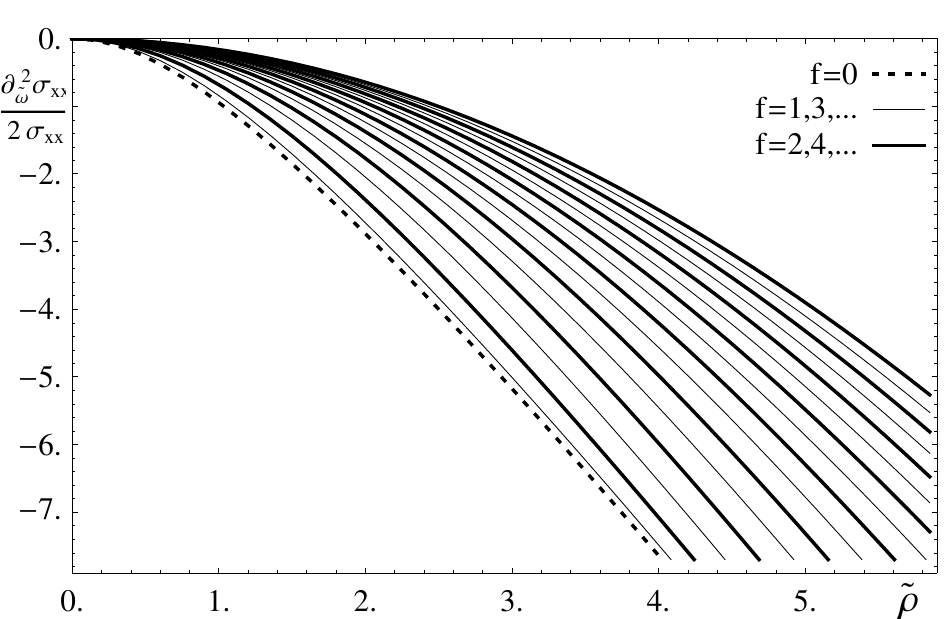}{The quadratic factor in the small-frequency expansion of the conductivity $\frac{\partial^2_\tom \sigma}{2 \sigma}$ as a function of the density for various values of $f$.}{fruits_mag_stuff}
The proportionality coefficient is approximately $\tau \sim 2.6 \pi \sqrt{\frac{ 2\varepsilon_0}{\rho_0}}$. The $f${}-dependence is not surprising, as increasing $f$ appears to increase the relaxation time, which is consistent with a decreasing effective temperature that was a recurring theme in ref. \refcite{baredef}. It is interesting though that at large densities, the effect of $f$ is only to shift the curves in fig. \mref{fruits_dens} and leaves the proportionality factor constant.

\DFIGURE{
\includegraphics[width=0.49\textwidth]{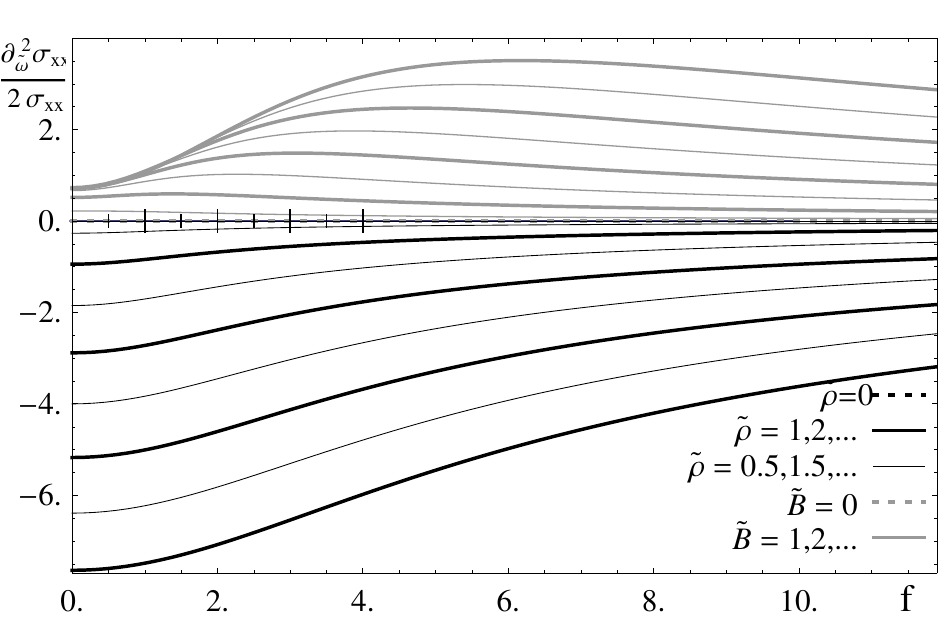}
\includegraphics[width=0.49\textwidth]{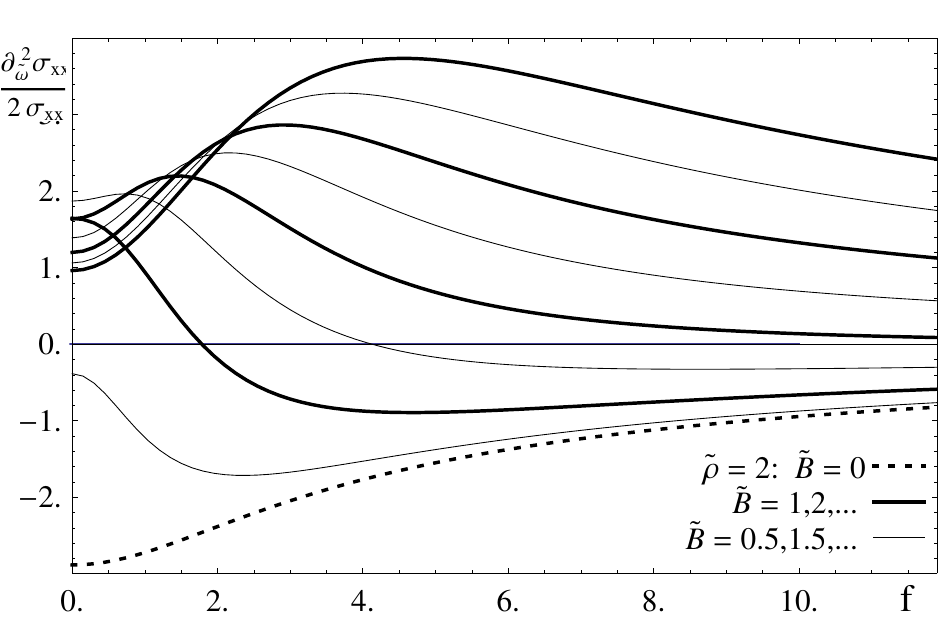}\\
\includegraphics[width=0.49\textwidth]{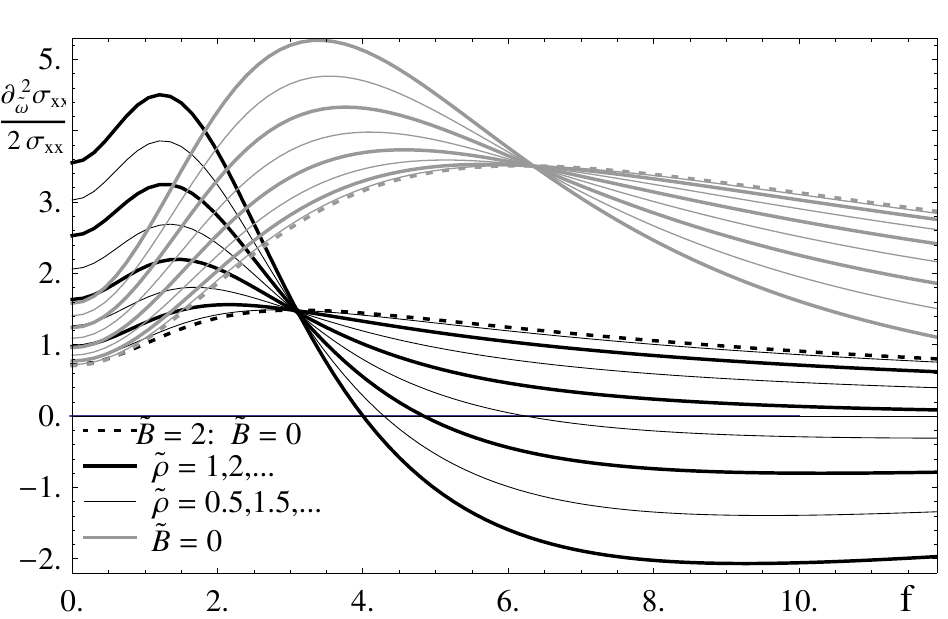}
\includegraphics[width=0.49\textwidth]{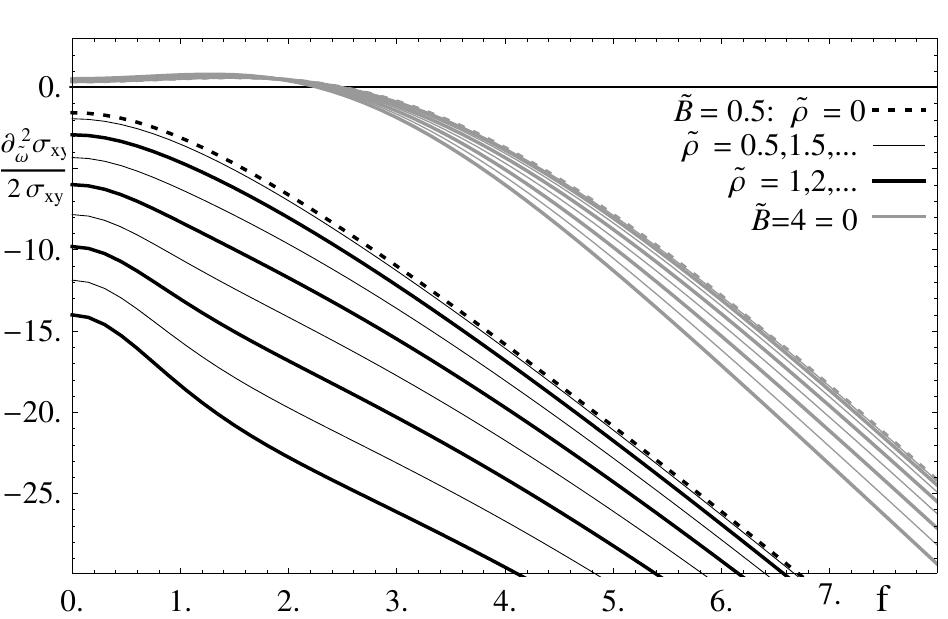}
\caption{The quadratic factor in the small-frequency expansion of the conductivity $\frac{\partial^2_\tom \sigma}{2 \sigma}$ as a function of $f$. Top left: The coefficient of the diagonal conductivity for various values of the density and the magnetic field, right: For various values of the magnetic field at $\tlrho =1 $. Bottom left: The coefficient of the diagonal conductivity for various values of the density at $\tlb \in \{2,4 \}$, right: The coefficient of the Hall conductivity for various values of the density at $\tlb \in \{0.5,4 \}$.}\mlabel{fruits_f}}
Looking at the coefficients as a function of $f$ in fig. \mref{fruits_f} shows our observations from a different perspective. Essentially, the effect of $f$ is to increase the relaxation time, and to decrease $\omega_c \tau$ at fixed $\tlb$. The most striking feature is the observation that we had above, that the coefficient in the Hall conductivity is proportional to $ f$ in regimes where it is negative, i.e. the ``Hall peak'' becomes narrower.

\DFIGURE{
\includegraphics[width=0.49\textwidth]{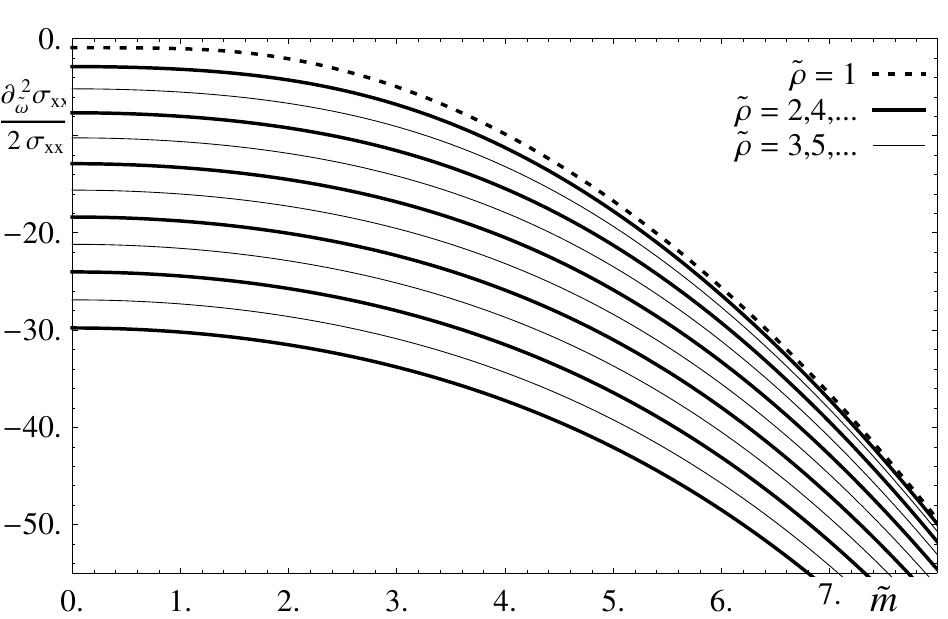}
\includegraphics[width=0.49\textwidth]{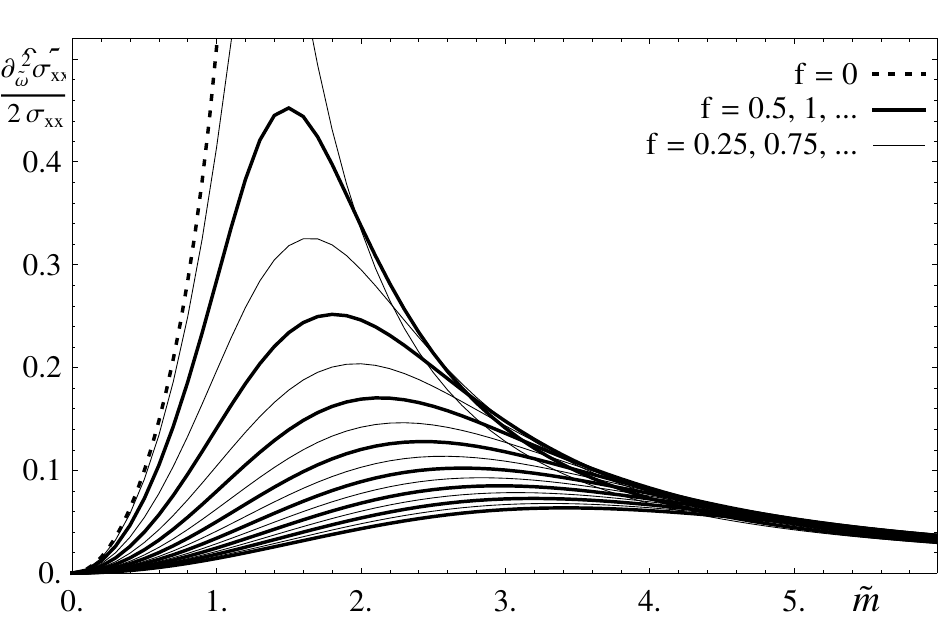}\\
\includegraphics[width=0.49\textwidth]{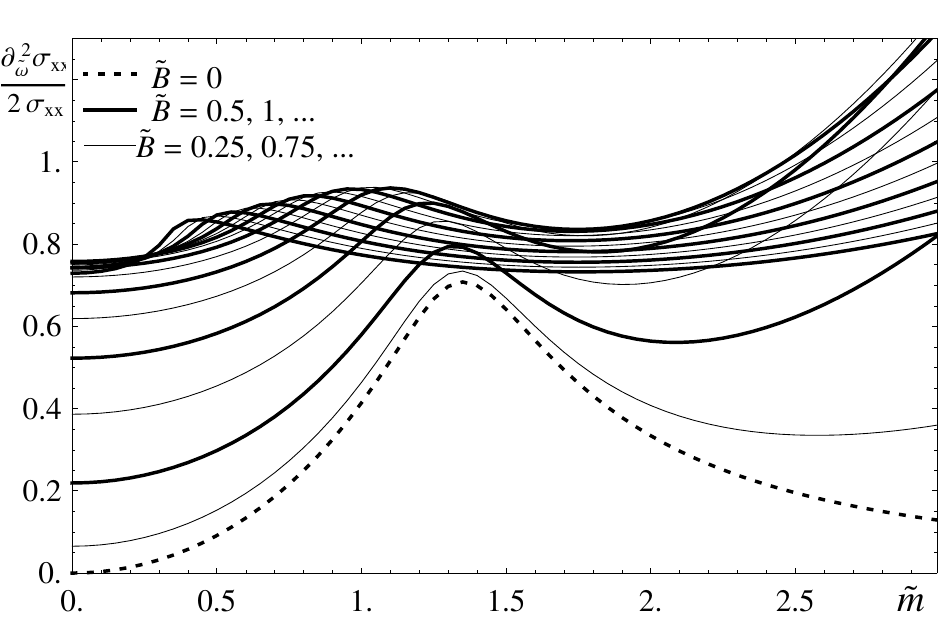}
\includegraphics[width=0.49\textwidth]{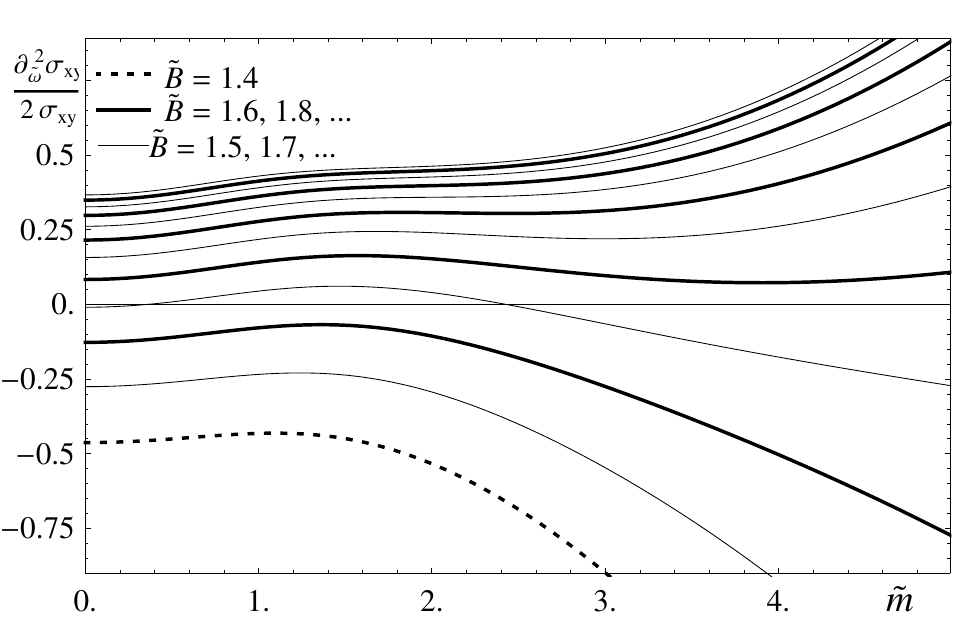}
\caption{The quadratic factor in the small-frequency expansion of the conductivity $\frac{\partial^2_\tom \sigma}{2 \sigma}$ as a function of $\tlm$. Top left: The coefficient of the diagonal conductivity for various values of the density, right: For various values of $f$. Bottom left: The coefficient of the diagonal conductivity for various values of the $\tlb$ at $f=0.25$, right: The coefficient of the Hall conductivity for various values of the magnetic field at $\trho 2$.}\mlabel{fruits_mass}}
Finally, we can look at the mass dependence in fig. \mref{fruits_mass}. The biggest surprise from the Drude picture view is the quadratic dependence of second the expansion coefficient on the mass. This indicates $\tau \propto \tlm^{-1}$, which is somewhat counterintuitive since one would have thought that the relaxation time increases with increasing mass. If one considers the Drude peak however to be a quasiparticle resonance, this is what one does classically expect since it means that the quasi particle becomes more stable at larger quark mass due to slower thermal motion and hence reduced collision rates. At vanishing density and different values of $f$, the result is also in contradiction with the free particle picture, since the DC conductivity is in a minimum at finite mass. There is an interesting maximum in the coefficient, which corresponds as $f\rightarrow 0$ to the critical quark mass of the phase transition discussed in ref. \refcite{thermpaper}. Hence, it occurs at the transition from the small-mass to the large-mass regime. This feature is even more apparent when plotting the coefficient against $\tlb$ for a small value of $f=0.25$, where there is a small maximum around the critical mass. Looking at the Hall conductivity the regimes in $\tlb$ in which there is a Drude peak and in which there is a magnetoresistance minimum behave approximately like the pure Drude peak and magnetoresistance effects. It is an interesting curiosity, that the transition between those regimes receives a very small mass dependence.

\subsection{Large Temperatures: Diffusion limit}\mlabel{diffanal}
In the diffusion limit, i.e. at $\tom \ll \tlq \ll 1$, we expect to be able to predict the transport properties from the diffusion behavior, i.e. from the diffusion constant $D$ and the susceptibility $\varepsilon$  because we expect the ``mean free path'' to be set by the temperature scale.

The diffusion constant was computed e.g. in ref. \refcite{Kovtun:2003wp} by studying the equations of motion of the gauge field in the gravity side to obtain Fick's Law,
\begin{equation}
\vec{j}(t,\vec{x}) \ = \ - D \, \vec{\nabla} j_0(t,\vec{x})  \ ,
\end{equation}
on the field theory side.
The derivation in ref. \refcite{Kovtun:2003wp} is very instructive and can be followed also in our case in the presence of background fields. The expression for the diffusion constant is then slightly modified and yields 
\begin{equation}
D \ = \ \frac{1}{\pi T}\left(-G \sqrt{-G^{tt} G^{uu} }G^{xx}\right)_{\! u\rightarrow 1} \, \int_0^1 \frac{d \, u}{\sqrt{-G} G^{tt} G^{uu}} \ ,
\end{equation}
where we keep in mind that in our notation $\sqrt{-G}$ contains a factor of the $u$ dependent coupling $g^{-2}_{eff}(u) =\sqrt{f^2 + (1-\Psi(u)^2)^2}$.

At $M_q = 0$, this can be evaluated analytically and expressed in terms of hypergeometric functions as:
\begin{eqnarray}
 D & = & \frac{(1\!+\!f^2)\sqrt{1+\! f^2\!+\tlb^2\!+\tlrho^2}}{\pi T\,(1+f^2+\tlb^2)}\!\int_0^1\! d u  \frac{1+f^2+b^2 u^4}{\!\left(1\!+\!f^2\! +\!(\tlb^2\! + \tlrho^2)u^4 \right)\! \sqrt{1+\!(f^2\!+\tlb^2\!+\tlrho^2)u^4\!}} \nonumber \\
& = & \frac{(1+f^2)\sqrt{1+ f^2+\tlb^2+\tlrho^2}}{\pi T\,(1+f^2+\tlb^2)} \times \\ \nonumber
& &\!\!\!\!\! \left(\!{}_2 F_1\!\left(\textstyle{\frac{1}{4}, \frac{1}{2}; \frac{5}{4};} -(f^2\!+\tlb^2\!+\tlrho^2)\right)   - \tlrho^2 \frac{ F_1 \left(\frac{5}{4};\frac{1}{2} ,1 ;\frac{9}{4} ; -(f^2\!+\tlb^2\!+\tlrho^2),-\frac{\tlrho^2+\tlb^2}{1+f^2} \right)}{5(1+f^2)}\right) .
\end{eqnarray}
Here ${}_2F_1$ is the Gauss hypergeometric function, which is asymptotically in our case $\sim \frac{\Gamma(1/4)^2}{4 \sqrt{\pi}(f^2+\tlb^2+\tlrho^2)^{1/4}}$ and $F_1$ is an Appell hypergeometric function that is here at $f=0$ asymptotically $\sim \frac{9\Gamma(5/4)^2}{5 \sqrt{\pi}(\tlb^2+\tlrho^2)^{5/4}}$. At $F\neq 0$, the decay will be with a smaller, non-rational, power.
\DFIGURE{
\includegraphics[width=0.49\textwidth]{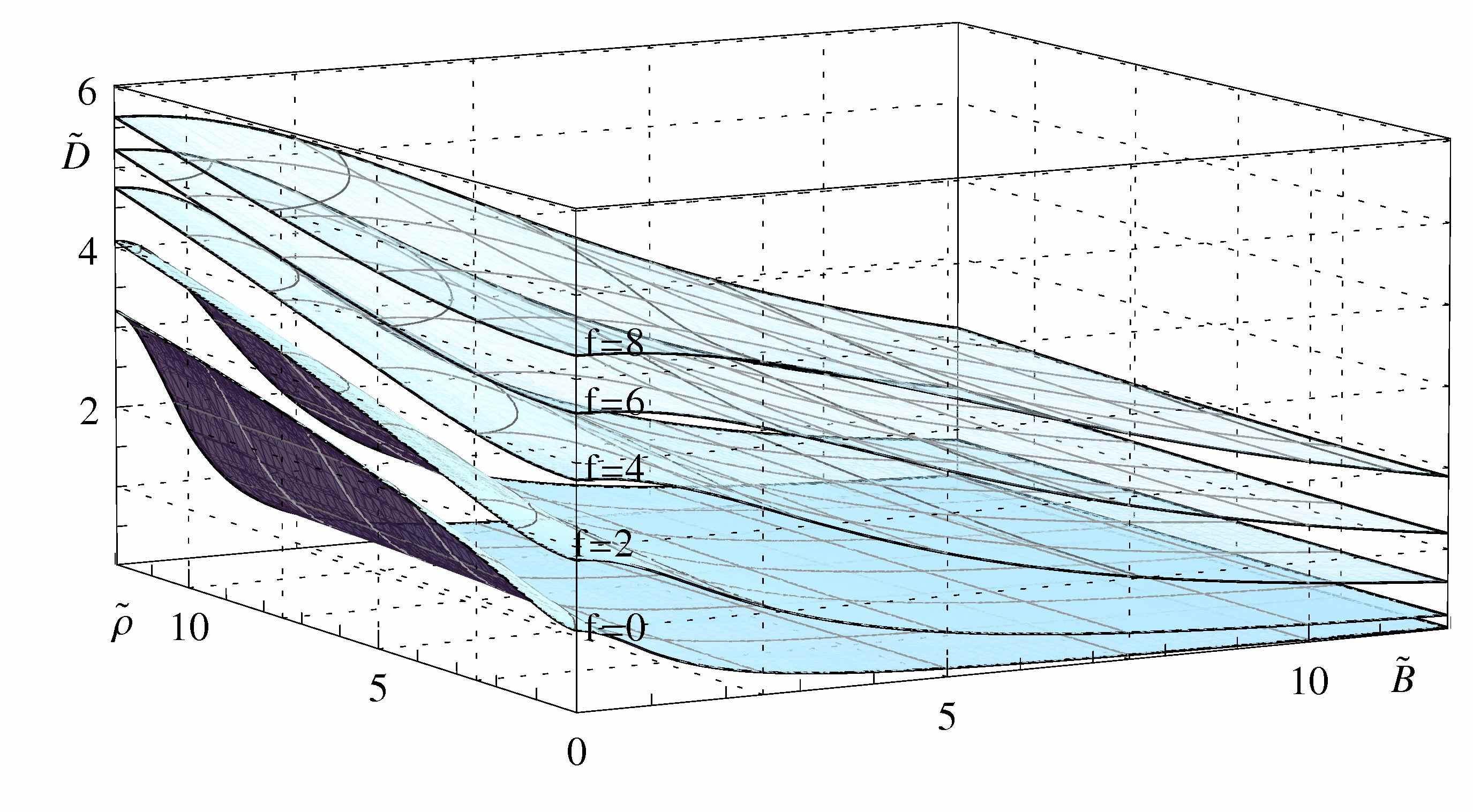}
\includegraphics[width=0.49\textwidth]{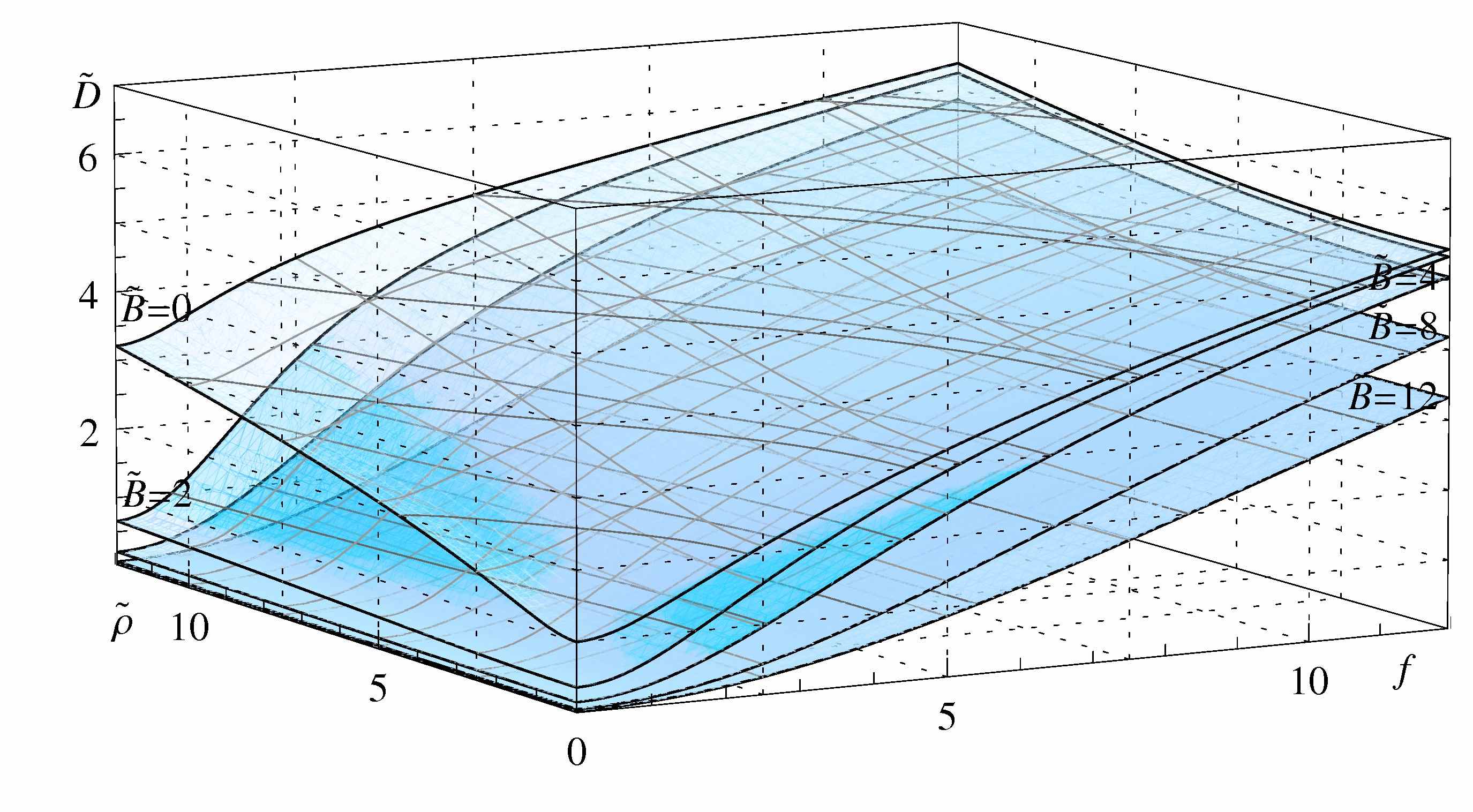}
\caption{The diffusion constant $\pi T \, D$ as a function of the magnetic field and density for different values of $f$ (left) and as a function of $f$ and the density for different values of the magnetic field (right).}
\mlabel{diffcplot}}
In fig. \mref{diffcplot}, we see that the diffusion constant is for small $f$ approximately proportional to $\sqrt{\tlrho}$ whereas for large $f$ the dependence is approximately linear. This may be due to the strong coupling because the usual classical geometric result for the diffusion constant is proportional to the mean free path -- which one expects to be inversely related to the density -- and the mean free path should dominated by the baryon density at large baryon density, at least in weak-coupling intuition. However if we are for example in a superfluid, this intuition does obviously not apply anymore.

At small $f$, the diffusion constant decays inversely proportional to the magnetic field, which represents the fact that charged particles in magnetic fields receive extra ``drag'' and become localized. At larger $f$, this decay slows down. Looking at the $f$-dependence, we see that the diffusion constant is approximately proportional to $f$, with an asymptotic slope that is independent from $\tlb$. This contrasts to the dependence on $\tlrho$, which disappears at large $\tlb$.

Obtaining the permittivity is similarly straightforward.
By definition (see e.g. \refcite{Chaikin1,Landau1})
\begin{equation}
\varepsilon \ = \ \lim_{\omega,q \rightarrow 0} C_{tt} \ , 
\end{equation}
where it is understood that the limit $\omega \rightarrow 0$ is to be taken first.
Taking the limit $\omega, q \rightarrow 0$ of the equation of motion for $A_t$, \reef{fteq}, gives us a Poisson equation
\begin{equation}\labell{diffpoisson}
\left(\sqrt{-G}G^{tt}G^{uu} A'_t(u)\right)' \ = \ 0 \ .
\end{equation}
We note that this equation does not yield an appropriate infalling wave behavior near the horizon, but it is easy to see from the full equations for $A_t$ and $A_x$ that for very small but finite $\omega \ll q \ll 1$, the behavior will be appropriately resolved near the horizon. Near the horizon, $A_t$ and $A_x$ are strongly coupled, with $A_t \sim \frac{\omega}{q} A_{x}$, and $A_x$ follows an oscillatory behavior.
To solve for $A_t$, we then simply integrate \reef{diffpoisson} with $A_t = 0$ as a boundary condition at $u \rightarrow 1$, which gives us readily the permittivity
\begin{equation}
\varepsilon  \ = \  \varepsilon_0 \left.\frac{A_t'}{A_t}\right|_{u \rightarrow 1} \ = \ \varepsilon_0 \frac{(\sqrt{-G}G^{tt}G^{uu})_{u=0}}{\int_0^1 d\, u \, \sqrt{-G}G^{tt}G^{uu}} \ =: \ \varepsilon_0 \varepsilon_r \ ,
\end{equation}
where $(\sqrt{-G}G^{tt}G^{uu})_{u=0} = 1$.
We can now see immediately, that the isotropic DC conductivity in section \mref{DClimit} is given by the diffusion result $\sigma_{yy} = \sigma_{xx} = \varepsilon D$ and in the DC limit there is no contribution from other modes, as expected. However, in contrast to the remarkable result in the conformal case in refs. \refcite{baredef,pavel} where the conductivity was at all frequencies determined precisely by the diffusion behavior, the diffusion behavior is now only valid at small frequencies and receives corrections as we move away from $\omega = 0$ as outlined in section \mref{smallfreq}.
Since the integral is the same as the one for the diffusion constant, we find that for $M_q = 0$, we obtain the relative permittivity
\begin{equation}
\varepsilon_r^{-1}\!  = \! {}_2 F_1\left(\textstyle{\frac{1}{4}, \frac{1}{2}; \frac{5}{4};} -(f^2+\tlb^2+\tlrho^2)\right)   - \tlrho^2 \frac{ F_1 \left(\frac{5}{4};\frac{1}{2} ,1 ;\frac{9}{4} ; -(f^2+\tlb^2+\tlrho^2),-\frac{\tlrho^2+\tlb^2}{1+f^2} \right)}{5(1+f^2)} \ .
\end{equation}

\DFIGURE{
\includegraphics[width=0.49\textwidth]{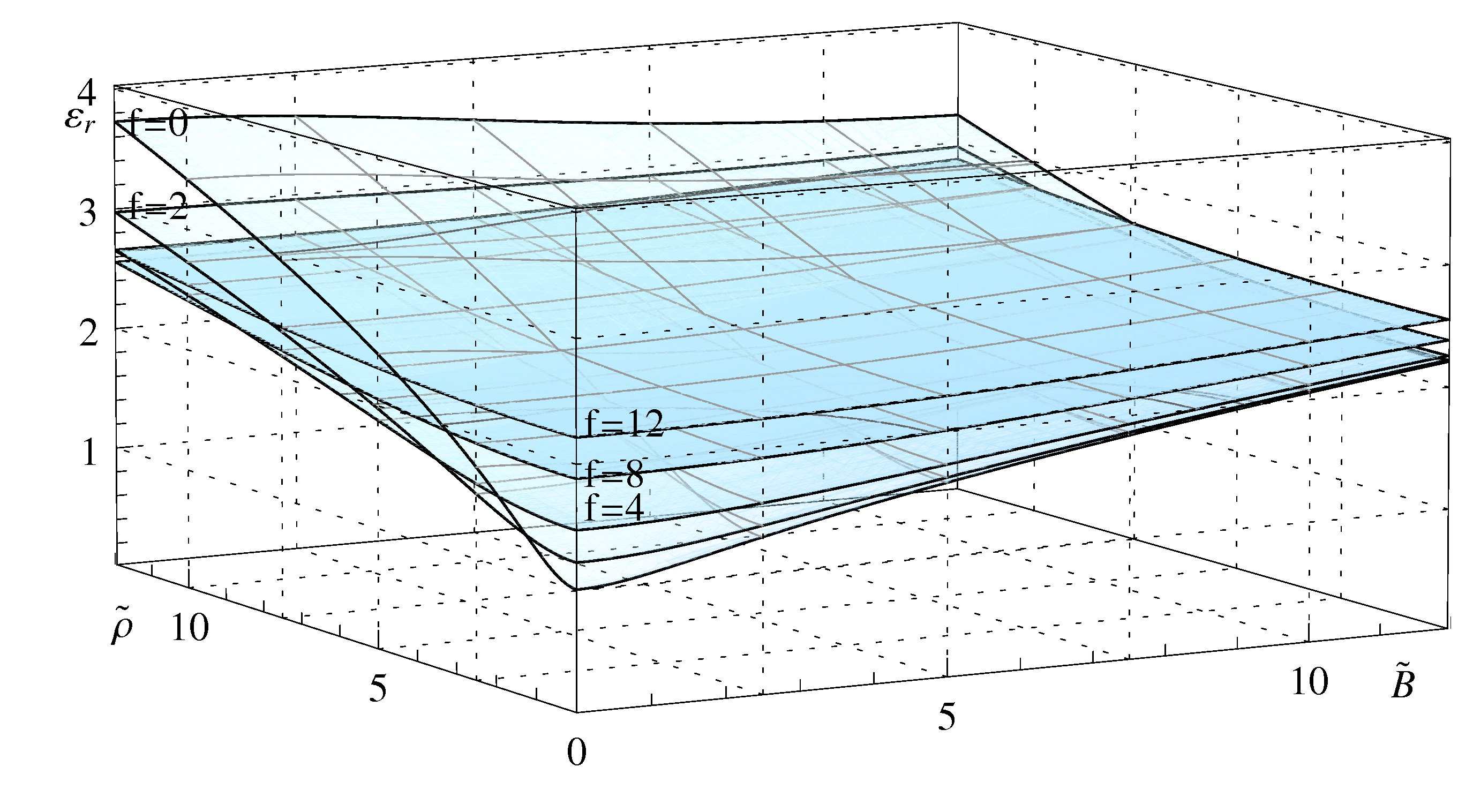}
\includegraphics[width=0.49\textwidth]{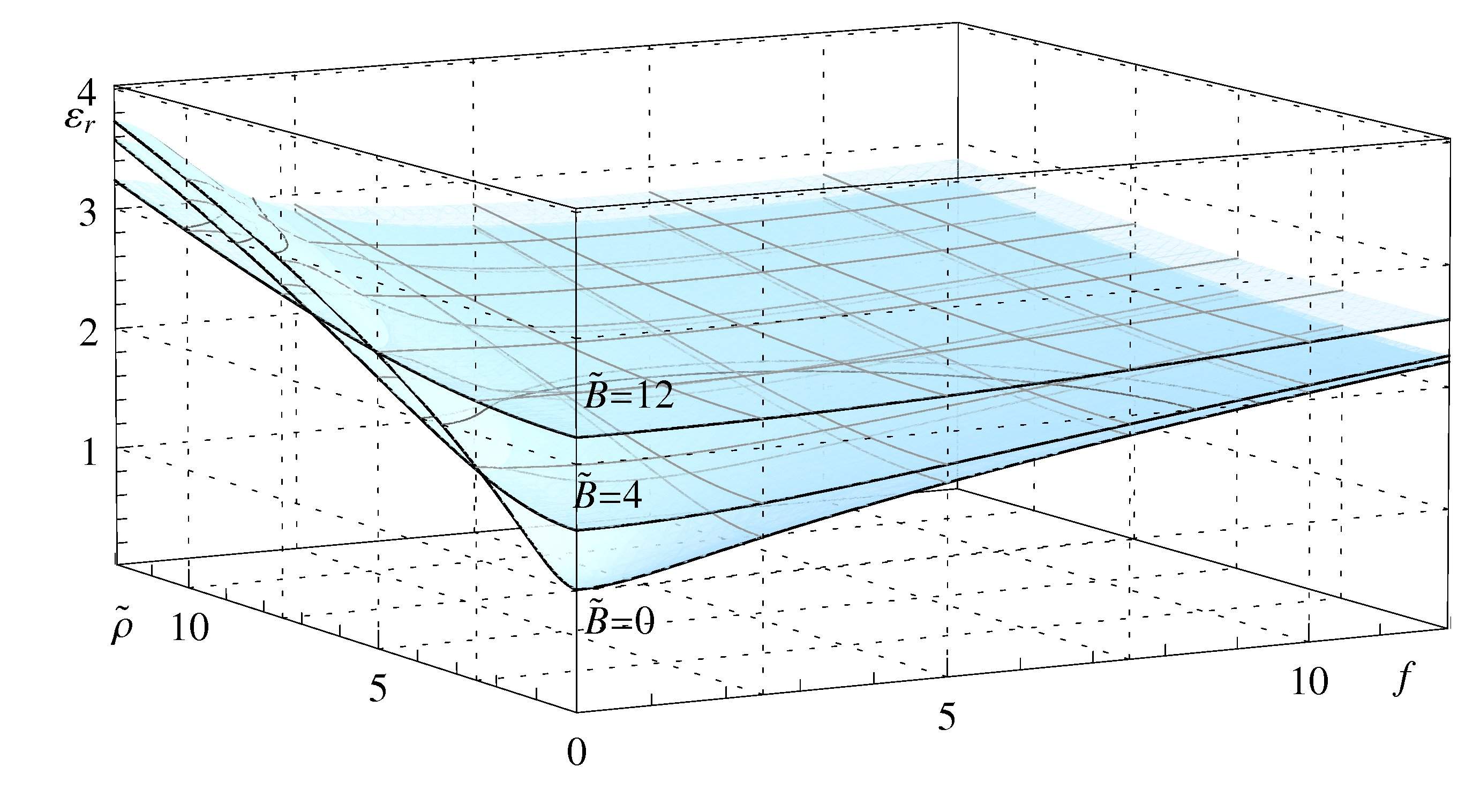}
\caption{The relative permittivity $\varepsilon_r$ as a function of the magnetic field and density for different values of $f$ (left) and as a function of $f$ and the density for different values of the magnetic field (right).}
\mlabel{epsrplot}}
In figure \mref{epsrplot}, we see the interesting fact that at large $f$, the relative permittivity becomes approximately constant. While one does not generically expect any specific dependence on the magnetic field, one would expect in a simple solid state model $\varepsilon_r \propto \qnn$ and hence  $\varepsilon_r \propto \rho_0$ at large $\tlrho$, which is realized here at large $f$, but at small $f$ it is proportional to $\sqrt{\rho_0}$ at large values of $\tlrho$. 

If we compare the results of this section to the Drude model reviewed in section \mref{metalcon}, we can identify
from the Einstein relation \reef{einsteindrude}
\begin{equation}
\mu \, = \, \frac{D}{T} \ \ \mathrm{and} \ \ \ 
\qnn \, = \, \pi T^2 \varepsilon_0 \varepsilon_r \ .
\end{equation}

\subsection{$T\rightarrow 0$ limit}\label{tzero}
Next, let us look at the low temperature limit of $\tlq, \tom \gg 1$. Here, we are interested in the  equations near $u=0$. The equations for $A_y$ and $\mathcal{A}_x$ \reef{fyyeq},\reef{fxxeq} are identical in this limit, and become
\begin{equation}
A_y'' \ + \ \left(\tom^2 - \tlq^2 \right) A_y \ = \ 0
\end{equation}
as in the ``conformal limit'' in ref. \refcite{baredef}, up to order $\tlm^2 \frac{u^2}{\tom^2}$ or $(e^2,b^2)\frac{u^4}{\tom^2}$. The appropriate solution gives us the diagonal conductivity $\tilde{\sigma}_{yy} = \varepsilon_0 \sqrt{1-\tlq^2/\tom^2}$ and $\tilde{\sigma}_{xx} = \frac{\varepsilon_0}{\sqrt{1-\tlq^2/\tom^2}}$. 
\subsubsection{Exponentially suppressed regime, $\tlq^2 \gg 1 \gg \tom^2$, at small backgrounds, $|\tlq| \gg |\tlrho|,|\tlb|$}\mlabel{conefftf}
To study the low temperature limit more in detail, we start with the regime  $|\tlrho|,|\tlb|\ll |\tlq|$ which is similar to the approximation in ref. \refcite{baredef}. 

It is straightforward to see that the dominating term in the solution at finite $h=1-u^2$ will still be $A_{y},\mathcal{A}_x \sim A^0_{y},\mathcal{A}_x e^{\pm \tlq \int \sqrt{-G^{xx}/G^{uu}}}$. Using this to estimate the contribution of the cross-terms in \reef{fyyeq},\reef{fxxeq}, we find that they are suppressed by a factor of $\tlq^{-1}$ with respect to the dominant diagonal terms. In the near-horizon regime at $\omega^2/q^2 \ll h \ll 1$, they are suppressed by a factor of $h/\tlq$, and in the regime $h\ll \omega^2/q^2 \ll 1$, they are suppressed by $h^2 \tlq/\tom^2$. Hence, we can proceed as follows: First we will obtain the diagonal conductivity $\tilde{\sigma}_{yy}$ ($\tilde{\sigma}_{xx}$ follows similarly) by solving the homogeneous part, because the contribution from the cross-terms to the diagonal conductivity will be suppressed by the order of the square of the suppression of the cross-terms and can hence be safely ignored. Then we will compute the Hall conductivity from the inhomogeneous part.

Again, let us use the Ansatz $A_y  = A_y^0 e^{\int^u \zeta} $, which gives us 
\begin{equation} \labell{zetaeq}
\zeta^2\, +\, \zeta' \, + \, \frac{\left(\sqrt{-G}G^{uu} G^{yy}\right)'}{\sqrt{-G}G^{uu }G^{yy}}\zeta \, + \, \frac{\left(\left(\sqrt{-G}\, G^{tu}G^{xy}\right)'\right)^2}{-G\, G^{tt} G^{uu} G^{xx} G^{yy}} \, + \, 
\frac{G^{xx}}{G^{uu}}\left(\frac{G^{tt}}{G^{xx}} \tom^2 \, - \, \tlq^2 \right) \ = \ 0 \ ,
\end{equation}
with the approximate result at $ h \gg \omega^2/q^2$ up to $\order (1)$, $\zeta  = - \frac{\left(\sqrt{-G}G^{uu} G^{yy}\right)'}{2 \sqrt{-G}G^{uu }G^{yy}}\pm \frac{\zeta_0'}{2\zeta_0}\pm \zeta_0$, $\zeta_0 :=  \tlq \sqrt{\frac{G^{xx}}{G^{uu}}}$, where we pick the negative sign corresponding to a solution that decays towards the horizon.
Next, we take $\zeta = - \zeta_0- \frac{\left(\sqrt{-G}G^{uu} G^{yy}\right)'}{2 \sqrt{-G}G^{uu }G^{yy}}- \frac{\zeta_0'}{2\zeta_0} + \epsilon$ and  gather the remaining terms up to linear order in $\epsilon$
\begin{eqnarray}
0 & = & \epsilon' \, - \, \epsilon \left(\frac{\zeta_0'}{\zeta_0}  \, + \, 2\zeta_0   \right) \, + \, \frac{\left(\left(\sqrt{g}\, G^{tu}G^{xy}\right)'\right)^2}{-G\, G^{tt} G^{uu} G^{xx} G^{yy}} \nonumber \\
& & + \, \left(\frac{\left(\sqrt{-G}G^{uu} G^{yy}\right)'}{2 \sqrt{-G}G^{uu }G^{yy}}+ \frac{\zeta_0'}{2\zeta_0} \right) \frac{\zeta_0'}{2 \zeta_0}  \, - \, \left(\frac{\left(\sqrt{-G}G^{uu} G^{yy}\right)'}{2 \sqrt{-G}G^{uu }G^{yy}}+ \frac{\zeta_0'}{2\zeta_0} \right)' \, + \, \frac{G^{tt}}{G^{uu}} \tom^2 \nonumber \\
& =: &  \epsilon' \, - \, \epsilon \alpha(u) \, - \, \beta(u) \ . \labell{epseq}
\end{eqnarray}
The general solution to this equation is 
\begin{equation}\labell{epssol}
\epsilon \, = \, e^{\int_0^u d\bar{u}\alpha(\bar{u})} \left(\epsilon_0 + \int_0^u d\tilde{u}e^{-\int_0^{\tilde{u}} d\bar{u}\alpha(\bar{u})} \beta(\tilde{u}) \right) \ .
\end{equation}
The second part is a small contribution $\in \mathbb{R}$ that is at most of order $\tlq^{-2}$, so we are only interested in the first part that evaluates to $\epsilon \, = \, \epsilon_0 \zeta_0 e^{2 \int_0^u d\bar{u}\zeta_0(\bar{u})}$, or $\epsilon \, = \, \epsilon_H \zeta_0 e^{-2 \int_u^1 d\bar{u}\zeta_0(\bar{u})}$.  $\epsilon_H$ will be fixed in the region $\tom^2/\tlq^2 \ll h \ll 1$, where there is an overlap between the asymptotic and near horizon solutions.

At $h \ll 1$, the equation becomes:
\begin{equation}
-4\partial_h \zeta\, +\, \zeta^2 \, - \, \frac{4}{h}\zeta \, + \, \frac{\tom^2}{h^2} \, - \, \frac{1}{1 + \frac{\tlb^2}{1+f^2} + \frac{\tlrho^2}{(1-\Psi_0^2)^2 + f^2}}\frac{\tlq^2}{h} \ = \ 0 \ .
\end{equation}
As in ref. \refcite{baredef}, this can be solved analytically in terms of hypergeometric functions and then be expanded for $\frac{\tlq^2}{1 + \frac{\tlb^2}{1+f^2} + \frac{\tlrho^2}{(1-\Psi_0^2)^2 + f^2}} h \gg 1$, giving us in the overlap region
\begin{eqnarray}
\zeta & \sim & -\, \frac{\tlq^2}{\sqrt{h} \sqrt{1 + \frac{\tlb^2}{1+f^2} + \frac{\tlrho^2}{(1-\Psi_0^2)^2 + f^2}}} \, + \, \frac{1}{h} \ + \, \ldots  \nonumber \\ 
& & ~~~~ - \,  \frac{ \pi i \tom \tlq}{\sqrt{h}\sqrt{1 + \frac{\tlb^2}{1+f^2} + \frac{\tlrho^2}{(1-\Psi_0^2)^2 + f^2}}}\, e^{- \tlq \frac{\sqrt{h}}{\sqrt{1 + \frac{\tlb^2}{1+f^2} + \frac{\tlrho^2}{(1-\Psi_0^2)^2 + f^2}}}} \ + \, \ldots \ .
\end{eqnarray}
This solution connects nicely to the asymptotic region, even matching subleading terms in the overlap region, to give us
 \begin{eqnarray}
 \zeta & = & - \tlq \sqrt{\frac{G^{yy}}{G^{uu}}}\, -\,\frac{\left(\sqrt{-G}G^{uu} G^{yy}\right)'}{2 \sqrt{-G}G^{uu }G^{yy}}\, -\,\frac{\sqrt{\frac{G^{yy}}{G^{uu}}}'}{2\sqrt{\frac{G^{yy}}{G^{uu}}}} 
 \, + \ \order (\tlq^{-2} ) \nonumber \\
 & & ~~~~
 -\,  i  \pi \tom \tlq \sqrt{\frac{G^{yy}}{G^{uu}}} e^{-2 \tlq \int_u^1 d\bar{u}, \sqrt{\frac{G^{yy}}{G^{uu}}}} \, \left(1+\order (\tom^2,\tlq^{-2}) \right) \ .
 \end{eqnarray}
Hence, the dissipative part of the diagonal conductivity reads  to leading order
\begin{equation}\labell{exconn}
\Re \tilde{\sigma}_{yy} \ = \ \varepsilon_0 \pi \tlq e^{-2 \tlq \int_0^1 d\, u \, \sqrt{\frac{G^{yy}}{G^{uu}}}} \ ,
\end{equation}
where one could again interpret the result as having an ``effective temperature'' scale of
\begin{equation}\labell{conefft}
\frac{T}{T_{eff}} \ = \ \frac{2}{\pi}  \int_0^1 d\, u \sqrt{\frac{G^{yy}}{G^{uu}}} \ .
\end{equation}
In the massless limit we can, as usual, find an analytic expression, which evaluates to
\begin{equation}
\frac{T}{T_{eff}} \ = \ \frac{2 \Gamma(5/4)}{\sqrt{\pi} \Gamma(3/4)} \sqrt{1+f^2} {\, }_2F_1\left(\frac{1}{4},\frac{1}{2};\frac{3}{4};-(f^2+\tlrho^2+\tlb^2)\right) \ .
\end{equation}
If we were to describe this result qualitatively as the behavior of a semiconductor, then the edge density of states would correspond to total density of baryons and anti-baryons in thermal equilibrium, and the difference $N_c-N_v$ would correspond to the baryon density $\rho_0$.

\DFIGURE{
\includegraphics[width=0.49\textwidth]{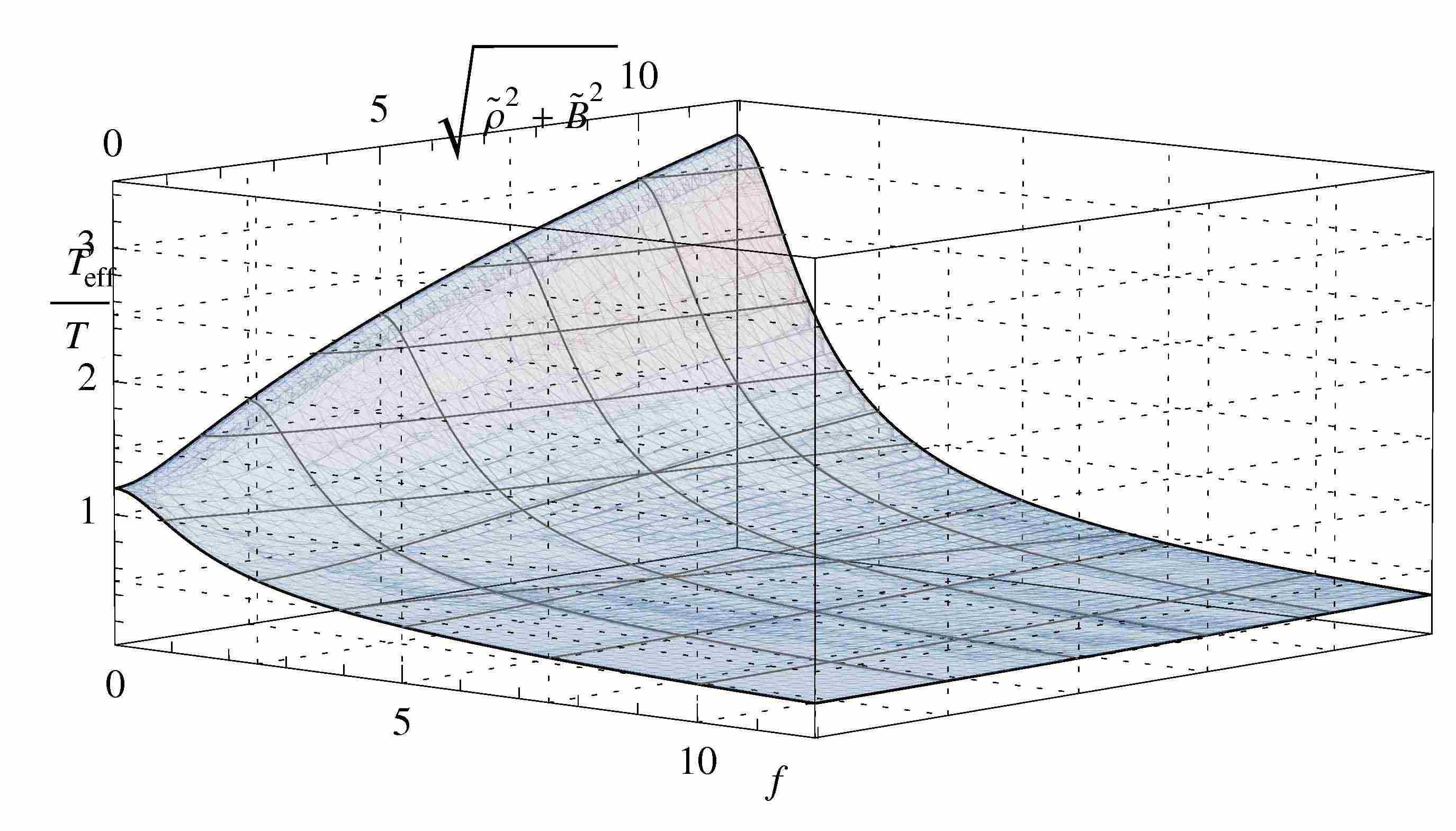}
\includegraphics[width=0.49\textwidth]{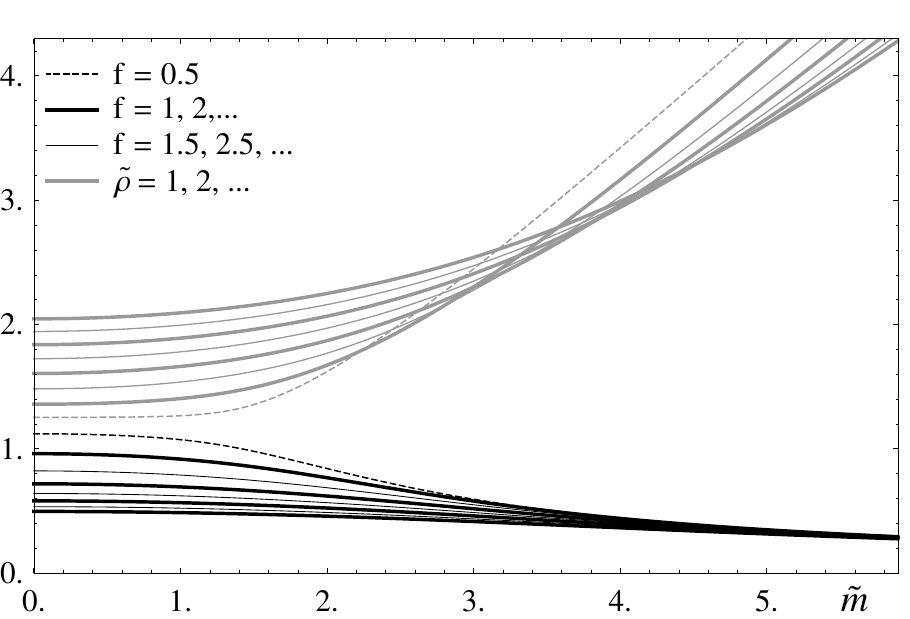}
\caption{The ``effective temperature'' $T_{eff}/T$. Left: As a function of $f$ and $\sqrt{\tlrho^2 + \tlb^2}$. Right: As a function of $\tlm$ for various values of $\tlrho$ and $\tlb$.}\mlabel{efftf_plot}}
In fig. \mref{efftf_plot}, we show how the effective temperature depends on the parameters of the defect. In addition to the dependence on $f$, that was previously found in ref. \refcite{baredef}, turning on a magnetic field or a finite density now raises the effective temperature approximately $\propto (\tlb^2 + \tlrho^2)^{1/4}$.
Furthermore, we find that turning on a finite mass in some sense ``enhances'' the effect of the density and of $f$ but the dependence on the mass in the presence of only $\tlb$ is not very significant.

Next, let us look at the off-diagonal terms. Do do so, we first need to write out the homogeneous part of the equation of motion for $\mathcal{A}_x =: \mathcal{A}_x^0 e^{\int^u \zeta_x}$:
\begin{equation}
\zeta_x^2\, +\, \zeta_x' \, +- \, \frac{\left(\sqrt{-G}G^{tt} G^{xx}\right)'}{\sqrt{-G}G^{tt }G^{xx}}\zeta_x   \, + \, 
\frac{G^{xx}}{G^{uu}}\left(\frac{G^{tt}}{G^{xx}} \tom^2 \, - \, \tlq^2 \right) \ = \ 0 \ ,
\end{equation}
which has the solution up to $\order (1)$, $\zeta_x  =  \frac{\left(\sqrt{-G}G^{tt} G^{xx}\right)'}{2 \sqrt{-G}G^{tt}G^{xx}}\pm \frac{\zeta_0'}{2\zeta_0}\pm \zeta_0$, where we again pick the negative sign. The dominant terms in the full homogeneous solutions are then 
\begin{equation}\labell{aahom}
\mathcal{A}_x \, = \, \mathcal{A}_x^0 \sqrt{\frac{\sqrt{-G}G^{tt} G^{xx}}{\zeta_0}}e^{-\int_0^u \zeta_0} ~~~~\mathrm{and} ~~~~
A_y \, = \, A_y^0 \frac{1}{\sqrt{\sqrt{-G}G^{uu} G^{yy} \zeta_0}} e^{-\int_0^u \zeta_0} \ .
\end{equation} 
There are now two ways to determine the perturbative contribution coming from the cross terms. Either we again solve for the exponents $\zeta$ -- which would then contain factors of $A^0_y/A^0_x$ -- or we can directly take a pertubation for $A_y$. Even though the latter one may seem most natural, in particular since the system is linear, we will use the first method since it gives us the result in a very neat way.
Substituting \reef{aahom} into the equation of motion for $A_y$ \reef{fyyeq}, we see that
the equation for $\epsilon$ \reef{epseq} receives now an additional term
\begin{equation}
\beta(u) \ \rightarrow \ \beta(u) \ - \ \zeta_x \left(\sqrt{-G} G^{tu}G^{xy}\right)'\sqrt{\frac{G^{uu}}{G^{tt}}}\frac{\mathcal{A}_x^0}{A_y^0} \ ,
\end{equation}
such that we obtain an extra contribution to $\epsilon$, taking only the leading term in $\beta$ $\propto \tlq$
\begin{equation}
\epsilon \ = \ e^{2 \int_0^u \zeta}\left(\epsilon_0 - \int_0^u d\tilde{u}\left(\zeta_x\left(\sqrt{-G} G^{tu}G^{xy}\right)'\sqrt{\frac{G^{uu}}{G^{tt}}}e^{-2 \int_0^{\tilde{u}} \zeta}\right) \frac{\mathcal{A}_x^0}{A_y^0}\right) \ .
\end{equation}
Now, if we look at the equations of motion \reef{fyyeq},\reef{fxxeq}, we remind ourselves that in the near horizon geometry, the equations of motion for $A_y$ and $\mathcal{A}_x$ look the same and the cross terms are suppressed by a factor of $h/\tlq$ with respect to the dominant terms. Hence the coupling occurs over the range $u\in ]0,1-\varepsilon]$ for small $\varepsilon$ and not in the near-horizon region. We keep $\varepsilon$ to regulate the asymptotic solution in the near horizon region, in which it is not valid. To capture the mixing then correctly, we fix $\epsilon_H$ at the horizon as above, such that we find
\begin{eqnarray}
\epsilon & \simeq & \sqrt{\frac{G^{yy}}{G^{uu}}}e^{\!-2\! \int_u^{1-\varepsilon}\! \zeta_0}\!\left(\epsilon_H\! +\! \int_u^{1-\varepsilon}\!\!d\tilde{u}\! \left(\zeta_x\!\left(\sqrt{-G} G^{tu}G^{xy}\right)'\!\frac{G^{uu}}{\sqrt{-G^{tt}G^{xx}}}e^{2\! \int_{\tilde{u}}^{1-\varepsilon}\! \zeta_0}\!\right)\frac{\mathcal{A}_x^0}{A_y^0}\right) \, \nonumber \\
& =: & \epsilon_{hom.} \, + \, \epsilon_{inh.} \ 
\end{eqnarray}
where we absorbed a factor of $\left.\sqrt{\frac{G^{uu}}{G^{xx}}}\right|_{u = 1-\varepsilon}$ into $\epsilon_H$.
Because of the exponential factor in the second term, the integral will be dominated around small $\tilde{u}$. The appropriate expansion gives us to leading order at $u \ll 1$
\begin{eqnarray}\labell{hallexpapprox}
\epsilon_{inh.}& = & - 4 \, \frac{\tlrho \tlb}{1+f^2} \, \tlq \, \frac{\mathcal{A}_x^0}{A_y^0} \, e^{2 \sqrt{1+f^2} \tlq u}\int_u^1 d\tilde{u} \, u^3 e^{-2 \sqrt{1+f^2} \tlq \tilde{u}}  \nonumber \\
& = & 4 \,  \frac{\tlrho \tlb}{1+f^2} \, \tlq \, \frac{\mathcal{A}_x^0}{A_y^0} \, e^{2 \sqrt{1+f^2} \tlq u}\, \left[\frac{\tilde{u}^4}{16 \tlq^4 (1+f^2)^2}\, \Gamma(4,2 \tlq \tilde{u} )\right]_{\tilde{u}=u}^1 \ 
\end{eqnarray}
and in the limit $u\rightarrow 0$, we find 
\begin{equation}
\lim_{u\rightarrow 0} \epsilon_{inh.} \ = \ - \frac{3}{2}\,  \frac{\tlrho \tlb}{\left(1+f^2\right)^{5/2}\, \tlq^3}\,  \frac{\mathcal{A}_x^0}{A_y^0} \ + \ \order (\tlm/\tlq^5)\ + \ \order (1/\tlq^7) \ .
\end{equation}
Finally, keeping $A_y|_{u=0} = A_y^0$ fixed, we get the leading terms (ignoring the exponentially suppressed terms)
\begin{equation}
\partial_u A_y \ = \ - \sqrt{1+f^2}\,  \tlq\, A_y^0 \ - \  \frac{3}{2} \, \frac{\tlrho \tlb}{\left(1+f^2\right)^{5/2}  \, \tlq^3} \, \mathcal{A}_x^0 \  + \ \order (M_q^2/\tlq^5)\ + \ \order (1/\tlq^7) \ ,
\end{equation}
and hence we can compute the Hall conductivity
\begin{equation}
\tilde{\sigma}_{yx} \ = \ - \varepsilon_0 \sqrt{1+f^2} \left.\frac{\delta A_y'}{\delta \mathcal{A}_x'}\right|_{u\rightarrow 0} \ = \ - \varepsilon_0\, \frac{3}{2}\, \frac{\tlrho \tlb}{\left(1+f^2\right)^{5/2}\, \tlq^4} \  + \ \order (M_q^2/\tlq^6)\ + \ \order (1/\tlq^8) \ .
\end{equation}
Following through the analysis attentively, one can also see that the imaginary part of the Hall conductivity is exponentially suppressed by a factor of $e^{-q/T_{eff}}$.

This result is remarkable, since the diagonal part of the dissipative conductivity is heavily suppressed with a factor $e^{-q/T_{eff}}$, while the off-diagonal part is only suppressed by a factor of $T^4/q^4$. This reflects the fact that at small temperatures we approach conformal symmetry in the field theory and hence the form of the conductivity in ref. \refcite{sl2z} that was discussed in ref. \refcite{baredef}. Having a purely off-diagonal conductivity is not surprising as it is for example the case on the Hall plateaus in the quantum Hall effect or as we demonstrated above in intrinsic semiconductors at small temperatures. 

Interestingly it occurs also in intrinsic semiconductors at low temperatures. There the absence of defects and the highly suppressed charge carrier density cause the relaxation time $\tau$ to diverge, whilst the carrier mobility remains approximately unchanged. Hence the diagonal conductivity \reef{drudecon} is suppressed, while the factor $\omega_c \tau$ in the Hall conductivity \reef{drudehall} causes the Hall conductivity to remain finite.
\subsubsection{Exponentially suppressed regime at large backgrounds, $\tlq^2 \gg |\tlrho|,|\tlb|\gg |\tlq|\gg 1$}
This regime is slightly more non-trivial, because now $\zeta_0$ splits into three regimes (for simplicity at $M_q = 0$):
\begin{eqnarray}
u \lesssim \frac{1}{\sqrt{\tlb}}, \frac{1}{\sqrt{\tlrho}} & : \ \  & \zeta_0 \, \sim \, \tlq \sqrt{1+f^2} \ , \\
\frac{1}{\sqrt{\tlb}}, \frac{1}{\sqrt{\tlrho}} , \sqrt{\tlrho} \lesssim u  \ll 1 & : \ \  &  \zeta_0 \, \sim \frac{q\sqrt{1+f^2}}{\sqrt{f^2+\tlb^2+\tlrho^2} u^2 } \ \ \ \mathrm{and}\\
h \ll 1 & : \ \  & \zeta_0 \, \sim \frac{q\sqrt{1+f^2}}{\sqrt{f^2+\tlb^2+\tlrho^2} h } \ .
\end{eqnarray}
In the asymptotic region, the solution is dominated by the decaying exponential, whereas in the near horizon region, it can be written in terms of the coordinate $s$ from section \mref{DClimit} as 
\begin{equation}
A_y \ = \ A_y^0 e^{i \nu\, s} \ = \ A_y^0 e^{i \nu\, s\int_0^u \frac{1}{\sqrt{-G}G^{uu}G^{yy}}} \ ,
\end{equation}
where  $\nu = \tom 
\sqrt{1+f^2}\frac{\sqrt{\tlb^2\left(f^2+(1-\Psi_0^2)^2\right) +(1+f^2)(\tlrho^2+ f^2 +(1-\Psi_0^2)^2)}}{1+\tlb^2+f^2}$.
If this solution were to overlap with the ``tail'' of the asymptotic solution, $u^4 > f^2+\tlrho^2+\tlb^2$, we could match them at some $1 \gg u_H \gtrsim (f^2+\tlrho^2+\tlb^2)^{1/4}$.
The fact that they do not overlap, however, can be simply seen from the different $u$ dependence of $\partial_u s = \frac{1}{\sqrt{-G} G^{uu} G^{yy}}$ and $\zeta_0= \tlq \sqrt{\frac{G^{xx}}{G^{uu}}} $. Hence, whatever we try now, the conductivity will disagree by some finite factor. One way to pretend that they do overlap is to simply set $h=1-u^4 \rightarrow 1$ and $u \rightarrow 1$, which corresponds to extending the $\frac{1}{\sqrt{\tlb}}, \frac{1}{\sqrt{\tlrho}} \lesssim u  \ll 1$ region towards the horizon and the near-horizon limit into the intermediate region. Matching the solutions under these conditions and ignoring the second part of the solution for $\epsilon$ \reef{epssol} gives us
\begin{equation}
\epsilon \, = \, - i\tom \sqrt{\frac{1+f^2+\tlrho^2+\tlb^2}{1+f^2}}\sqrt{\frac{G^{yy}}{G^{uu}}} e^{-2 \tlq \int_u^1 \sqrt{\frac{G^{yy}}{G^{uu}}}} 
\end{equation}
with the corresponding conductivity
\begin{equation}
 \tilde{\sigma}_{yy} \, = \, - i\varepsilon_0  \frac{\tlq}{\tom} \, + \ldots  \,+ \, \varepsilon_0  \sqrt{\frac{1+f^2+\tlrho^2+\tlb^2}{1+f^2}} e^{-2 \tlq \int_0^1 \sqrt{\frac{G^{yy}}{G^{uu}}}} \, + \ldots  \ .
\end{equation}
Certainly we cannot trust any $\order (1)$ and polynomial factors, but the point to make here is that we should still expect the exponential suppression from the effective temperature \reef{conefft}. This is because the integral in the exponent is dominated by the region in which $\zeta_0$ does indeed dominate the solution, we took care of the deep near horizon region and elsewhere there are no ``large'' terms in the equations of motion.

In principle one can also try to find a solution in the regime $\frac{1}{\sqrt{\tlb}}, \frac{1}{\sqrt{\tlrho}} \lesssim u  \lesssim 1$ and then ``glue'' it to the near horizon and asymptotic solutions to gain more accuracy, but there is limited insight to be learned from this and it would be much more tedious than the calculations in \mref{conefftf}.

The Hall conductivity will still be dominated by the asymptotic regime since the mixing from the near horizon region is exponentially suppressed.
In principle, it is then still $\tilde{\sigma}_{yx}  =  - \varepsilon_0 \frac{3}{2} \frac{\tlrho \tlb}{(1+f^2)^{5/2}\tlq^4}$, however we need to note that the integral that was computed in section \mref{conefftf} is dominated around the maximum of $u^3 e^{-2 \tlq \sqrt{1+f^2} u }$ at $u_{max} = \frac{3}{2 \tlq \sqrt{1+f^2}}$ and decays then also on the scale $\delta u = \frac{1}{2 \tlq \sqrt{1+f^2}}$. $\zeta$ and hence also the exponential suppression however start to change around $u_{stop} \sim (f^2 + \tlrho^2 + \tlb^2)^{-1/4}$. To properly evaluate the integral in the approximation \reef{hallexpapprox}, $u_{stop}$ need to be significantly larger than $u_{max}$ - otherwise the ``tail'' of the polynomial term in the integral will not be sufficiently suppressed  and will give a finite contribution to the result, which will greatly overestimate the Hall conductivity. 
Certainly one can always use the full integral
\begin{equation}
\tilde{\sigma}_{yx} \, = \, 
-\varepsilon_0 \frac{\sqrt{1+f^2}}{\tlq}e^{-2 \int_0^{1} \zeta_0}
\int_0^{1}d\tilde{u}\, \zeta_x\left(\sqrt{-G} G^{tu}G^{xy}\right)'\frac{G^{uu}}{\sqrt{-G^{tt}G^{xx}}}e^{2 \int_{\tilde{u}}^{1} \zeta_0}
\end{equation}
but this is certainly a somewhat less insightful result and cannot be computed analytically.
\DFIGURE{
\includegraphics[width=0.49\textwidth]{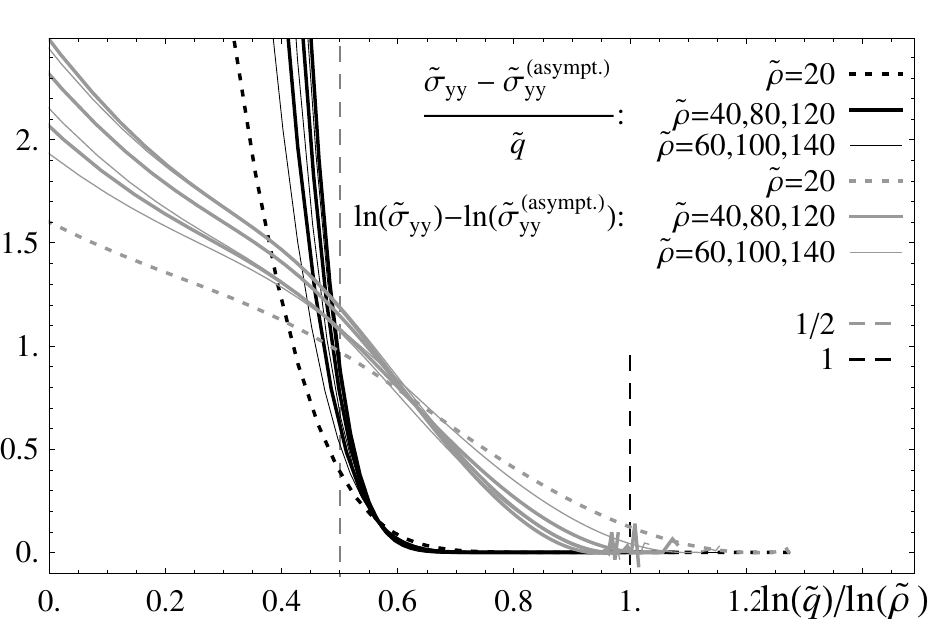}
\includegraphics[width=0.49\textwidth]{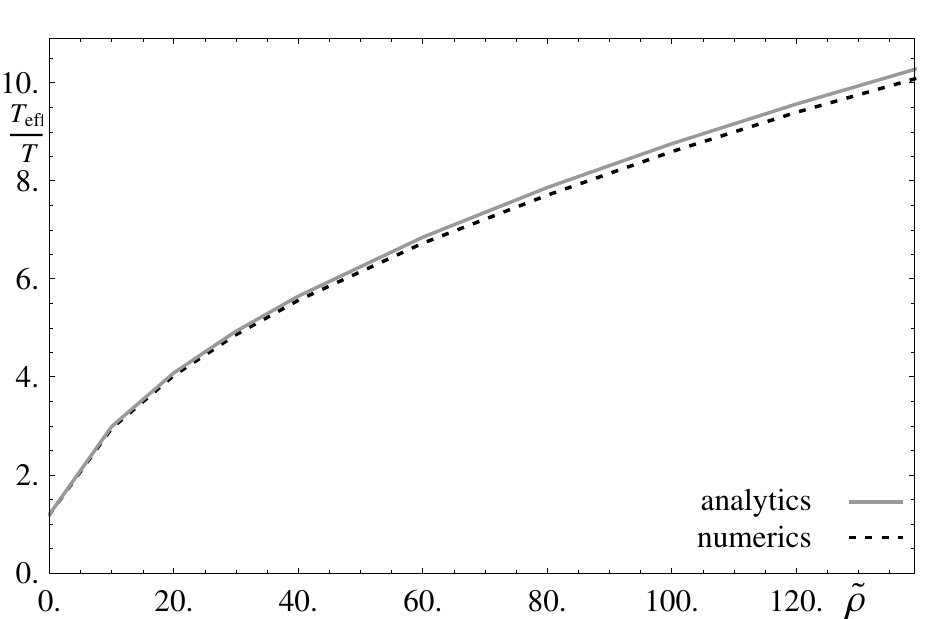}
\caption{Left: The check to the deviation from the $T\rightarrow 0$ limit as described in the text. Right: The numerical estimate $\frac{T}{T_{eff}} \sim -\pi \partial_{\tlq} \ln \left(\frac{\tilde{\sigma}_yy}{\tlq}\right)$ compared to the analytical result.}
\mlabel{efftf_checkr}}
In figure \mref{efftf_checkr}, we demonstrate the boundaries between the different regimes. To see how far the regime of the previous section reaches, we plot $\ln \left( \sigma_{yy}\right)-\ln \left( \sigma^{(asym.)}_{yy}\right)$ against $\frac{\ln \, \tlq}{\ln \, \tlrho}$. $ \sigma^{(asym.)}_{yy}$ is defined taking $\ln \left( \frac{\sigma^{(asym.)}_{yy}}{\tlq}\right)$ to be the linear expansion of $\ln \left( \frac{\sigma^{}_{yy}}{\tlq}\right)$ at large values of $\tlq$ approximately where $\tilde{\sigma}_{yy}\sim 10^{-15} \varepsilon_0$, shortly before the numerics fail. $\frac{\ln \, \tlq}{\ln \, \tlrho} = n$ corresponds to $\tlq = \tlrho^n$. This is sensitive to changes in the factor in front of the suppressed conductivities at large $\tlq$. 
To check for the overall limit of the exponentially suppressed regime, we look directly at $ \sigma_{yy}- \sigma^{(asym.)}_{yy}$. It is easy to see that the boundaries at approximately $\tlrho,\tlb \sim \tlq$ and $\tlrho,\tlb \sim \tlq^2 $ are verified. Using that data, we also looked at the $\order (1)$ factor $\pi$ in front of the exponential term in the conductivity in \reef{exconn}. It turns out that for our values of $\frac{q}{T_{eff}}\sim 35$, the numerically estimated factor varied from $\approx 3.8$ at $\tlrho=0$ to $15$ at $\tlrho = 140$, where the numerics carried us only up to $\tlq\sim\tlrho$. At the latter values, we could not expect close agreement because we were outside the regime that we considered in section \mref{conefftf} -- and the value of $3.8$ seems reasonably close to $\pi$.

\subsubsection{Dominantly large backgrounds $\tlrho,\tlb \gg \tlq^2 \gg  1$}
As $\tlrho,\tlb \gg \tlq^2$, we find that $\int_0^1 \zeta_0 \sim 2 \tlq (f^2 + \tlrho^2 + \tlb^2)^{-1/4}\ll 1$, and hence the exponential suppression factor disappears. Furthermore, as indicated above, our estimate for the Hall conductivity does not apply anymore, as it will be dominated by the region of $u$ in which the assumption $\zeta_0\gg 1$ does not apply anymore.

Looking at the problem in another way in terms of the coordinate $s$ from section \ref{DClimit} for the DC conductivity with the equation for $A_y$ \reef{DC_Ay} gives us the relevant term 
\begin{equation}
\partial_u \partial_s A_y = \ldots + \sqrt{-G}G^{yy}G^{xx} \tlq^2 A_{y} \ .
\end{equation}
Integrating this analytically in the massless case and for for sufficiently slowly varying $\partial_s A_y \ll A_y$, i.e. $\nu \ll 1$, we find $\delta \, \partial_s A_y \int_0^1 d u \, \partial_s A_y \lesssim \frac{(1+f^2) \tlq^2}{(f^2 + \tlb^2 +\tlrho^2)^{1/4}}\in\mathbb{R}$. A second integration gives $\frac{\delta A_y}{A_y} \lesssim \frac{(1+f^2)\tlq^2}{\sqrt{f^2 + \tlb^2 +\tlrho^2}}$ and for larger frequencies $\nu$, the oscillatory behavior of $A_y$ implies that the integral is further suppressed by a factor $1/\tom$.

Hence, for $\tlrho,\tlb \gg \tlq^2$, the real part of the conductivity is dominated by the results of the isotropic case in section \mref{DCsmall}.
The inductive (imaginary) part of the conductivity obviously still receives the term $\tilde{\sigma}_yy \sim -i\varepsilon_0 \frac{\tlq}{\tom}$.
%
%
%
\section{Numerical Results}\mlabel{numcon}
In this section, we study information that can be derived from computing the correlators numerically, in particular the overview of the frequency-dependent conductivity (i.e. the spectral curves), the diffusion and relaxation behavior in the hydrodynamic regime, and the spectrum of quasi-particles.
\subsection{Spectral Curves}\mlabel{fancyspectral}
In this section, we present the conductivity spectrum in the presence of various background quantities.

%
\DFIGURE{
\includegraphics[width=0.49\textwidth]{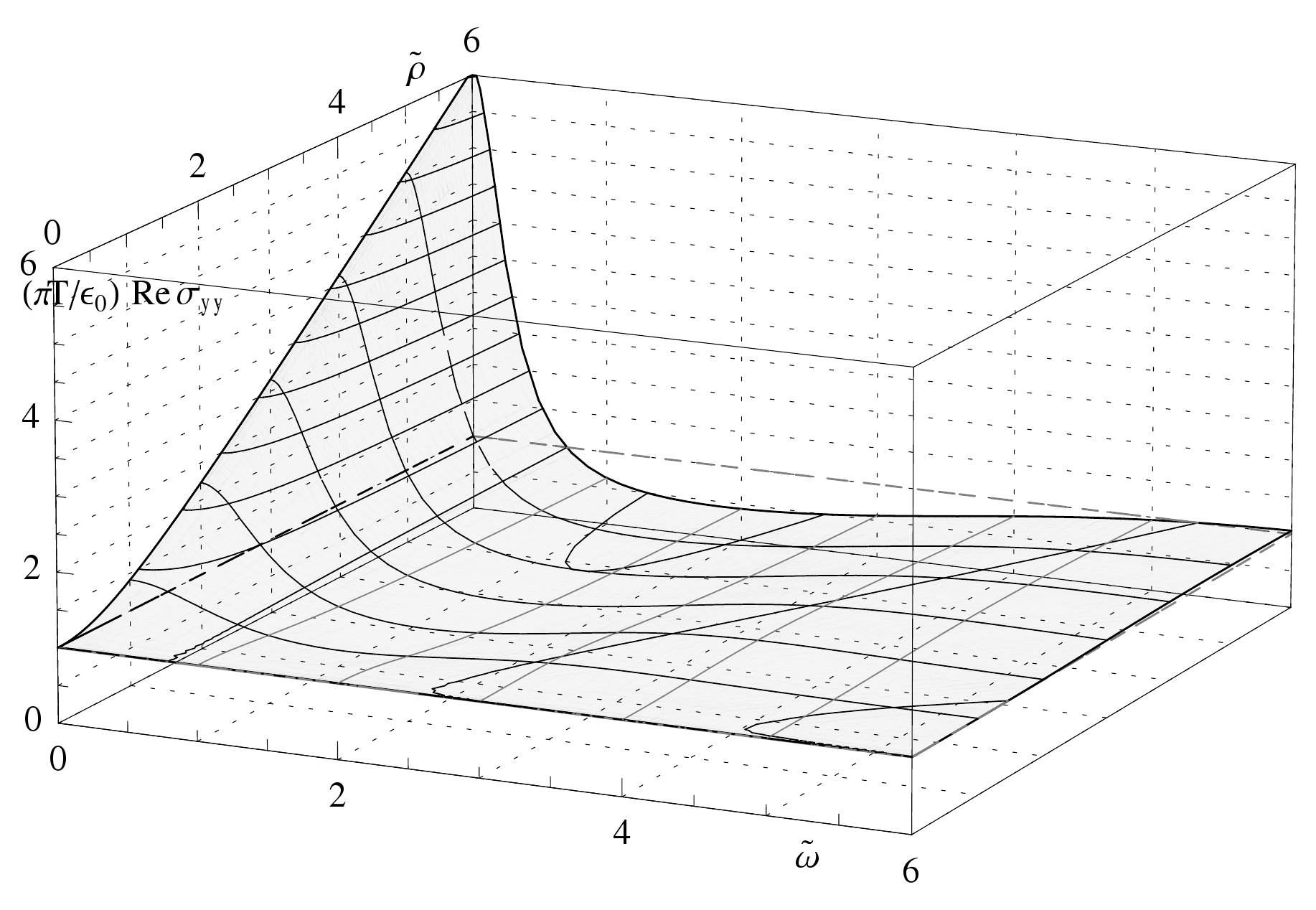}
\includegraphics[width=0.49\textwidth]{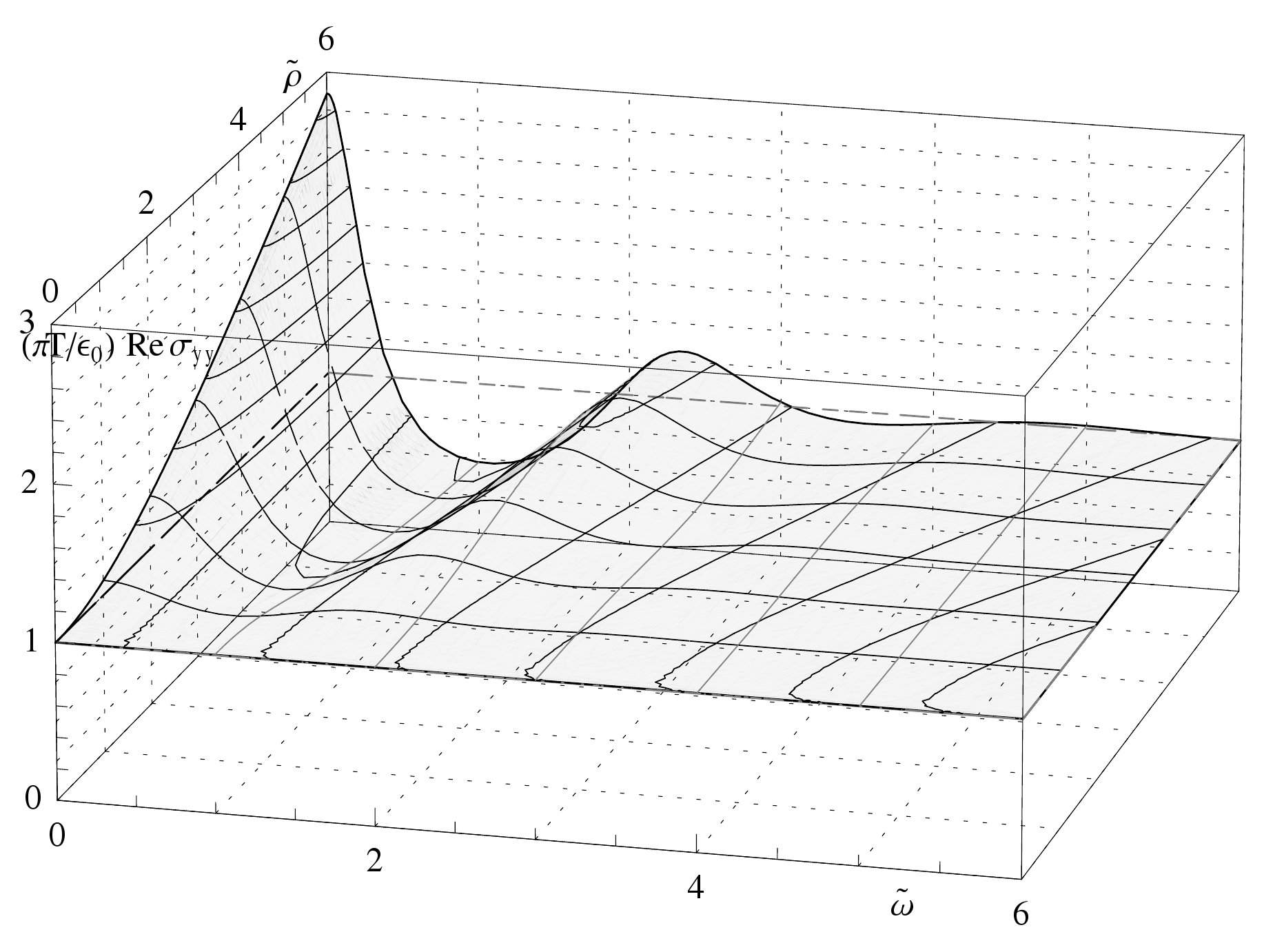}\\
\includegraphics[width=0.49\textwidth]{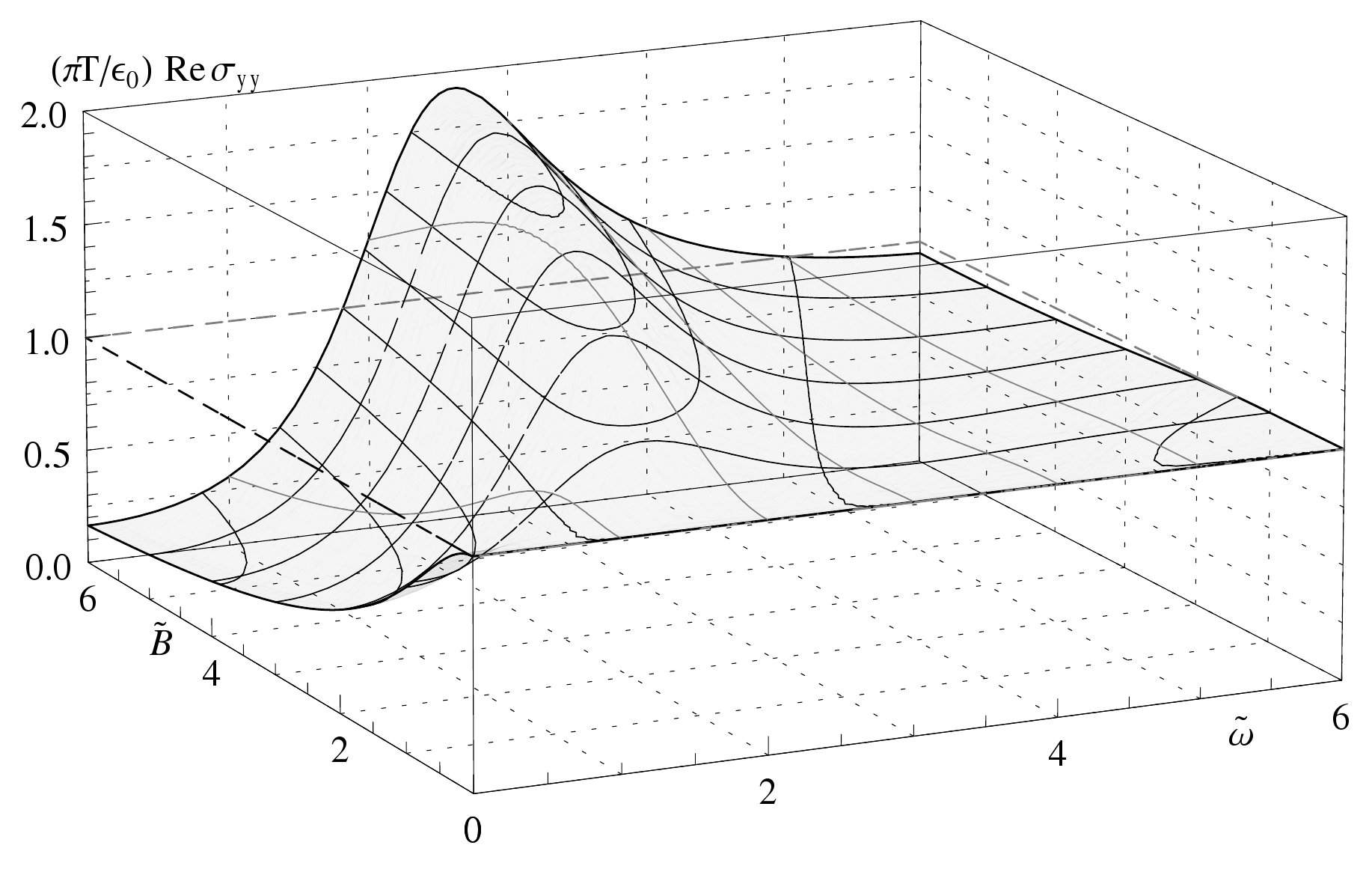}
\includegraphics[width=0.49\textwidth]{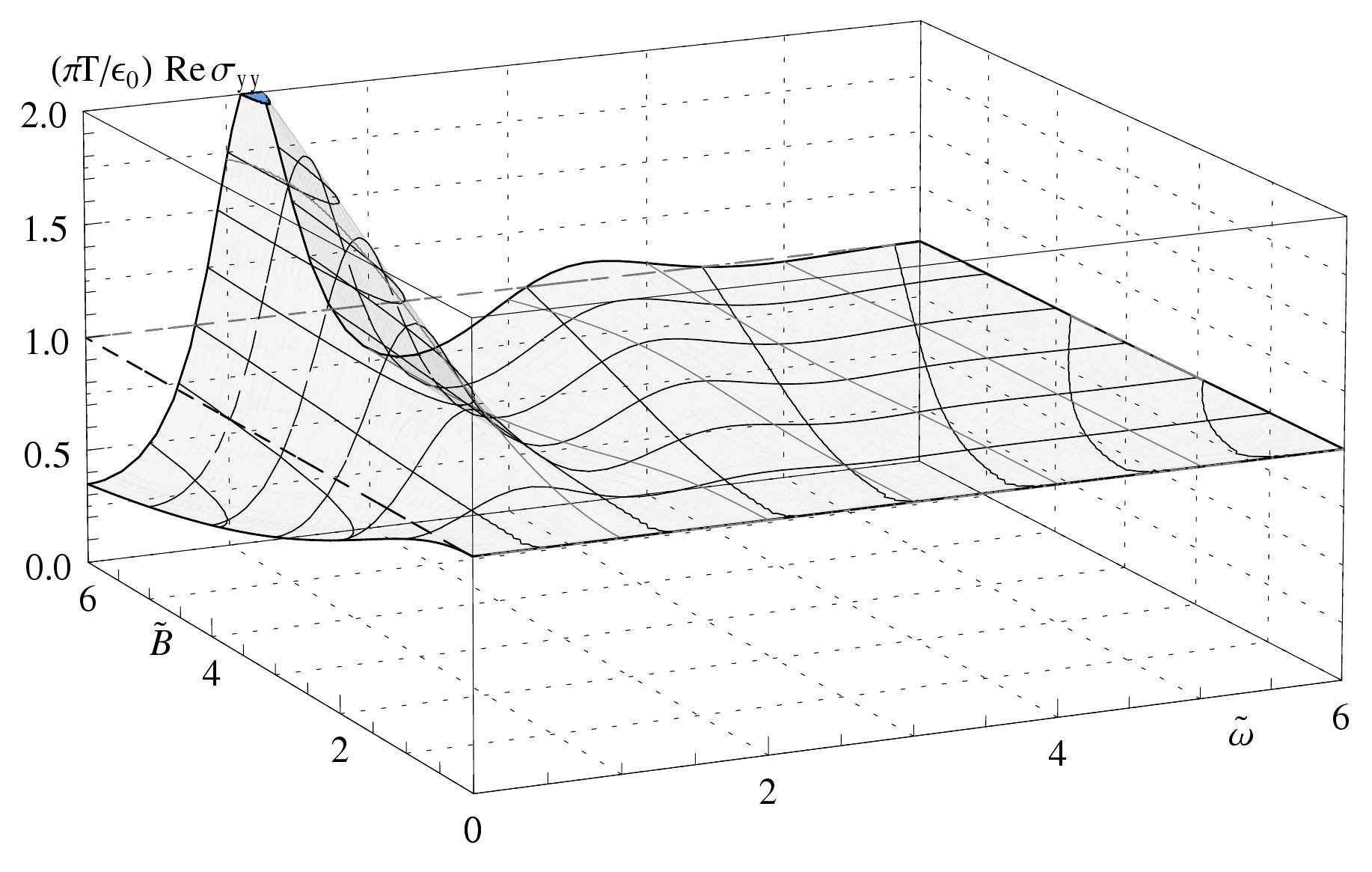}
\caption{The real, diagonal part of the isotropic conductivity at $\tlq=0$ as a function of frequency and net baryon density (top) or magnetic field (bottom) for $f=0$ (left) and $f=2$ (right).}
\mlabel{con3d_b}}
First, let us look at the case of finite density and magnetic field alone in fig. \mref{con3d_b}. From the result of the electromagnetic duality in section \mref{femdual}, and the very generic results for the hall conductivity in section \mref{resonances}, we expect to see a sequence of resonances, in which maxima and minima are exchanged between the case of finite net baryon density and magnetic field. The fact that ``plasmon'' (finite-density) resonances are relatively strong is not surprising since this is a strongly coupled system -- and plasmons are a finite coupling effect. The small-frequency regime reflects very well the classical Drude model expectations and the small-frequency expansion from section $\mref{DCsmall}$ -- with the Drude peak and magnetoresistance.
Looking at the resonances, we find that they are approximately equally spaced at $n \omega_p$ or $(n-1/2)\omega_c$, respectively -- and they decay quickly. Comparing this to what we learned in section \mref{resonances} reveals interesting information about the quasi-particles that carry the current: They a) must be massive and b) do not consist of chiral fermions, in sharp contrast to graphene \cite{graphenehall}.
It is also interesting to see that the amplitude seems to be again decaying exponentially as in the resonance on the width of the defect that were studied in ref. \refcite{baredef}. The frequencies
$\omega_c$ and $\omega_p$ are, however, not proportional to $\tlb$ and, respectively, $\tlrho$ and even start off at a finite value. In terms of generic weak-coupling intuition, this would need to be explained  by a non-linear magnetization behavior and non-linear chemical potential, and in terms of a changing mass of the quasi-particles that carry the current. 
%

%
\DFIGURE{
\includegraphics[width=0.49\textwidth]{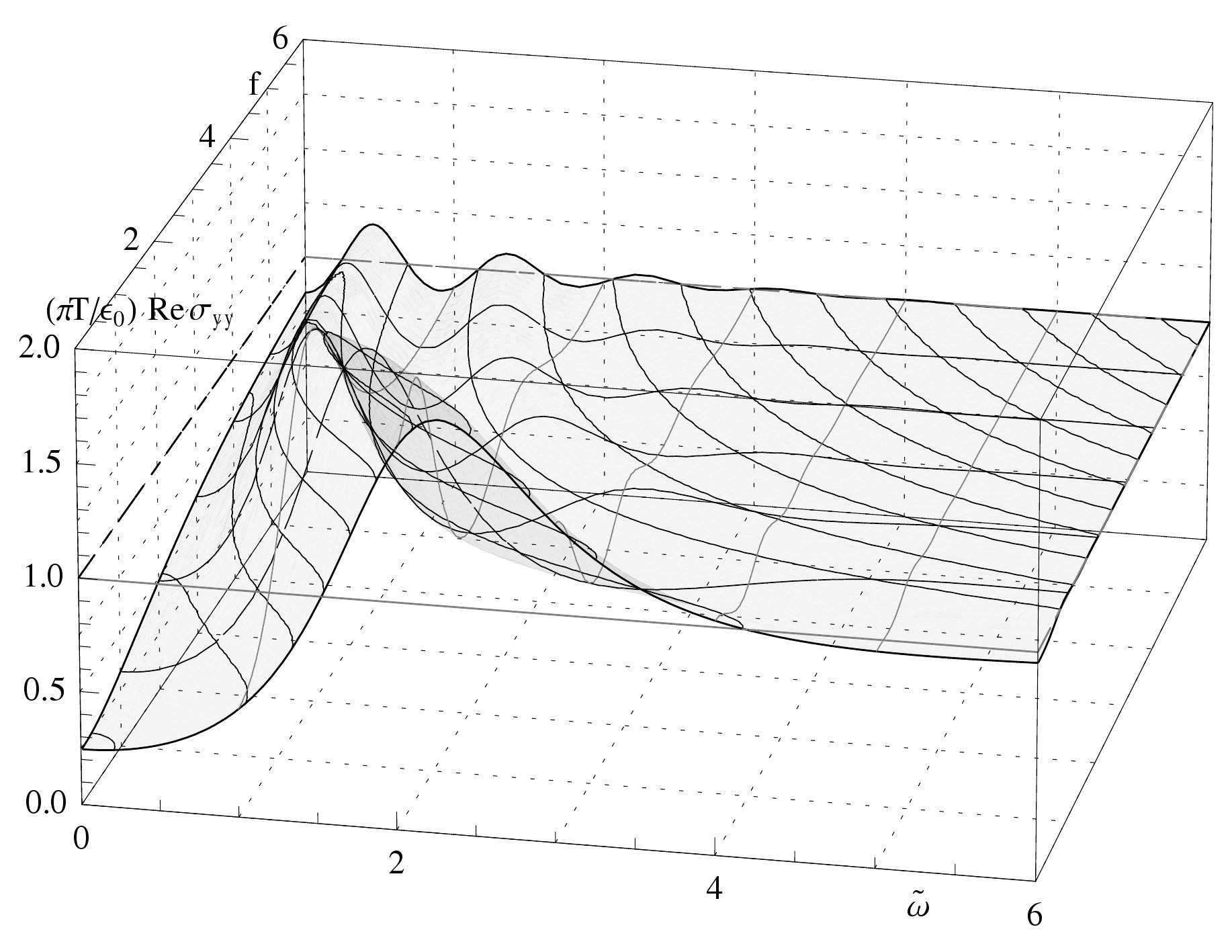}
\includegraphics[width=0.49\textwidth]{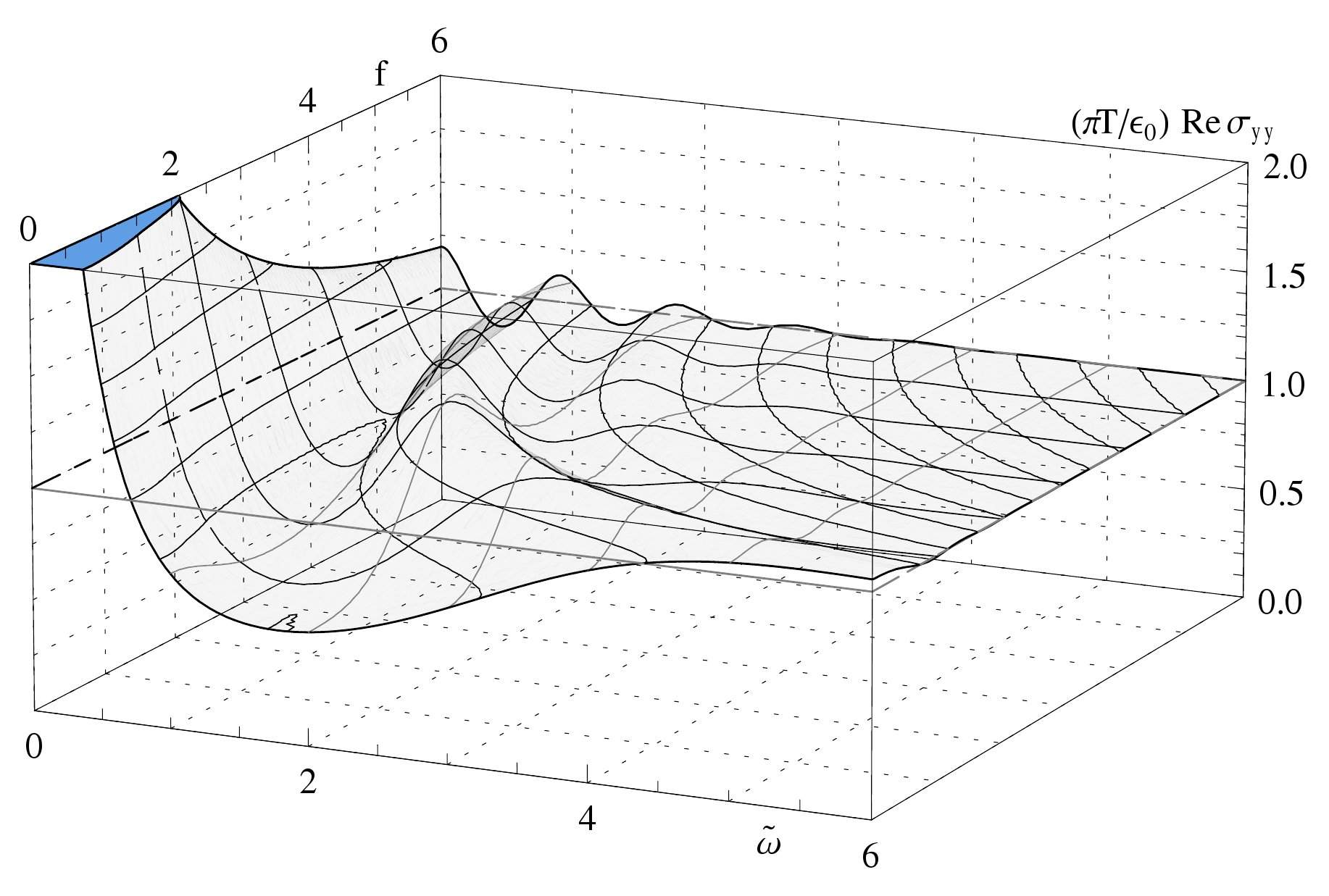}
\caption{The real, diagonal part of the isotropic conductivity at $\tlq=0$ as a function of frequency and magnetic field for $\tlb=4$ (left) and $\tlrho = 4$ (right).}
\mlabel{con3d_f}}
Looking at the $f${}-dependence in figure \mref{con3d_f}, we find that on the one hand, increasing $f$, i.e. an increasing width $z_{max}$ or stronger ``confining potential'', reduces the amplitude of the resonances  at small frequencies. This is consistent with the value that we found for the DC conductivity \reef{DCtensor}. In contrast to this, we find that the suppression of the resonances with increasing frequencies decreases with increasing $f$ and we can see the tower of modes, that is at small amplitudes hinted at by the $\tilde{\sigma} = \varepsilon_0$ lines in the plot. This agrees with the effective temperature \reef{conefft} that decreases proportionally to $ f^{-1/2}$. 
Secondly, we find that the parameters $\omega_c$ and $\omega_p$ decrease with increasing $f$, which we can again explain by a non-linear behavior of the response functions or by an $f${}-dependent quasiparticle mass.
%
%
%
%
%
%
%
%

\DFIGURE{
\includegraphics[width=0.49\textwidth]{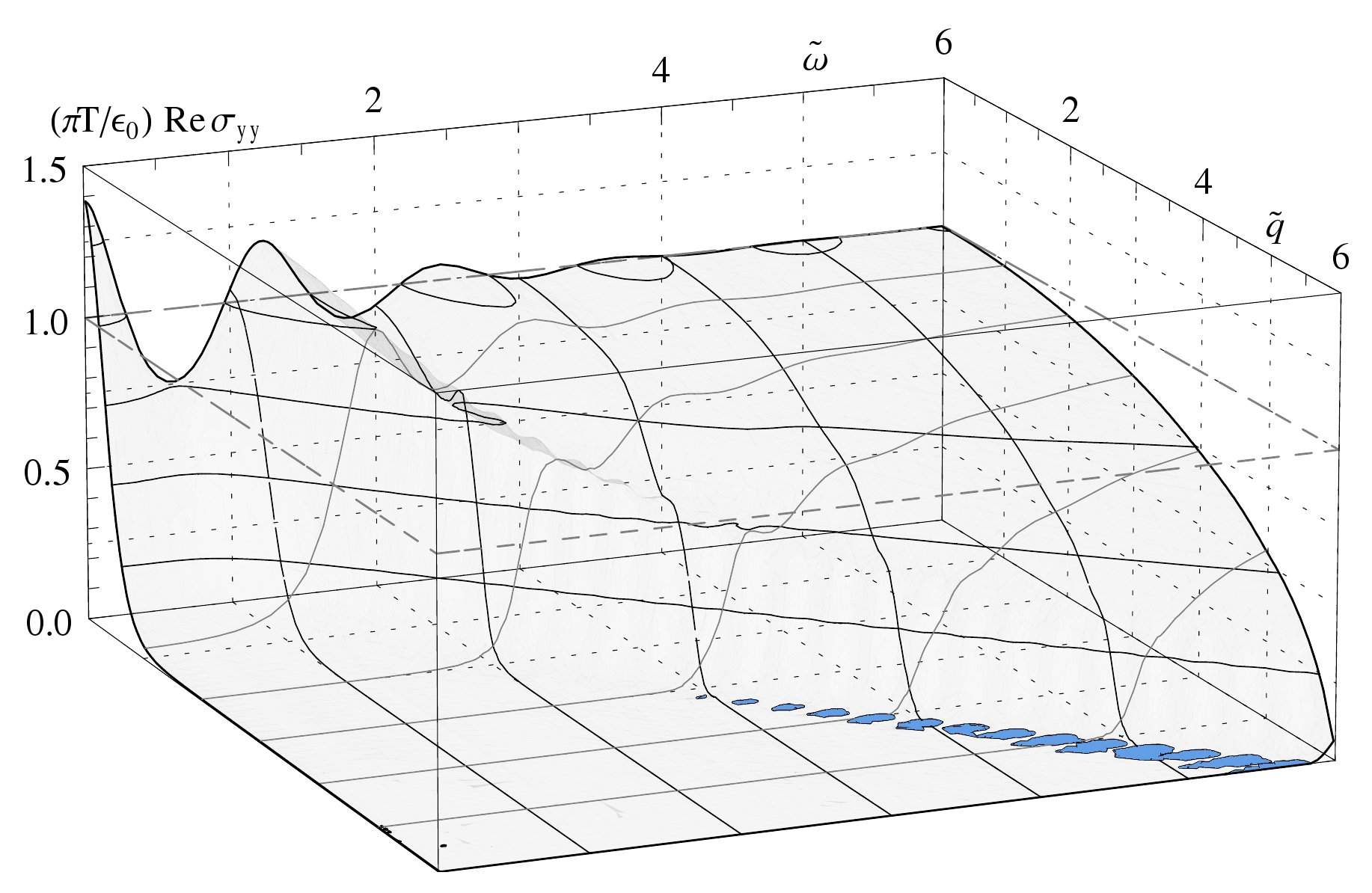}
\includegraphics[width=0.49\textwidth]{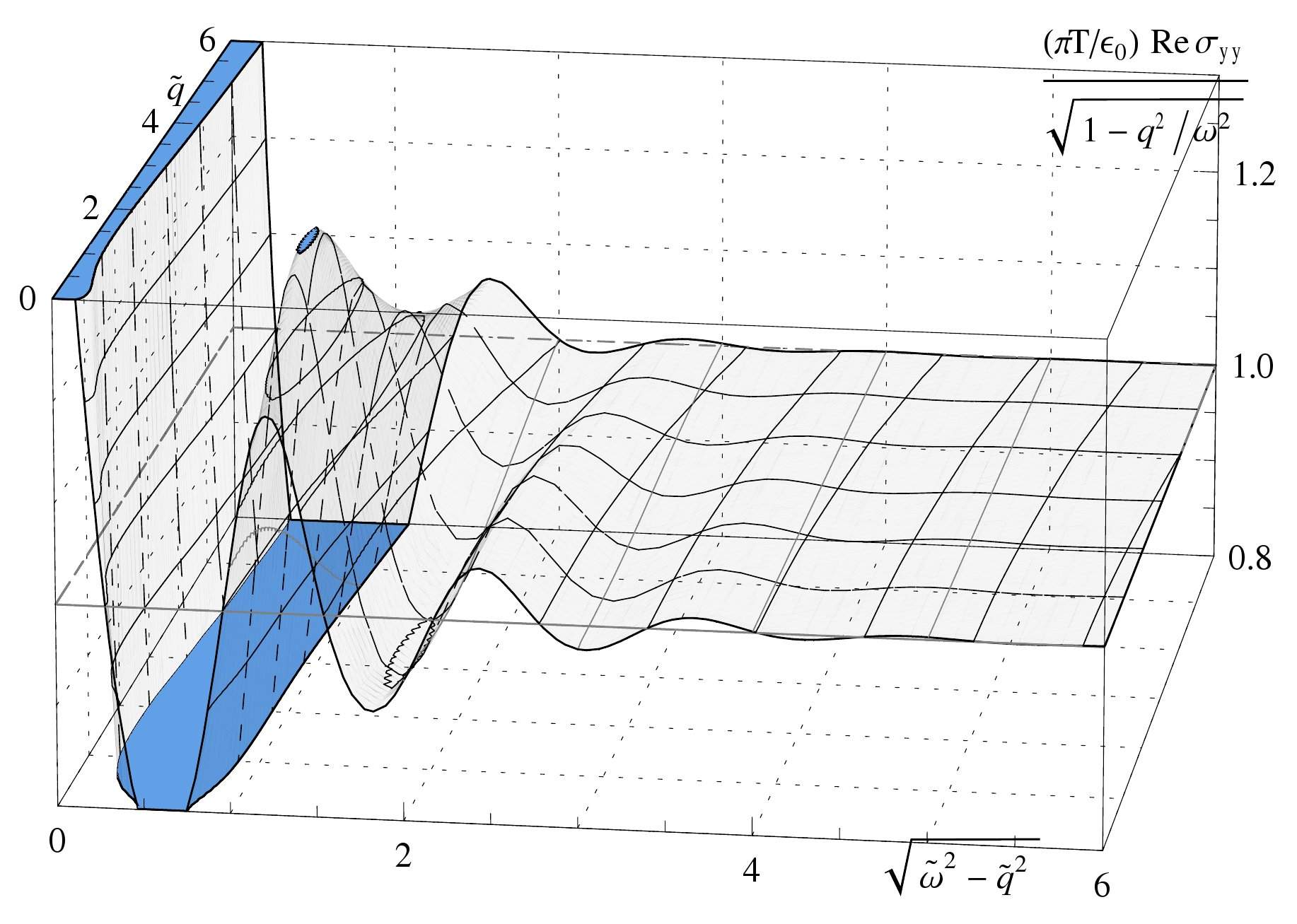}\\
\includegraphics[width=0.49\textwidth]{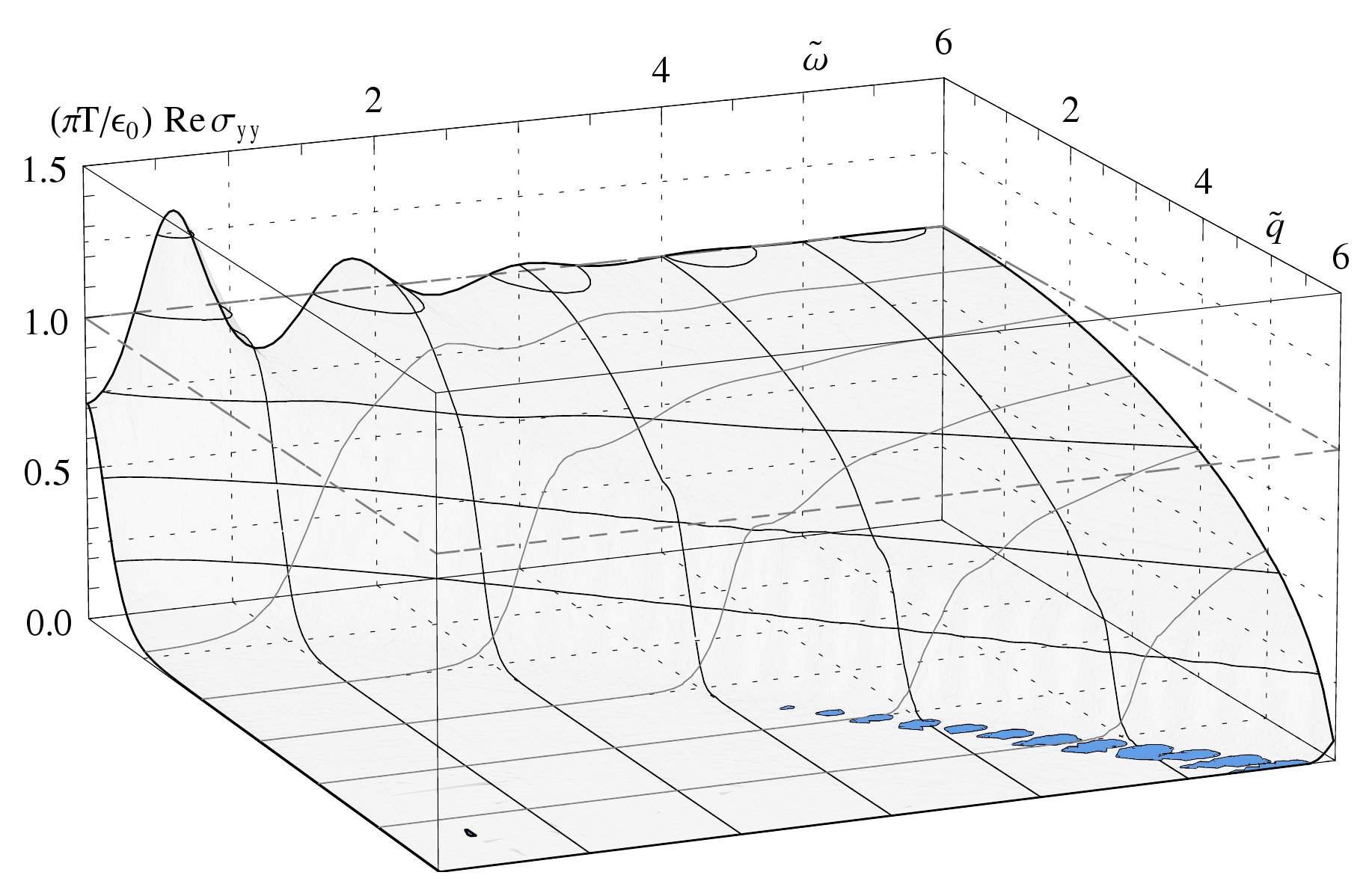}
\includegraphics[width=0.49\textwidth]{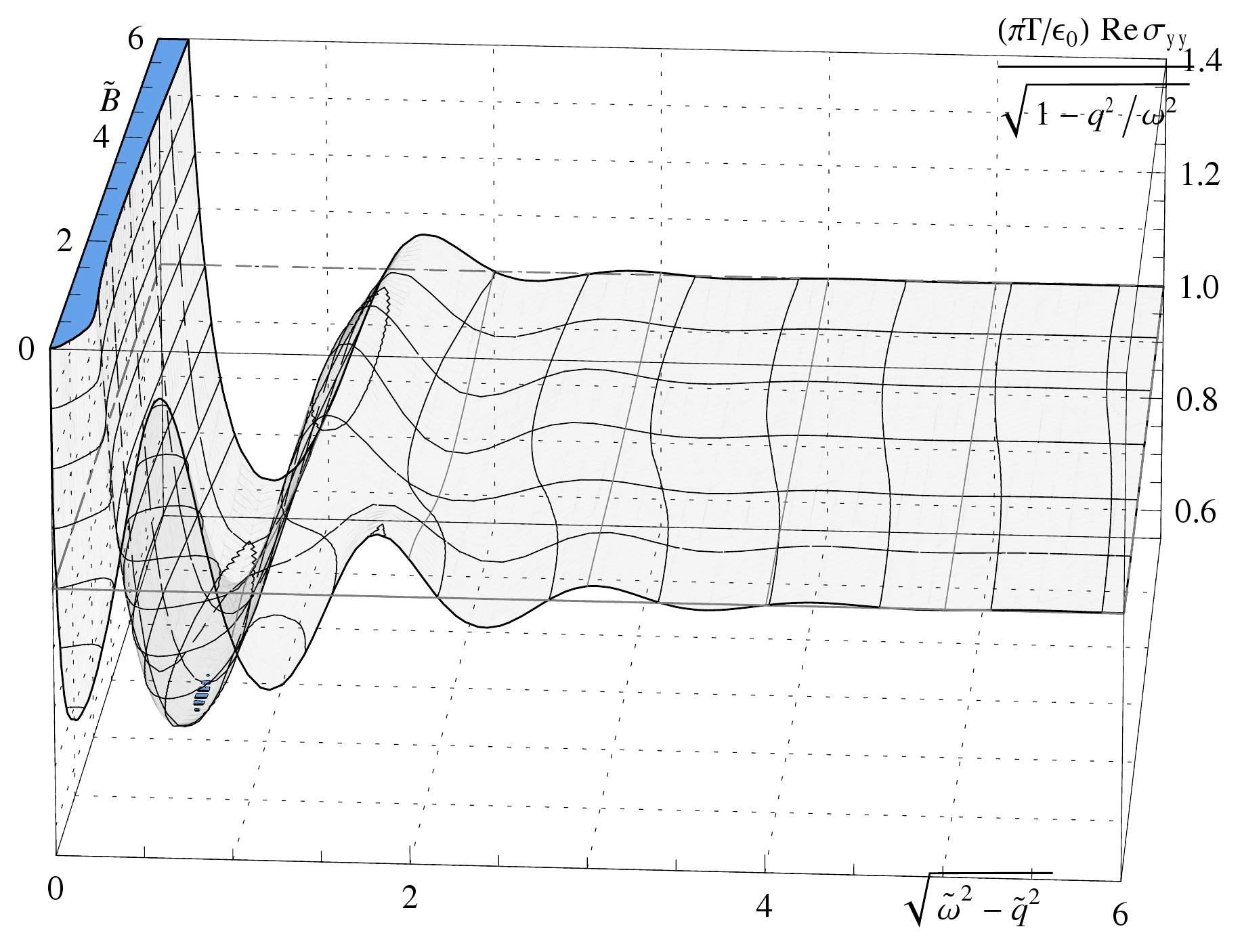}
\caption{The real part of the conductivity $\sigma_{yy}$ at varying $\tlq$, $f=2$ and $\tlrho = 4$ (top) or $\tlb = 4$ (bottom) 
 as a function of $\tom$ (left) and rescaled as   $\frac{\tilde{\sigma}_{yy}}{\varepsilon_0 \sqrt{1-\tlq^2/ \tom^2}}$ as a function of the ``rest-frame'' frequency $\sqrt{\tom^2 -\tlq^2}$ (right). }
\mlabel{con3d_k}}
To see what happens when we turn on a finite wavenumber of the perturbations, we look at fig. \mref{con3d_k}, where we show a few of the higher resonances at $f=4$ in order to see  how they  depend on the wavenumber $\tlq$. Looking at the plots on the left, there seems to be only a small difference between the behavior of the Landau levels and plasmons. This difference becomes however very significant when one plots the ``normalized'' conductivity, $\frac{\tilde{\sigma}}{\sqrt{1-\tlq^2/\tom^2}}$ as a function of the ``rest-frame frequency'' $\sqrt{\tom^2-\tlq^2}$ as it was done in ref. \refcite{baredef}. Then, we see that the density resonances connect smoothly to the resonances in the optical regime (i.e. above the conduction threshold $\tom=\tlq$) in the ``semiconductor'' case at $\tlq \gg 1$. Certainly the statement about the continuity of the pole or ``resonance'' at $\sqrt{\tom^2-\tlq^2}=0$ is somewhat meaningless, since this arises always due to the rescaling (at finite temperatures), but only says that the correlator is finite at $\omega = q$ and does not imply a pole in the correlator. The magnetic resonances, however, seem to be discontinuous -- the $n=0$ Landau level seems to disappear, when the $\sqrt{\tom^2-\tlq^2}=0$ pole arises, and the higher resonances behave in a non-monotonic way.
\DFIGURE{
\includegraphics[width=0.49\textwidth]{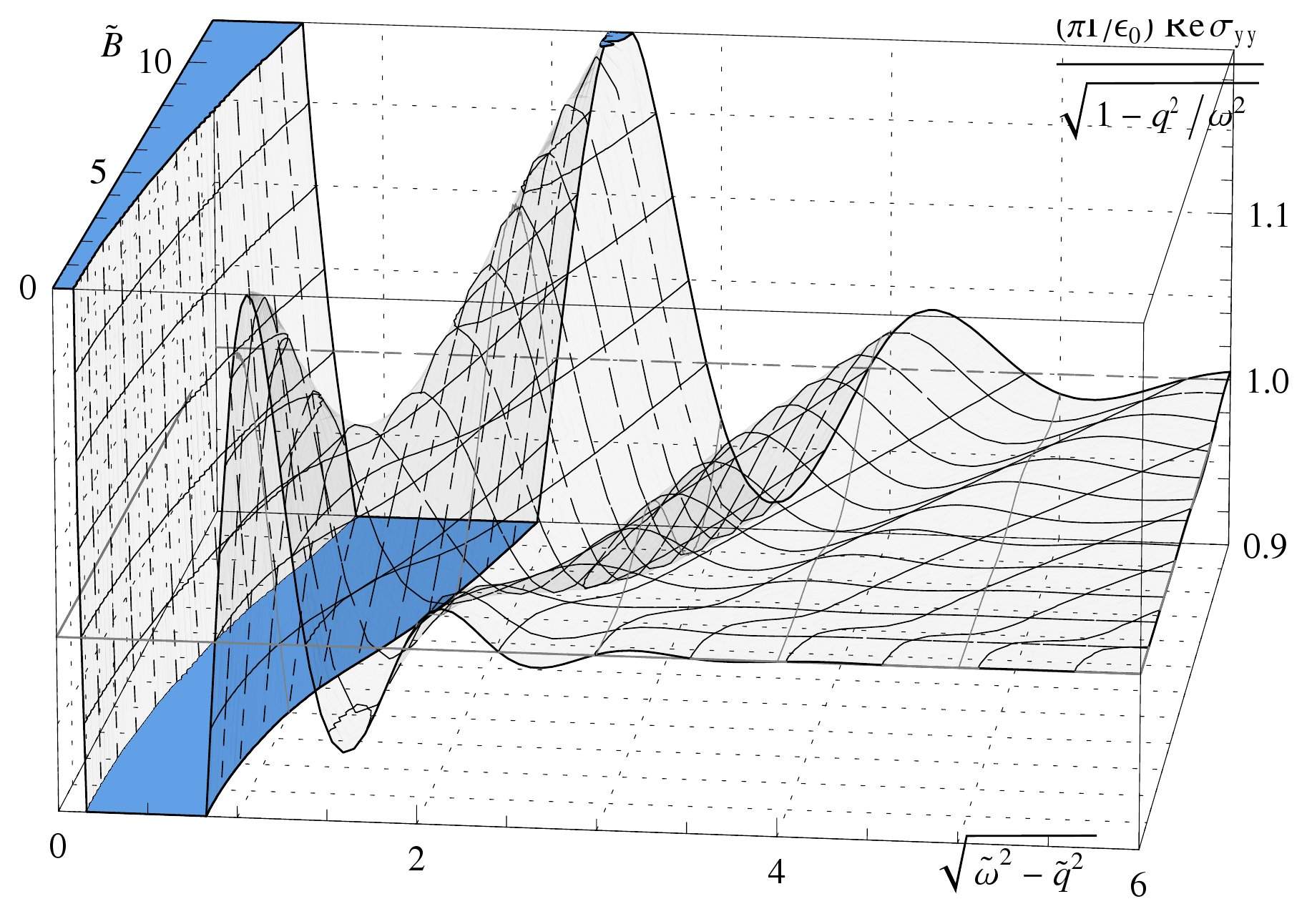}
\includegraphics[width=0.49\textwidth]{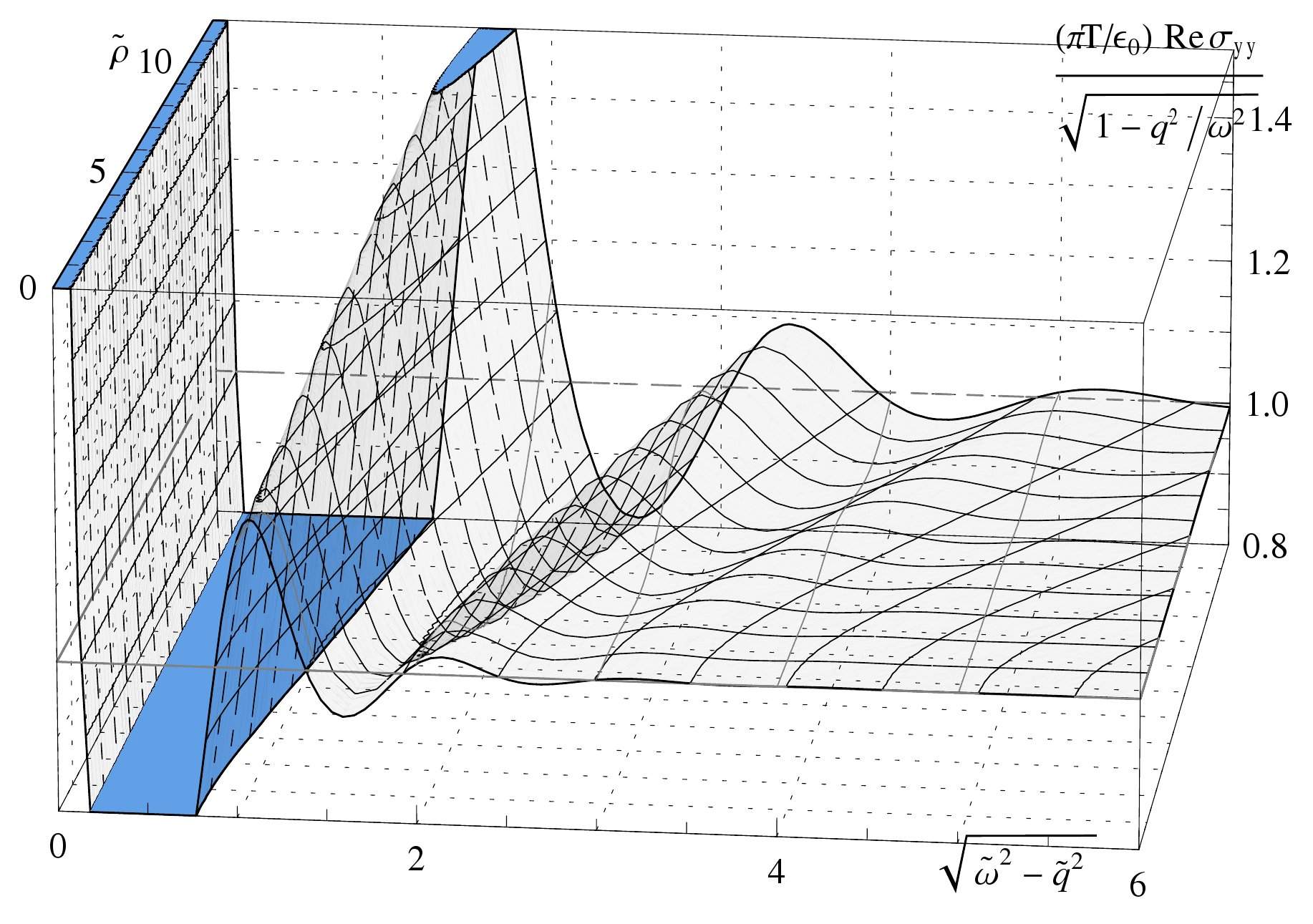}
\caption{ $\frac{\Re \tilde{\sigma}_{yy}}{\varepsilon_0 \sqrt{1-\tlq^2/\tom^2}}$ at $\tlq=\pi/2$ and $f=4$ as a function of the ``rest-frame'' frequency $\sqrt{\tom^2 -\tlq^2}$ . Left: As a function of the magnetic field. Right: As a function of the density.}
\mlabel{con3d_fk}}
In order to see more in detail where this discontinuity comes from, we can look at the $\tlb$ and $\tlrho$-dependence at a finite wavenumber $\tlq=\pi/2$ and finite $f=4$ in figure \mref{con3d_fk}. There we see that we start off with the ``bare defect'' and its finite-width resonances, and as we turn on the net baryon density, they shift smoothly, as if we were to decrease the width of the defect. 
As we turn on a magnetic field, while there is still no apparent splitting of resonances -- as one might expect if new kinds of resonances are turned on, they are not monotonically connected. This implies that there are some non-monotonous changes in the residue and location of the poles.
It can be easily seen from the electromagnetic duality in the plain defect, that at wavenumber $\tlq=0$, there can be only one pole, which is at $\tom=0$, and hence, assuming continuity, the residue of the poles from the finite-width resonances must be proportional to $\tlq$. On the gravity side, this corresponds to the fact that at $\tlq=0$ there is only one mode function in the gauge field and the equations for $A_y$ and $\mathcal{A}_x$ are the same, but at finite $\tlq$, the equations for $A_y$ \reef{fyyeq} and $\mathcal{A}_x$ \reef{fxxeq} become different. Hence we find two distinct mode functions. The same argument applies for turning on $\tlrho$ or $\tlb$. This also reflects the fact that generically, the density of states of Landau levels \reef{landens} is proportional to the magnetic field.

A rough, argument in the field theory is that turning on $\tlq$ corresponds to introducing an inhomogeneity in the $x$ direction. Hence, the $U(1)$ perturbations become localized in that direction, whereas they are not localized in the $y$ direction.
Plasmons are not generically localized, so they do not change this configuration. Landau levels however are intrinsically localized quasi-particles, so they break translation invariance also in the $y$ direction and change the pattern of the resonances less smoothly. From EM duality, we know that magnetic resonances in $\sigma_{xx}$ connect smoothly to the finite-width resonances and the density resonances connect less smoothly. This is precisely because in the $x$ direction the translation invariance of the plasmons becomes broken by finite $\tlq$, whereas the Landau levels were already localized.

\DFIGURE{
\includegraphics[width=0.49\textwidth]{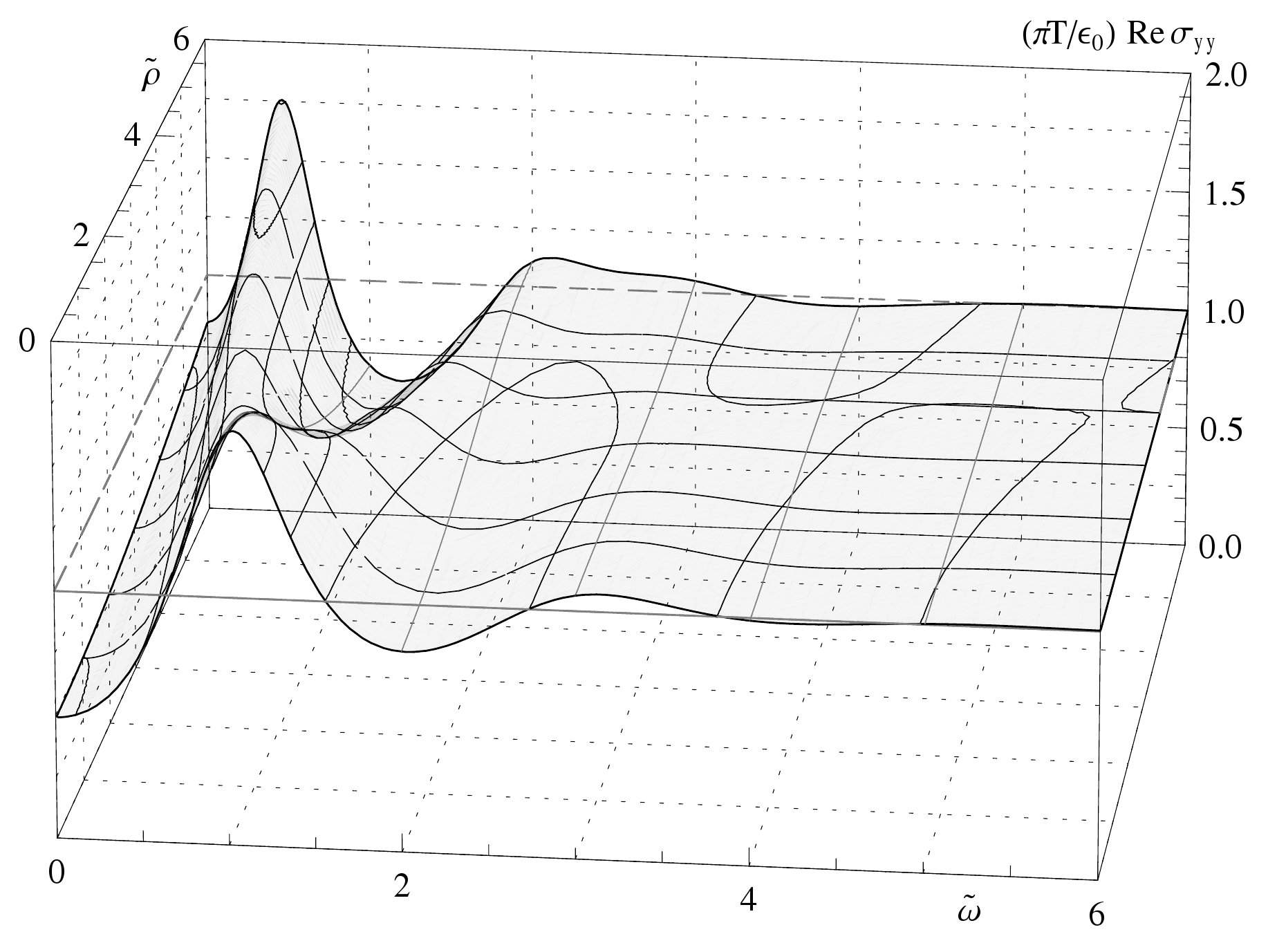}
\includegraphics[width=0.49\textwidth]{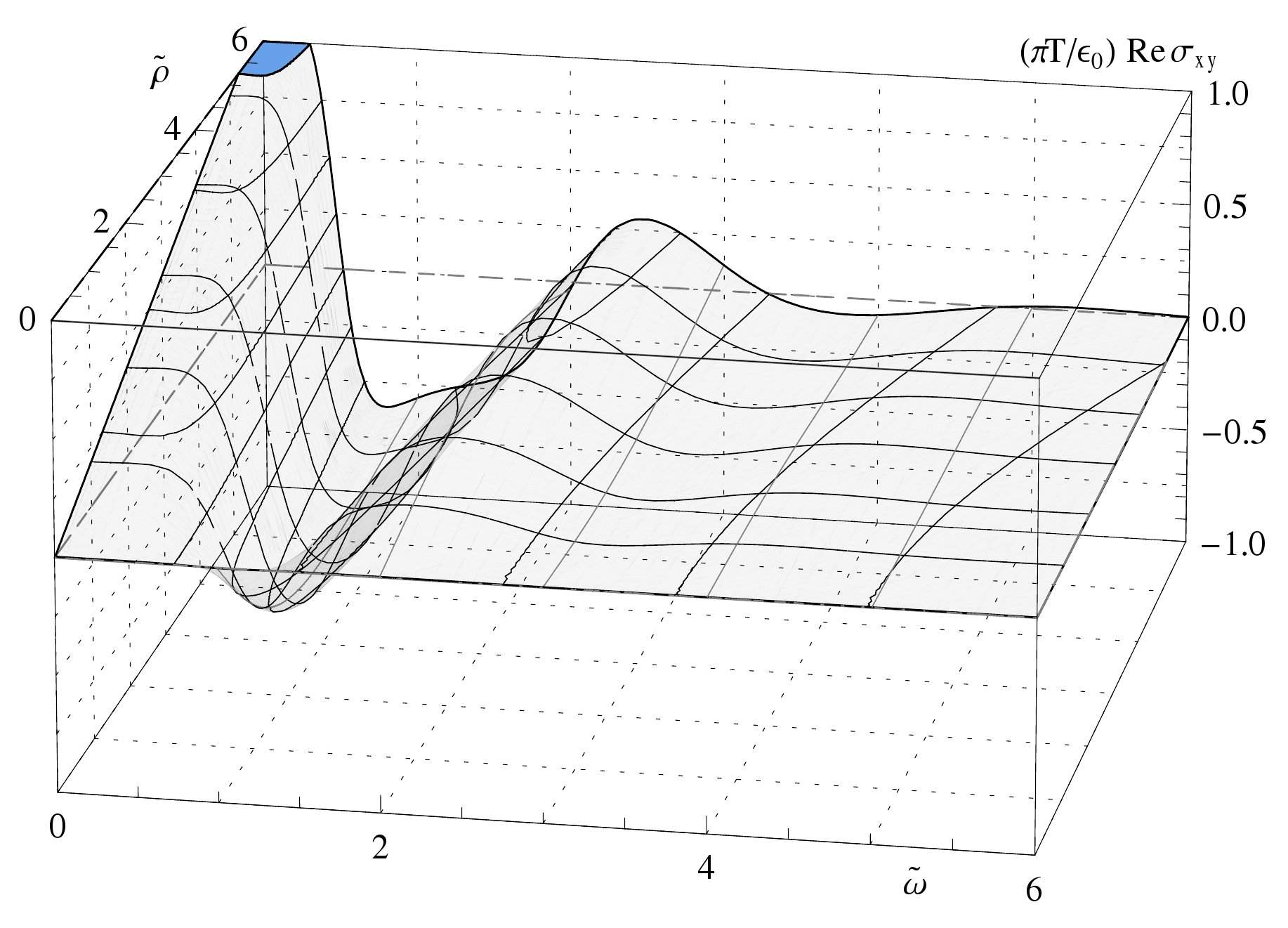}\\
\includegraphics[width=0.49\textwidth]{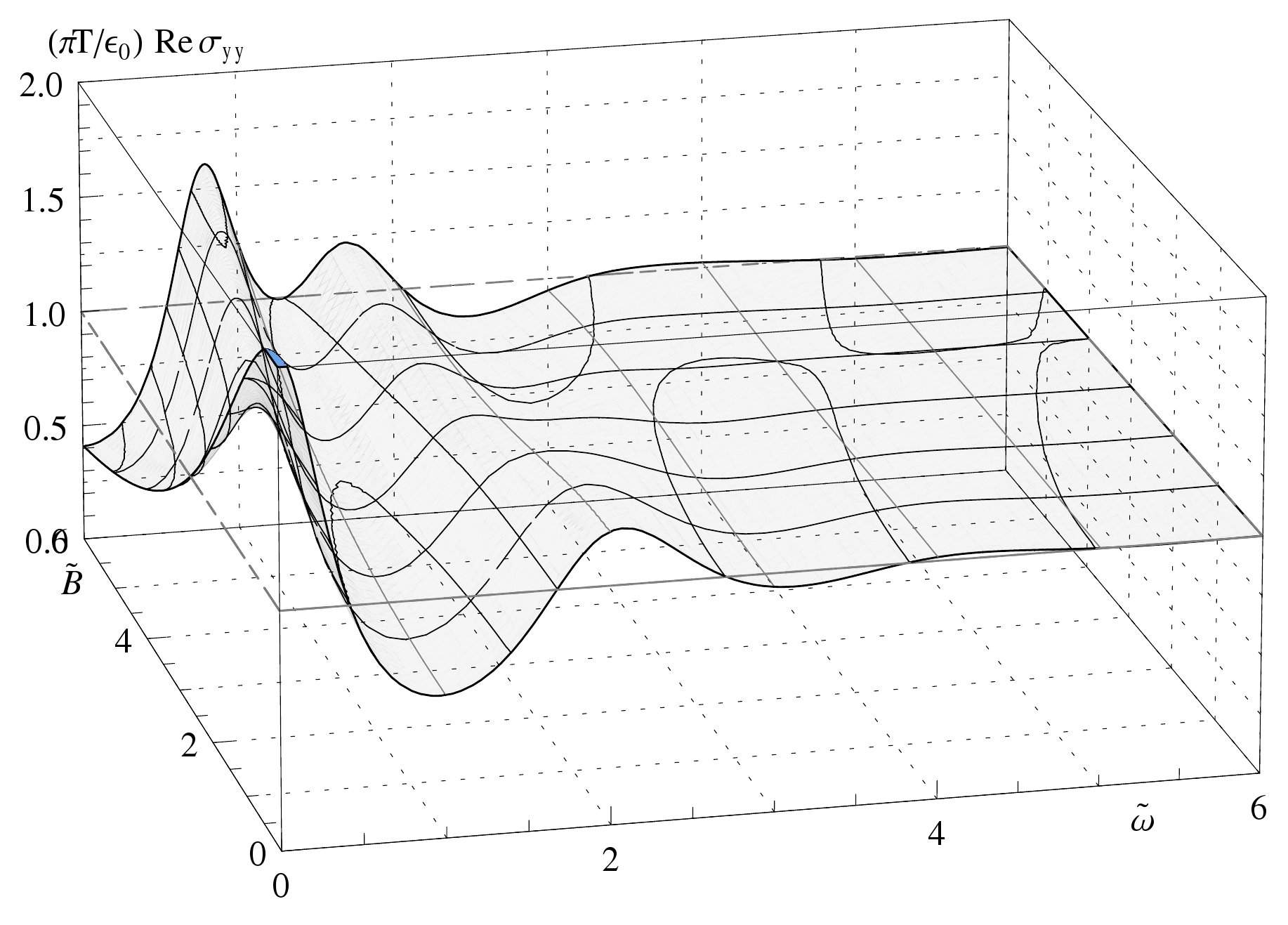}
\includegraphics[width=0.49\textwidth]{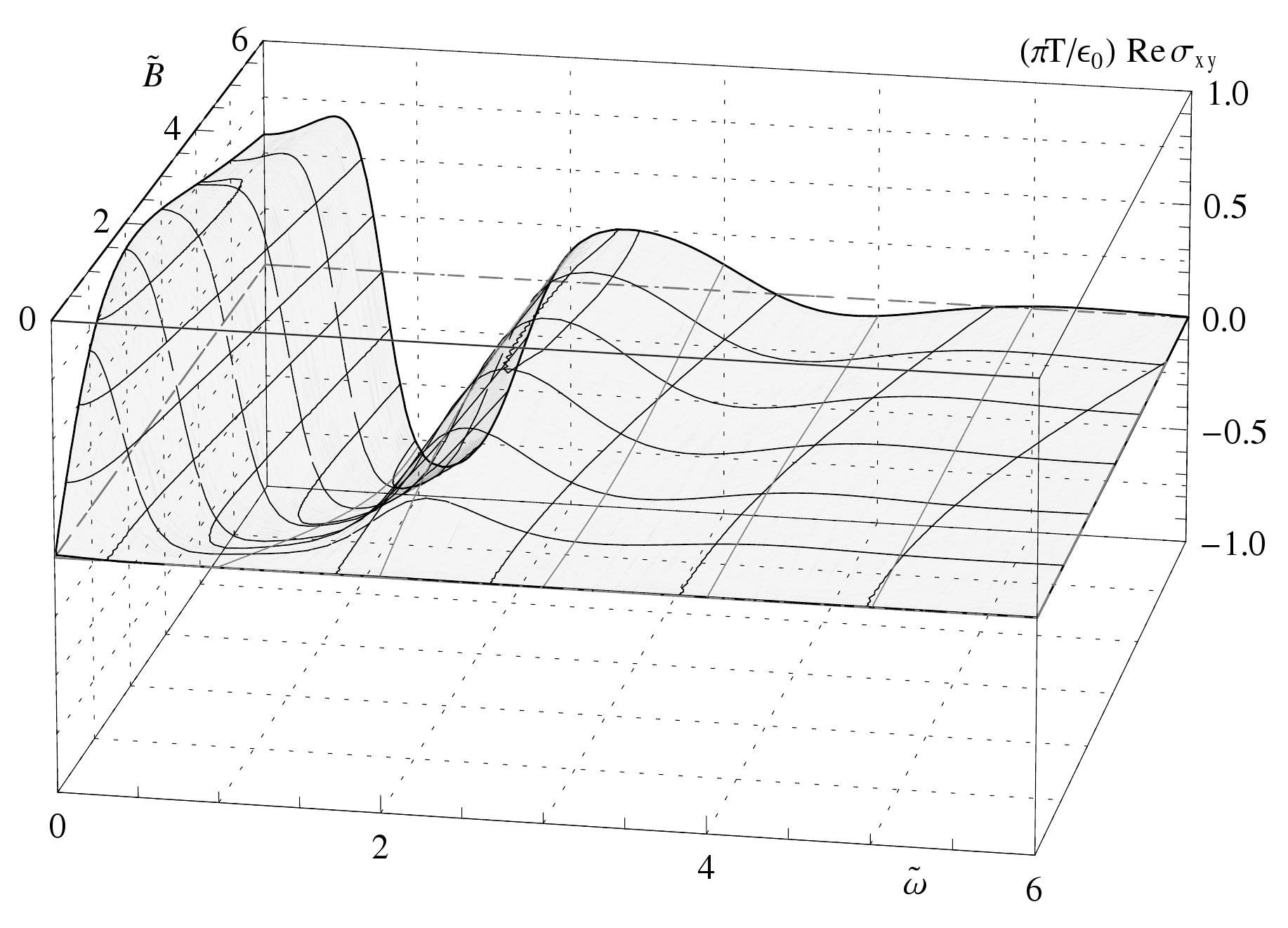}
\caption{The real part of the isotropic conductivity at $\tlq=0$, $\tlb=4$ and $f=2$ as a function of frequency and density (top) and at $\tlrho = 4$ as a function of the magnetic field (bottom). Left: diagonal part of the conductivity tensor. Right: Hall conductivity.}
\mlabel{con3d_hall}}
After studying the effects of having either $\tlb$ or $\tlrho$ turned on, let us look at the case when they appear simultaneously in fig. \mref{con3d_hall}. In the diagonal part of the conductivity tensor, we see that the magnetic- or density resonances split in two as we turn on a net density or magnetic field, respectively. It is interesting, that there is no ``tower'' of excitations splitting off each resonance and that the mean frequency of each ``split level'' charges only by a small amount.
Furthermore, we find that at each resonance in the diagonal conductivity, the Hall conductivity changes sign, at least for the first two resonances. This is just the continuation of what one expects classically for the first resonance as we saw in section \mref{metalcon}. It is also what one expects semi-classically, if the split states have either positive or negative magnetic moment, carrying a total net Hall current similar to the edge current in the Quantum Hall effect. By continuity, this implies that the plasmons (at zero magnetic field) and the Landau levels (at zero net ``charge'' density) have vanishing net magnetic moment and equal degeneracy  (2). It is also worth noting that  it is impossible to have any of the resonances cross $\omega=0$ no matter how much one tunes the parameters, which clearly indicates that the system has no Fermi level in the classical sense. Otherwise we would see Landau levels crossing $\omega = 0$.

\DFIGURE{
\includegraphics[width=0.49\textwidth]{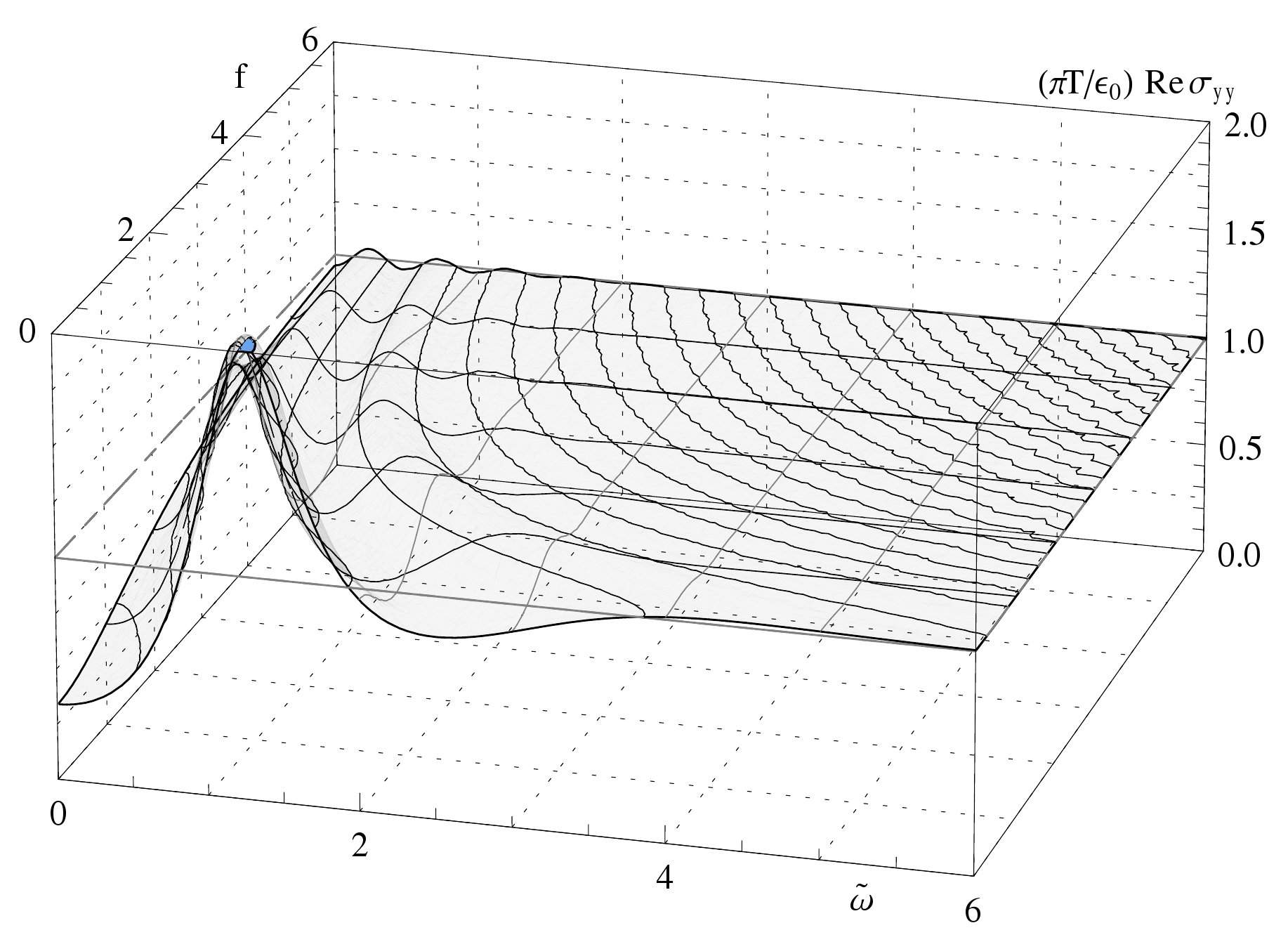}
\includegraphics[width=0.49\textwidth]{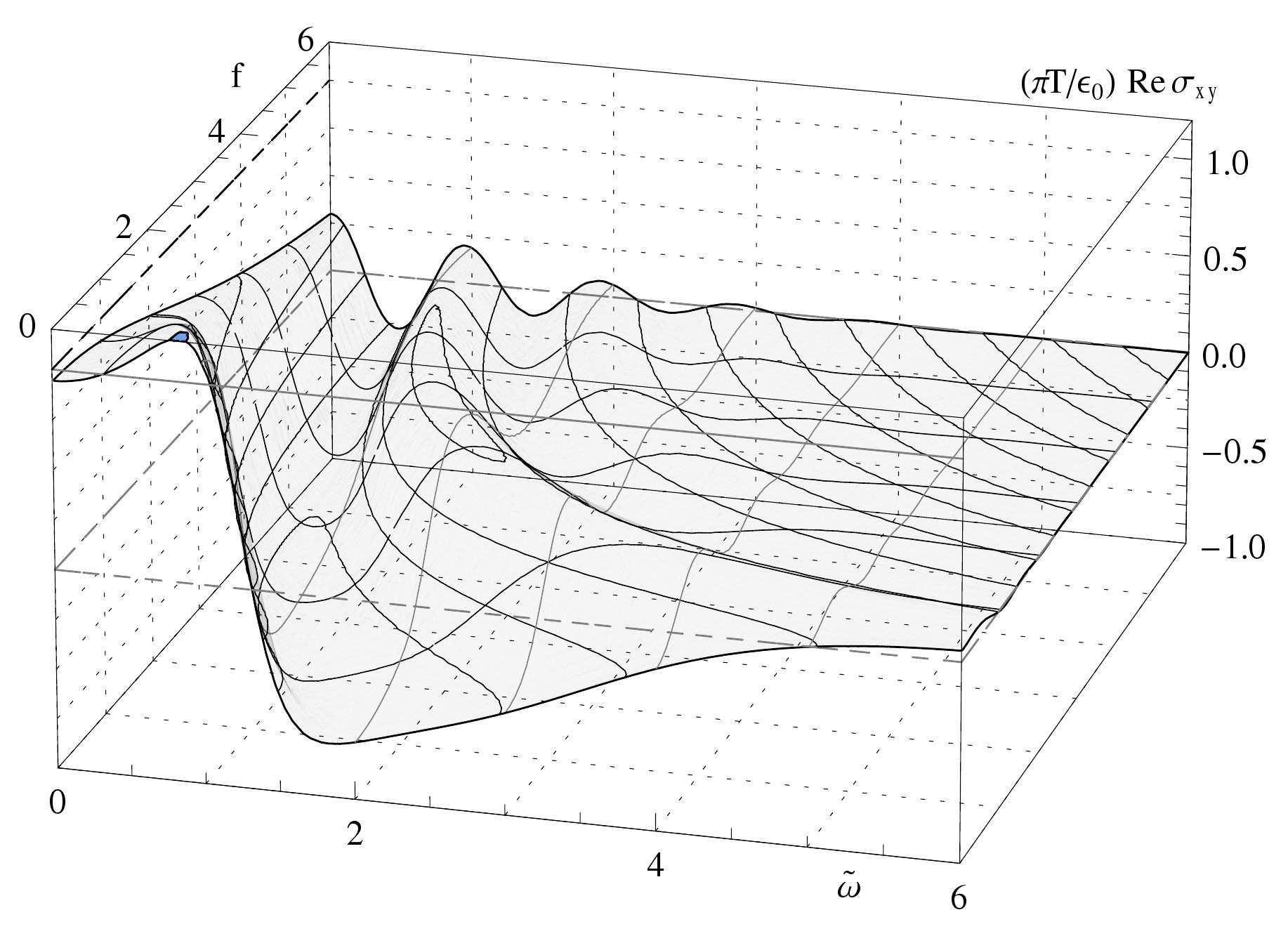}\\
\caption{The real part of the isotropic conductivity at $\tlq=0$, $\tlb=4$ and $\tlrho=4$ as a function of frequency and $f$. Left: diagonal part of the conductivity tensor. Right: Hall conductivity.}
\mlabel{con3d_hall_f}}
For completeness, we can look at the $f$-dependence of the Hall effect in figure \mref{con3d_hall_f}. This confirms our observations of the relation between the resonances in the diagonal part of the conductivity and the Hall conductivity. In the regime  of highly suppressed resonances this appears through their periodicity that differ by a factor of 2. We can also see that their frequencies roughly behave as the resonance frequencies of the plasmons and Landau levels.


%
%
%
\DFIGURE{
\includegraphics[width=0.49\textwidth]{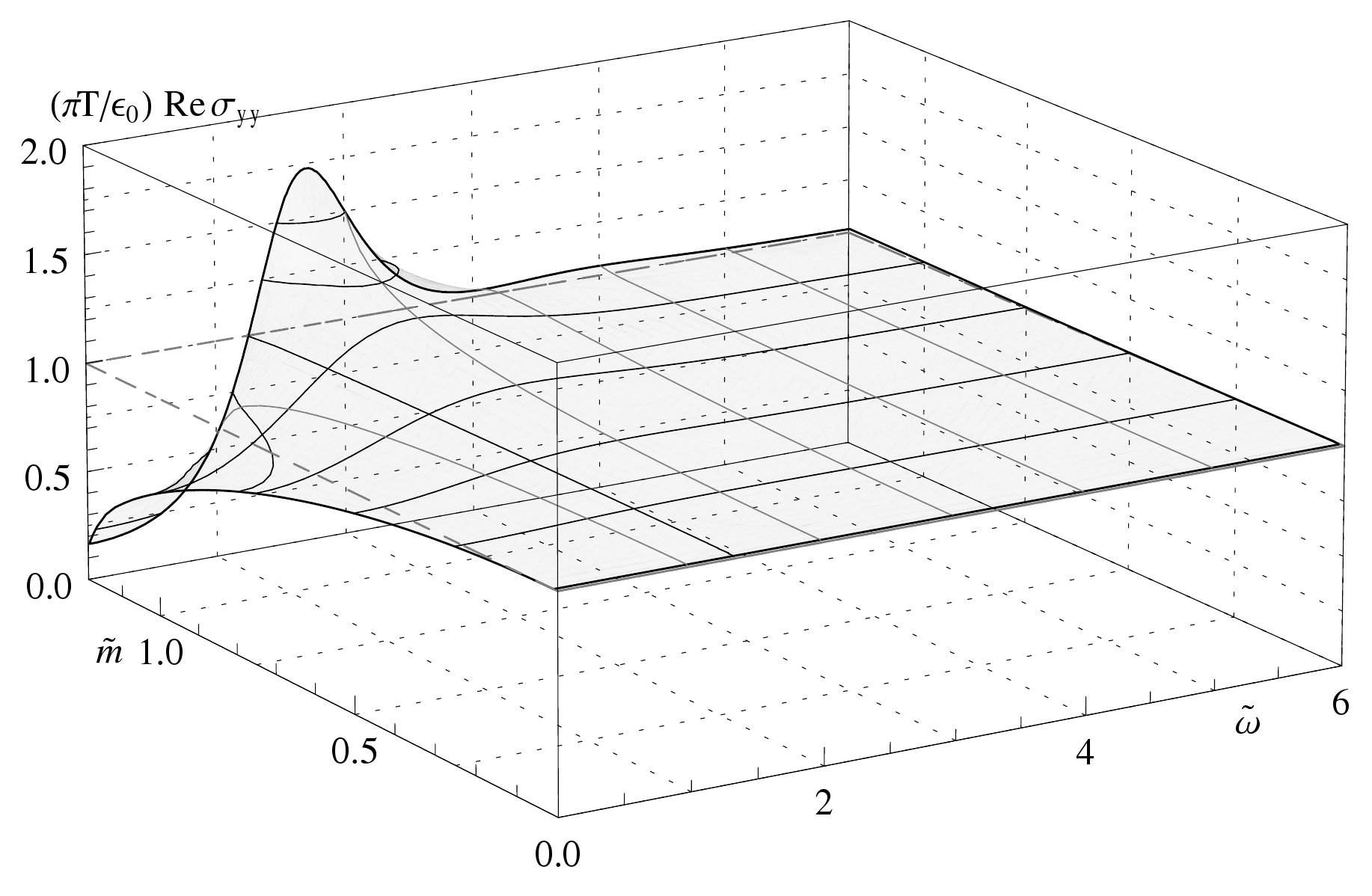}
\includegraphics[width=0.49\textwidth]{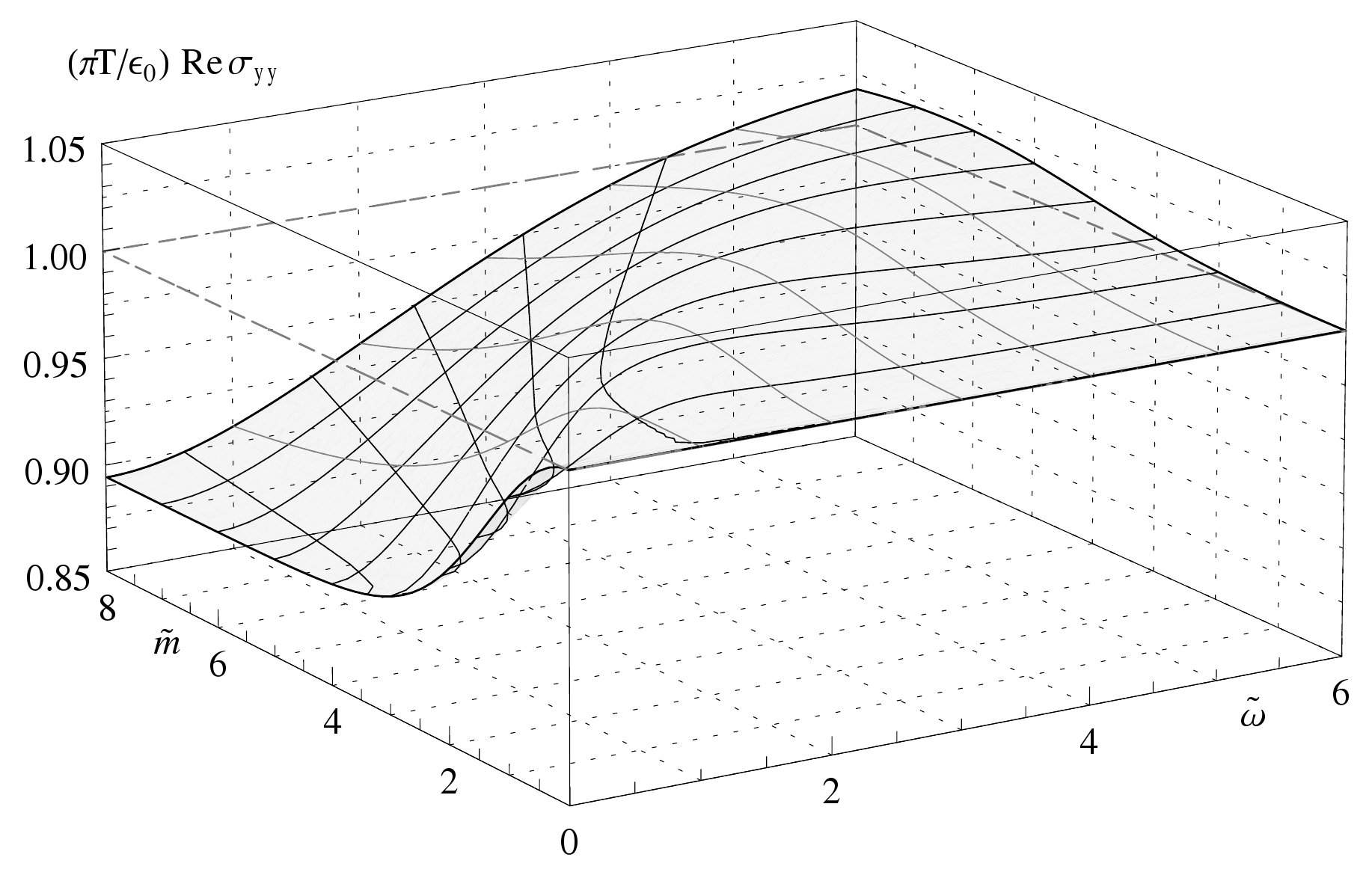}
\caption{The real, diagonal part of the isotropic conductivity at $\tlq=0$ as a function of frequency and quark mass for $f=0$, $\tlm\in [0,1.18]$ (left) and $f=0$, $\tlm\in [0,8]$ (right).}
\mlabel{con3d_m}}
Finally, let us look at the mass dependence. In fig. \mref{con3d_m}, we look at the conductivity at $\tlq=\tlrho=\tlb=0$, where we actually see the DC conductivity from \reef{DCtensor}. At $f=0$, we see a significant change of the conductivity with a resonance around $\tom \sim 1.7$ as $\tlm$ approaches the critical mass of the phase transition.
This $\tlm${}-dependence is suppressed at finite $f$, and at $f=2$, the most significant change takes place only over $\tlm\sim0\ldots4$ -- simply because it depends roughly on $f^2 + (1-\Psi^2)^2$ and not on the mass directly, such that the mass dependence becomes ``frozen'' as $\Psi_0$ becomes close to $1$. In contrast to this indirect mass dependence, the location of the very shallow maximum seems to be roughly proportional to $\tlm$. This gives some nice insight into the IR and UV dependence of the underlying physics. Processes that take place at small energies, i.e. in the IR will be dominated by gravity background near $u=1$, and hence depend on $\Psi_0$ and show most of their mass dependence in the regime of $\tlm \sim \order (1)$. Effects that depend on high energies, i.e. the UV, however depend on the background near $u=0$ and hence depend on $\tlm$ (and only to subleading order) on $\tlc$.
\DFIGURE{
\includegraphics[width=0.49\textwidth]{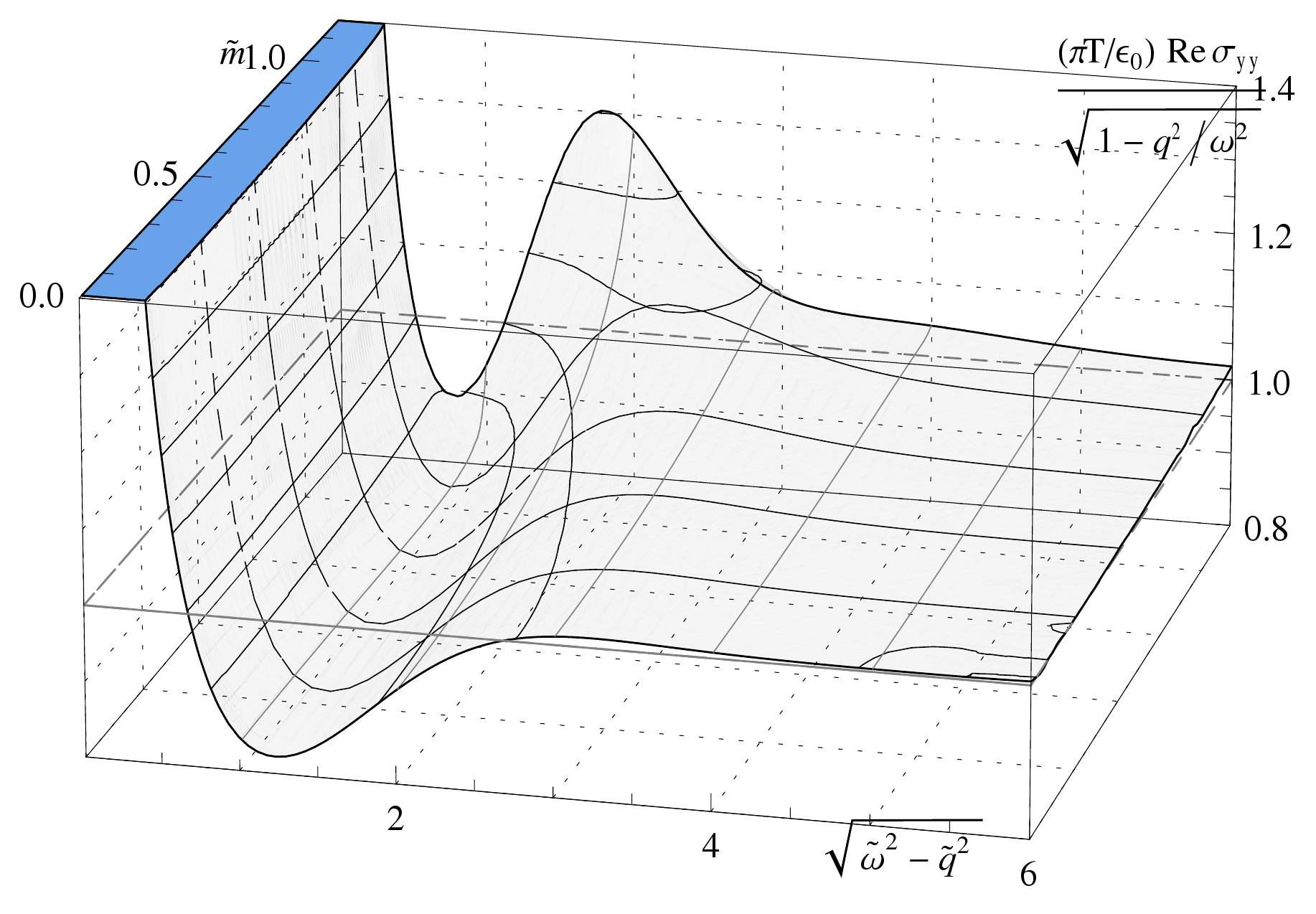}
\includegraphics[width=0.49\textwidth]{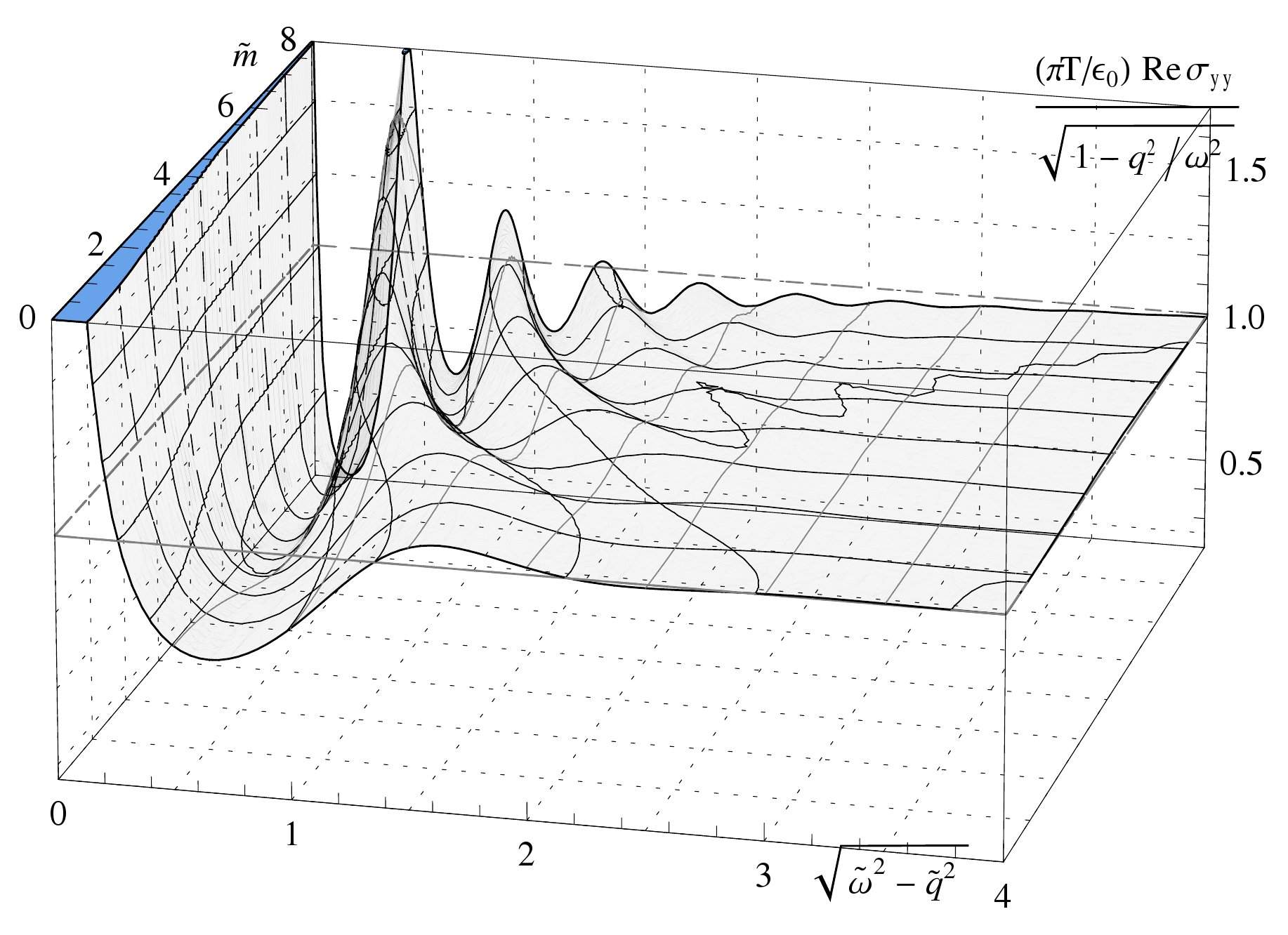}\\
\includegraphics[width=0.49\textwidth]{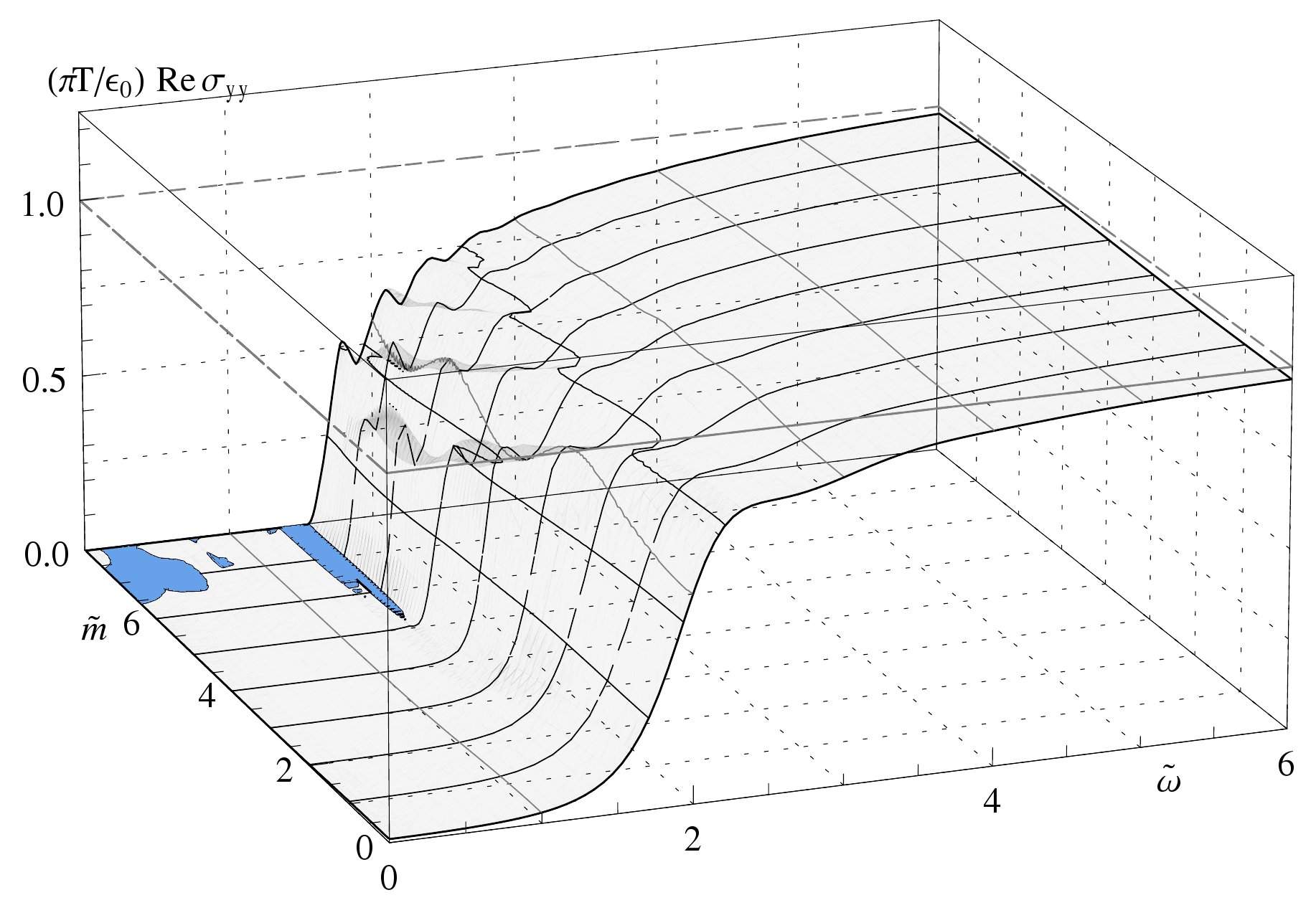}
\begin{minipage}[b]{0.49\textwidth}{\caption{Top: The real part of the normalized conductivity $\frac{\sigma_{yy}}{\sqrt{1-q^2/\omega^2}}$ at $\tlq=\pi/2$ as a function of ``rest-frame'' frequency $\sqrt{\tom^2-\tlq^2}$ and quark mass at $\tlq=\pi/2$ for $f=0$, $\tlm\in [0,1.18]$ (left) and  $f=2$, $\tlm\in [0,8]$ (right). Bottom: $\sigma_{yy}$ for $f=2$, $\tlm\in [0,8]$.}\mlabel{con3d_m_kpi2}}
\end{minipage}}
We can observe the influence of the quark mass on the finite-$\tlq$ resonances in figure \mref{con3d_m_kpi2}. There we see that the gap between the resonances is roughly proportional to $\tlm^{-1}$ at large $\tlm$ and the change starts $\propto \tlm^2$ at small $\tlm$ - as one does generically expect for a relativistic system. As naively expected, the resonances are also narrower at large $\tlm$ and their amplitude increases. If we look at the overall level of the conductivity (i.e. ignore the resonances) there seems to be the correction that we found at $\tlq=0$, now as a correction to the background around which the resonances take place at small $\sqrt{\tom^2 - \tlq^2}$. This also agrees with the picture that we see at $f=0$. Looking at the un-scaled conductivity, at the bottom in fig. \mref{con3d_m_kpi2}, we also see that the conductivity approaches the $t\rightarrow 0$ limit, $\tilde{\sigma}_{yy} = \varepsilon_0 \Re \sqrt{1-\tlq^2/\tom^2}$ as we increase the quark mass.

\DFIGURE{
\includegraphics[width=0.49\textwidth]{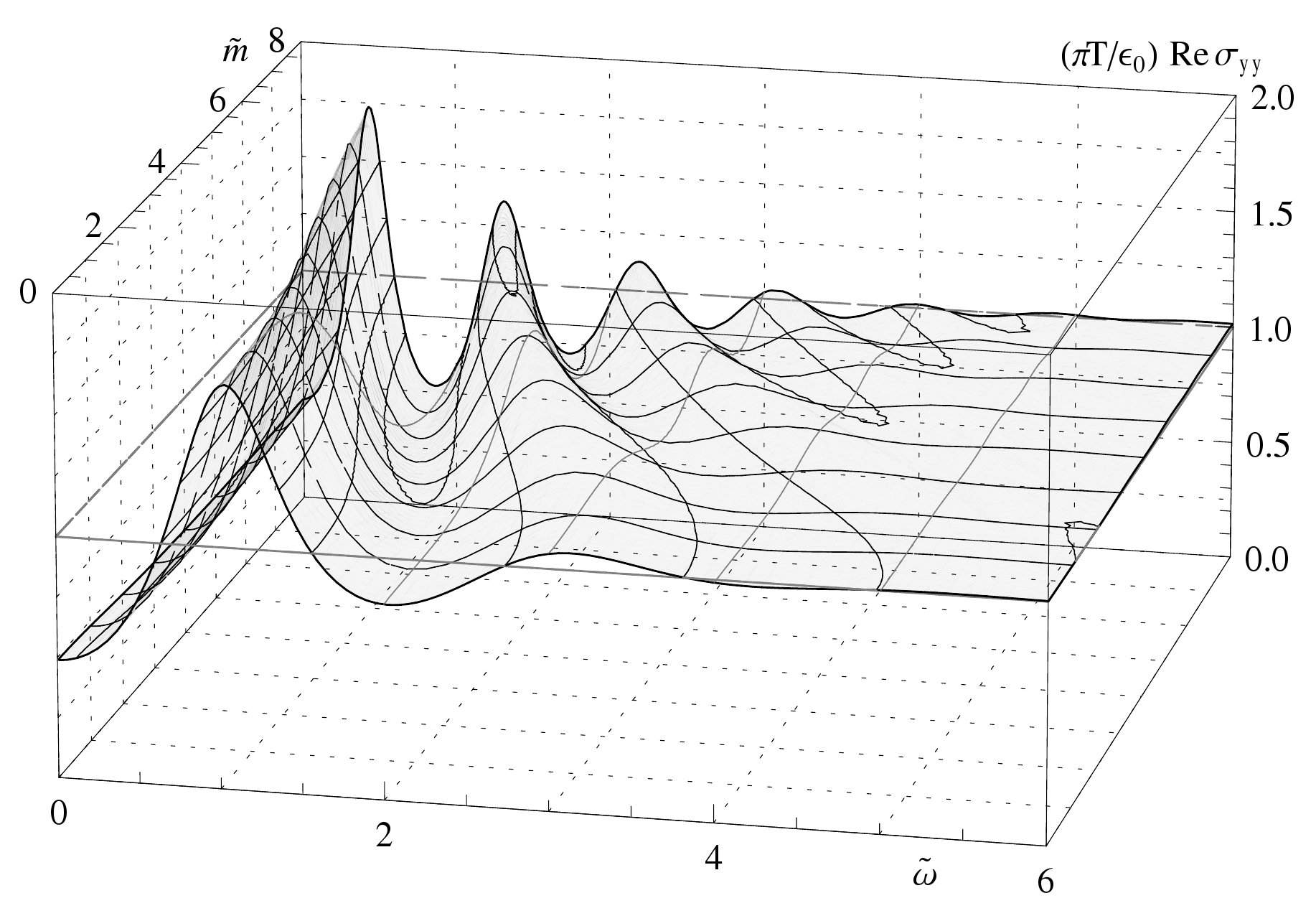}
\includegraphics[width=0.49\textwidth]{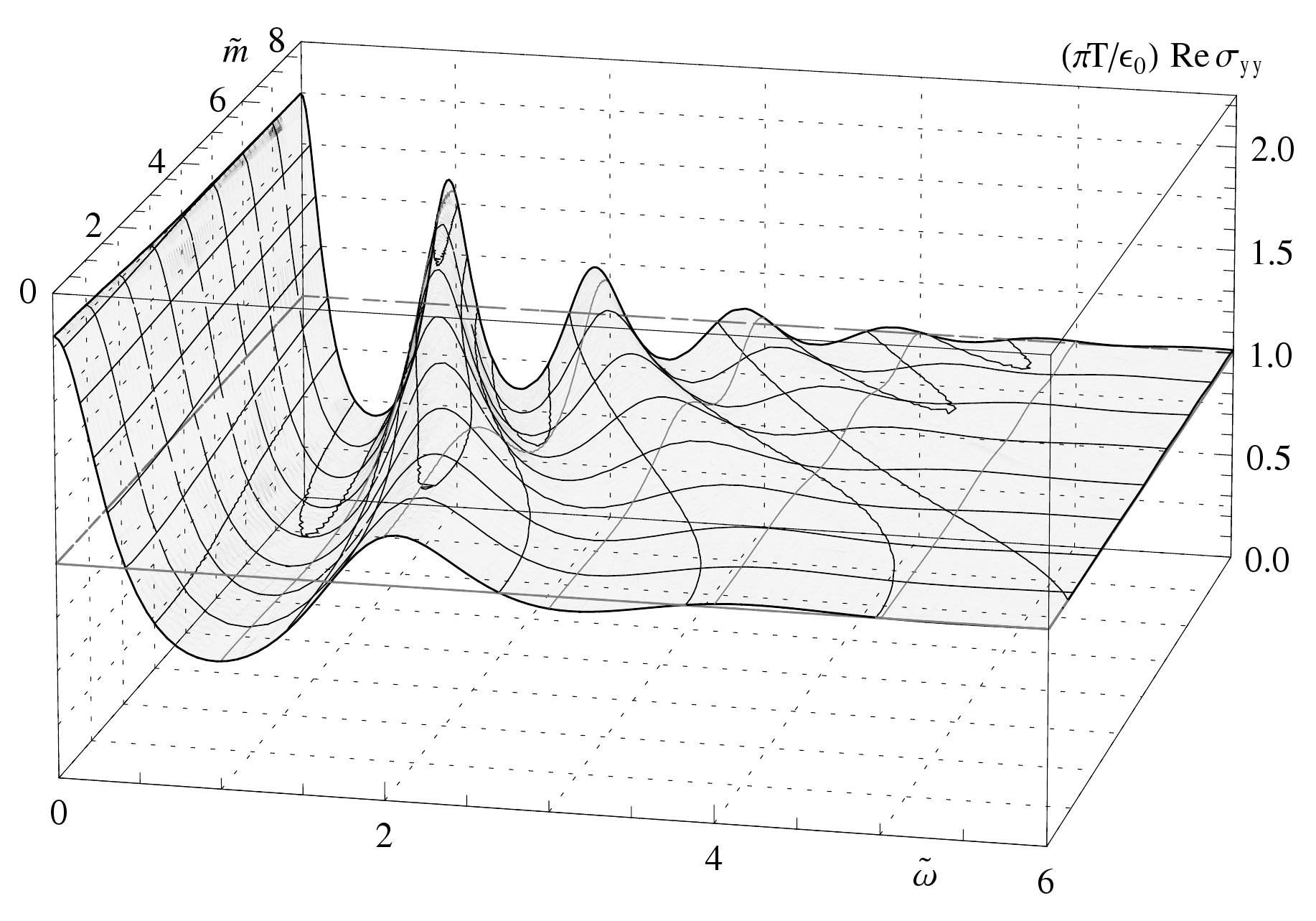}
\caption{The real, diagonal part of the isotropic conductivity at $\tlq=0$ as a function of frequency and quark mass at $f=2$ for $\tlb=4$ (left) and $\tlrho = 4$ (right).}
\mlabel{con3d_m_eb}}
\DFIGURE{
\includegraphics[width=0.49\textwidth]{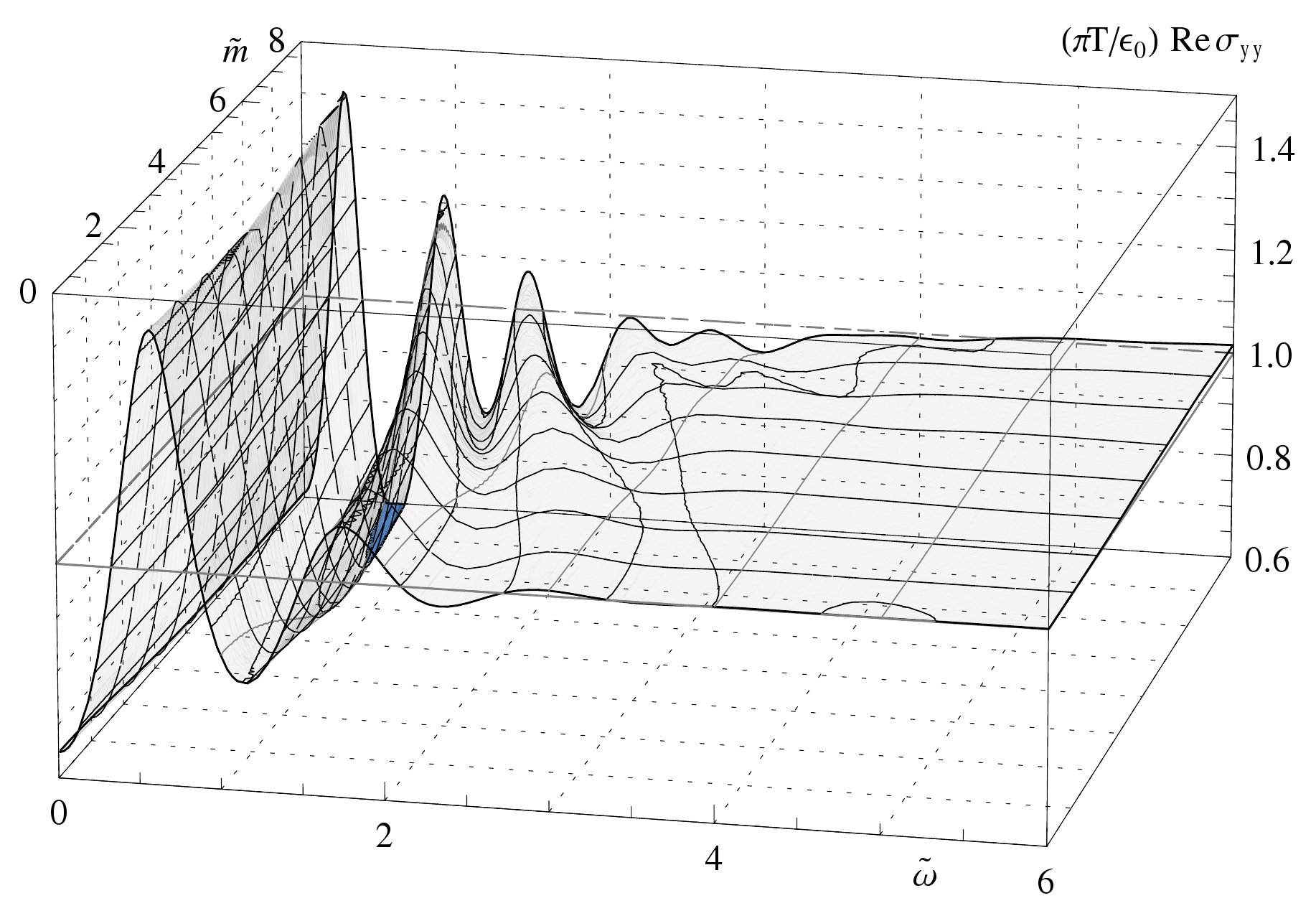}
\includegraphics[width=0.49\textwidth]{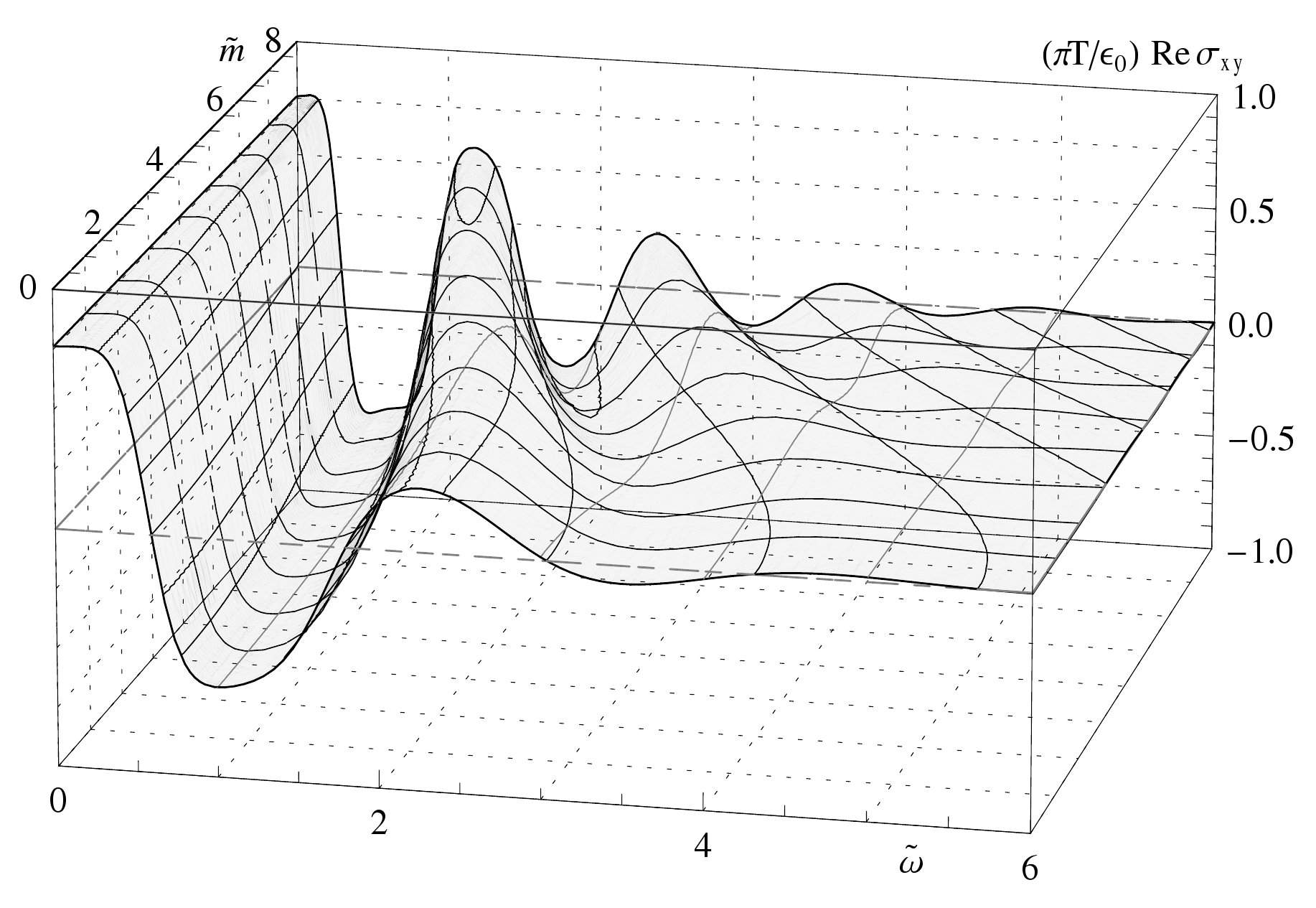}\\
\includegraphics[width=0.49\textwidth]{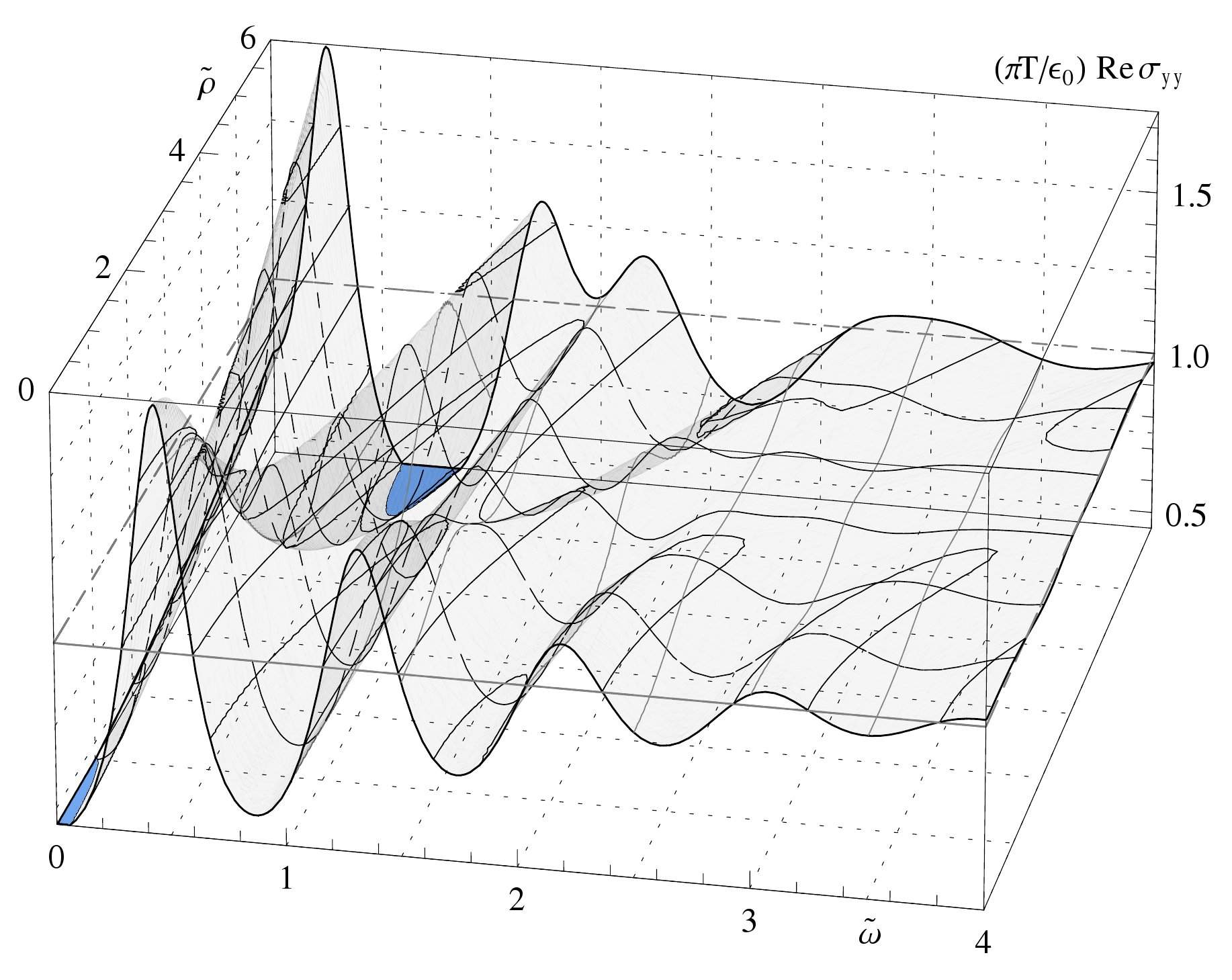}
\includegraphics[width=0.49\textwidth]{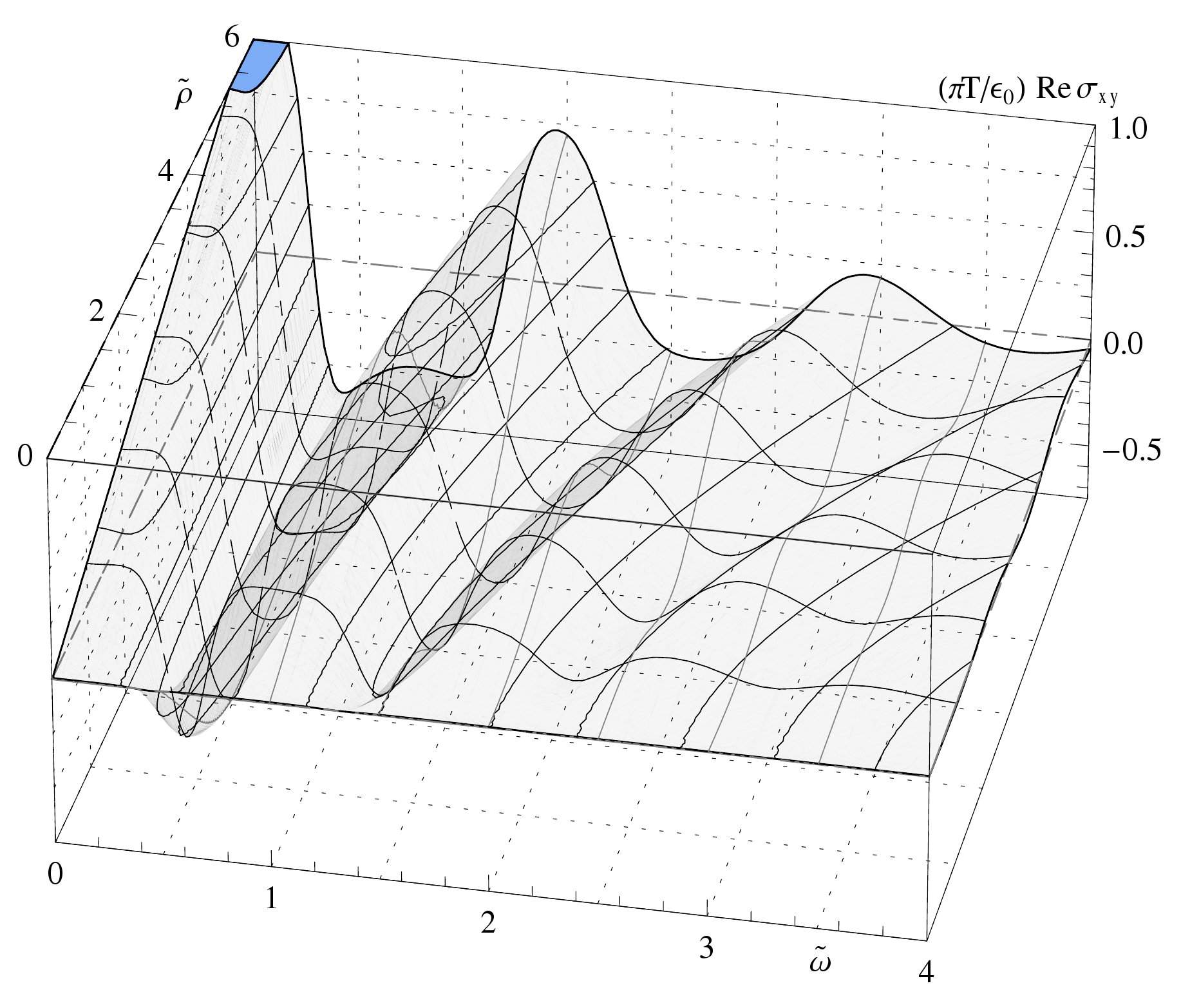}\\
\caption{The real, diagonal (left) and Hall (right) part of the isotropic conductivity at $\tlq=0$ and $f=2$. Top: As a function of $\tlm$ at $\tlb=4$, $\tlrho=4$. Bottom: As a function of $\tlrho$ at $\tlb=4$ and $\tlm=8$.}
\mlabel{con3d_m_hall}}
These generic effects of turning on $M_q$ can also be seen in the plasmons and Landau levels, and in the Hall effect, in figures \mref{con3d_m_eb} and \mref{con3d_m_hall}. Again, we see that on the one hand, the resonances become more stable at large $\tlm$, and on the other hand that the energy levels receive at small $\tlm$ a small correction $\propto \tlm^2$ and at large masses scale $\propto m^{-1}$, just like $\omega_c$ and $\omega_p$ do classically.

\HFIGURE{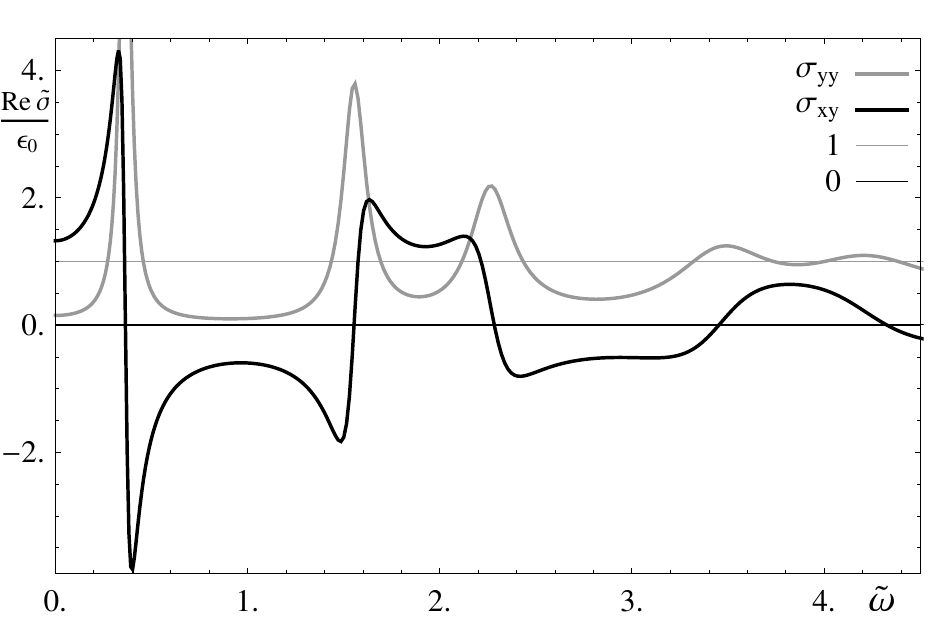}{$\sigma_{yy}=\sigma_{yy}$ and $\sigma_{xy}$ at $\tlq=0$, $\tlm=32$, $\tlrho=32$ and $\tlb=24$.}{hall_pole}
Finally, we can turn on a large mass (in this case $\tlm=32$) in order to study the structure of the Hall effect more rigorously. In fig. \mref{hall_pole}, we see that the Hall conductivity has a small overall positive (or negative if we rather look at $\sigma_{yx}$ or negative $\tlb \tlrho$) background, and there are poles with alternating residue, each precisely located at a maximum of the diagonal part of the conductivity. This supports exactly our suggestion above that the Hall current is carried collectively by localized states with net positive or negative magnetic moment.


\subsection{Small frequency regime}\mlabel{diffnrelax}
In this section, we look at the behavior of the purely dissipative poles of the correlator $C_{xx}$ on the imaginary axis, that dominate the conductivity at small frequencies and wavenumbers. Our particular interest is how they influence the DC conductivity and how the transition to ``semiconductor-like'' in the quasiparticle regime at larger wavenumbers occurs, i.e. how those poles disappear.

The numerical strategy behind locating the poles is reasonably straightforward. First, we divide the imaginary axis in three regions, based on an educated guess, and localize the poles in these regions in a simple recursive process at some initial wavenumber, magnetic field, quark mass and density. Then, we can identify regions around those poles that allow us to ``track'' them as we change the parameters, without having to scan the whole imaginary axis. One caveat though is, that it is numerically increasingly difficult to find the poles as their residue decreases, so we keep a minimum wavenumber (we will use $\tlq \geq 0.02$) to always find the ``middle'' pole. We may also ``lose track'' of poles if their residue becomes too small. The other caveat is that with our rudimentary method, we need to filter the result afterwards for whether a suspected pole is a pole or just a local extremum or noise. In most cases the distinction is obvious, but in some cases we will look at the value of the residue that we estimate. Furthermore, since this process is reasonably numerically intensive, we will limit the computations to a few examples.
%

Applying accurate numerics, we find that there are only at most three purely dissipative poles; the diffusion pole at small imaginary values of $\tom \ll 1$, and two more rapidly decaying poles at $\tom\, \order (1)$.
We recall that in ref. \refcite{baredef}, there were found the first two of those three poles -- the diffusion pole and one corresponding to decay on thermal scales. Those poles were found to move along the imaginary axis as we increase the wavenumber, meet at some critical wavenumber, and for more short scale excitations turn into massive quasi-particles.
Obviously, at $\tlq=\tlb=\tlrho = 0$, there can only be the diffusion pole because then the electromagnetic duality together with isotropy restricts $\varepsilon_0^{-1}\tilde{\sigma}\in \{-1,1\}$. As we tune these quantities to zero, the residue of both poles vanishes. One of them just disappears to a constant conductivity, while the other one turns into a unit step function of frequency in the conductivity.

\DFIGURE{\includegraphics[width=0.49\textwidth]{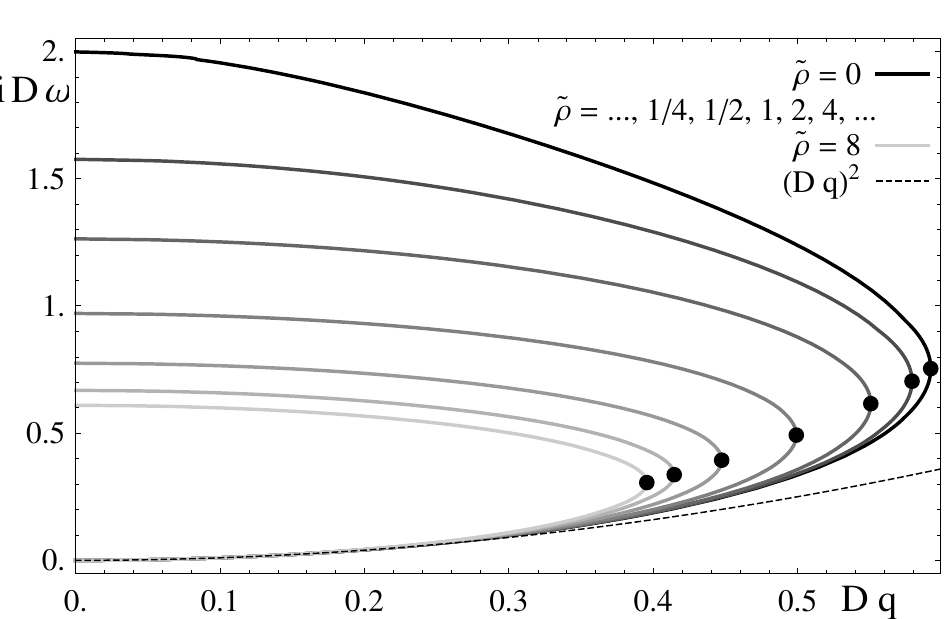}\includegraphics[width=0.49\textwidth]{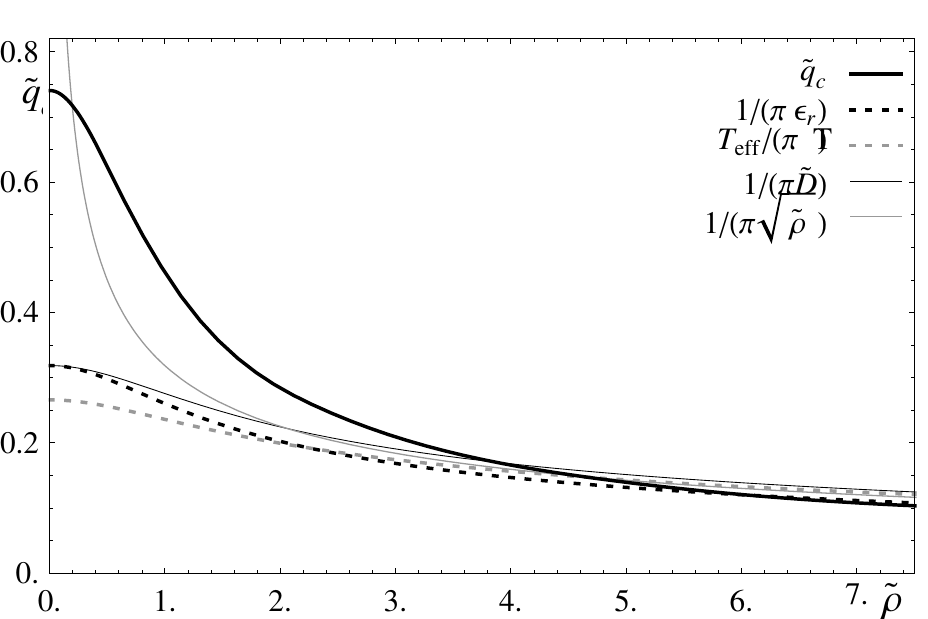}
\caption{Left: Imaginary frequency of the diffusion pole and the first higher pole as a function of $\tlq$ for different values of $\tlrho$. The frequency and wavenumber are scaled with $D$, such that the diffusion equation is $D\omega = -i (D q)^2$. The dot indicates the point where there is the branch cut, and the mode frequency of the poles gain a real part to become quasi-particles. Right: The critical wavenumber as a function of $\tlrho$, compared to various length scales of the problem: the effective temperature, diffusion constant and the electric permittivity.}\mlabel{diffu_e}}
First, let us look at the poles in the presence of a finite density. In fig. \mref{diffu_e}, we show the frequency of the diffusion pole and the second pole as a function of $\tlrho$, and we see that the lower pole follows, at small $\tlq$, the diffusion behavior, and then, at some critical wavenumber, they merge and we have again the branch cut with the transition from the dissipative to quasiparticle behavior. It is interesting to note that even beyond the diffusion behavior, the curves agree reasonably closely upon the appropriate rescaling with the diffusion constant. In fig. \mref{diffu_e}, we also plot the critical wavenumber as a function of $\tlrho$. We see that there is no length scale in our system that fits it particularly well compared to a simple $\pi^{-1}\tlrho^{-1/2}$ approximation -- even though $\frac{1}{\pi \tilde{D}}$ seems to fit best asymptotically. As in the case of the diffusion constant, we note that the critical wavenumber is proportional to $1/\sqrt{\rho}$, and $\sqrt{\rho}$ is approximately the mean separation of the quarks. So at a small net quark density, $q_c$ is dominated by scattering off gluons and quarks from the thermal equilibrium, and at a large net quark density, it is the baryon density that sets this scale.

The imaginary frequency of the third pole is slightly increasing with increasing $\tlq$, but it has a very small residue that decreases with increasing $\tlq$. Hence it can only be seen at $\tlq \sim \order(0.1)$ and we do not plot it here.

\HFIGURE{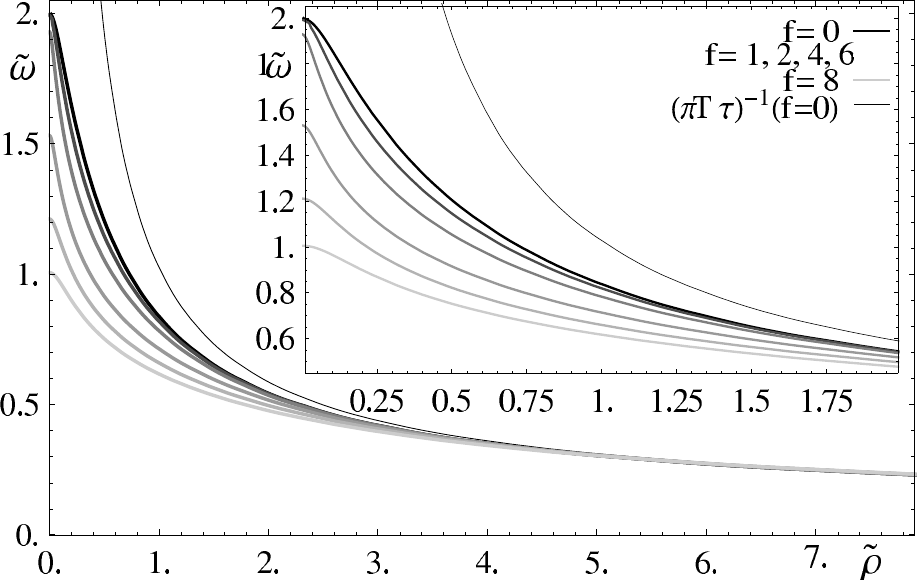}{The location of the second pole on the imaginary axis as a function of the net baryon density for various values of $f$. For comparison, we show the inverse correlation time $\tau^{-1}$ computed from the conductivity for $f=0$ only. The other curves for the relaxation time will be in worse agreement at small frequencies.}{relaxplot}

To study the nature of the second pole, we look in figure \mref{relaxplot} at how the location of that pole depends on the net baryon density. We find that this too is proportional to $\tlrho^{-1/2}$ at large $\tlrho$, and approaches some finite value at small $\tlrho$. Also, the dependence on $f$ dominates only the small-$\tlrho$ regime. In that figure, we also compare the location of this pole to the inverse of the relaxation time $\tau^{-1}$ that we obtain from $\left.\frac{\partial_{\omega}^2 \sigma_{yy}}{\sigma_{yy}}\right|_{B,\omega \rightarrow 0}$ as defined in section \mref{metalcon} and computed in \mref{smallfreq}. We find that for large $\tlrho$, they are in perfect agreement, whereas for small $\tlrho$, the inverse relaxation time diverges. For clarity, we show only the relaxation time for $f=0$. The relaxation time for larger $f$ is in worse agreement ($\tau^{-1}$ becomes larger at small $\tlrho$), however the agreement at large $\tlrho$ is equally good. This disagreement at small $\tlrho$ reflects the special conformal nature of the system at $\rho=0$, with the constant conductivity from the electromagnetic duality in ref. \refcite{baredef}. This causes the relaxation time that we computed from the conductivity to vanish, while we can obviously expect that any excitation still decays on a finite timescale as dictated for example by causality. The reason why we do not see this relaxation time in the conductivity at $\tlq = \tlrho = 0$ and hence why the constant conductivity does not violate causality is that the residue of this relaxation pole vanishes in the isotropic limit at vanishing density. Furthermore, we notice that this theme of exact convergence to the Drude model (\reef{drudecon} has precisely a pole at $\omega = - i \tau^{-1}$) at large $\tlrho$ is recurrent and was already seen in figure \mref{fruits_mag_stuff}  in section \mref{smallfreq}.
From naive intuition about weak coupling, one might be puzzled as to why the relaxation time scales as $\sqrt{\tlrho}$, i.e. proportional to the inverse of mean separation between the quarks whereas in a simple geometric weakly coupled model, the relaxation time is proportional to the mean free path, that is proportional to the density. Because this system is strongly coupled and there are long-range correlations, however, this intuition breaks down.

\DFIGURE{
\includegraphics[width=0.49\textwidth]{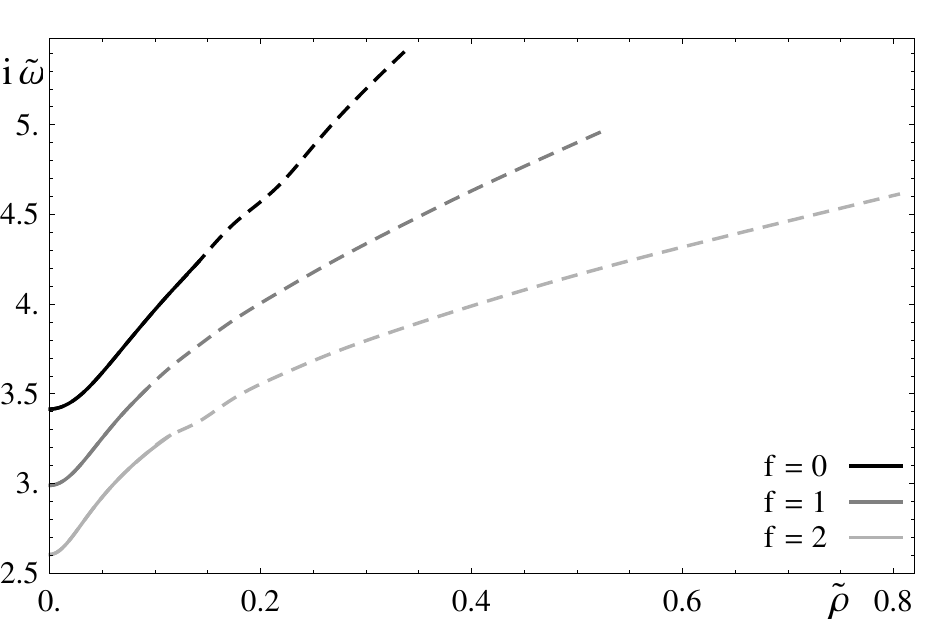}
\includegraphics[width=0.49\textwidth]{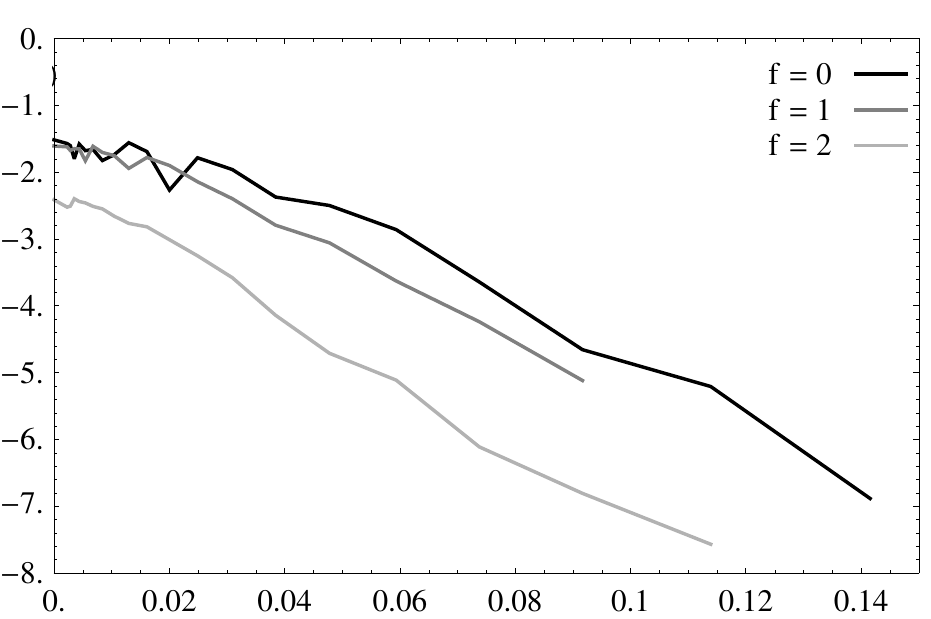}
\caption{Left: The location of the second relaxation pole as a function of the net baryon density for various values of $f$. Right: The logarithm of the residue of the second relaxation pole.}
\mlabel{hipole_e}}
This case of very small $\tlq=0.02$, varying $\tlrho$ and $f$ is also a good example to study what happens to the third pole. To do that, we can look at fig. \mref{hipole_e}, which shows the location of this higher relaxation pole and (the logarithm of) its residue as a function of $\tlrho$, at various values of $f$. We see that the residue decreases exponentially with increasing $\tlrho$, while the poles move to higher imaginary frequencies, until they can't be tracked anymore. To illustrate this better, the frequency  in the region of $\tlrho$ in which there is no reliable residue information anymore is plotted dashed. This is because the accuracy of the location of the pole that is necessary to determine the residue with our methods is of the order of the residue. Obviously, there is no guarantee that the extremum that one finds actually is a pole if there is not good enough data to compute the residue. One could argue that we could go to much higher accuracy since the accuracy grows exponentially with the number of steps. However 
the number of steps needed to track the poles grows with the inverse of the residue, i.e. exponentially with increasing $\tlrho$ or $\tlq$.
If we look at the density dependence of this pole, we find that its imaginary frequency increases with increasing density -- which is just what we expect for a naive model of weakly-coupled particles with finite cross sections $\sigma$. In particular, for small cross sections in $d$ dimensions, $\sigma^{1/(d-1)} \ll \rho^{-1/d}$, one expects classically $\tau_{cl.}^{-1}\sim v \sigma \rho$, where $v$ is some characteristic speed. Hence for larger densities or larger cross sections, the scaling would naively approach $\sqrt{\rho}$. This is just what we see in fig. \mref{hipole_e}. Hence, despite the limited accuracy and reliability in tracking these poles, we can safely associate this pole with a classical, weakly-coupled relaxation behavior. 


\DFIGURE{\includegraphics[width=0.49\textwidth]{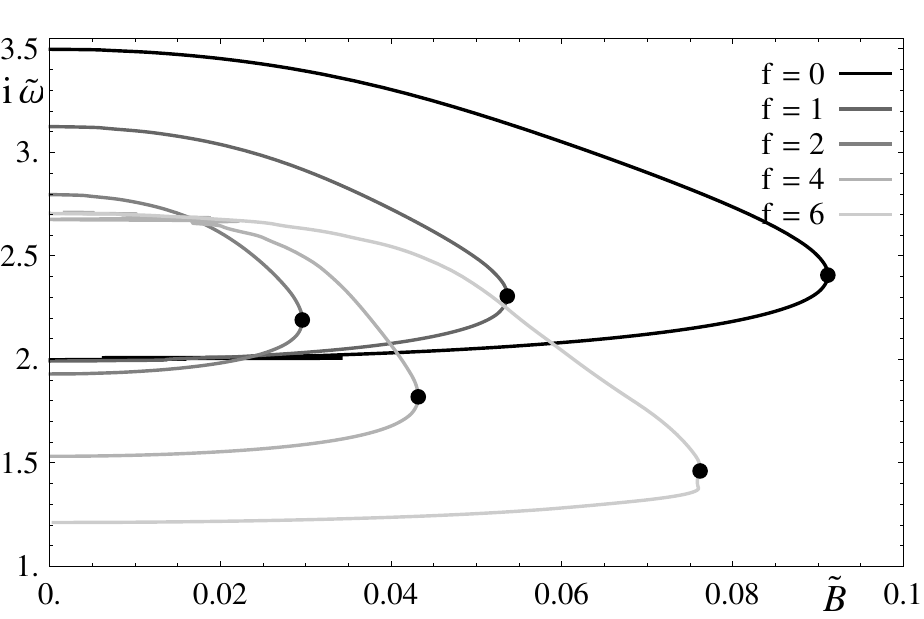}\includegraphics[width=0.49\textwidth]{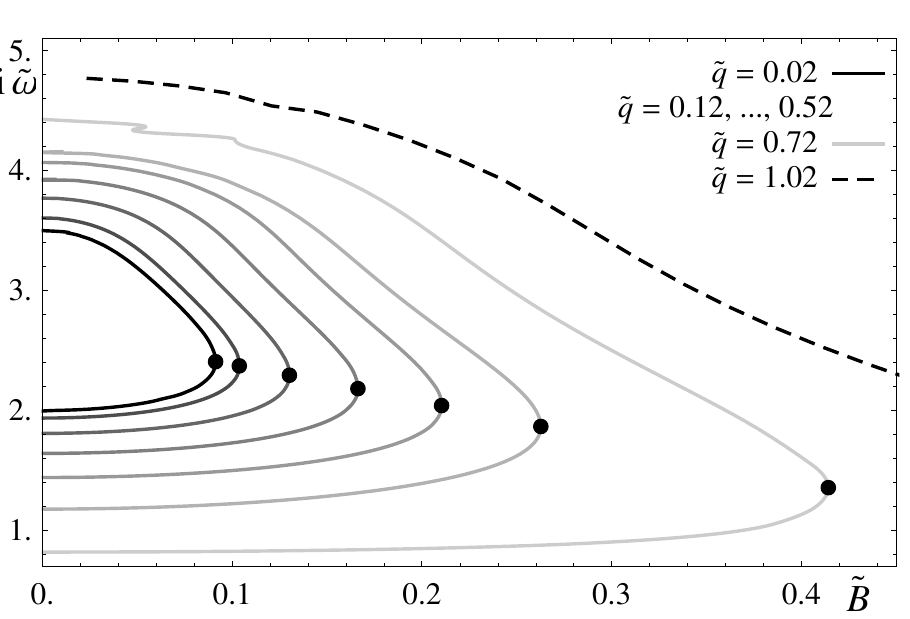}\caption{Left: The location of the relaxation poles as a function of the magnetic field for various values of $f$. Right: For different wavenumbers.}\mlabel{hipolescomm}}
A behavior very much in contrast to the case of turning on the finite density can be found in in the left of fig. \mref{hipolescomm}, where we show how the relaxation poles behave at small wavenumber in the presence of a magnetic field.
We see that with increasing magnetic field, the relaxation poles merge at some critical magnetic field, $\tlb_c$, and then turn into the first Landau level. This contrasts to the classical Drude-model analysis, where the magnetic pole moves away from the imaginary axis as soon as the magnetic field is turned on. Essentially what is happening is that the creation of the first Landau level is inhibited below $\tlb_c$ because of the strong coupling.
If we assume a crude model, in which the frequencies are given by $-i \frac{\tau^{-1}_{cl.}+ \tau^{-1}}{2} \pm\sqrt{\tom_c^2 - \frac{(\tau^{-1}_{cl.}- \tau^{-1})^2}{4}}$, then we see the reason for the dependence of the curves on $f$. Now, remember that we found in section \mref{DClimit}, that $\tom_c \tau \sim \frac{\tlb}{\sqrt{1+f^2}}$, such that $\tlb_c$ is approximately given by the ratio of the relaxation times at vanishing magnetic field, $\tlb_c \sim\frac{ \sqrt{1+f^2}}{2}\left(\frac{\tau^{-1}_{cl.}}{\tau^{-1}}-1\right)_{B=0}$. This ratio depends non-monotonically on $f$, because apparently the location of the classical relaxation pole is not closely related to the location of the first relaxation pole. Obviously, this generic behavior is not exact, but provides a rather qualitative description.
%

This magnetic dependence is consistent with the dependence of the relaxation poles on the wavenumber that we show in the right of fig. \mref{hipolescomm}. In fact, now it is most apparent that the behavior is reasonably well-described in terms of $\tau_{cl.}/\tau$ only. This can be seen by computing  $\frac{1}{2\tlb_c}\left(\frac{\tau^{-1}_{cl}}{\tau^{-1}}-1\right)_{B=0}$. For the data in fig. \mref{hipolescomm}, this ratio is approximately constant, ranging from $4.2$ at $\tlq = 0.02$ to $5.3$ at $\tlq = 1.2$. Certainly it is not close to 1, but we could not expect this, as we chose our expression only as an example of how a branch cut in the solution for the location of the poles as a function of the background parameters can look like. 
This shows us however how suppressing the first Landau level is related to strong coupling, as the critical magnetic field is proportional in some approximation to the difference between the classical and strong-coupling inverse relaxation times.

\DFIGURE{
\includegraphics[width=0.99\textwidth]{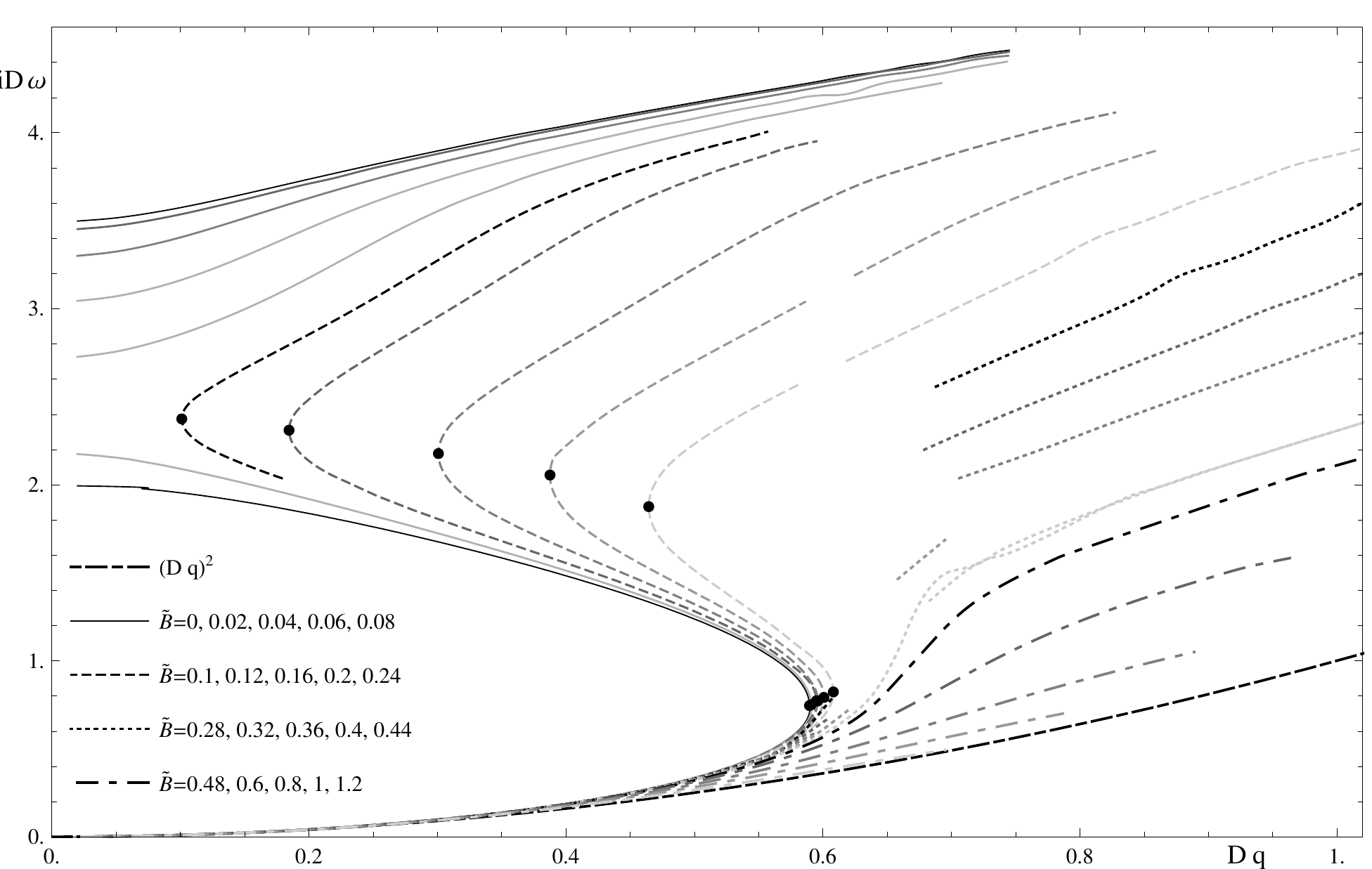}\
\caption{The location of the poles on the imaginary axis as a function of the wavenumber for different values of the magnetic fields. The frequency and wavenumber are scaled with the diffusion constant, as the diffusion behavior can be written as $i D \omega = (D\, q)^2$.}
\mlabel{big_easy}}
Finally, we can look in fig. \mref{big_easy} at the dependence of the location of all three poles on the wavenumber at varying magnetic field. At vanishing magnetic field, we start off with the system in which there is the hydrodynamic to quasiparticle transition, and there exists always the classical relaxation pole, that moves towards larger $\tau_{cl.}^{-1}$ with increasing $\tlq$. Going beyond the critical magnetic field, we see that the relaxation poles re-appear at some wavenumber $q_B < q_c$ that increases with increasing magnetic field. Beyond some second critical magnetic field, at which $\tlq_B = \tlq_c$, there is only one imaginary pole, that starts off at small $\tlq$ as a diffusion pole and turns at large $\tlq$ into the classical relaxation pole. Interestingly, at large wavenumbers, we always see only the classical relaxation pole, so the ``strong'' relaxation pole is an effect that arises only  at small wavenumbers, i.e. long-distance perturbations, which is what one actually expects, because of the diverging correlation length. The effect that the magnetic field inhibits relaxation on large length scales is precisely what we expect because of the localizing effect of the magnetic field on charged particles. This behavior between the critical magnetic fields, in which a pair quasiparticle poles (with positive and negative real part of the frequency) turns into a pair of relaxation poles and then into a different quasiparticle pole nicely reflects the transition between the regime dominated by Landau levels and the regime dominated by resonances on the width of the defect, that we observed in the previous section. Furthermore, we see that under the appropriate rescaling with the diffusion constant, the curves agree even beyond the actual behavior proportional to $\tlq^2$ -- indicating that $\tlq_c$ is reasonably well-described by the lengthscale from the diffusion constant.
The attentive reader will notice the ``hole'' in the plot near $\tlq_c$. This arises because it is difficult to track the poles in this regime using our method of scanning the frequency for each value of $\tlq$ in this case. Obviously, swapping that order should allow us to determine the location of the poles also in this regime.

\DFIGURE{
\includegraphics[width=0.49\textwidth]{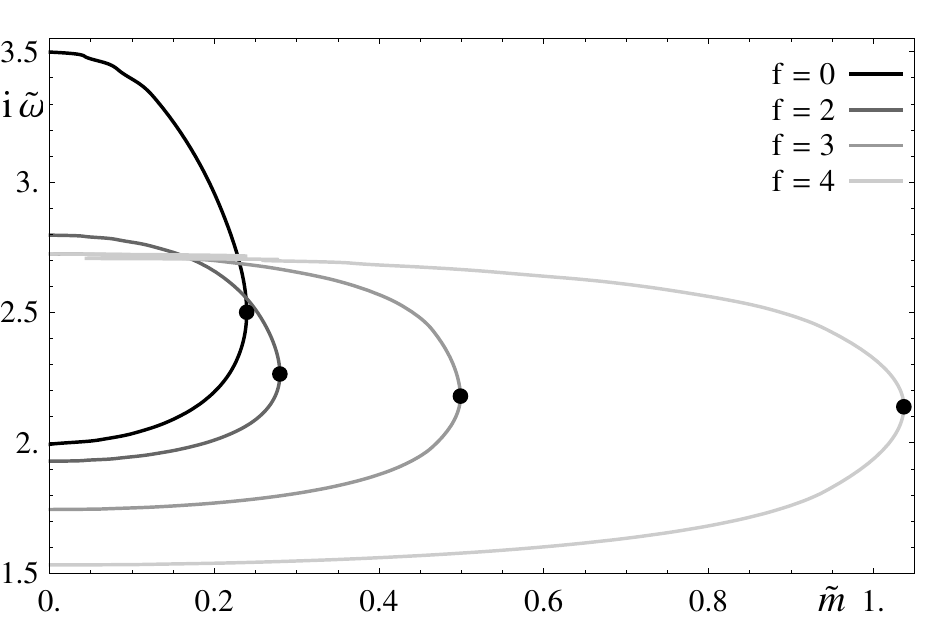}
\includegraphics[width=0.49\textwidth]{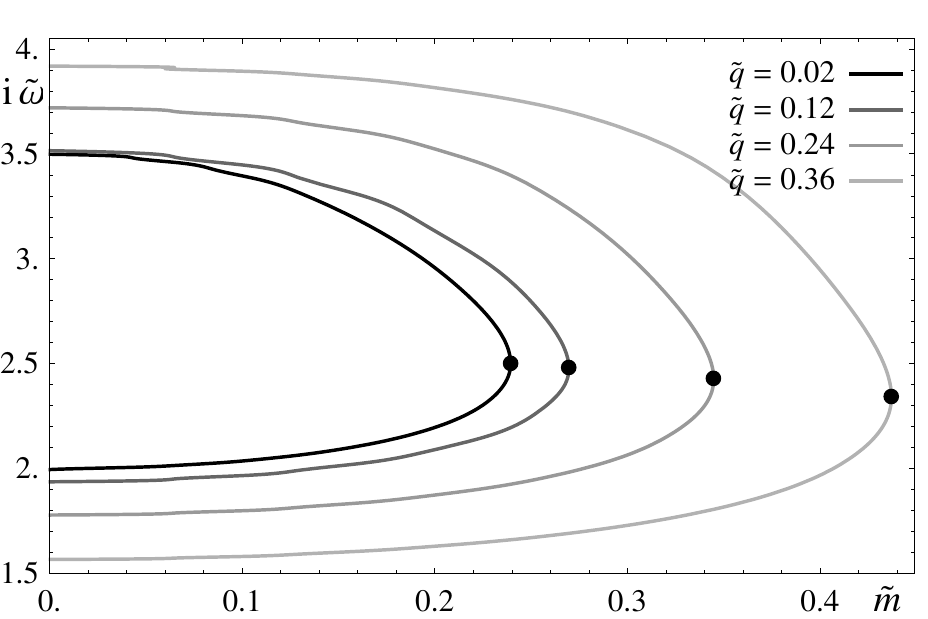}
\caption{The imaginary frequencies of the relaxation poles as a function of the quark mass. Left: For various values of $f$. Right: For various values of $\tlq$.}
\mlabel{relax_m}}
The effect of turning on a finite quark mass is very similar to the case of the background magnetic field, as we see in fig. \mref{relax_m}. The most significant difference is in the $f${}-dependence, as the critical mass at which the relaxation frequencies receive a real part increases quickly as we increase $f$. This is due to the fact that the hydrodynamics is dominated by physics in the IR, i.e. at small radii and hence depends on $\Psi_0$ rather than the mass directly and the quark mass as a function of $\Psi_0$ increases with increasing $f$.

\HFIGURE{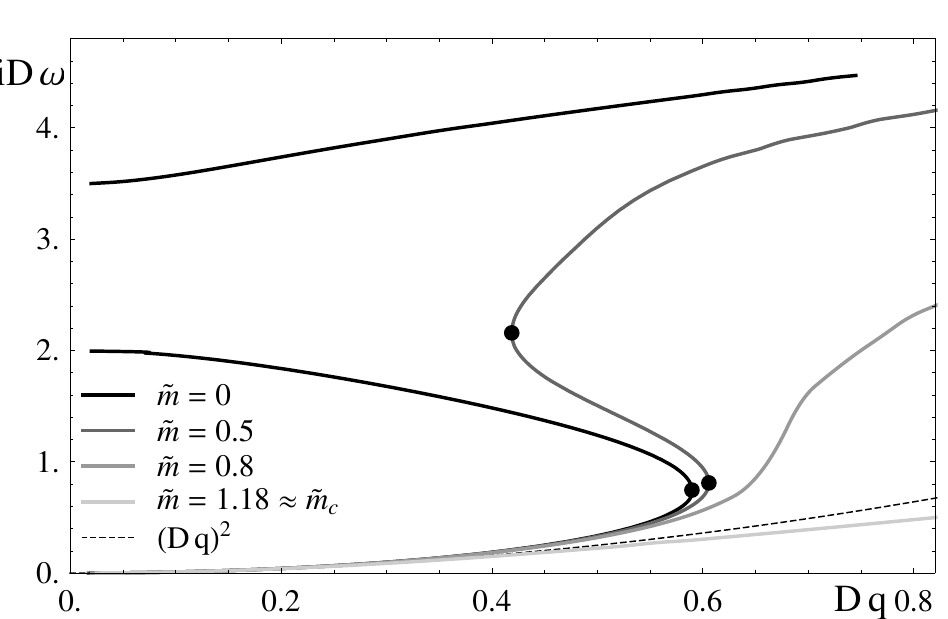}{The location of the poles on the imaginary axis as a function of the wavenumber for different values of the quark mass. The frequency and wavenumber are scaled with the diffusion constant. For comparison, the critical mass for the embedding is $\tlm \sim 1.196$.}{all_q_m}
Looking at the picture of the location of the poles as a function of the wavenumber in fig. \mref{all_q_m},
we find again that the system at small imaginary frequencies and small wavenumbers is well-described by the diffusion behavior, however not up to as high wave numbers as in the case of the background magnetic field. The critical mass above which there is always only the classical relaxation pole is just below $\tlm \sim 0.75$.
\subsection{Landau Levels and Plasmons}\label{llplasmon}
In this subsection, we will try to shed some light on the nature of the magnetic and density resonances that we observed in section \mref{fancyspectral}, by studying the corresponding quasiparticle poles. 
See e.g. also discussion in
ref. \refcite{RobSpec} or ref. \refcite{fast} for why, in general, the thermal correlators will have poles in the lower
half of the complex frequency plane.
For simplicity, we will focus on the isotropic ($\tlq = 0$) case. 
In principle, there are again different methods of estimating the location of the poles. The least reliable method is simply fitting Lorentzians to the resonances. However it cannot give the right answer, as we expect a sequence of infinitely many poles with separation $\nu_0 - i \gamma_0$, and if we consider the $n^{th}$ resonance and provided $\nu_0$ and $\gamma_0$ are of the same order, we need to consider more than $\order(n)$ neighboring poles. The more precise method involving only the data on the real (frequency) axis is then as in ref. \refcite{baredef} based on assuming an appropriate sequence of poles, summing it (ideally analytically), and fitting the parameters locally around each maximum - assuming they vary slowly enough, such that the ``backreaction'' from the varying parameters is sufficiently suppressed. The third method is simply trying to fit the poles by scanning an area in the complex frequency plane using an appropriate guess obtained from the data on the real axis. Then, we can use the usual recursion to find the poles. The most time consuming step in that method is to scan the search area for the first time, since we can not assume that the poles are the only local extrema.
For the former technique, we use the Ansatz
 \beq
C_{y y} \, = \, - \varepsilon_0 \, \sum_{n \ge 1} \frac{1}{\pi}\left(
\frac{n (\tlnu_0 + i \tlgam_0)^2}{\tlnu + n (\tlnu_0 + i \tlgam_0)}
- (\tlnu_0 + i \tlgam_0)   \, + \,      \frac{n (\tlnu_0 -  i
\tlgam_0)^2}{\tlnu - n ( \tlnu_0 -i \tlgam_0)} + (\tlnu_0 - i
\tlgam_0)  \right) \ , \labell{gen_ansatz}
 \eeq
 that was also used in ref. \refcite{baredef}.
This Ansatz basically says that all the poles are located with equal spacing on a straight line at $\nu_n = n (\pm\nu_0 - i \gamma_0)$ (or $\nu_n = (n-1/2) (\pm\nu_0 - i \gamma_0)$ for the Landau levels), with residues $\frac{\varepsilon_0}{\pi} \frac{\nu_n}{n}$ (or $\frac{\varepsilon_0}{\pi} \frac{\nu_n}{n-1/2}$).

However, it turns out that methods to try to fit the resulting analytic expression 
 \bea
\!\!\!\!\!\!\!\!\!\Im C_{y y} \, = \, \varepsilon_0\, \Im \tlnu \cot\left(\pi
\frac{\tlnu}{\tlnu_0 + i \tlgam_0}\right) \,
=  \, \varepsilon_0\, \Im \tlnu \frac{\sin \frac{2 \pi \tlnu
\tlnu_0}{\tlnu_0^2 + \tlgam_0^2} \, -\, i \sinh \frac{2 \pi \tlnu
\tlgam_0}{\tlnu_0^2 + \tlgam_0^2}}{\cosh \frac{2 \pi \tlnu
\tlgam_0}{\tlnu_0^2 + \tlgam_0^2} \, -\, \cos \frac{2 \pi \tlnu
\tlnu_0}{\tlnu_0^2 + \tlgam_0^2}} \ ,
 \labell{lin_ansatz}
 \eea
 i.e.
  \beq\labell{anscon}
\Re \tilde{\sigma}_{yy} \, = \,\varepsilon_0\, \frac{\tlnu}{\tom} \frac{
\sinh \frac{2 \pi \tlnu \tlgam_0}{\tlnu_0^2 + \tlgam_0^2}}{\cosh
\frac{2 \pi \tlnu \tlgam_0}{\tlnu_0^2 + \tlgam_0^2} \, -\, \cos
\frac{2 \pi \tlnu \tlnu_0}{\tlnu_0^2 + \tlgam_0^2}} \, ,
 \eeq
 are unreliable in this case, mainly because of the periodicity of the expression.

From \reef{anscon}, we can then obtain the parameters for the poles from the $n^{th}$ local maximum, $\sigma_{n}$, $\omega_{n}$ and its neighboring minima $\sigma_{n-1/2}$, $\sigma_{n+1/2}$. Assuming $\sigma_{n-1/2} = \sigma_{n+1/2} = \sigma_{min}$, the exact result is in the case of plasmons
\begin{equation}
\sigma_{n} - \sigma_{min} \, = \, - 2 \frac{\varepsilon_0}{\sinh  \frac{2 \pi \tom_{n} \tlgam_0}{\tlnu_0^2 + \tlgam_0^2}} \ \ , \ \ \tom_n \, = \, n\frac{\tlnu_0^2 + \tlgam_0^2}{\tlnu_0}
\end{equation}
and similar for Landau levels. This expression can be trivially inverted to obtain $\tlnu_0$ and $\tlgam_0$. Taking into account that $\sigma_{n-1/2} \neq \sigma_{n+1/2}$, the correction when using $\sigma_{min} = \frac{1}{2} \left(\sigma_{n-1/2} + \sigma_{n+1/2}\right)$ will be of order $\frac{\pi^2\tlgam_0^2}{\tlnu_0^2} \left(\cosh \frac{n \tlgam_0}{\tlnu_0} \right)^{-1}$, i.e. it will only be significant for the first few poles. However this seems to be a less bothersome shortcoming to take than ``misfitting'' poles because of the periodicity. Note that the two terms in the error term, make sure that beyond $n=1$ always either of them gives us a good suppression.

\DFIGURE{
\includegraphics[width=0.49\textwidth]{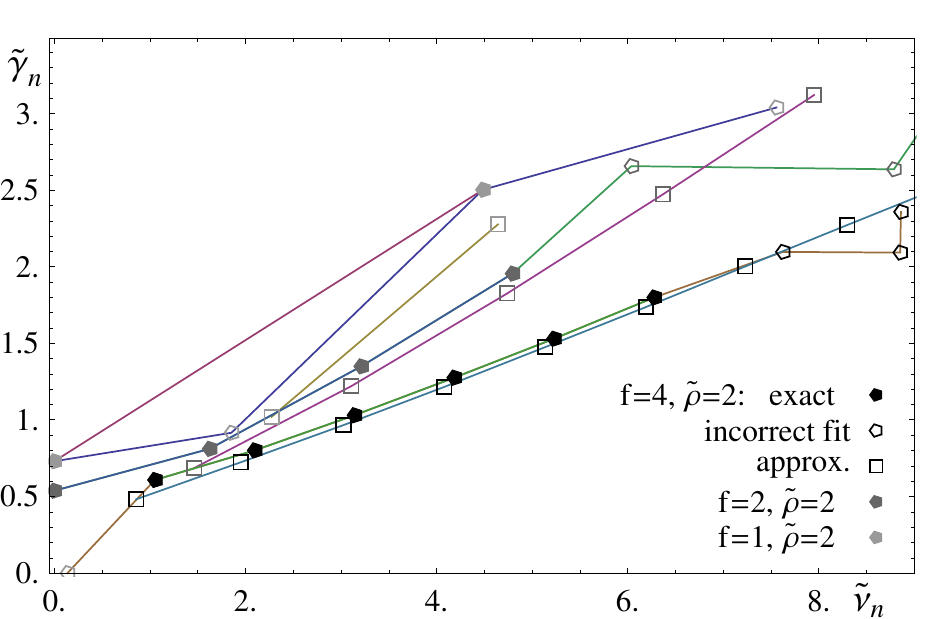}
\includegraphics[width=0.49\textwidth]{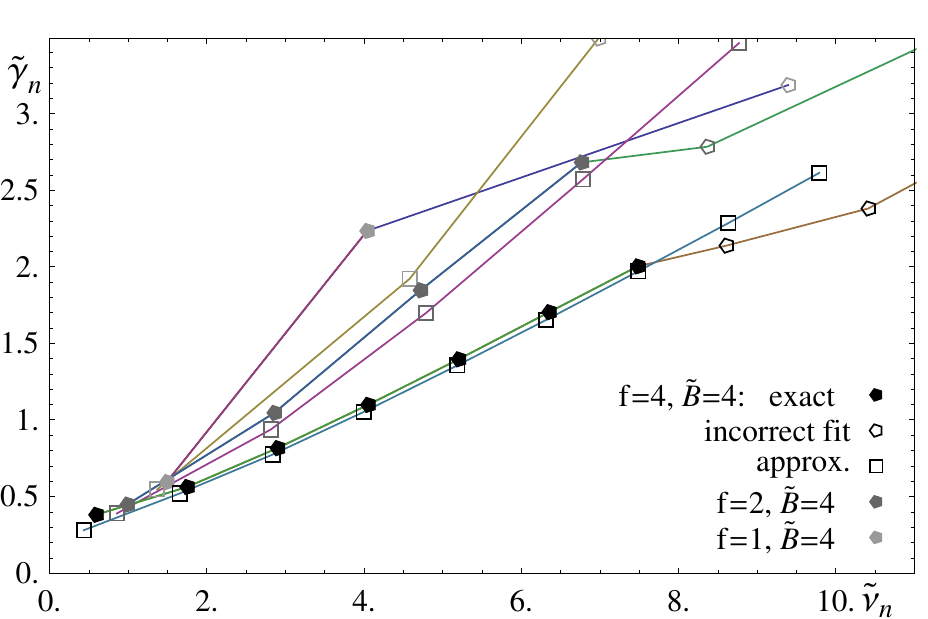}
\caption{The location of the poles $\omega_n = \nu_n - i\gamma_n$  in the complex frequency plane from the different estimates as described in the text. For orientation purposes, the lines connect the poles of different order in the same background. Left: Finite $\tlrho=2$, Right: Finite $\tlb=4$.}
\mlabel{comppolesfancy}}

In fig. \mref{comppolesfancy}, we compare the two methods to obtain the poles -- the ``exact'' direct search and the ``approximate'' result from fitting to the Ansatz -- for various values of $f$, $\tlb$ and $\tlrho$. In the direct search, we used the maximum value of the spectral function in the last recursion as an indicator as to whether the pole has been found, with a threshold at $\frac{\tilde{\sigma} (\tlnu)}{\varepsilon_0} = 100$. In practice the value will be either much larger or much smaller than this number. In the plot, we still show the ``misfitted'' poles for reference. 
Overall, we see that the agreement of $\tlnu_0$ is very good at finite $f$, and the estimates from the spectral curves are much more reliable in the sense that there are no poles that are ``not found'' or have large displacements. In fact, even if there is some disagreement, the spacing $\tlnu_0$ between the poles
is more accurate than the overall shift of the poles.
While there may be larger disagreements for the first pole and at small values of $f$, we should note that in those cases  the direct fitting also fails frequently.
There is no clear trend for the dependence on $\tlrho$ and $\tlb$. Typically the approximation is slightly worse for very large values and very small values, because in the former case the decay is more rapid, causing the resonances to be more asymmetric. In the latter case, we also see only the first few resonances and are not in the slowly decaying oscillatory regime -- this time because of the larger spacing between resonances. We will not demonstrate this in the plots in fig. \mref{comppolesfancy}, simply because $\tlgam_0/\tlnu_0$ is approximately constant when varying $\tlb$ or $\tlrho$ at fixed $f$, and hence this is difficult to display in a clear fashion.
The agreement in the imaginary direction is slightly worse, which is not unexpected because it depends more on the conjectured form of the residues for the Ansatz.

\DFIGURE{
\includegraphics[width=0.49 \textwidth]{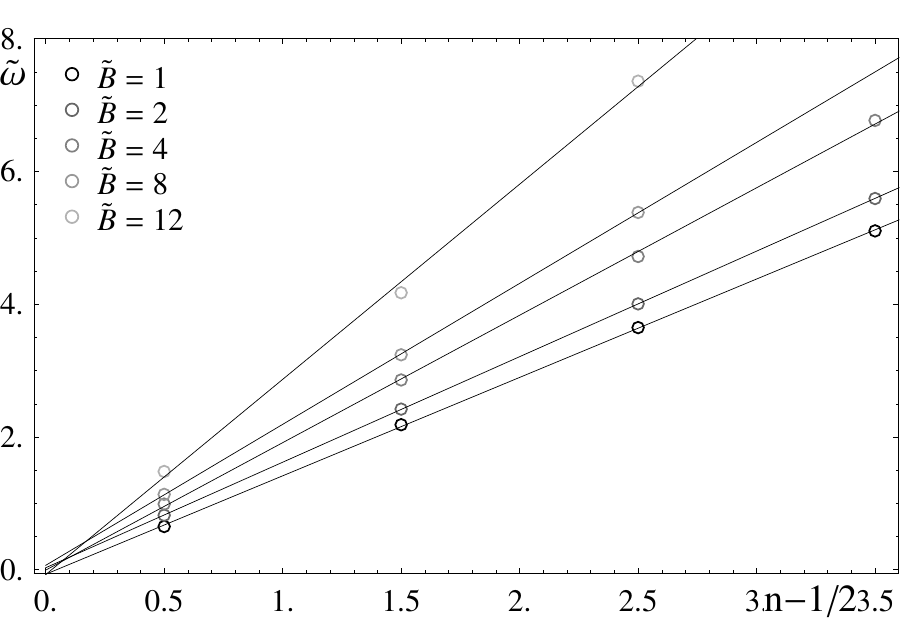}
\includegraphics[width=0.49 \textwidth]{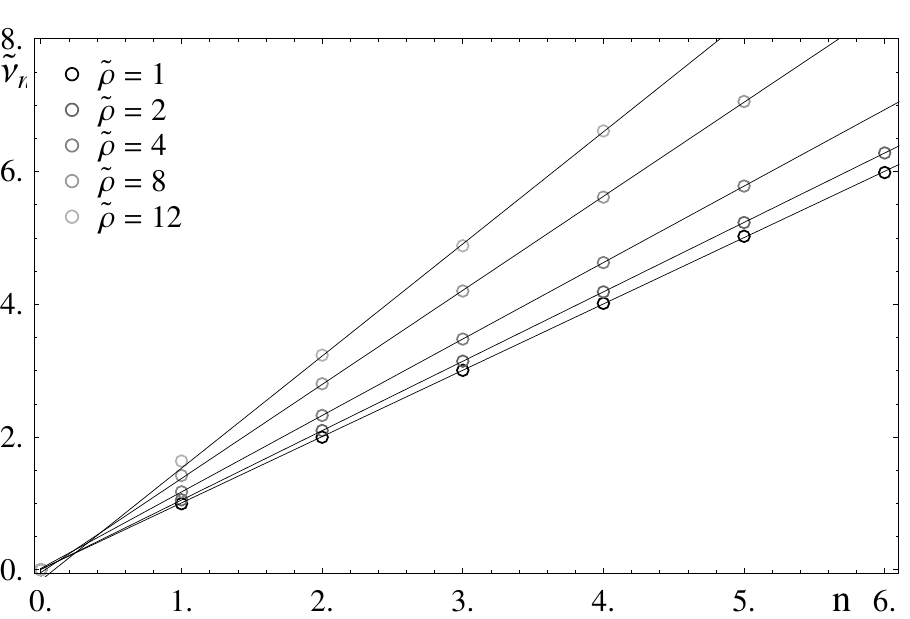}
\caption{The real part $\tlnu_n$ of the location of the poles in the complex frequency plane as a function of the level number. The lines are linear fits. Left: Various values for $\tlb$ at $f=2$. Right: Varying $\tlrho$ at $f=4$.}
\mlabel{accupoles_real}}
Next, let us look at the spectrum of the resonances as a function of the resonance level $n$. In figure \mref{accupoles_real}, we show the real part of the poles $\tlnu_n$ for various choices of $\tlrho$ and $\tlb$ at $M_q = 0$. Because it is important to distinguish between behaviors that are, for example, of the kind $\sqrt{n(1+n)}$ from strictly linear behaviors, we used the direct search for poles in order to obtain the highest accuracy. It turns out that the poles follow extremely closely a linear behavior $\tlnu_n = n \tlnu_0 + \delta \tlnu$ (or $\tlnu_n = (n -1/2) \tlnu_0 + \delta \tlnu$ in the case of a magnetic background field), with a small negative value for $\delta\tlnu$. The latter is easily explained from the behavior of the first pole that we found in the hydrodynamic regime. The constant separation implies that the spectrum of Landau levels is indeed the classical one for the finite mass case. This is slightly puzzling though, because one would have thought that if the collective excitations that carry the current have an effective mass, this mass would be finite. Then we should see a transition from the massive to the massless behavior at some frequency. Hence, either that mass must be large or frequency dependent, or some unusual mechanism gives rise to the Landau levels.

\DFIGURE{
\includegraphics[width=0.49 \textwidth]{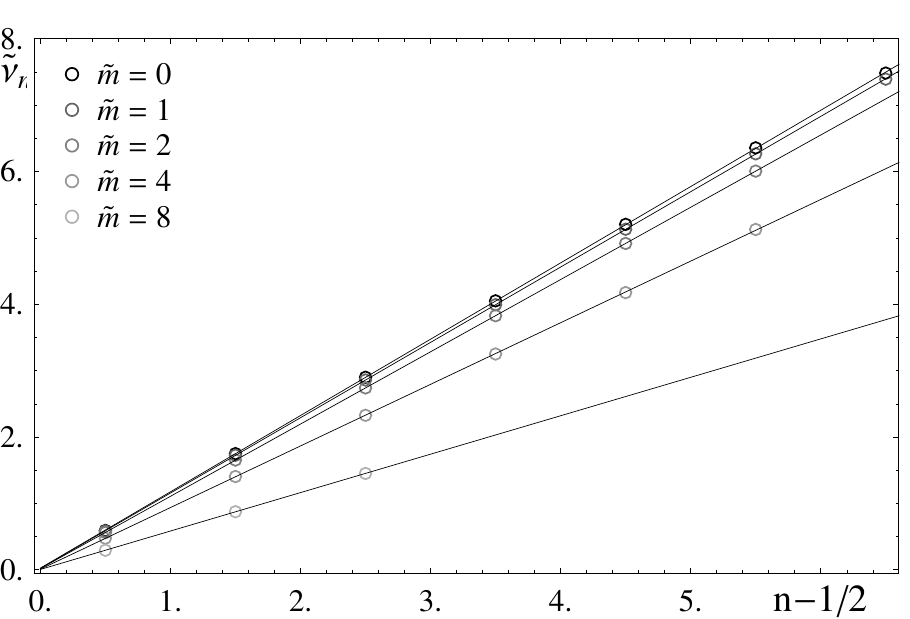}
\includegraphics[width=0.49 \textwidth]{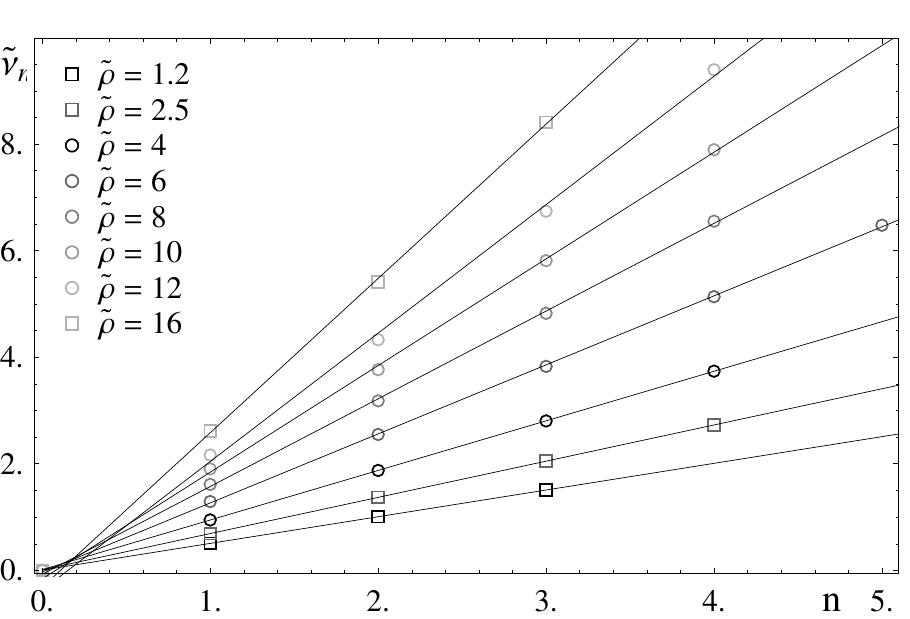}
\caption{The real part $\tlnu_n$ of the location of the poles in the complex frequency plane as a function of the level number at finite quark mass. Left: Varying $\tlm$ at $\tlb=4$ and $f=4$. Right: Different values of $\tlrho$ at $\tlm = 8$ and $f=2$.}
\mlabel{accupoles_mass}}
Looking at the case of finite quark mass in fig. \mref{accupoles_mass}, we find again no sign of a non-constant spacing between the poles, surprisingly even around the level of the quark mass. There is however a transition in the value of $\tlnu_0$ around that region, which we will follow up on later.

\DFIGURE{
\includegraphics[width=0.49 \textwidth]{accupoles_re_E_F4.pdf}
\includegraphics[width=0.49 \textwidth]{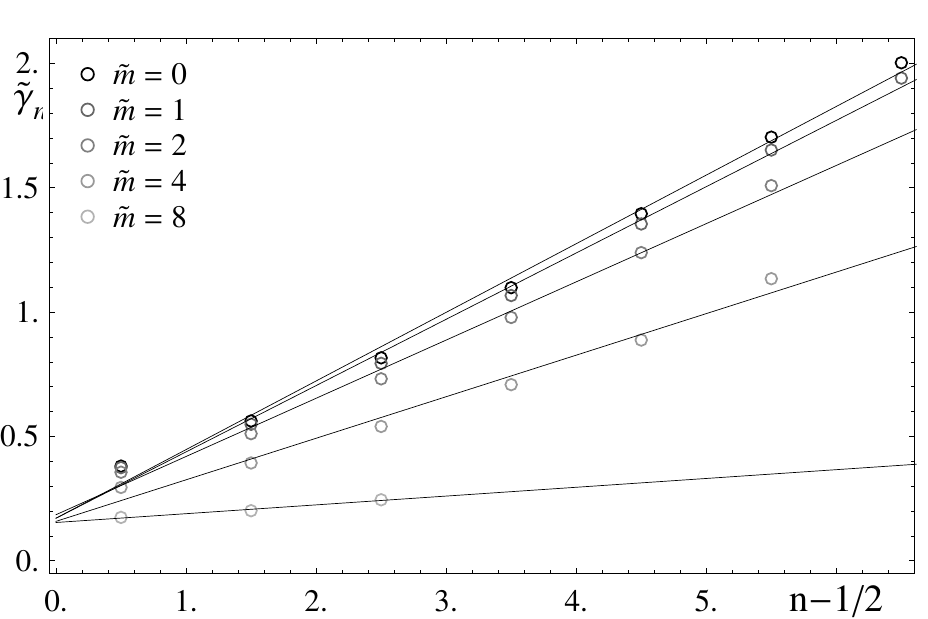}
\caption{The imaginary parameter in the location of the poles, $\tom_n = \tlnu_n - i \tlgam_n$. Left: Varying values of $\tlrho$ at $f=4$. Right: Varying mass $\tlm$ at $f=4$ and $\tlb = 4$.}
\mlabel{accupoles_imag}}
Finally looking at the behavior of the imaginary part of the poles, which reflects the inverse lifetime, in fig. \mref{accupoles_imag}, we find a small but significant deviation from the linear relation $\tlgam_n = n \tlgam_0 + \delta \tlgam$ (and accordingly for the Landau levels). This small drift towards larger $\tlgam_n$ for small $n$ can again be explained from the non-trivial behavior of the first (hydrodynamic) poles and from the fact that we are in a finite temperature background, which renders the first few resonances that are close to the temperature scale less stable.

\HFIGURE{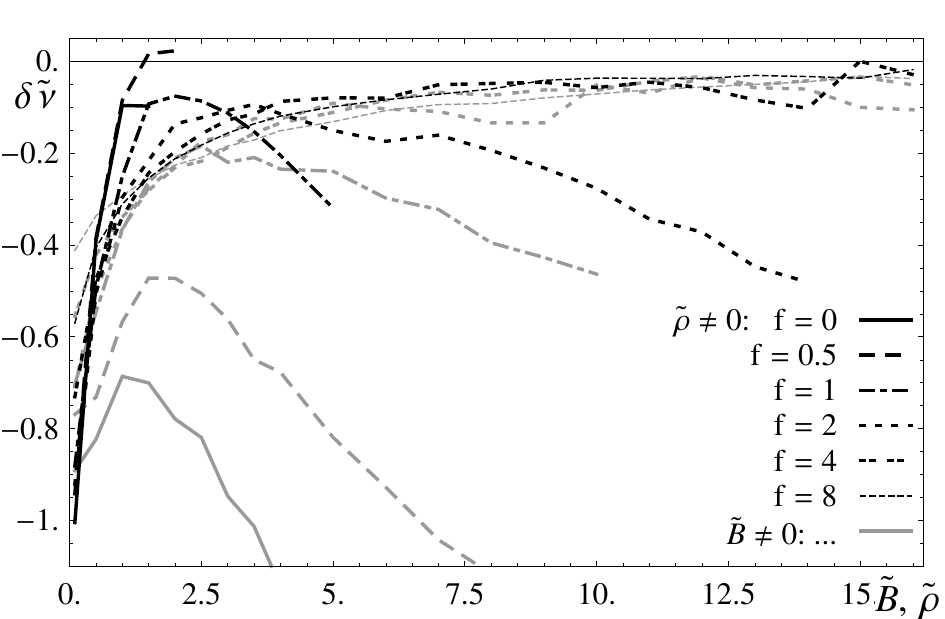}{The overall shift in the poles $\delta \tlnu$ as a function of $\tlb$ or $\tlrho$ for various values of $f$.}{delnu}
Finally, we can look at the resonances. Before looking at $\tlnu_0$, let us study the shift $\delta \tlnu$ in fig. \mref{delnu}. We find that for large $f$, there is a universal behavior $\delta \tlnu \propto \tlb^{-1},\tlrho^{-1}$ with a proportionality constant that seems independent of $f$. A significantly different behavior exists only for small values of $f$ and large values of $\tlb$ or $\tlrho$. This may be simply due to the worse fitting because in those cases we found only the first two poles, and the second one has already a very low amplitude such that it is at the limit of what can be recognized as a resonance above the background. This overall behavior may be simply due to the fact that the first quasiparticle pole originates from the relaxation poles on the imaginary axis as we found in the previous section. 

\DFIGURE{
\includegraphics[width=0.49 \textwidth]{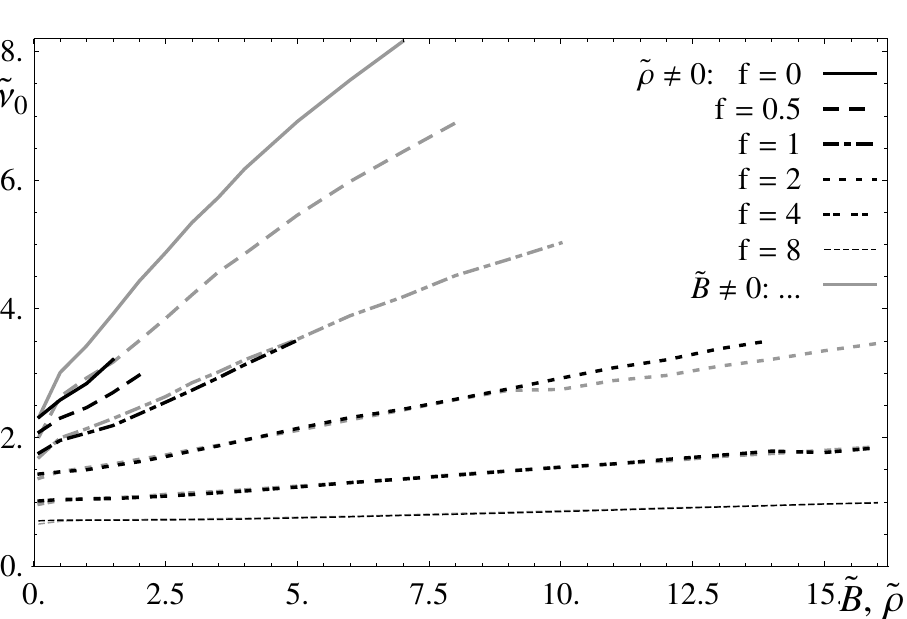}
\includegraphics[width=0.49 \textwidth]{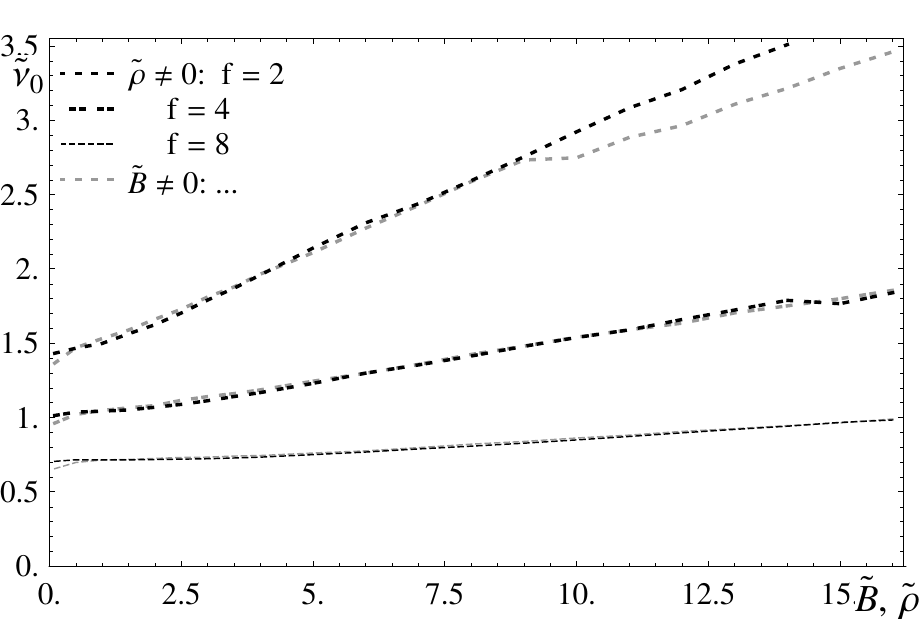}
\caption{The separation $\tlnu_0$ in the spectrum of ``plasmons'' and ``Landau levels'' as a function of $\tlb$ or $\tlrho$ for various values of $f$. The plot on a right only shows the highest $f$ curves.}
\mlabel{tlnu_0}}
Next, let us look at the value of $\tlnu_0$ in fig. \mref{tlnu_0}. There are two features to notice: Firstly the finite value of $\tlnu_0$ at vanishing magnetic field or density, and secondly the non-trivial dependence on $\tlb$ or $\tlnu$ -- both of which are quite different from what we would have naively expected for plasmons and Landau levels.
It turns out that there are two equivalent ways to interpret this situation.

\DFIGURE{
\includegraphics[width=0.49 \textwidth]{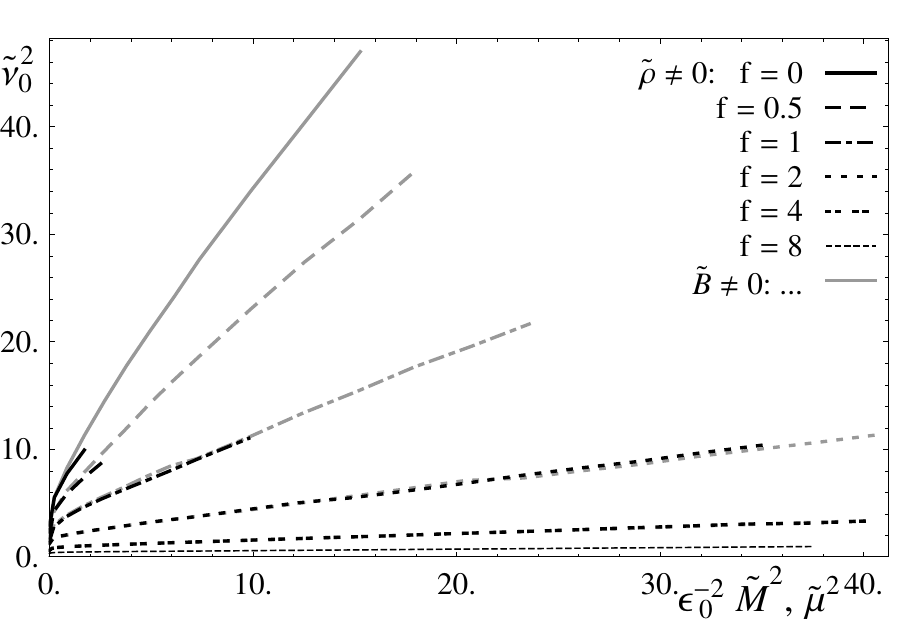}
\includegraphics[width=0.49 \textwidth]{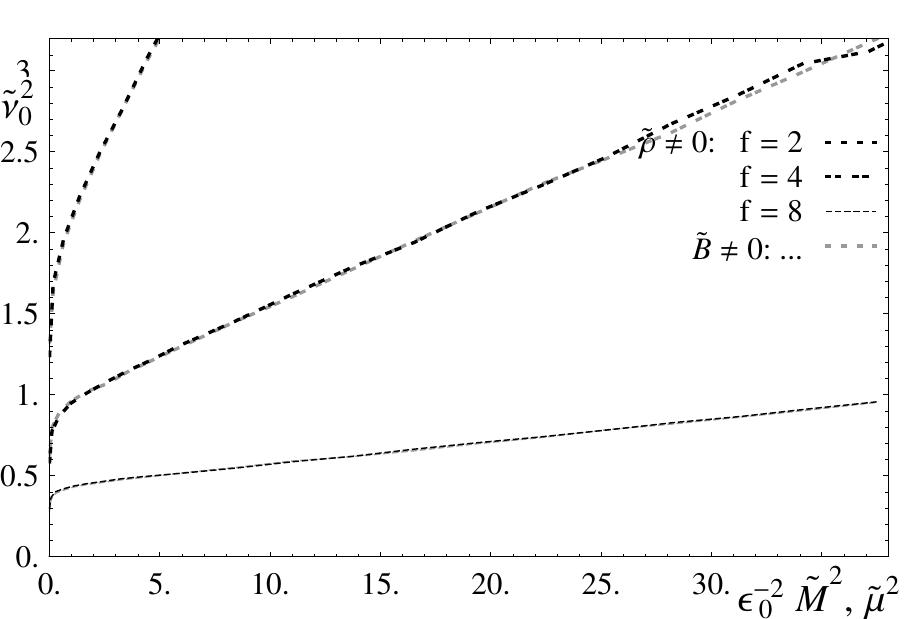}
\caption{$\tlnu_0$ presented as $\tlnu^2$ as a function of the square of the magnetization or chemical potential  $\tilde{M}^2$ or $\tilde{\mu}^2$ for various values of $f$. The plot on a right only shows the highest $f$ curves.}
\mlabel{magmagres}}
An interesting physical picture can be obtained by representing the data as $\tlnu_0^2$ as a function of the square of the magnetization $\tilde{M}$ or chemical potential $\tilde{\mu}$ that are computed in ref. \refcite{thermpaper}. As we see in fig. \mref{magmagres} the result are perfectly straight lines, such that the resonances are given by $\omega_c^2 = \omega_0^2 + \frac{M^2}{\alpha(f)^2}$ or $\omega_p^2 = \omega_0^2 + \frac{\mu^2}{\alpha(f)^2}$. Obviously, we could interpret the function $\alpha(f)$ as some kind of a mass scale that depends on $f$ only, i.e. not on the width of the defect, but only its ``topological'' property. If we were -- inappropriately -- to look at the corresponding classical Schr\"odinger equation, we would see that then the magnetic and density perturbations are not independent, but mixes with some other potential. This is  in contrast to the fact that the overall ``amplitude'' of the resonances, and hence the residue of the poles is at least for small fields proportional to the magnetic field. This is just what happens in the case of the classical Hall effect as discussed in section \mref{metalcon}. In fact, we can check a few values for $\omega_0$, 
and compare them to the resonances at finite $f$ due to a finite ``width'' of the defect, that were found in ref. \refcite{baredef}
to find that they are identical to the $f\rightarrow 0$ limit of those resonances.
%

Inspired by this, we can try to check the relation between the resonances and the length scale given by the effective temperature that we observed in fig. \mref{magmagres}. To do so, we can plot the ratio $\frac{\tlnu_0 T}{T_{eff}}$ in fig. \mref{tlnuteff}, where we see that it approaches within errors $\frac{\tlnu_0 T}{T_{eff}} \sim 2$. 
\HFIGURE{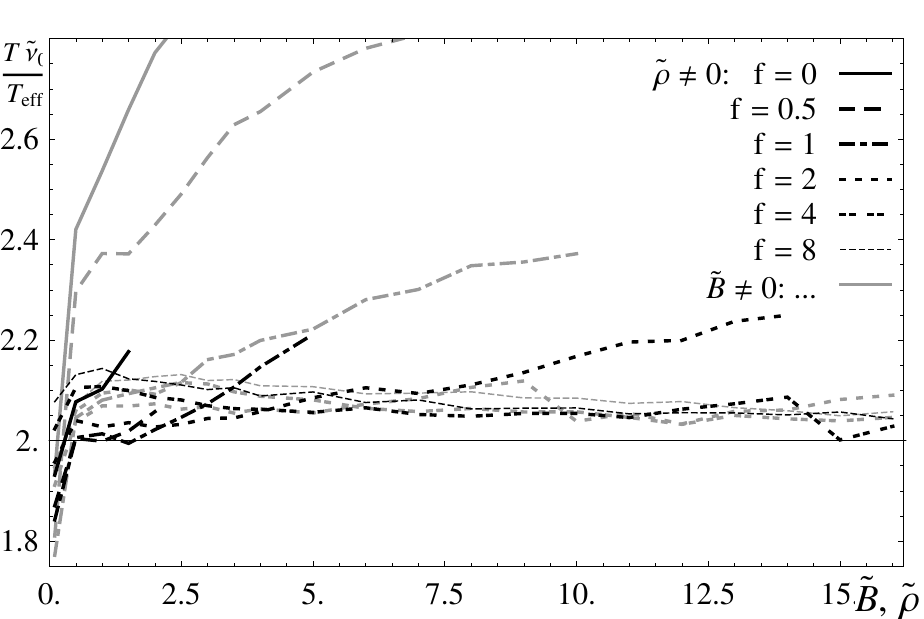}{The ratio $\frac{\tlnu_0 T}{T_{eff}}$ as a function of $\tlb$ or $\tlrho$ for various values of $f$.}{tlnuteff}
This is not a big surprise, as there is an underlying exact relation between $\frac{T_{eff}}{T}$ and $\mu$ (or $M$), that we can easily verify numerically and may in principle be able to derive analytically.

\DFIGURE{
\includegraphics[width=0.49 \textwidth]{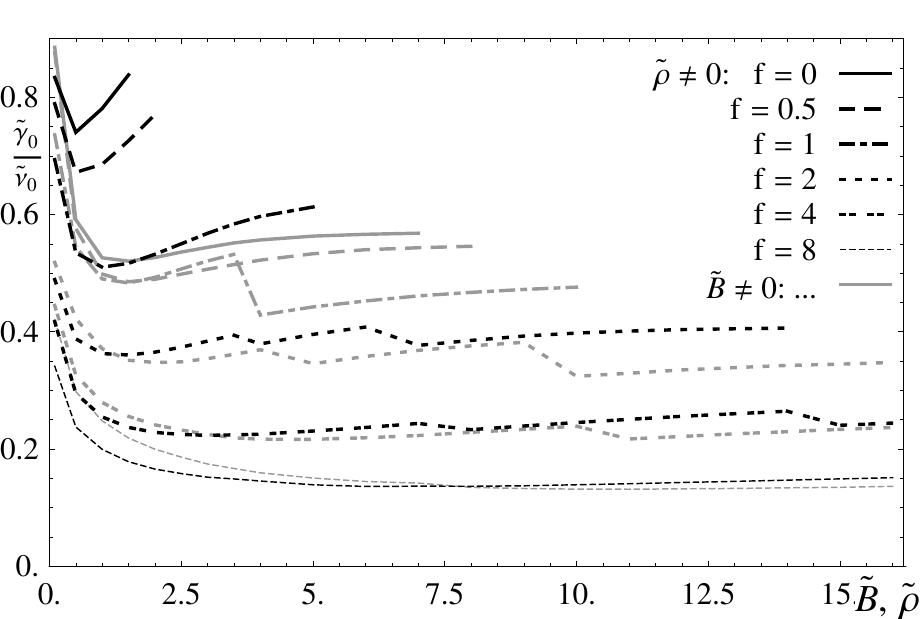}
\includegraphics[width=0.49 \textwidth]{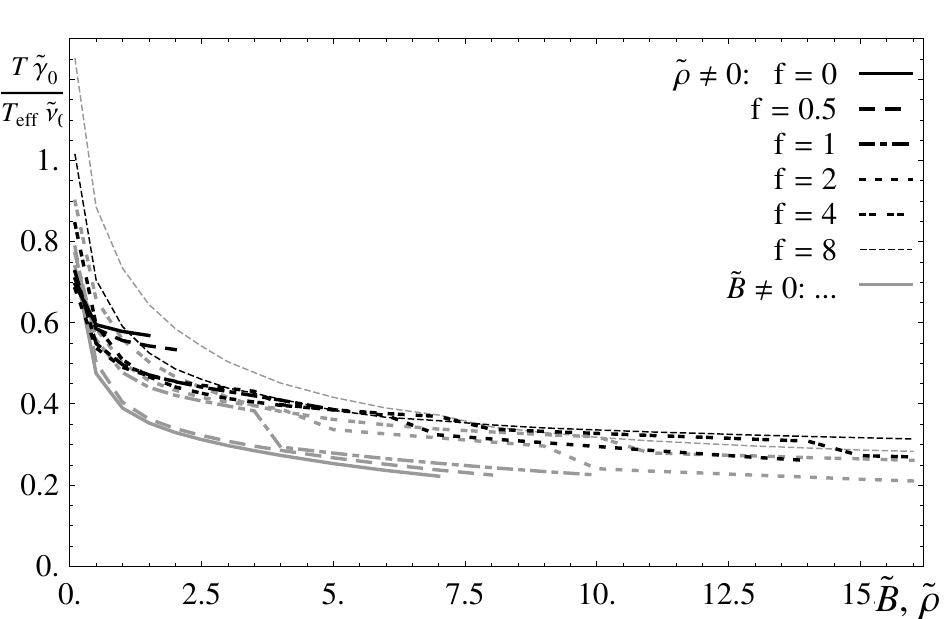}
\caption{The ``inverse lifetime to mass ratio'' $\frac{\tlgam_0}{\tlnu_0}$ for the magnetic and density resonances as a function of $\tlb$ or $\tlrho$ for various values of $f$.}
\mlabel{gamnaughts}}
Before concluding the study of the parameters of the resonances, let us look at the ratio $\frac{\tlgam_0}{\tlnu_0}$ in fig. \mref{gamnaughts}. Firstly, we notice the steps in the curves, that happen to coincide with data sets at which the highest pole in the sequence of resonances drops out of the fit because of its decreasing amplitude. This indicates the limitations in fitting the parameters (that we expect to converge only asymptotically) accurately.
%
Obviously, we could try to account for the non-linearity in $\tlgam_n$ as it was done in ref. \refcite{baredef}, but this has the downside that fitting with more parameters makes the result less reliable and may simply hide the limitations of the numerical result. Secondly, we see that within those limitations and even though $\frac{\tlgam_0}{\tlnu_0}$ itself seems to approach constants in $\tlnu$ or $\tlb$, it seems to be best described as being proportional to $\frac{T_{eff}}{T}$. Within the errors, it seems that the appropriate ratio is independent of $f$ and becomes also approximately constant in $\tlnu$ and $\tlb$.

\DFIGURE{
\includegraphics[width=0.49 \textwidth]{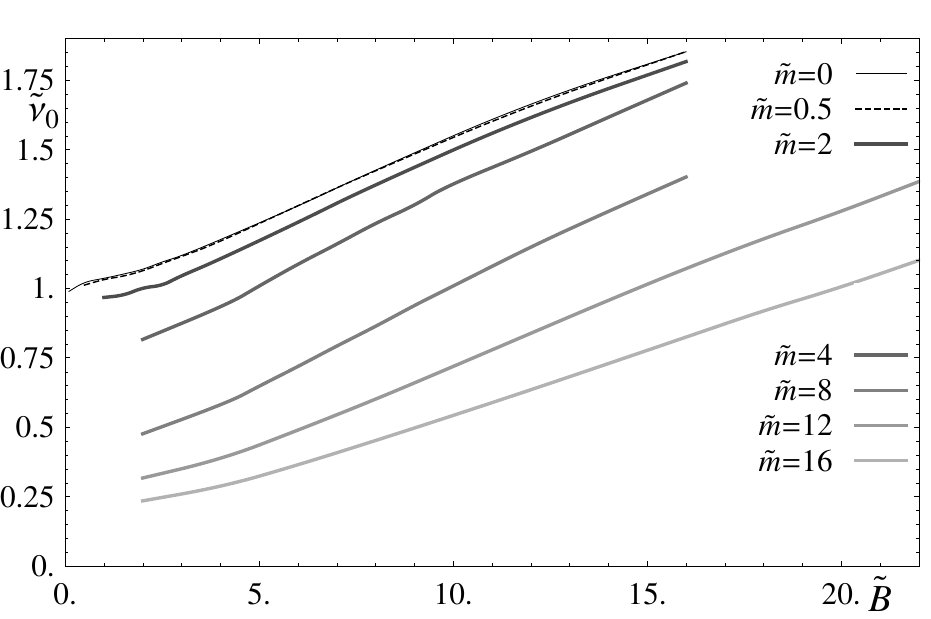}
\includegraphics[width=0.49 \textwidth]{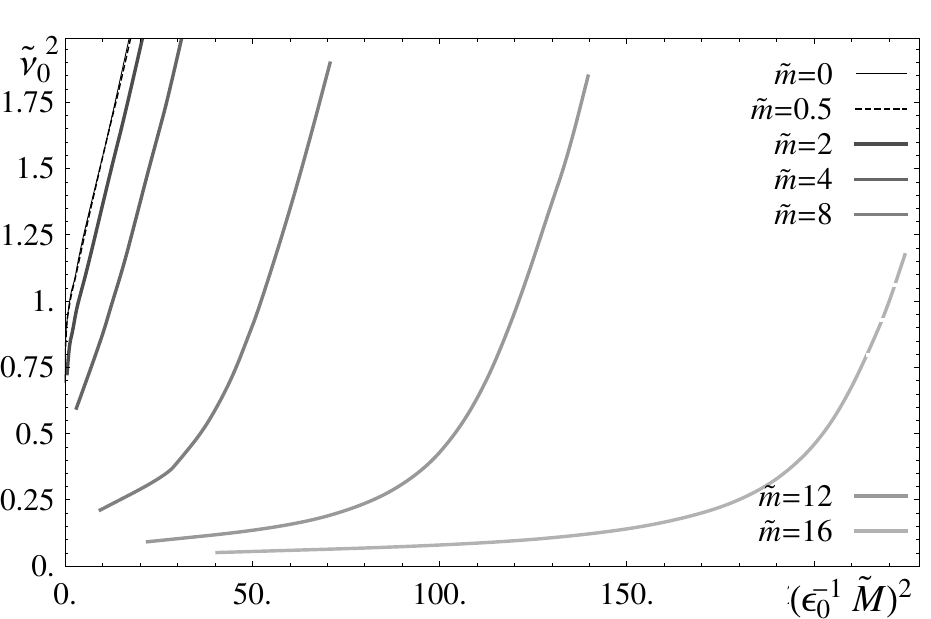}
\caption{$\tlnu_0$ for various values of the quark mass parameter $\tlm$ at $f=4$. Left: As a function of the magnetic field. Right: $\tlnu_0^2$ as a function of the magnetization $\tilde{M}^2$.}
\mlabel{magpoles_real}}
Next, we can look at the location of the poles at a finite quark mass in fig. \mref{magpoles_real}. If we look only at $\tlnu_0$ as a function of $\tlb$, we see a surprise, as there is only an overall shift in the curves depending on the quark mass. On the other hand, if we plot $\tlnu^2_0$ as a function of the square of the magnetization $\tilde{M}^2$, we see clearly the scale $\tilde{M}\sim \tlm$ that separates the massive and massless regime. Below $\tilde{M}\sim \tlm$, $\tlnu_0$ is suppressed with increasing mass and above $\tilde{M}\sim \tlm$, the behavior is similar to the massless case. If we compare $\tlnu_0$ to the effective temperature, we find that $\frac{\tlnu_0 T}{T_{eff}}$ is approximately constant around $2.04 \ldots 2.06$ with no apparent systematic trend, so even in this case the effective temperature sets the appropriate scale for the quasiparticle energy spectrum.
In principle, it would be interesting to study also  smaller values of $f$, however at smaller $f$ and large masses, it is not possible to find reliably at least the first two or three poles.

%
\DFIGURE{\includegraphics[width=0.49 \textwidth]{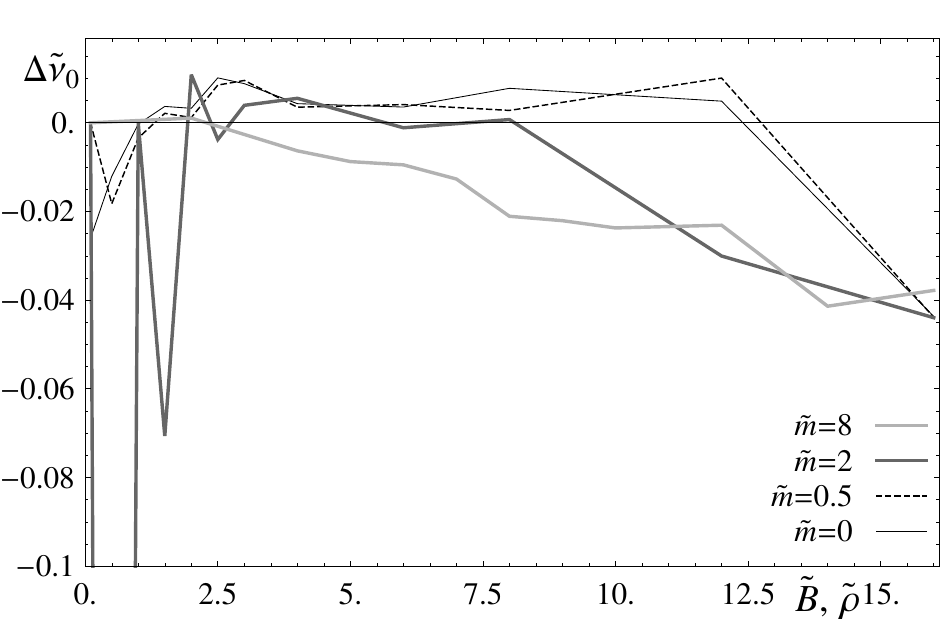}\includegraphics[width=0.49 \textwidth]{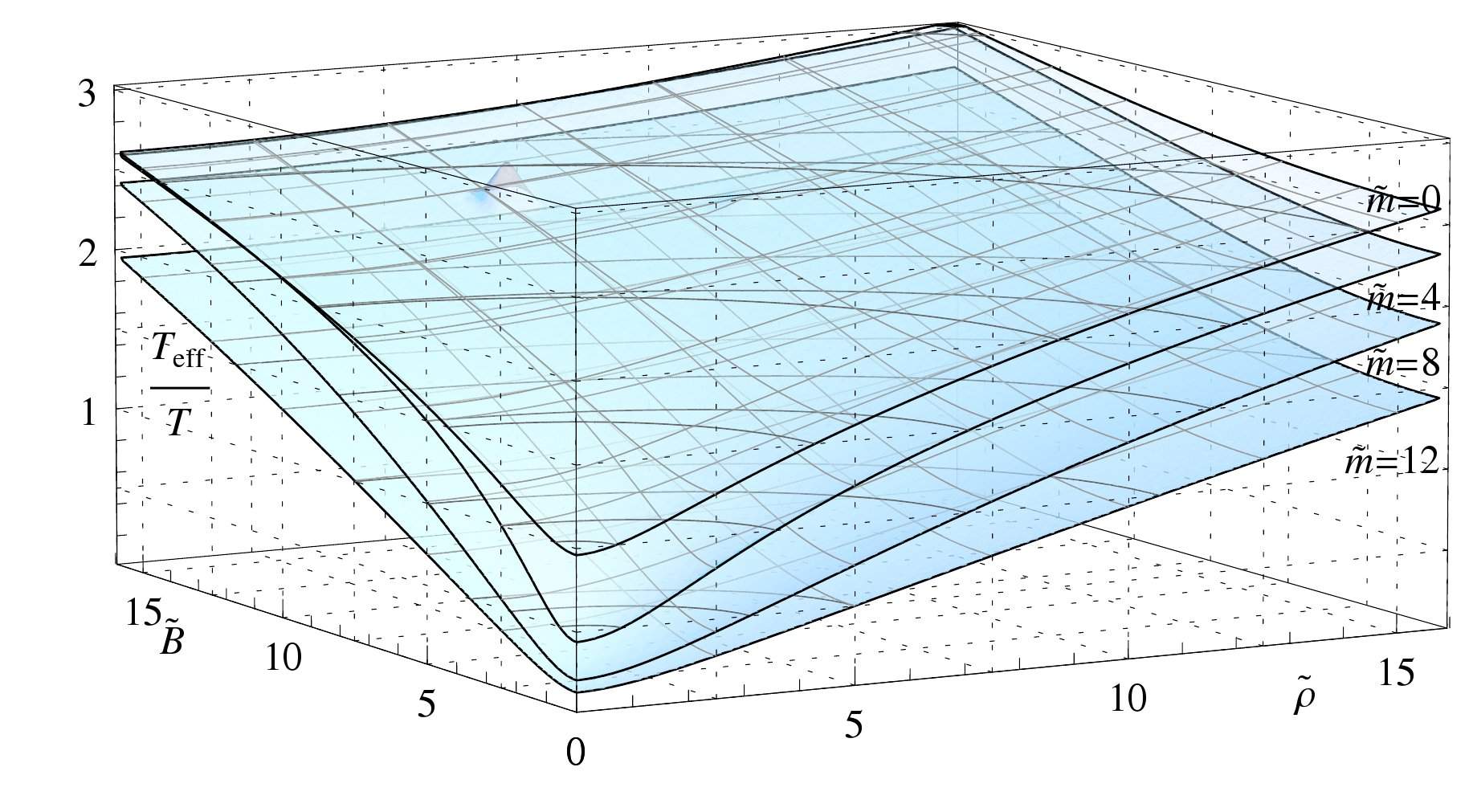}\caption{Left: The difference in $\tlnu_0$ between the case of the magnetic field and density at $f=4$ for various values of the quark mass as a function of  $\tlrho =\tlb$. Right: The scale $T_{eff}/T$ at $f=1$ and various values of the quark mass as a function of the net baryon density and magnetic field.}\mlabel{effcomm}}
In principle, we expect that the results in the presence of the finite density and the magnetic field are different as we turn on the finite mass. Comparing $\tlnu$ for these cases in the left fig. \mref{effcomm}, we see that the difference is very small, even as $\tlm > \tlb$. We are uncertain as to whether these deviations are significant. From the behavior at $\tlm = 8$, it seems that there may be a small effect, which is suppressed in the quasiparticle regime.
This question is resolved on the right in \mref{effcomm}, where we plot $T_{eff}/T$ at finite mass. We see that there is indeed a small difference between the dependence on $\tlrho$ and $\tlb$ at finite quark mass. Upon close inspection we also notice the separation between the regimes above and below the mass scale with the scalings $\frac{T_{eff}}{T} \propto \tlb^{1/2}$ or $\frac{T_{eff}}{T} \propto \tlb$, respectively, at least in the case of the magnetic field. It seems that there, the scale is $\tlb \sim \tlm$, whereas for the density it is $\tlrho \sim \tlm^2$.

\DFIGURE{
\includegraphics[width=0.49 \textwidth]{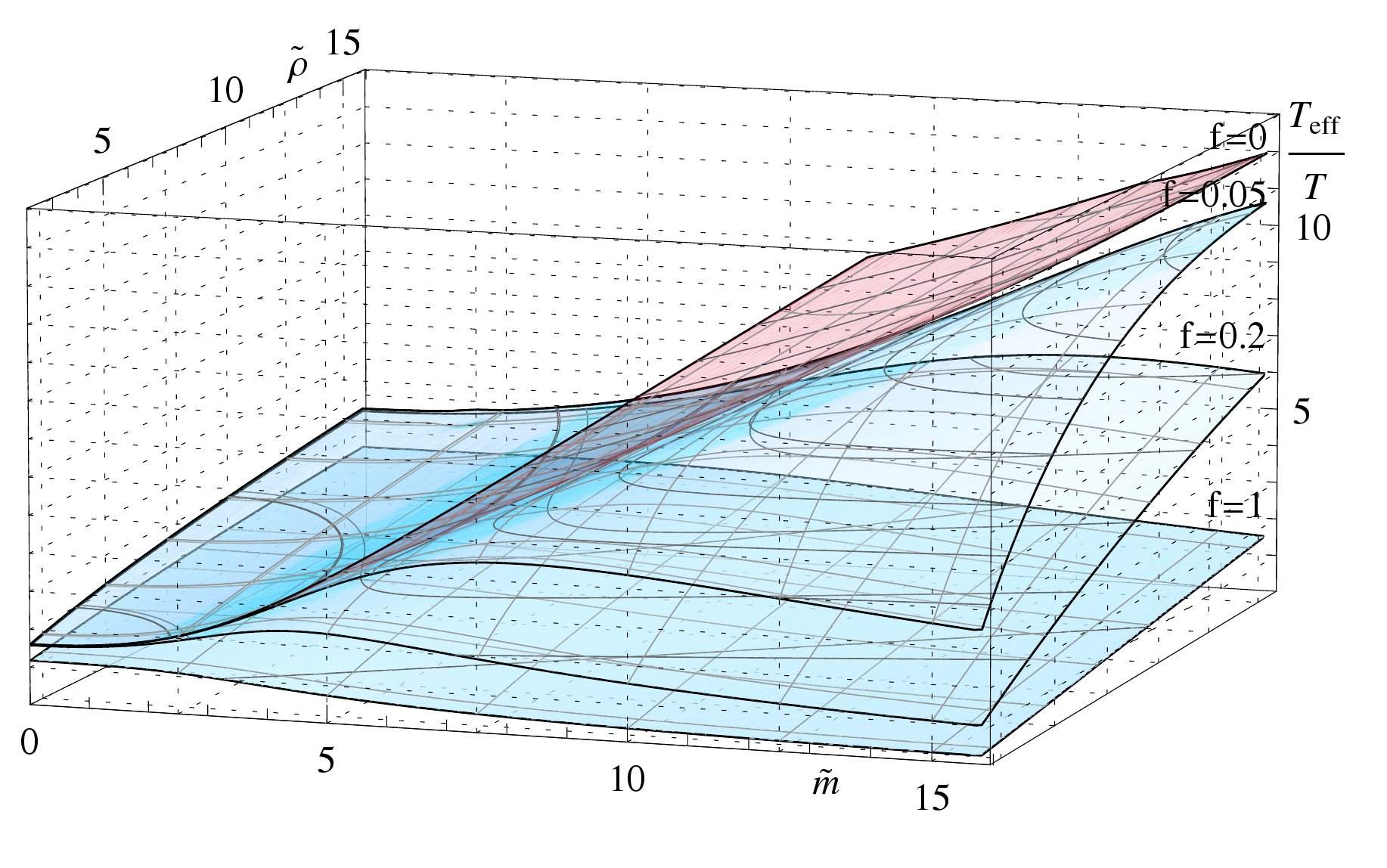}
\includegraphics[width=0.49 \textwidth]{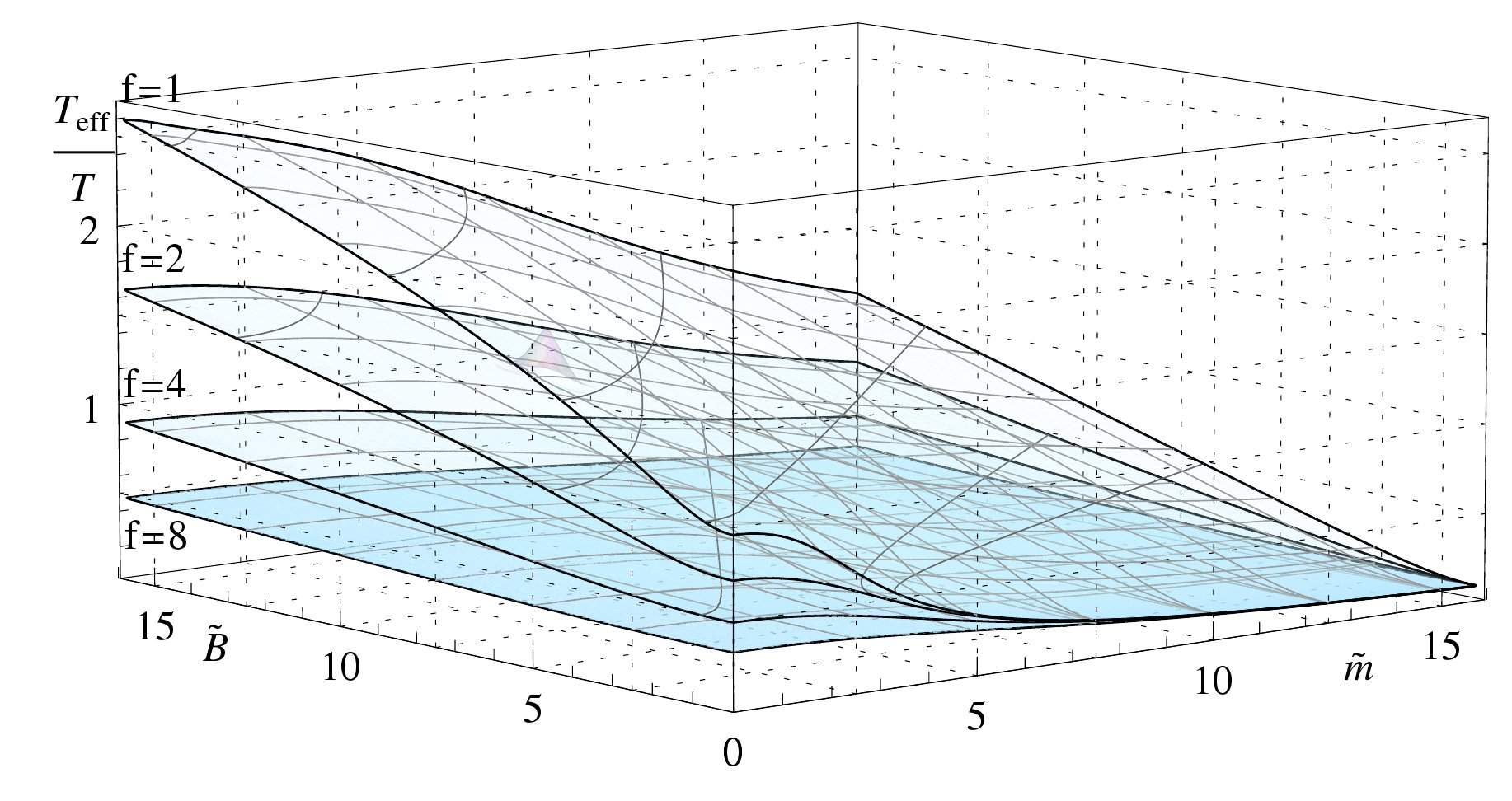}
\caption{Left: $\frac{T_{eff}}{T}$ as a function of the mass and the net baryon density for various (small) values of $f$. We choose the lower bound of the range $2\le \tlrho \le 16$ to avoid the phase transition at the critical mass and numerical problems near it. Right: $\frac{T_{eff}}{T}$ as a function of the mass and the magnetic field for various values of $f$.}
\mlabel{efftfmore}}
Because of the very good agreement between the quasiparticle spectrum and $T_{eff}$, let us have one more look at the mass-dependence of $\frac{T_{eff}}{T}$ in fig. \mref{efftfmore}. Generically, the dependence on $\tlrho$ and $\tlb$ is very similar at large values of $f$ and differs very significantly at small values, hence our choice of plots. At small values of $f$, we see a very interesting behavior. It seems that at vanishing $f$, $T_{eff}$ is essentially given by the mass, with only a subleading dependence on $\tlrho$ at large masses, which actually reduces $T_{eff}$ with increasing $\tlrho$. As we turn on $f$, this behavior turns over into a more common behavior, starting first at small $f$ at large masses and small densities. This may be some transition from a purely 2-dimensional system to a system that extends also in the third dimension. What makes this behavior so surprising is that normally, both in the mass dependence and also at $M_q=0$, any dependence on $f$ is subleading at large $\tlrho$. We should be careful with conclusions however, because we have not tested the dependence of the resonances on $T_{eff}$ in this regime.
Looking at the $\tlb$ dependence at large $f$, we find that the behavior is more generic, with a $\sim \frac{\tlb}{\tlm}$ scaling of $T_{eff}$ at $\tlm \gg \tlb$ and dependence approximately proportional to $\sqrt{\tlb}$ at $\tlm \ll \tlb$. We also note that The transition between the two regimes becomes clearer at increasing $f$, i.e. as we widen the defect. Large $f$ also suppress the $\tlb$ dependence of $T_{eff}$.

\DFIGURE{
\includegraphics[width=0.49 \textwidth]{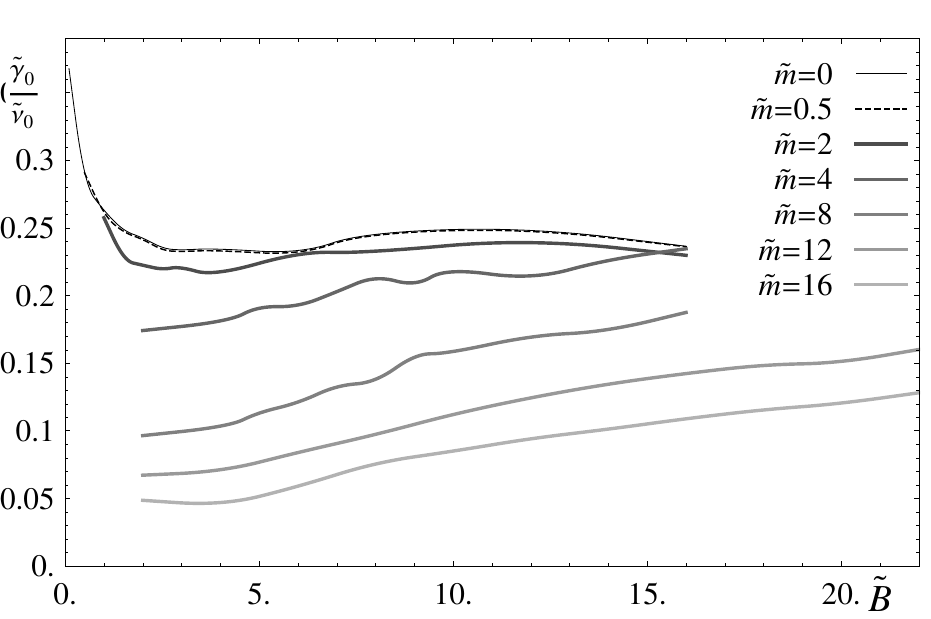}
\includegraphics[width=0.49 \textwidth]{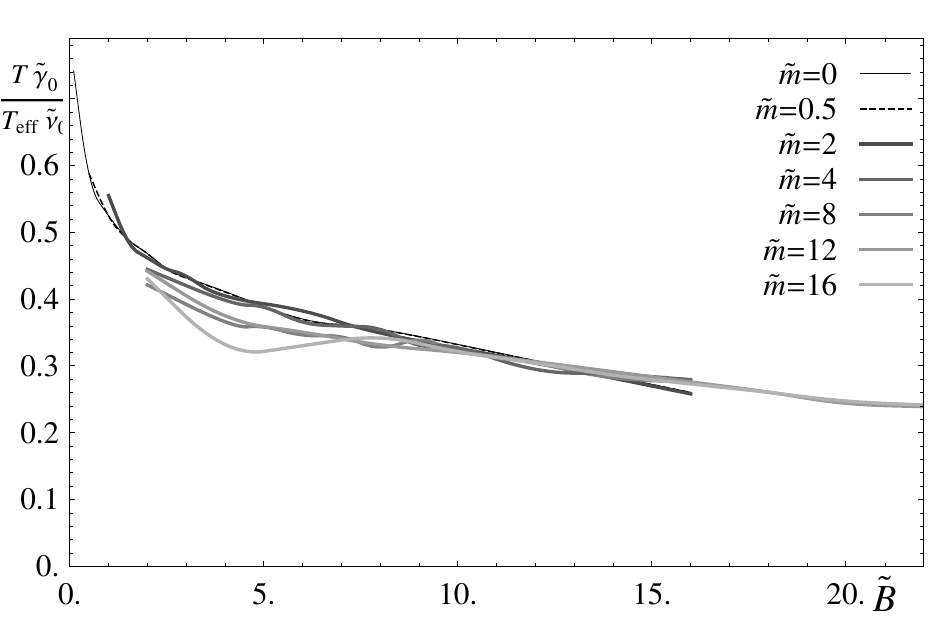}
\caption{The ``inverse lifetime to mass ratio'' $\frac{\tlgam_0}{\tlnu_0}$ for the magnetic resonances as a function of $\tlb$ for $f=4$ and various values of $\tlm$.}
\mlabel{tlnumass}}
To conclude, let us also in this case look at the ratio $\frac{\tlnu_0}{\tlgam_0}$ which we show  in fig. \mref{tlnumass}. Again, we see that $\frac{\tlnu_0}{\tlgam_0}$ is approximately constant in $\tlb$, however it seems to depend on the quark mass. If we divide by the effective temperature, the dependence on the quark mass is removed, however there seems to be some dependence on $\tlb$.

\DFIGURE{
\includegraphics[width=0.99 \textwidth]{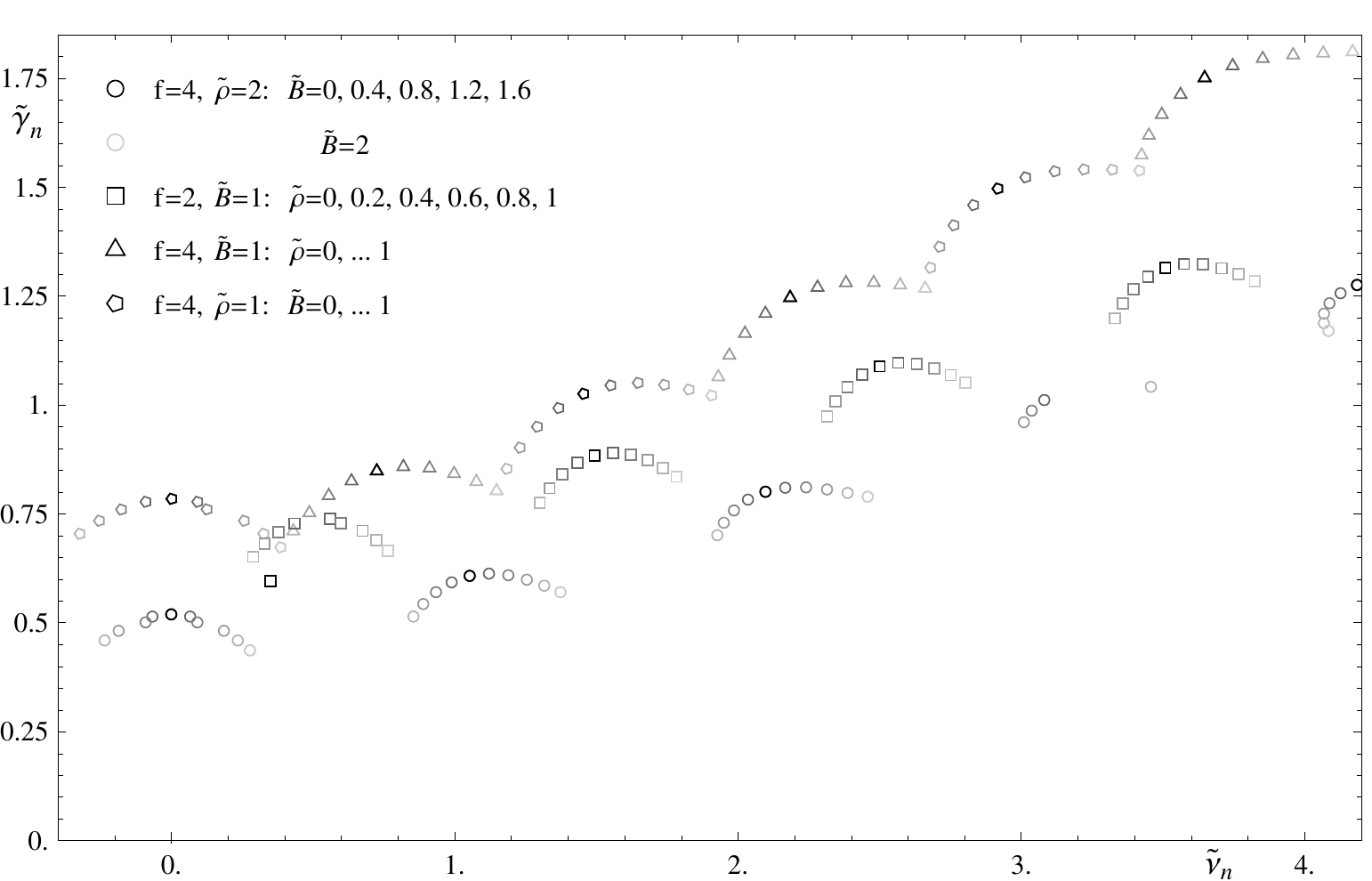}
\caption{The splitting of the poles due to the Hall effect at various values of $\tlrho$, $\tlb$ and $f$. The black symbols indicate the sequence of poles in the absence of the Hall effect for some choice of $f$ and $\tlb$ or $\tlrho$. With decreasing gray shade, we turn on the $\tlrho$ or $\tlb$, respectively, causing the original pole to split in two poles.}\mlabel{hallpoles}}
Finally, we can look at how the poles split if we turn on the Hall effect. In fig. \mref{hallpoles}, we show the location of the poles for various values of $f$, $\tlrho$ and $\tlb$. In black, we show the poles in the absence of the Hall effect, and then we show the sequence of poles as we gradually turn on the ``other'' parameter. The most surprising result is that the total spacing $\tlnu_0$ of the pairs of poles remains approximately unchanged, with quadratic dependence on the ``smaller'' background quantity $\delta \tlnu_0(\tlb) := \sqrt{\tlnu_0^2-\tlnu_0(\tlb=0)^2} \propto \tlb$ that is within errors consistent with the behavior of the effective temperature. The splitting of the poles, $\tlnu_\Delta$ depends approximately linearly on the magnetic field or density that we turn on, however there seems to be no simple dependence of the proportionality coefficient on the obvious candidates such as the specific magnetic moment $\frac{\partial M}{\partial \rho_0}$, the density of states or the magnetic susceptibility. We show this in fig. \mref{hallfitplots}
\DFIGURE{
\includegraphics[width=0.49 \textwidth]{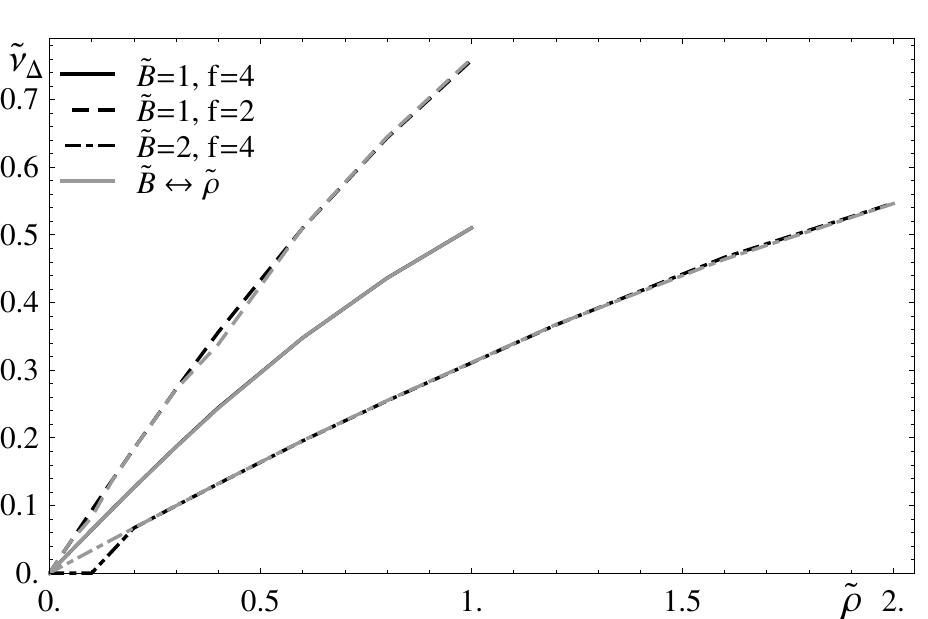}
\includegraphics[width=0.49 \textwidth]{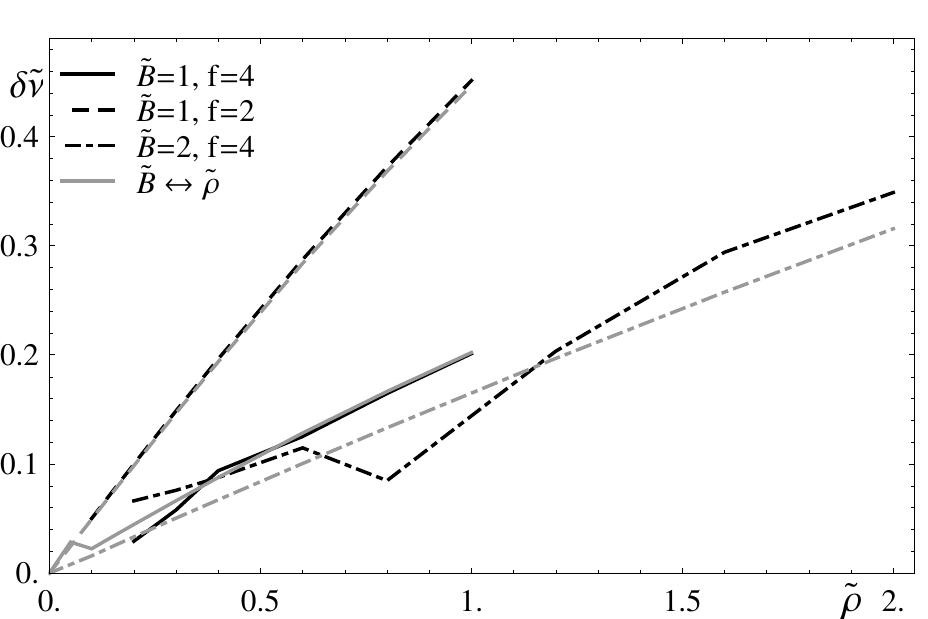}
\caption{Left: The splitting of the energy of the quasiparticle poles due to the Hall effect, $\tlnu_\Delta$. Right: The shift in $\tlnu_0$ from turning on the second parameter, written as $\delta \tlnu_0(\tlb) := \sqrt{\tlnu_0^2-\tlnu_0(\tlb=0)^2} \propto \tlb$ (and $\tlrho \leftrightarrow \tlb$).}
\mlabel{hallfitplots}}

The origin of the resonances on the gravity side is again straightforwardly explained in terms of quasinormal modes on the brane -- but rather than re-writing the equations as a Schr\"odinger equation and studying the potential as in refs. \refcite{fast,sugra1,sugra2}, let us look at the modes more directly. 
Taking the same Ansatz $A_y=A_0 e^{\int \! ds \zeta}$ as for the effective temperature, but now in terms of the variable $s$ that we used in section \mref{DClimit} and computing the equation of motion for $\zeta$ 
as in \reef{zetaeq}, we obtain
\begin{equation} \labell{zetaeqq}
\zeta^2\, +\, \dot{\zeta} \, +  \, \left(\left(\sqrt{-G}\, G^{tu}G^{xy}\right)'\right)^2\frac{G^{uu} }{ G^{tt} } \, - \, 
G G^{tt} G^{uu}G^{xx} G^{yy} \tom^2\ = \ 0 \ .
\end{equation}
Since we just want to have a brief picture, we will only work to leading order, i.e. we use the approximate solution $\zeta_0 :=  i \tom \sqrt{G G^{tt} G^{uu}G^{xx} G^{yy}}$ and the pertubation $\zeta = \zeta_0 + \epsilon$.
This gives us the linearized equation of motion for $\epsilon$
%
\begin{equation}\labell{shitt}
0 \ = \ \dot{\epsilon} \, - \, 2 \epsilon \zeta_0   \, + \,\dot{\zeta}_0
\ =: \  \dot{\epsilon} \, - \, \epsilon \alpha(s) \, - \, \beta(s) \ ,
\end{equation}
again with the general solution 
$\epsilon \, = \, e^{\int_0^s d\bar{s}\, \alpha(\bar{s})} \left(\epsilon_0 + \int_0^s d\tilde{s}\, e^{-\int_0^{\tilde{s}} d\bar{s}\, \alpha(\bar{s})} \beta(\tilde{s}) \right) $. 
Essentially what happens now is that the resonances arise from the inhomogeneous term. If we imagine that the source term $\beta$ were a delta function at some position $s_0$ with amplitude $\xi$, then the contribution from this term would be $\epsilon \, = \, \xi e^{\int_s^{s_0} d\bar{s}\alpha(\bar{s})}$. Setting $s=0$ and taking the real part gives us the resonances we want as $\delta \sigma \sim \xi \cos 2\omega\int_0^{s_0} \sqrt{G G^{tt} G^{uu}G^{xx} G^{yy}}$. 

To see where the resonances originate from in the geometry, we consider the high frequency limit $\tom \gg 1$ and remind ourselves that the integral of a periodic function vanishes, however the contribution to an integral of the type $\int F(x) e^{i\omega x}$ from some region around $x_0$ will be of the order $\frac{F'(x_0)}{\omega}$.
In our case, the frequency in the exponent is also not constant, but we can take care of that by a coordinate change $s \rightarrow \int \zeta_0$. After taking the derivative and changing back to $s$, we find that the term that gives us the contribution in the integral is $ds \, \partial_s \frac{\dot{\zeta}(s)}{\zeta(s)}$. This expression already shows us straightforwardly that these resonances appear only in the presence of the background fields, since otherwise $\sqrt{-G G^{tt} G^{uu}G^{xx} G^{yy}} = 1$.
%
In fig. \mref{finalgoddamnplot}, we show this term for various choices of the magnetic field, $f$ and the density. We see  how this length scale arises, and we see also the structure that gives rise to the line splitting in the Hall effect. It also demonstrates how the different phase between the magnetic and plasma resonances arises, essentially through flipping the sign in this term. The shape of this contribution gives rise to the amplitude of the oscillations, and the positive and negative sections tell us that there is a higher suppression by $\tom$. Note that at small $u \ll \sqrt{\tlb,\, \tlrho}$ we have $s\sim u$ and $\zeta \sim \sqrt{1+f^2}$. Hence at small or $\order(1)$ values for $\tlb$ and $\tlrho$, the spacing between resonances will be mostly controlled by the value of $\zeta$, and not through $s$. The curves in the plot however control through their shape the nature and the stability of the resonances.
\HFIGURE{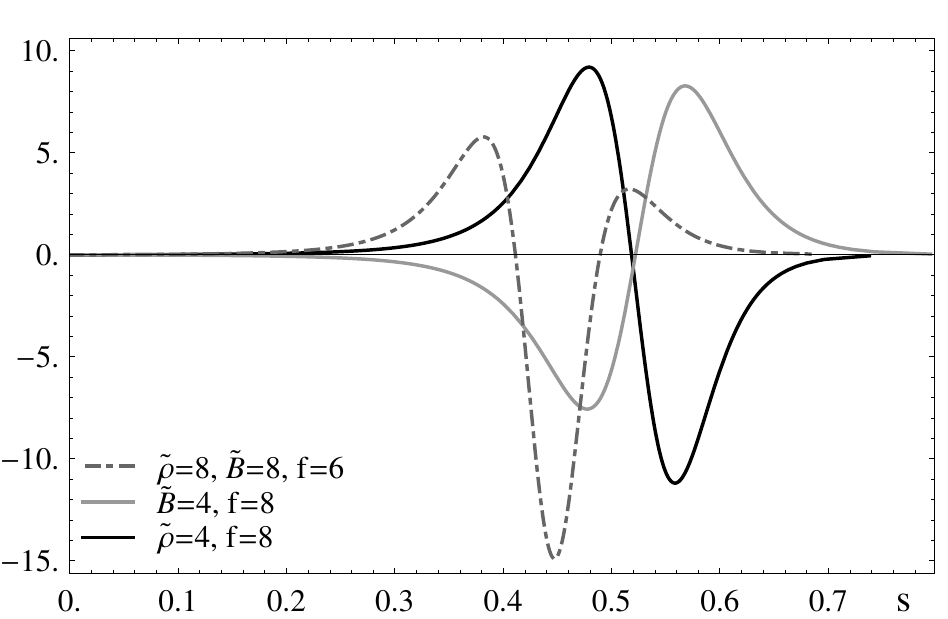}{Coefficient giving an estimate of the contribution to the resonances as described in the text.}{finalgoddamnplot}
Essentially, we can interpret this as scattering off a potential step, as it was discussed in ref. \refcite{baredef} and in a similar context also in refs. \refcite{fast,RobSpec}--\refcite{sugra2}.
%
%
%
%
%
%
%
%
%
%
%
%
%
%
%
%
\section{Discussion and Conclusions} \mlabel{discuss}
In this paper, we used holographic techniques to investigate the
transport properties of certain defect CFT's. In particular, we
studied matter on a $(2+1)$-dimensional defect emersed in a heat bath of 3+1 dimensional $\mathcal{N} = 4$ $SU(N_c)$ SYM theory.
Compared to the analysis in ref. \refcite{baredef}, we added to our analysis the presence of a finite background magnetic field, finite quark mass and net baryon density. This allowed us on the one hand to enlarge the class of theories that we are studying, and on the other hand it allowed us to compare the rich structure of our results to known phenomena in condensed matter physics, that are based either on generic considerations or on physics in the weakly coupled regime.  We tried to distinguish between a) generic properties that seem to be independent of physical details of particular models, b) intuition that carries over from the weak coupling regime and c) properties that are specific to the strong coupling regime and allow us to build some new intuition that is generic in the strong coupling regime.

Most of our analysis covers two distinct defect CFTs. The first was realized by
embedding $\nf$ probe D5-branes in an AdS$_5\times S^5$ background and in this case, the system (at $T=0$)
preserves eight supersymmetries. The second system involves
embedding $\nf$ probe D7-branes in the AdS$_5\times S^5$
and the resulting defect CFT preserves no supersymmetries. 
As it contains only fermions at the massless level, it is of particular interest to condensed matter physics.
In section \mref{fancysetup}, we looked at the gravitational setup that corresponds to turning on the various parameters in the field theory side. 
In both
cases, the theory could be deformed by introducing 
a flux in the AdS part of the world volume corresponding to a finite net baryon number density and chemical potential in the dual CFT,
a flux on the sphere corresponding to a shift $N_c \rightarrow N_c + \delta N_c$ in the level of the gauge group on one side of the defect
and
a deformation of the sphere
corresponding to a finite quark mass in the D5 case.  
In the D7 case, the internal flux was needed to stabilize the setup, and the operator dual to the deformation of the sphere seems to have non-integer conformal dimension as described in section \mref{adscmt_nonsusy}.
It turns out however that it is  not quite sure in how far the D7 setup persists in the light of gravitational backreaction. 

Perhaps surprisingly, the transport
properties of both defect CFT's were essentially identical in the massless case. Certainly, higher order effects might not be identical anymore.  
Furthermore it is curious that the only change is a Chern-Simons term with non-constant coupling at finite quark mass. However, we did not pursue this avenue at the massive level, since the results would be unreliable. 
%
%
%
%
%
%

We then went on to provide a few general results in section \mref{fancygetcon}. In \mref{femdual}, we studied how the field theory outcome of the EM duality in gravity side changes in the case of the extra parameters. We found the very interesting result, that the transport properties are now related under a simultaneous exchange of the dimensionless magnetic field and density and the transverse and longitudinal coordinate, while interchanging the 2-dimensional conductivity tensor with its inverse. From a condensed matter point of view, this related completely distinct parameters of the theory for a large class of 2+1 dimensional theories whose gravity dual obeys EM duality. 
Throughout section \mref{numcon},
 we saw how this duality gets broken as we consider a finite mass. However, still, this breaking appears gradually, with the parameter $M_q/T$.
This duality seems to be part of the $SL(2,\mathbb{Z})$ duality discussed in the field theory \cite{sl2z,dolan}, extended to finite frequencies and complex conductivities. The second generator may then be obtained from the theta term, \ie from the ``topological'' Hall conductivity discussed in ref. \refcite{baredef}. Certainly, it would be an interesting point to address in how far strongly coupled systems observed in nature display such a duality and possess a self-dual point in phase space -- in our system at the conformal point at vanishing net density and magnetic field.

In section \mref{analcon}, we then discussed analytic results in several regimes. In the DC limit, we found the Drude conductivity at finite density, a magnetoresistance effect and a Hall conductivity. They can be parametrized under the Drude model, obviously giving a new description to the ``Drude-parameters'' in terms of the parameters of the theory, as the underlying microscopic physics is different. However, the overall scalings in the limit of large density and magnetic field are the familiar ones.
Coincidentally, if we assume that the form of this DC conductivity has a high degree of generality, this could address the minimum quantum conductivity in Graphene \cite{naturegrev}. There, it is known that at the neutrality point, which corresponds in our case to vanishing net baryon density, there is a minimum in the conductivity of $e^2/h$ per carrier type, and this has apparently been of significant interest in the community \cite{pie}. Under an appropriate translation of the parameters, this is precisely what we observe in our case, as we also find, an increasing conductivity as we move away from the neutrality point, which goes beyond the results in ref. \refcite{pavel}. However, for example from the observed magnetic resonances, our defects seem to be quite different from the chiral nature of graphene \cite{graphenehall}.
We then studied the small-frequency limit, in which we reproduce the existence of a Drude peak and the minimum at $\omega=0$ in the case of the magnetoresitance. We could also identify a relaxation time, and accurately reproduce a relation between the frequency dependence of the diagonal conductivity and the hall conductivity, in the limit of large densities. In general, the structure of the frequency dependence resembled the generic prediction from the Drude model. We also found however, that the behavior of the specific values of parameters that depend on a particular model changes compared to the Drude model -- for example the relaxation time receives a dependence on the magnetic field. 

When we compared the relaxation time to the location of the purely dissipative poles in the hydrodynamic regime that we studied numerically in section \mref{diffnrelax}, we found that there was a disagreement with the dominant ``relaxation pole'' only at small densities. This can be explained from the remarkable constant DC conductivity from the EM duality at vanishing density found in refs. \refcite{baredef,pavel}, that is obviously not considered in the Drude model. In general we observed, in this regime and elsewhere, that there is always a total finite quark density in thermal equilibrium (which we can't control) that influences the transport properties even at vanishing net density. Beyond the dominant relaxation pole, that showed an unusual dependence on the density, which we attributed to the strong coupling, there was a second relaxation pole, that shows a more classical behavior, but has no significant contribution to the charge transport. We were able to reproduce the transition to the quasiparticle regime that was found in ref. \refcite{baredef}, and also found another transition in which the relaxation poles merge and turn into the first Landau level at some critical magnetic field, rather than the diffusion pole merging with the first relaxation pole as in ref. \refcite{baredef}. This transition can be attributed to strong coupling and is absent in free particle models. Overall, there is an interesting interplay between those relaxation poles and the diffusion pole as we tune the parameters, and there is a common theme that before poles move to large (imaginary) frequencies or leave their regime of validity, they either merge into quasiparticle poles or have decaying residue. 

The diffusion pole also has an unusual dependence on the density, which can be motivated from the strong coupling properties. We verified that the numerically obtained diffusion constant agrees with the one obtained analytically from the membrane paradigm. We also computed the permittivity. This gives us what we called the ``relative'' permittivity, that depends on the mass, magnetic field, internal flux and density. The diffusion constant and permittivity reproduce from the diffusion behavior, i.e. from the Einstein relation, the precise value of DC conductivity.

In the opposite regime, i.e. in the low temperature limit, our analytical approximations were concentrated at the exponentially suppressed conductivity at small frequencies and at large wavenumbers, $q \gg T$. There, we again found the conduction threshold at $\omega = q$ and extended the result of the exponentially suppressed diagonal conductivity that was found in ref. \refcite{baredef}. Now, the ``effective temperature'' that controls the ``Boltzmann factor'' however also depends on the other parameters of the theory. In particular, the density and magnetic field now raise that factor, i.e. reduce the exponential suppression. This is however not to be misunderstood as doping a semiconductor. At very large values of those parameters, of the order $\tlb,\tlrho \gg \tlq^2$, we were able to demonstrate that the conductivity turns into the DC result. 
Computing the Hall conductivity gave an interesting result as we obtained a finite value of the Hall conductivity even in the regime where the diagonal conductivity vanishes. This is however a common theme in condensed matter physics, for example in semiconductors at low temperatures or on Hall plateaus in the quantum Hall effect.

The rest of the work in the quasiparticle regime was mostly numerics-based. In section \mref{fancyspectral}, we gave an overview over the spectral curves, where we found the appearance of what one could describe as the strong-coupling equivalent of Landau levels and plasmons, and line splitting in the Hall effect. We also noticed that increasing the mass or $f$ makes the results approach the low temperature limit, as we expect from our results on the effective temperature. Furthermore, we looked at how the resonances from the isotropic regime carry over to the finite-wavenumber regime and connect to the resonances in the conformal case, which were discovered in ref. \refcite{baredef}. In particular, we found that for the transverse correlator, the density resonances connect smoothly whereas the magnetic resonances connected less smoothly -- which can be explained in terms of the localizing property of the magnetic field.
The other approach to the quasiparticle regime was to extract the location of the poles in the correlator in order to obtain the quasiparticle spectrum. We found that the poles are exactly equally spaced, indicating that the mechanism underlying the magnetic resonances is just a quantum harmonic oscillator as in the classical generation of Landau levels.
The length scales corresponding to the spectrum can be explained in two manners: On the one hand, they are just given by  approximately 2 times the inverse of the effective temperature, over essentially all the parameter range including the mass. On the other hand they can be related to the magnetization or chemical potential, which splits in the massive case into regimes below and above the quark mass. In this scenario, however, it seems that the Landau levels or plasmons are strongly coupled to the resonances over the ``width'' of the defect, as apparent from the minimum spacing of the resonances. This fits in nicely with the line splitting in the Hall effect, where each pole splits in two, indicating that there are overall two types of resonances in the system. Because of this and because of the unusual spacing, it seems that the magnetic and density resonances are not Landau levels or plasmons in the classical sense.
In a qualitative description, we also discussed how the quasiparticles arise from quasinormal modes in the scattering off a step in a potential in the gravity side.

As discussed in section \ref{checkapp}, we ignored the coupling between the scalars and the vectors in the pertubation of the worldvolume theory of the probe branes. This arises, however, only if we consider simultaneously finite scalar backgrounds (the ``width'' $\delta z$ arising from finite $\delta N_c$ and the compact embedding at finite quark mass) and finite vector background, i.e. the density and magnetic field. Furthermore, it turns out to be relevant for the transport properties only at finite wavenumber, in particular for the transverse conductivity in the presence of a magnetic field and for the longitudinal conductivity at finite net density. Hence, in these cases some details may not be captured by our analysis. However, the features resulting from the UV (asymptotic region), e.g. resonances or exponential suppression, and in the case of finite $\delta N_c$ also features arising from the IR (near-horizon region). Particular cases that should be taken with caution are the Hall effect at finite mass and finite wavenumber and the discussion of the breakdown of EM duality when turning on the finite mass -- in case it is done at finite wavenumber. Also, certainly there may be some interesting new physics hidden in the mixing of the fields (i.e. the operators in the field theory side).

Comparing our results to those obtained from field theory methods \cite{subir}, we found that there were a few similarities as a resonance or threshold at $\omega = q$ is also generically obtained using field theory methods. Furthermore, our results can be expressed in terms of a universal function that depends on $\omega/T$. However, it turns out that expressed in this way, this universal function depends on quantities like $\rho_0/T^2$, and hence also depends on the temperature. Overall, it seems that using AdS/CFT, we could more straightforwardly obtain a very rich and complex behavior of this universal function. Also, it seems that the methods in ref. \refcite{subir} do not find quasiparticle resonances, that seem to be an integral part of the defect that we studied.

For directions of future research, it would certainly be interesting gain a better interpretation of our results in terms of the microscopic theory beyond what we have attempted in this paper. This reveals a big weakness in using the AdS/CFT correspondence, as it is in practice to some degree like performing an experiment; and the ``microscopic'' theory in AdS/CFT is not the field theory, but the gravitational configuration. Another interesting direction would be to study the problems of the D3-D7 defect more in detail, as having a purely fermionic system is very appealing, even if it is only in the sector of the fundamental representation on the defect and the 3+1 bulk is still SYM. It would be interesting to see what kind of effects may then appear in the ``massive'' case. Certainly, it would also be interesting to study the consequences of the mixing between the gauge fields and the scalars in the probe brane action.
%
%
%
%
%

\section*{Acknowledgments} 
I would like to thank Rob Myers for many helpful discussions and suggestions and for proofreading an earlier draft of this paper and
Alexander Abanov, Mohammed Ansari, Brian Dolan, Johanna Erdmenger, Jaume Gomis, Troels Harmark,
Matt Headrick, Doug Hoover, Gary Horowitz, Matthias Kaminski, Pavel Kovtun, Per Kraus, Karl Landsteiner, David Mateos,  Volodya Miransky, Markus
M\"uller, Marta Orselli,  Andrei Parnachev, Mukund Rangamani, Subir Sachdev, Sang-Jin Sin, Aninda Sinha, Kostas Skenderis, and Yidun Wan 
for helpful
discussions and useful comments.
Research at the Perimeter Institute
is supported in part by the Government of Canada through NSERC and
by the Province of Ontario through MRI. This research is also supported
from an NSERC Discovery grant, from the Canadian Institute for
Advanced Research and by the Korea Science and Engineering Foundation (KOSEF) grant funded by the Korea Government (MEST) through the Center for Quantum Spacetime (CQUeST) of Sogang University with grant number R11-2005-021.
%
%
%
\appendix
\section{Components of the metric}\mlabel{gform}
In this appendix, we write out the components of the metric in the most convenient combination, including the implicit effective coupling in $\sqrt{-G} := \frac{\sqrt{(1-\Psi^2)^2 + f^2}}{\sqrt{1+f^2}} \sqrt{-\det G}$:
\begin{eqnarray}\nonumber
\!\!\!\!\! &\!\!\!\!\!\!\!\!\!\! &\textstyle{ \sqrt{-G}G^{xx}G^{yy} \ = \ \frac{\left(f^2 + (1-\Psi(u)^2)^2\right)\sqrt{(1-\Psi^2)+u^2(1-u^4)\Psi'(u)^2}}{\left(1+\frac{\tlb^2 u^4}{1+f^2} \right)\sqrt{1-\Psi(u)^2}\sqrt{ (f^2+\tlrho^2 + \tlb^2)u^4 +1 - \left(1+\frac{\tlb^2 u^4}{1+f^2} \right)\left(1-(1-\Psi(u)^2)^2 \right) } }}
\end{eqnarray}\\[-9mm]
\begin{eqnarray}\nonumber
\!\!\!\!\! &\!\!\!\!\!\!\!\!\!\! &\textstyle{ \sqrt{-G}G^{uu}G^{yy} \ = \ \frac{(1-u^4)\sqrt{1-\Psi(u)^2}\sqrt{ (f^2+\tlrho^2 + \tlb^2)u^4 +1 - \left(1+\frac{\tlb^2 u^4}{1+f^2} \right)\left(1-(1-\Psi(u)^2)^2 \right) } }{\left(1+\frac{\tlb^2 u^4}{1+f^2} \right) \sqrt{(1-\Psi^2)+u^2(1-u^4)\Psi'(u)^2}}}
\end{eqnarray}\\[-9mm]
\begin{eqnarray}\nonumber
\!\!\!\!\! &\!\!\!\!\!\!\!\!\!\! &\textstyle{ \sqrt{-G}G^{tt}G^{yy} \ = \ } \nonumber \\ \nonumber
\!\!\!\!\! &\!\!\!\!\!\!\!\!\!\! &~~~~~~~~~~~\textstyle{ - \frac{\left((1-\Psi(u)^2)^2 + f^2+(\tlrho^2 + \tlb^2)u^4  - \frac{\tlb^2 u^4}{1+f^2} \left(1-(1-\Psi(u)^2)^2 \right)\right)\sqrt{(1-\Psi^2)+u^2(1-u^4)\Psi'(u)^2}}{(1-u^4)\left(1+\frac{\tlb^2 u^4}{1+f^2} \right)\sqrt{1-\Psi(u)^2}\sqrt{ (f^2+\tlrho^2 + \tlb^2)u^4 +1 - \left(1+\frac{\tlb^2 u^4}{1+f^2} \right)\left(1-(1-\Psi(u)^2)^2 \right) }}}
\end{eqnarray}\\[-9mm]
\begin{eqnarray}\nonumber
\!\!\!\!\! &\!\!\!\!\!\!\!\!\!\! &\textstyle{\sqrt{-G}G^{tt}G^{uu} \ = \-\sqrt{ (f^2+\tlrho^2 + \tlb^2)u^4 +1 - \left(1+\frac{\tlb^2 u^4}{1+f^2} \right)\left(1-(1-\Psi(u)^2)^2 \right) }  \ \times }\nonumber \\ \nonumber
\!\!\!\!\! &\!\!\!\!\!\!\!\!\!\! & ~~~~~~~~~~~~~~~~~~~~~~~~~~\textstyle{\frac{\sqrt{1-\Psi(u)^2}\left((1-\Psi(u)^2)^2 + f^2+(\tlrho^2 + \tlb^2)u^4  - \frac{\tlb^2 u^4}{1+f^2} \left(1-(1-\Psi(u)^2)^2 \right)\right)}{\left(1+\frac{\tlb^2 u^4}{1+f^2} \right)\left(f^2 + (1-\Psi(u)^2)^2\right) \sqrt{(1-\Psi^2)+u^2(1-u^4)\Psi'(u)^2}}} 
\end{eqnarray}\\[-9mm]
\begin{eqnarray}\nonumber
\!\!\!\!\! &\!\!\!\!\!\!\!\!\!\! & \textstyle{\sqrt{-G}G^{tu}G^{xy} \ = \ \frac{\tlb \tlrho}{1+\frac{\tlb^2 u^4}{1+f^2}}} \ . 
\end{eqnarray}
At $M_q = 0$, i.e. $\Psi(u) = 0$, these simplify to:
\begin{eqnarray}\nonumber
\sqrt{-G}G^{xx}G^{yy} & = & \frac{1+f^2}{\left(1+\frac{\tlb^2 u^4}{1+f^2} \right)\sqrt{1+(f^1+\tlrho^2 +\tlb^2)u^4}}
\end{eqnarray}\\[-9mm]
\begin{eqnarray}\nonumber
\sqrt{-G}G^{uu}G^{yy} & = & \frac{(1-u^4)\sqrt{1+(f^1+\tlrho^2 +\tlb^2)u^4}}{1+\frac{\tlb^2 u^4}{1+f^2} }
\end{eqnarray}\\[-9mm]
\begin{eqnarray}\nonumber
\sqrt{-G}G^{tt}G^{yy} & = & -\frac{1+f^2+(\tlrho^2 + \tlb^2)u^4}{(1-u^4)\left(1+\frac{\tlb^2 u^4}{1+f^2} \right)\sqrt{1+(f^1+\tlrho^2 +\tlb^2)u^4}}
\end{eqnarray}\\[-9mm]
\begin{eqnarray}\nonumber
\sqrt{-G}G^{tt}G^{uu} & = & - \frac{\left(1+f^2+(\tlrho^2 + \tlb^2)u^4\right)\sqrt{1+(f^1+\tlrho^2 +\tlb^2)u^4}}{\left(1+\frac{\tlb^2 u^4}{1+f^2} \right)}
\end{eqnarray}\\[-9mm]
\begin{eqnarray}\nonumber
\sqrt{-G}G^{tu}G^{xy} & = & \frac{\tlb \tlrho}{1+\frac{\tlb^2 u^4}{1+f^2}} \ .
\end{eqnarray}

\section{Weak-Coupling Condensed Matter Physics}\mlabel{solidrev}
Even though we are interested in the strong coupling regime which one expects to be quite different from the free electron gas picture, some intuition and generic properties can be learned from this very straightforward limit and it can serve as a phenomenological description. Also, weak coupling is the only reference regime over which we have good control and where there are readily available textbook-type results. Hence, we remind the reader of the very basic model, which can be found in standard textbooks e.g. \refcite{SolidState1}--\refcite{SolidState3}.
\subsection{Metals}\mlabel{metalcon}
In the Drude model, we assume a gas of non-interacting charge carriers with finite (effective) mass $m_{eff}$ and charge $\qe$, which we will write out explicitly. Eventually, it will turn out, however, that using the coefficients that do not involve $m_{eff}$ is a suitable parametrization also in the relativistic case. To obtain the conductivity, one then considers a small electromagnetic background field, to which the charge carriers are coupled classically via the Lorentz force $m_{eff}(\partial_t {\vec{p}} - \tau^{-1}\vec{p})= \qe \vec{E} + \qe \vec{v}\times \vec{B}$. As the charge carriers are massive, they have a finite net velocity $\vec{v}$, which is assumed to be neutralized on the time scale of a relaxation time $\tau$. Classically, one has then a mean velocity $\vec{v}= \frac{ \qe \, \tau}{m_{eff} }\vec{E} = : \mu \vec{E}$, where we defined the charge carrier mobility $\mu$.
The charge carrier mobility is related to the Diffusion constant by the Einstein relation
\begin{equation}\labell{einsteindrude}
D \ = \ \frac{\mu T}{\qe} \ .
\end{equation}
 Further, for our massive case, the magnetic field can be rewritten in terms of the cyclotron frequency as $\omega_c  =  \frac{B \, \qe}{m_{eff}} = \mu B/\tau$.
Now, let us consider two species of charge carriers with equal mass and relaxation time, but opposite charge $\pm \qe$, such that we have a total density of charge carriers $\qnn = \qnn_+ + \qnn_-$  and a net charge density $\Delta \qnn = \qnn_+ - \qnn_-$. This is relevant in our case, since even at vanishing net baryon number density, $\rho_0=0$, at finite temperature $T \gtrsim M_q$ we will always have a finite total baryon density.

To obtain the conductivity, one then assumes an oscillatory electric field $\vec{E} = \vec{E}_0 e^{-i\omega t}$ and current $\vec{j} = \vec{j}_0 e^{-i\omega t}$, but constant magnetic field $\vec{B}$ and obtains the diagonal conductivity
\begin{equation}\labell{drudecon}
\sigma^{\parallel} \! =  \frac{\qe \qnn \mu (1-i\omega \tau)}{(1-i\omega \tau)^2+ \omega_c^2 \tau^2}  =  \frac{\qe \qnn \mu }{1+ \omega_c^2 \tau^2}\left(\!1\! + i\omega\tau \frac{1- \omega_c^2 \tau^2}{1+ \omega_c^2 \tau^2} - \omega^2\!\tau^2\! \frac{1- 3\omega_c^2 \tau^2}{(1+ \omega_c^2 \tau^2)^2}\!\right) +  \order(\omega \tau)^3  .
\end{equation}
Taking into account the positive and negative charges, the Hall conductivity becomes
\begin{equation}\labell{drudehall}
\sigma^{\perp} \! =  \frac{\qe \Delta \qnn \mu \omega_c \tau}{(1-i\omega \tau)^2+ \omega_c^2 \tau^2} =  \frac{\qe \Delta \qnn \mu \omega_c \tau}{1+ \omega_c^2 \tau^2}\left(1 + \! \frac{2 i\omega\tau}{1+ \omega_c^2 \tau^2} - \omega^2\tau^2 \frac{3- \omega_c^2 \tau^2}{(1+ \omega_c^2 \tau^2)^2}\right) + \order(\omega \tau)^3 \ .
\end{equation}
The dissipative part of the conductivity is then
\begin{eqnarray}
\Re \sigma^{\parallel} & = & \qe \qnn \mu \frac{1+ \omega^2 \tau^2+ \omega_c^2 \tau^2}{(1+\omega_c^2 \tau^2-\omega^2 \tau^2)^2+ 4\omega_c^2 \tau^2} \ \ \mathrm{and} \ \ \ \\
\Re \sigma^{\perp} & = & \qe \Delta \qnn \mu \omega_c \tau \frac{1 - \omega^2 \tau^2+ \omega_c^2 \tau^2}{(1+\omega_c^2 \tau^2-\omega^2 \tau^2)^2+ 4\omega_c^2 \tau^2} \ .
\end{eqnarray}
%
%
%
%

The DC conductivity at $B=0$ is commonly referred to as the Drude conductivity, and at small frequencies it is also called the Drude peak, due to the small value of $\tau$ in metals at room temperature. 
Similarly, the fact that the DC conductivity is suppressed at finite magnetic fields is referred to as the magnetoresistance effect. 

It is also interesting to notice that at the magnetic resonance around $\omega_c$, the Hall conductivity changes sign, and this turns into a pole at large $\tau$, i.e. in practice at very small temperatures or in very ``clean'' semiconductors. 

We can also observe a few generic properties of the frequency dependence. For example, at $B=0$, $\frac{\partial^2_\omega \sigma^\parallel }{\sigma^\parallel} \, = \,\frac{2}{ \tau^{2}}$ and, provided the relaxation time is independent of the magnetic field, the behavior at large frequencies is  $\frac{\partial^2_\omega \sigma^\parallel }{\sigma^\parallel} \, = \, -\frac{6}{\omega_c^2 \tau^{4}}$.
An interesting relation is also $\frac{\partial^2_\omega \sigma^\parallel \ \sigma^\perp}{\sigma^\parallel \ \partial^2_\omega \sigma^\perp} \, = \, \frac{1}{3}$ at vanishing magnetic field and $3$ at large magnetic fields. Also, the magnetic field at which  $\frac{\partial^2_\omega \sigma}{\sigma} $ changes sign is $\tau^2\omega_c^2 = \frac{1}{3}$ for $\sigma^\parallel$ and $3$ for $\sigma^\perp$. 
The fact that the quadratic term changes sign implies that the Drude peak moves away from the real axis and becomes a magnetic resonance.

We already see a limitation of the free electron gas picture, because -- as we will see in section \mref{resonances} -- the quantum mechanical treatment implies that the first magnetic resonance is at $\frac{1}{2}\omega_c$ and taking into account a finite coupling implies that there will be plasma density resonances.
\subsection{Semiconductors}\mlabel{semicon}
Semiconductors are somewhat less generic than the Drude model of conductivity, i.e. we must assume that we are dealing with fermions, but the discussion obviously carries over to any system of charge carriers with an excitation gap.
In a semiconductor, we assume that the valence and conduction bands are separated, $E_v < E_c$, where the valence band is, as the name says, filled such that the chemical potential lies between the bands $E_v < \mu < E_c$.

At small temperatures, the dissipative conductivity will be dominated by $\qnn$, such that we are most interested in obtaining the density of conduction ``electrons'' $\qnn_c$ and valence ``holes'' $\qnn_v$ assuming some density of states $g_c$ and $g_v$:
\begin{equation}
\qnn_c \, = \, \int_{E_c}^\infty \! d E\, g_c(E) \frac{1}{e^{(E-\mu)/T}+1}
\end{equation}
and
\begin{equation}
\qnn_v \, = \, \int_{-\infty}^{E_v} \! d E \, g_v(E)\left(1- \frac{1}{e^{(E-\mu)/T}+1}\right)\, = \, \int_{-\infty}^{E_v} \! d E \, g_v(E) \frac{1}{e^{(\mu-E)/T}+1} \ .
\end{equation}
Assuming that we are dealing with low temperatures $E_c -\mu \gg T$ and $\mu- E_v\gg T$, we can re-write this as
\begin{equation}\labell{semicon2}
\qnn_c(T) \, = \, N_c(T) e^{-(E_c-\mu)/T} \ , \ \ \ \qnn_v(T) \, = \, N_v(T) e^{-(\mu- E_v)/T}
\end{equation}
where we defined the edge densities of states as
\begin{equation}
N_c(T) \, := \, \int_{E_c}^{\infty} \! dE \, g_c(E) e^{-(E-E_c)/T} \ , \ \ N_v(T) \, := \, \int^{E_v}_{\infty} \! dE \, g_v(E) e^{-(E_v-E)/T} \ .
\end{equation}

Using those definitions, we can combine the equations \reef{semicon2} to write down the ``law of mass action''
\begin{equation}
\qnn_c \qnn_v \, = \, N_c N_v e^{-(E_c-E_v)/T} \, = \, \qnn_i(T)^2
\end{equation}
which determines the charge carrier density of an intrinsic (undoped) semiconductor, $\qnn_i = \qnn_v  = \qnn_c$.
This small charge carrier density implies (together with the purity of the crystal) that $\tau$ may diverge at small temperatures, as it is usually (in a metal or metallic phase with high charge carrier density) dominated by the charge carriers and their thermal motion (independent of the purity). Hence semiconductors may show, for example, a finite Hall conductivity even if the material is effectively an insulator.

The chemical potential can then be obtained from the edge densities of state
\begin{equation}
\mu \, = \, \frac{E_v+E_c}{2} \, + \, \frac{1}{2} T \ln  \frac{N_v}{N_c} \ .
\end{equation}
One can see that in practice in an intrinsic semiconductor at $E_c-E_v \gg T$, the edge density of states has only a very small influence on the chemical potential, and hence on the conduction threshold and the suppression of the conductivity.
%
%
\subsection{Resonances}\mlabel{resonances}
In the ``optical'' regime at larger frequencies, we are interested in quasiparticle resonances, in particular in Landau levels and plasmons.
In the case of massive charged particles, the derivation of Landau levels is straightforward. Assuming coordinates $\{t,x,y,z\}$ with the magnetic field in the $z$ direction, the gauge potential can be written as $A= B\, x \, dy$. Substituting this into the Schr\"odinger equation gives rise to a quantum harmonic oscillator with a frequency $\omega_c = \frac{\qe B}{m}$, with the solution
\begin{equation}
E_n \, = \, E_{kin.}^{(z)} \, + \, \omega_c (n -\frac{1}{2}) \ , \ \ n\in \mathbb{Z}^+ \ .
\end{equation}
Obviously, our charge carries are constrained to the $z=0$ plane, so $E_{kin.}^{(z)} = 0$.
Using the same naive strategy for a scalar in the Klein-Gordon equation, we find that 
\begin{equation}
E_n^2 \, = \, p_z^2\, + \, m^2\, + \,  \omega_M^2 (2n -1) \ , \ \ \omega_M^2 \, = \, \qe B \ ,
\end{equation}
putting the Landau levels in the massless case at $E_n =  \pm \omega_M \sqrt{2 n -1}$. In either case, one can apply a simple argument by assuming a finite sample size to derive the density of states (per unit area) of 
\begin{equation}\labell{landens}
\frac{N_{Lan.}}{A} \ = \ \qe \, B \ .
\end{equation}
In a similar fashion, Landau levels can also be obtained from the Dirac equation (see ref. \refcite{graphenehall} and references therein):
For massless chiral fermions e.g in graphene, the result is  \cite{graphenehall}
\begin{equation}
E_n  \ = \ \pm \omega_M |v_F|\sqrt{| n|}\ , \ \ n\in \mathbb{Z}
\end{equation}
where $v_F$ is the Fermi speed ($v_f = 1$ for a ``real'' relativistic system) and for a system of chiral fermions with finite effective mass, as in multi-layered graphene, \cite{graphenehall}
\begin{equation}
E_n  \ = \ \pm \omega_c  \sqrt{|n|(|n|+1)}\ , \ \ n\in \mathbb{Z} \ .
\end{equation}
The zero-energy, field-independent Landau level is unique to chiral fermions, which makes them very easy to identify \cite{graphenehall}.

The other quasi-particles that we are interested in are plasmons. These are collective excitations resulting from density perturbations of the (electron) gas. They can be derived in several ways. Classically it can be derived simply from the continuity equation $\nabla \cdot \vec{j} = \partial_t \rho$ and from Gauss' law $\nabla \cdot \vec{E} = 4\pi \rho$. Using $\sigma^\parallel \vec{E} = \vec{j}$, we can arrive with an equation for $\omega$, 
$ 4\pi \qnn \mu \tau^{-1} = \omega (\omega +i \tau^{-1})$. 
Another classical derivation is to assume a neutral gas of positive and negative charges in which the charges are displaced in, say the $x$ direction, leaving two strips of density $\pm \Delta \qnn$ with a width $d$ that obeys the classical equation of motion $\partial_t^2 d = -\frac{4\pi \qe^2 |\Delta \qnn|}{m}$.
Finally, a proper derivation is based on computing the Green's function in a gas of weakly interacting fermions (of equal charge in magnitude and sign) with the Coulomb potential. A very instructive derivation can be found in ref. \refcite{Nagaosa} and yields 
\begin{equation}
\omega_p \ = \ \frac{8\pi}{3} \frac{k_F^3}{m} \ = \ \frac{4\pi \qe^2 \qnn}{m} \ .
\end{equation}
where $k_F$ is the momentum corresponding to the Fermi energy. The Green's function is at small momenta $|q| \ll |\omega|$
\begin{equation}
D \ = \ \frac{4 \pi}{q^2} \frac{1}{1-\frac{\omega_p^2}{(\omega+ i \tau^{-1})^2}} \ .
\end{equation}
The derivation follows with only small changes also in the massless case, where we find $\omega_p \ = \ k_F^2  \, 8\pi/3 $ .  Plasmons are usually (i.e. in weakly coupled systems) observed through optical scattering, where one can observe spectra from multiple plasmon excitations. 
Studying plasmons and surface plasmons in various materials an using them for photonic devices seems to be a very active field of research. 

%
%
%
%
%

%
%
%
%

\end{document}